# AUTOMATING ABSTRACT INTERPRETATION OF ABSTRACT MACHINES

JAMES IAN JOHNSON

April 2015

*Submitted in partial fulfillment of the requirements for the degree of Doctor of Philosophy*

*to the*

*Faculty of the*
*College of Computer and Information Science*
*Northeastern University*
*Boston, Massachussetts, USA*

**NORTHEASTERN UNIVERSITY**
**GRADUATE SCHOOL OF COMPUTER SCIENCE**
**Ph.D. THESIS APPROVAL FORM**

*THESIS TITLE:* Automating Abstract Interpretation of Abstract Machines

*AUTHOR:* J. Ian Johnson

*Ph.D. Thesis Approved to complete all degree requirements for the Ph.D. Degree in Computer Science.*

| | |
|---|---|
| *Thesis Advisor* | 2015/4/22 *Date* |
| *Thesis Reader* | 3/30/2015 *Date* |
| *Thesis Reader* | 3/30/15 *Date* |
| *Thesis Reader* | 4/22/2015 *Date* |

*GRADUATE SCHOOL APPROVAL:*

| | |
|---|---|
| *Director, Graduate School* | 4/23/2015 *Date* |

*COPY RECEIVED IN GRADUATE SCHOOL OFFICE:*

| | |
|---|---|
| *Recipient's Signature* | 4/23/2015 *Date* |

*Distribution: Once completed, this form should be scanned and attached to the front of the electronic dissertation document (page 1). An electronic version of the document can then be uploaded to the Northeastern University-UMI website.*

## COLOPHON

This document was typeset using the typographical look-and-feel `classicthesis` developed by André Miede. The style was inspired by Robert Bringhurst's seminal book on typography *"The Elements of Typographic Style"*. `classicthesis` is available for both LaTeX and LyX:

http://code.google.com/p/classicthesis/

The PDF construction used pdfTeX. `pdflatex -v`:

```
pdfTeX 3.1415926-2.4-1.40.13 (TeX Live 2012/Debian)
kpathsea version 6.1.0
Copyright 2012 Peter Breitenlohner (eTeX)/Han The Thanh (pdfTeX).
There is NO warranty.  Redistribution of this software is
covered by the terms of both the pdfTeX copyright and
the Lesser GNU General Public License.
For more information about these matters, see the file
named COPYING and the pdfTeX source.
Primary author of pdfTeX:
  Peter Breitenlohner (eTeX)
  Han The Thanh (pdfTeX).
Compiled with libpng 1.2.49; using libpng 1.2.49
Compiled with zlib 1.2.7; using zlib 1.2.7
Compiled with poppler version 0.20.4
```

Figures were constructed with Racket 6's `plot` library, graphviz's `dot` version 2.26.3 (20100126.1600), and Inkscape 0.48.4 r9939 (Jan 22 2014).

*Final Version* as of May 1, 2015 (`classicthesis` version 1.0).

*Dedicated to my grandparents,*

*Mary Jane & Robert Johnson, and*
*Imogene & Willard Speck*

# ABSTRACT

Static program analysis is a valuable tool for any programming language that people write programs in. The prevalence of scripting languages in the world suggests programming language interpreters are relatively easy to write. Users of these languages lament their inability to analyze their code, therefore programming language analyzers (abstract interpreters) are not easy to write. This thesis more deeply investigates a systematic method of creating abstract interpreters from traditional interpreters, called Abstracting Abstract Machines.

Abstract interpreters are difficult to develop due to technical, theoretical, and pragmatic problems. Technical problems include engineering data structures and algorithms. I show that modest and simple changes to the mathematical presentation of abstract machines result in 1000 times better running time - just seconds for moderately sized programs.

In the theoretical realm, abstraction can make correctness difficult to ascertain. Analysis techniques need a reason to trust them. Previous analysis techniques, if they have a correctness proof, will have to bridge multiple formulations of a language's semantics to prove correct. I provide proof techniques for proving the correctness of regular, pushdown, and stack-inspecting pushdown models of abstract computation by leaving computational power to an external factor: allocation. Each model is equivalent to the concrete (Turing-complete) semantics when the allocator creates fresh addresses. Even if we don't trust the proof, we can run models concretely against test suites to better trust them. If the allocator reuses addresses from a finite pool, then the structure of the semantics collapses to one of these three sound automata models, without any foray into automata theory.

In the pragmatic realm, I show that the systematic process of abstracting abstract machines is automatable. I develop a meta-language for expressing abstract machines similar to other semantics engineering languages. The language's special feature is that it provides an interface to abstract allocation. The semantics guarantees that if allocation is finite, then the semantics is a sound and computable approximation of the concrete semantics. I demonstrate the language's expressiveness by formalizing the semantics of a Scheme-like language with temporal higher-order contracts, and automatically deriving a computable abstract semantics for it.

# ACKNOWLEDGMENTS

Readers unfamiliar with Jorge Cham's *PhD comics* are likely not PhDs or PhD students. For those not in the know: the trials, tribulations, trivialities and sometimes moral turpitude of the PhD as depicted in these works of comedy really happen all the time. I know. I'm a data point. Nothing in my life has ever been as difficult as these six years, and I could not have done it without the help I received from faculty, colleagues, friends and of course family. First of all, I thank my committee:

- David Van Horn, my advisor. Our relationship started with him as a postdoc with some cool ideas and great presentation skills. His philosophy and approach to research are both fundamentally pedagogical and progressive: everything he does, however complicated it was before, becomes easy, obvious, and better. Sure, this makes publishing difficult (I think reviewers get off on being confused), but I found this way of operating enviable. His focus on the long game calmed my indignation of rejection. His willingness to hear me out with a half-baked idea kept me from censoring my creativity. And his sense of humor kept our conversations enjoyable.
- Olin Shivers, my co-advisor. Previously my advisor, Olin gives his students room to explore and grow as researchers. He's famously entertaining, and always has his eyes on a shiny future.
- Mitchell Wand (Mitch) introduced me to programming languages research and a new way of thinking about proof. His ability to cut through arguments forced me to think more precisely, to get to the heart of the matter. I thought I had mathematical maturity before I met Mitch, but after working with him for a year on hygienic macros, well... This man knows semantics, and since I had the privilege of our time together, I feel I know too.
- Cormac Flanagan has done a vast amount of work in practical program analysis. I appreciate the effort he's put in to reviewing this dissertation.

My co-authors' help and support improved our publications more than I could have: thanks to Matthew Might, Ilya Sergey, and again, David Van Horn.

I thank the other Northeastern faculty who helped me in this process: Matthias Felleisen, for having my back; Amal Ahmed, for her help with the harder correctness arguments in this document; Panagiotis Manolios (Pete), for first teaching me formal methods; and Thomas Wahl for his perspective from the model-checking community. I would also like to thank J Strother Moore for welcoming me at the ACL2 seminar in my final year at the University of Texas, and generously funding my attending the 2009 ACL2 Workshop. I could not have done so much of my work without the development team for Racket, most notably Matthew Flatt. Thanks for all the bugfixes.

My colleagues in the lab are by far my greatest learning asset. We spent countless hours together working, learning, complaining and joking. My time in the PhD was immensely humbling, not because of the difficulty of the work, but knowing all of you tremendously talented people.

- Claire Alvis: *undefined amount of fun*
- Dan Brown: *categorically helpful*
- Harsh Raju Chamarthi: *theorem disprover*
- Stephen Chang: *father, proof-reader, friend*
- Ryan Culpepper: *master macrologist*
- Christos Dimoulas: *keeping us honest with contracts*
- Carl Eastlund: *macro logician*
- Tony Garnock-Jones: *happy to subscribe to your conversation topics*
- Evgeny (Eugene) Goldberg: *satisfying conversationalist*
- Dave Herman: *web freedom fighter and rusty macrologist*
- Mitesh Jain: *reimagining correctness*
- Jamie Perconti: *very cool if true*
- Tim Smith: *my go-to for obscure automata theory*
- Vincent St-Amour: *telling us how we're doing it wrong*
- Paul Stansifer: *fun partner and macro advocate*
- Asumu Takikawa: *affable, helpful and* 全て上手
- Sam Tobin-Hochstadt: *inspired crazy macro hacker turned professor*
- Aaron Turon: *concurrently brilliant and a good person*
- Dimitrios Vardoulakis: *full stack analyst*

I thank Neil Toronto for his `plot` library in Racket, and for all the time he spent helping me use it to produce the plots in this dissertation.

I thank my friends for keeping me from floating off into jargon-land every time I open my mouth.

- Matthew Martinez (Mattousai): you're my best college buddy. Best wishes for your life in Ireland.

- Nicholas Marquez (Alex): may you and Alex find other Alexes to happily Alex your Alex while you Alex with Alex.

- Daniel Davee (Mage): I know you're the physicist, but quantum mechanics will not make undecidable problems decidable.

Finally and most importantly, I thank my family for all their love and support.

- Shaunie: I love you and your ability to put up with me.
- V: I hope you never ever have to read this document.
- Mom & Dad: the condo will appreciate, and we appreciate the condo.
- Grandpa Johnson: for all the stories and ego-boosting.

# CONTENTS

# 1 INTRODUCTION AND CONTRIBUTIONS

*"What we hope ever to do with ease, we must first learn to do with diligence."*
*~Samuel Johnson*

## 1.1 MY THESIS

*Precise and performant analyses for higher-order languages can be systematically and algorithmically constructed from their semantics.*

Higher-order languages are pervasive, take many forms, and in several cases have complicated semantics (*e.g.*, Python, PHP, JavaScript). Static analyses are useful tools for programmers to detect problems before programs run, prove the safety of program transformations to improve performance, and even prove correctness properties. Static analyses are also *black magic* that only experts can perform. Principles of Program Analysis (POPA) [71] for instance uses heavy formalism to define analyses in a different way than the language itself is formalized. Worse, it requires additional external knowledge to efficiently implement. There is a gap between language implementation and language analysis.

There are several reasons that this gap exists. Compiler textbooks [67, 3], even modern ones [5], only cover the oldest framework for analysis [48], which presumes the language has a strict separation of data and control flow. Perhaps this is because overcoming the strict separation took a PhD [87]. The dissertation is highly cited, well-written, and solves the data-flow problem that previously held back the efficient implementation of functional languages that can be expressed in a specialized form. The oft-cited book for standard analyses which includes a functional language analysis (POPA) uses machinery that is distant from standard language implementation techniques. There is a gap in the literature for designing and implementing analyses for languages of different shapes and sizes. In this dissertation, I bridge the gap with a collection of techniques for constructing analyses from little more than a language's interpreter written as an abstract machine.

Another word for "gap" we might choose is "pitfall." Analysis construction has many pitfalls.

- **Unsoundness:** Keeping analyses true to the semantics of a programming language can be a difficult task when the two are separate artifacts. One can introduce bugs in any re-implementation effort. Analyses further make design harder since abstractions can easily leave out behavior.

- **Imprecision and state-space explosion:** Increased precision is a double-edged sword. On the one hand, high precision distinguishes values enough to rule out some execution paths. On the other hand, too many distinctions can lead to an explosion of states either due to either inherent complexity of the computation space, or accidental complexity from redundant state representations. Imprecision means there are too few distinctions to rule out impossible behavior.

  Not ruling out bad behavior means that bad executions are also explored – more computation. In both cases of high and low precision, analysis performance can be adversely affected.

- **Non-termination:** It is all too easy to introduce sources of non-termination into a program analysis. The slightest misstep with data representation can make an unbounded state space, not just intractably large.

This dissertation describes systematic techniques to protect analysis designers from the above problems. Furthermore, these systematic techniques can be made algorithmic for a metalanguage that expresses programming language semantics. Simply write down your language's semantics as a typical abstract machine, press a button, and get a sound analysis back.

## 1.2 STRUCTURE OF THE DISSERTATION

This dissertation is split into two parts:

1. Some specific "by hand" but still systematic construction techniques for program analyses are exposited.
   - This chapter discusses why I focus on abstract machines instead of a popular intermediate representation as the vehicle for analysis. The end of the chapter summarizes my previously published material and how it relates to my thesis.
   - Chapter 2 recounts the origin of the technique employed in this dissertation, known as "abstracting abstract machines." (AAM)
   - Chapter 3 shows simple semantics transformations that are effective for producing efficient analyzers.
   - Chapter 4 shows that AAM can be refined to construct pushdown models instead of finite ones. The technique employed gives the semantics access to the whole stack, so features like garbage collection and stack inspection are easy to express.

2. The lessons learned from the first part's constructions are distilled into a language and its semantics.
   - Chapter 5 gives the concrete semantics of a core language for expressing abstract machines.
   - Chapter 6 gives the abstract semantics of the previous chapter's language by building off the notions of abstraction used in AAM.
   - Chapter 7 presents a case study using the prior chapter's language to express a semantics of temporal higher-order contracts.

## 1.3 THE CASE FOR ABSTRACT MACHINES

First, what *is* an abstract machine? The term is overloaded to mean anything from an arbitrary automaton to a specified but unimplemented microprocessor design. What I mean by abstract machine is a construct along the lines of the popular SECD [57] or CESK machines [31]. Informally, an abstract machine[1] is space of *machine states* that have an execution behavior defined by a finite set of *reduction rules*. Machine states, or just *states*, have well-defined structure that reduction rules match on to rewrite into new states. Reduction rules govern how a machine transitions between states, and need not be 1-to-1 for input and output states. This means that one starting state can lead the machine to travel many execution paths, not just one.

[1] *The term "abstract machine" will be made most formal in Chapter 5.*

Second, why not choose a popular compiler intermediate representations like *single static assignment* (SSA) or *continuation-passing style* (CPS)? The fact of the matter is that SSA and CPS do not express the entire machine state, and thus cannot be the language of how a machine transitions between states. They are not what a programmer will typically write, so there is already a compilation step that may need some analysis. The goal of this dissertation is to bridge the gap by removing unnecessary detours. Indeed, both SSA and CPS can be given perfectly fine abstract machine semantics. The converse is not necessarily true: arbitrary abstract machines are not necessarily translatable to SSA or CPA without stretching what is actually meant by SSA or CPS.

Finally, why are abstract machines the appropriate target? An analysis is only correct with respect to a specified language. Specification is a *trusted process*, so we must be sure we get it right. Two ways to be sure of correctness are to visually audit the spec, and to test it. Abstract machines are high-level enough to express readable and understandable programming language semantics, but are low-level enough to provide a reasonable execution strategy for testing. The structure of abstract machines is flexible enough to elegantly express otherwise difficult semantic features like composable continuations.

The above benefits for abstract machines are also benefits of interpreters, an arguably more natural way to express the semantics of a programming language. An important strike against interpreters is that they have undefined behavior for divergent programs. Abstract machines take relatively *small steps*, and can represent intermediate computations. The definitions for these steps are terminating by design and do not conceal behavior in a metalanguage. Divergent programs do not diverge in an abstract machine, they just always have another step to take. **Programs with abstract components are much more likely to diverge.** Consider *factorial*'s execution on the abstract number $\mathbb{N}$:

$$\begin{aligned}
\mathit{factorial}(n) &= \texttt{if } n \stackrel{?}{=} 0 \texttt{ then} \\
&\quad\ \ 1 \\
&\ \ \texttt{else} n * \mathit{factorial}(n-1) \\
\mathit{factorial}(\mathbb{N}) &= \{1, \mathbb{N} * \mathit{factorial}(\mathbb{N}-1)\} \\
&= \{1, \mathbb{N} * \mathit{factorial}(\mathbb{N})\} \\
&= \{1, \mathbb{N} * \textbf{UH-OH}\}
\end{aligned}$$

Its termination condition is unclear, since $\mathbb{N} - 1 = \mathbb{N}$.

As such, a foundation that gives meaning to infinite executions is crucial. The framework of denotational semantics can give meaning to divergent programs, but in an extensional sense – two divergent programs are the same. Abstract machines' steps give us insight into just what a machine is doing – an intensional view of computation. A program's intension – *how* it computes what it computes – is what one might use an analysis to understand. Understanding comes in many flavors: optimization, finding security vulnerabilities, refactoring, semantic navigation, debugging laziness, adherence to style guidelines (linting), *etc.*.

## 1.4 PREVIOUSLY PUBLISHED MATERIAL

Two chapters of this dissertation are largely restated from my publications. Chapter 3 covers the implementation work that originally appeared in Johnson et al. [41]. The performance improvements I showed with that work demonstrate the performance aspect of my thesis. Chapter 4 covers the rephrasing of Vardoulakis and Shivers [103] in the AAM framework, and extends it with stack inspection and composable continuations, as originally appeared in Johnson and Van Horn [44] (with some bugfixes). The notion of *context* and *storable context* from this work improves on existing analysis technology's ability to precisely characterize first-class continuations, demonstrating the precision aspect of my thesis.

An additional publication, Johnson et al. [42], is related to the pushdown work in Chapter 4, but uses a technique outside of the AAM methodology that I explore in this dissertation. In that work, we defined and explored an entire class of automata, *regular introspective pushdown automata*, and its reachability problem. The reachability complexity in that class is intractible in general, so we specialized the machinery to solve the easier garbage collection problem. Having done that work, I'm convinced that the methodology in Chapter 4 is far simpler to explain, prove, implement and motivate.

Part I

# SYSTEMATIC CONSTRUCTIONS

# INTRODUCTION TO PART I: SYSTEMATIC CONSTRUCTIONS

*"A vocabulary of truth and simplicity will be of service throughout your life."*
*~Winston Churchill*

In some respects, everything in this part of the dissertation is not new. Higher-order control flow analysis has existed since the 1980's (Jones' flow analysis of lambda expressions [46] and Shivers' 0CFA [85]). Higher-order pushdown analysis is newer (CFA2 in 2010 [99]) but still precedes this work. What is new is how I formulate, prove, and implement them as abstract machines. Abstract machines give a simple and unified view of both concrete and abstract interpretation of programs.

The first chapter reviews the first foray into formulating analyses with abstract machines. It goes through a full derivation from the call-by-value lambda calculus expressed with a reduction semantics to an abstract abstract machine ($CESK_t^*$) that is a computable, sound approximation. The second chapter builds off the first by rigorously systematizing folklore implementation strategies step-by-step. The resulting abstract machines are directly translated to code for a 1000-fold performance improvement. The third chapter owes most of its machinery for pushdown analysis to CFA2. The lessons from that work are distilled into a new vocabulary and proof technique that allows both concrete and abstract interpretation in the same model. We take CFA2 further with abstract machines by adding garbage collection, stack inspection, and composable control operators.

# 2 ABSTRACTING ABSTRACT MACHINES

"Abstracting Abstract Machines" (AAM) is a technique conceived by Van Horn and Might [95] for constructing analyses of programming languages. Specifically, it makes finite-state approximations of programs by simply running them in a slightly modified semantics. AAM is founded on three ideas:

1. concrete and abstract semantics ideally should use the same code, for correctness and testing purposes,
2. the level of abstraction should be a tunable parameter,
3. both of the above are achievable with a slight change to the abstract machine's state representation.

The first two points are the philosophy of AAM: correctness through simplicity, reusability, and sanity checking with concrete semantics. The final point is the machinery that we recount in this chapter. The first point of simplicity emphasizes a "turn-the-crank" approach to analysis construction.

The slight modification is to restate all recursive data structures in a program's state to instead redirect their self-reference through some "address" in a store[2]. This way, the only source of new values is the space of addresses. If the space of addresses is made finite, the state space becomes finite. To remain a sound approximation, the store maps to *sets* of storable objects to not lose information. Store updates then become *weak*, so that an update to store $\sigma[a \mapsto v]$ becomes $\sigma[a \mapsto \sigma(a) \cup \{v\}]$[3], which may also be written as $\sigma \sqcup [a \mapsto \{v\}]$. Uses of the store, symmetrically, choose elements non-deterministically from the sets they access at some address.

[2] Sometimes called a "memory," or generally an "environment"

[3] Previously unmapped addresses are mapped to $\emptyset$

## 2.1 STANDARDIZING NON-STANDARD SEMANTICS: *alloc* AND *tick*

The term "non-standard semantics" originates from the context of abstract interpretation. A semantics is non-standard simply when it is not the standard semantics – it can gather extra information about execution, or just be structured differently. The open-endedness of this term is great for a broad framework, but AAM provides more structure to focus the design space while still remaining broad enough for most applications. The operational semantics for a language is, for lack of a better term, lightly decorated.

At any point that the semantics needs to construct some recursive data structure, AAM dictates that we appeal to a metafunction, *alloc*,

to provide an address at which to put the "recursive" part of the data structure in the store.[4] The allocated address takes the place where the recursive part of the data structure would have gone. For example, lists' recursive constructor

[4] *If a rule has multiple allocation sites and/or non-deterministic choices, the alloc function can take extra arguments to point at which allocation site it's allocating for, and the current environment of choices made before reaching the current allocation point.*

$$\texttt{cons} : \texttt{Value} \times \texttt{List} \rightarrow \texttt{List} \text{ becomes}$$
$$\texttt{cons} : \texttt{Value} \times \texttt{Addr} \rightarrow \texttt{List}$$

The question becomes what to give *alloc* as input? AAM suggests that each state just contain an extra component, $t$, that can guide *alloc*'s choice of addresses. Thus we have the following spaces and functions as parameters to an abstracted abstract machine:

$$\begin{aligned} \mathit{Addr} &\text{ an arbitrary set with decidable equality} \\ \mathit{Time} &\text{ an arbitrary set with decidable equality} \\ \mathit{alloc} &: \mathit{State} \times \mathit{Time} \rightarrow \mathit{Addr} \\ \mathit{tick} &: \mathit{State} \times \mathit{Time} \rightarrow \mathit{Time} \end{aligned}$$

The open-endedness of abstract interpretation's notion of abstract semantics is boiled down to the choice and representation of addresses with a "helper" parameter. The name for the helper is instructive as it is historically a generalization of a notion of "binding time" from the kCFA family of control-flow analyses. The *Time* domain in this case would be a list of function application *labels* (expecting expressions to be uniquely labeled) with length at most $k$. The *tick* function would then extend and truncate this list when encountering a function application state. Finally, addresses are pairs of binder and binding time, so *alloc* destructs a function application state to find the binder, and pairs it with the current time.

There are of course infinitely many other strategies that are waiting to be found. A paper at VMCAI [36] suggests a number of *Time* and *tick* constructions that abstract the execution trace in interesting ways. The authors used this work to build a JavaScript analysis with a multitude of allocation strategies that they evaluated for precision and performance [47].

### 2.1.1 *The lambda calculus to the* $\mathsf{CESK}^*_t$ *machine*

This section is a succession of refinements to the representation and implementation the semantics for the lambda calculus, where the result is simultaneously a correct implementation of the lambda calculus and a sound and computable approximation of lambda calculus expressions' evaluation. All but the final semantics are adaptations from Felleisen [31].

THE LAMBDA CALCULUS is the canonical simple language to expound ideas for functional languages:

$$\begin{aligned} e \in \mathit{Expr} &::= x \mid (e\ e) \mid \lambda x.\ e \\ x \in \mathit{Var} &\text{ a set} \end{aligned}$$

There is one rule of computation: function application substitutes the argument for the variable "bound" by the function (β-reduction).

$$C[(\lambda x.\ e\ e')] \longmapsto_\beta C[[^{e'}/_x]e]$$

Here C is a *context*, or simply an expression with a hole in it.

$$C \in \mathit{Context} ::= [] \mid (C\ e) \mid (e\ C) \mid \lambda x.\ C$$

$[^{e'}/_x]e$ is "capture-avoiding substitution" of $e'$ for $x$ in the expression $e$. A lambda expression (a function) is also called an *abstraction*; it is a term with a named hole that is plugged in by substitution; the name of the hole cannot leak out by coincidentally being the same as one in an expression being substituted in. This "leaking" is also called "name capture," and is avoidable by renaming a binder to something *fresh*: a name that does not exist in the expression being substituted.

$$\begin{aligned} [^e/_x]y &= x \stackrel{?}{=} y \to e, y \\ [^e/_x](e_0\ e_1) &= ([^e/_x]e_0\ [^e/_x]e_1) \\ [^e/_x]\lambda y.\ e' &= \lambda y.\ [^e/_x]e' \qquad \text{if } y \notin \mathit{fv}(e) \end{aligned}$$

A variable is "free" in an expression if it appears out of the scope of a binder:

$$\begin{aligned} \mathit{fv}(x) &= \{x\} \\ \mathit{fv}((e_0\ e_1)) &= \mathit{fv}(e_0) \cup \mathit{fv}(e_1) \\ \mathit{fv}(\lambda x.\ e) &= \mathit{fv}(e) \setminus \{x\} \end{aligned}$$

Its β reduction rule for reducible expressions (redexes) is often applicable in many places at once, and it is impossible to algorithmically determine the "best," *i.e.*, shortest, reduction sequence. It is necessary to therefore determine a *reduction strategy*, with popular choices being "by name," "by value," and "by need."

THE CALL-BY-VALUE LAMBDA CALCULUS is very common and can express the others with reasonable language extensions (like state). The choice to decompose an application on the left or right is removed by a specialized context, E, called an "evaluation context." In call-by-value, the first expression in an application is evaluated to a lambda expression before continuing on the right. Additionally, reduction cannot happen within a lambda expression.

$$E \in \mathit{EvalContext} ::= [] \mid (E\ e) \mid (v\ E)$$
$$v \in \mathit{Value} ::= \lambda x.\ e$$

With this kind of context instead of C above, reduction is deterministic.

THE CK MACHINE represents this kind of evaluation context so that there is no need to decompose the expression into a context and a redex, substitute, plug in the hole, rinse and repeat. Instead, the evaluation context can be represented in a more favorable way: an evaluation context can be seen as a sequence of composed functions.

$$\begin{aligned}
\mathit{decompose}([]) &= \mathit{identity\text{-}function} \\
\mathit{decompose}((E\ e)) &= ([]\ e) \circ \mathit{decompose}(E) \\
\mathit{decompose}((v\ E)) &= (v\ []) \circ \mathit{decompose}(E)
\end{aligned}$$

These small functions we will call *frames* of the continuation. If we now represent these functions as data structures, with $\circ$ instead a list constructor, we get

$$\phi \in \mathit{Frame} ::= \mathbf{appL}(e) \mid \mathbf{appR}(v)$$
$$K \in \mathit{Kont} ::= \epsilon \mid \phi{:}K$$

Reduction can now use the top of the continuation to pivot and look for the next redex without a full plug / decompose step.

$$\begin{aligned}
(e_0\ e_1), K &\longmapsto e_0, \mathbf{appL}(e_1){:}K \\
v, \mathbf{appL}(e){:}K &\longmapsto e, \mathbf{appR}(v){:}K \\
v, \mathbf{appR}(\lambda x.\ e){:}K &\longmapsto_\beta [{}^{v}/_{x}]e, K
\end{aligned}$$

Substitution isn't really how one would implement the lambda calculus. Programs don't rewrite themselves as their method of running, they have data structures, registers, memory.

THE CEK MACHINE represents substitutions as data structures to be interpreted as execution proceeds. Names are not substituted away, but instead are added to a data structure for delaying a substitution called an *environment*. Expressions now have variables in them that aren't bound by a lambda expression, but instead by the environment. For this reason, an expression (with an "open" scope) and environment (that "closes" the scope) pair is called a "closure." The frame components are augmented to carry environments.

$$\rho \in \mathit{Env} = \mathit{Var} \underset{\text{fin}}{\rightharpoonup} (\mathit{Expr} \times \mathit{Env})$$
$$\phi \in \mathit{Frame} ::= \mathbf{appL}(e, \rho) \mid \mathbf{appR}(v, \rho)$$

The resulting semantics is the CEK machine.

$$\begin{aligned}
x, \rho, K &\longmapsto v, \rho', K \qquad \text{where } (v, \rho') = \rho(x) \\
(e_0\ e_1), \rho, K &\longmapsto e_0, \rho, \mathbf{appL}(e_1, \rho){:}K \\
v, \rho, \mathbf{appL}(e, \rho'){:}K &\longmapsto e, \rho', \mathbf{appR}(v, \rho){:}K \\
v, \rho, \mathbf{appR}(\lambda x.\ e, \rho'){:}K &\longmapsto_\beta e, \rho[x \mapsto (v, \rho)], K
\end{aligned}$$

This is looking more like a possible language implementation, except these environments can get big and full of garbage (substitutions that end up never happening). This can be fixed by treating substitutions as resources.

THE CESK MACHINE explicitly allocates a fresh address to store the substitution in a store that can be garbage-collected. Environments hold on to these addresses instead of the values themselves.

$$\begin{aligned}
&\rho = \mathit{Var} \underset{\text{fin}}{\rightharpoonup} \mathit{Addr} \\
&\mathit{Addr} \text{ a set} \\
&\sigma \in \mathit{Store} = \mathit{Addr} \to (\mathit{Value} \times \mathit{Env})
\end{aligned}$$

The resulting semantics is the CESK machine.

$$\begin{aligned}
x, \rho, \sigma, K &\longmapsto v, \rho', \sigma, K \qquad \text{where } (v, \rho') = \sigma(\rho(x)) \\
(e_0\ e_1), \rho, \sigma, K &\longmapsto e_0, \rho, \sigma, \mathbf{appL}(e_1, \rho){:}K \\
v, \rho, \sigma, \mathbf{appL}(e, \rho'){:}K &\longmapsto e, \rho', \sigma, \mathbf{appR}(v, \rho){:}K \\
v, \rho, \sigma, \mathbf{appR}(\lambda x.\ e, \rho'){:}K &\longmapsto_\beta e, \rho'[x \mapsto a], \sigma[a \mapsto (v, \rho)], K \qquad \text{where } a \notin \mathbf{dom}(\sigma)
\end{aligned}$$

Garbage collection finds a conservative set of addresses that must remain in the store for evaluation to continue normally, and drops the rest since they are useless. The typical way to compute this set is to first find what addresses the state originally refers to (touch), then iteratively find what the values touch that those addresses point to. The iteration step is called the reachability computation. A closure touches the addresses that it can possibly refer to: the free variables that are mapped in the environment.

$$\begin{aligned}
&\mathcal{T}(e, \rho) = \mathcal{T}(\mathbf{appL}(e, \rho)) = \mathcal{T}(\mathbf{appR}(e, \rho)) = \{\rho(x) \ : \ x \in \mathit{fv}(e)\} \\
&\mathcal{T}(\epsilon) = \emptyset \\
&\mathcal{T}(\phi{:}K) = \mathcal{T}(\phi) \cup \mathsf{touch}(K) \\
&\mathcal{R}(\mathit{root}, \sigma) = \{b \ : \ a \in \mathit{root}, a \leadsto_\sigma^* b\} \\
&\text{where } \frac{(v, \rho) = \sigma(a) \qquad b \in \mathcal{T}(v, \rho)}{a \leadsto_\sigma b}
\end{aligned}$$

Garbage collection is then the restriction of the store to reachable addresses from the state's touched addresses:

$$\begin{aligned}
&e, \rho, \sigma, K \longmapsto_\Gamma e, \rho, \sigma|_L, K \\
&\quad \text{where } L = \mathcal{R}(\mathcal{T}(e, \rho) \cup \mathcal{T}(K), \sigma)
\end{aligned}$$

| $\varsigma, t \longmapsto \varsigma', u$ | $a = alloc(\varsigma),\ u = tick(\varsigma, t)$ |
|---:|:---|
| $x, \rho, \sigma, K, _$ | $v, \rho', \sigma, K, _$ where $\mathbf{inl}\,(v, \rho') \in \sigma(\rho(x))$ |
| $(e_0\ e_1), \rho, \sigma, K, _$ | $e_0, \rho, \sigma \sqcup [a \mapsto \mathbf{inr}\,K], \mathbf{appL}(e_1, \rho){:}a, _$ |
| $v, \rho, \sigma, \mathbf{appL}(e, \rho'){:}b, _$ | $e, \rho', \sigma, \mathbf{appR}(v, \rho){:}b, _$ |
| $v, \rho, \sigma, \mathbf{appR}(\lambda x.\ e, \rho'){:}b, _$ | $e, \rho[x \mapsto a], \sigma \sqcup [a \mapsto \mathbf{inl}\,(v, \rho)], K, _$ |
| | where $\mathbf{inr}\,K \in \sigma(b)$ |

Figure 1: The $CESK_t^*$ machine

These machines all implement the full semantics of the lambda calculus, but with the CESK machine, we are very close to a representation that easily abstracts to a finite state automaton.

THE $CESK_t^*$ MACHINE represents recursive data structures that the semantics constructs[5] with "recursive parts" rerouted through the store. Fortunately, the only recursion left in the CESK machine is the list structure for the continuation. Thus, the tail of a continuation *cons* is replaced with an address. Second, we use the aforementioned *weak* update semantics for extending the store, and use the *alloc* and *tick* functions. Finally, all uses of the store non-deterministically resolve to one of the values stored in the given address.

[5] *expressions are no longer rewritten/built, so there are only finitely many of them*

$$
\begin{aligned}
\hat{\kappa} \in \widehat{Kont} &= \epsilon \mid \phi{:}a \\
s \in Storable &= (Value \times Env) + \widehat{Kont} \\
\sigma \in Store &= Addr \underset{\text{fin}}{\rightharpoonup} \wp(Storable) \\
\varsigma \in State &= Expr \times Env \times Store \times \widehat{Kont}
\end{aligned}
$$

Garbage collection is similar to previously, with only the following changes:

$$
\begin{aligned}
\mathcal{T}(\mathbf{inl}\,(v, \rho)) &= \mathcal{T}(v, \rho) \\
\mathcal{T}(\mathbf{inr}\,\epsilon) &= \emptyset \\
\mathcal{T}(\mathbf{inr}\,\phi{:}a) &= \mathcal{T}(\phi) \cup \{a\}
\end{aligned}
$$

$$
\frac{s \in \sigma(a) \qquad b \in \mathcal{T}(s)}{a \rightsquigarrow_\sigma b}
$$

The size of the state space is a function of the size of the analyzed expression, the address space's size, and the size of the *Time* space.

$$\begin{aligned}
\mathit{size}(e, A, T) &= |\mathit{Expr}| * |\mathit{Env}| * |\mathit{Store}| * |\mathit{Kont}| * T \\
\text{where } |\mathit{Expr}| &= \mathit{treesize}(e) \\
|\mathit{Var}| &= |\mathit{supp}(e)| \\
|\mathit{Value}| &= |\lambda s(e)| \\
|\mathit{Env}| &= A^{|\mathit{Var}|} \\
|\mathit{Frame}| &= |\mathit{Expr}| * |\mathit{Env}| + |\mathit{Value}| * |\mathit{Env}| \\
|\mathit{Kont}| &= 1 + |\mathit{Frame}| * A \\
|\mathit{Storable}| &= |\mathit{Kont}| + |\mathit{Value}| * |\mathit{Env}| \\
|\mathit{Store}| &= (2^{|\mathit{Storable}|})^A
\end{aligned}$$

The auxiliary functions are simple tree-walks:

$$\begin{aligned}
\mathit{treesize}(x) &= 1 \\
\mathit{treesize}((e_0\ e_1)) &= 1 + \mathit{treesize}(e_0) + \mathit{treesize}(e_1) \\
\mathit{treesize}(\lambda x.\ e) &= 1 + \mathit{treesize}(e) \\
\mathit{supp}(x) &= \{x\} \\
\mathit{supp}((e_0\ e_1)) &= \mathit{supp}(e_0) \cup \mathit{supp}(e_1) \\
\mathit{supp}(\lambda x.\ e) &= \{x\} \cup \mathit{supp}(e) \\
\lambda s(x) &= \emptyset \\
\lambda s((e_0\ e_1)) &= \lambda s(e_0) \cup \lambda s(e_1) \\
\lambda s(\lambda x.\ e) &= \lambda s(e) \cup \{\lambda x.\ e\}
\end{aligned}$$

The second two functions have size at most $\mathit{treesize}(e)$, which is what we refer to as $n$: the "input size" for complexity analysis.

The case where *alloc* produces fresh addresses and thus implements the CESK semantics exactly means that $A = \omega^6$, and the state space is unbounded. The case where *alloc* produces addresses from a finite pool means that the state space is finite, and furthermore $\longmapsto^*$ (reachable states) is finite and effectively computable. The exponents makes this technical use of "effective" not so practically effective, however.

[6] *The first limit ordinal, isomorphic to the set of natural numbers*

## 2.2 WIDENING FOR POLYNOMIAL COMPLEXITY

Provided that the address and time spaces are polynomially sized, the state exploration algorithm for the $\mathit{CESK}_t^*$ machine can be accelerated to run in polynomiral time at the cost of precision. The key idea is to factor out the exponentially sized components to be shared amongst states. Instead of a state space that looks like $2^{\mathit{State}}$, we have a state space that looks like $\mathit{Large}^{\mathit{Small}}$. The large components grow monotonically for each small state. If the large components grow only polynomially many times, and *Small* is polynomially sized, the number of steps to compute is bounded by the product of those polynomials: still polynomial. Stepping a single state takes $\mathcal{O}(|\mathit{Storable}| * \log n)$ time due to environment lookups and extensions ($\log n$), the number of non-deterministic choices ($|\mathit{Storable}|$) and the assumption that the following are constant time operations:

- allocation of machine components
- consing output states to a result list
- *alloc* and *tick*

Concretely, the original AAM paper suggests a widening that shares the store amongst all states, and an allocation scheme that makes the $\rho$ component the identity function (0CFA). There are a couple of ways to do this:

1. treat the set of all seen states and the global store as one *big* state that steps (all states step, all modifying the global store) until it reaches a fixed point;
2. separate seen states and the stores they were seen at from *unexplored* states, and only step unexplored states, which all modify the global store.

I call the first kind a *whole-space* semantics, and the second a *frontier-based* semantics. The second is more precise, because states that are never again visited do not need to be processed again. However, since we determine if a state is unexplored by comparing stores, the comparison can be expensive. It turns out there is a simple fix for this that Shivers [87] originally discovered, and which I recount and modify in the next chapter. Essentially states are stored with the *age* of the store instead of the whole store.

COMPLEXITY the global store can be updated $|\mathit{Storable}| * |\mathit{Addr}|$ many times, and $|\mathit{Expr}| * |\mathit{Env}| * |\mathit{Kont}| * |\mathit{Time}|$ many states can be explored. With the given allocation strategy, these sizes are both $\mathcal{O}(n^3)$. Environment lookups are removable because $\mathit{Env} = \{\lambda x.x\}$, so that leaves store lookups. The store can be represented as a vector, and all names can be represented as unique indices into the vector, so we treat lookups and updates as $\mathcal{O}(1)$. The number of iterations for the whole-space semantics is in the worst case the number of states plus the number of store updates: $\mathcal{O}(n^3)$. The cost of a big step is the number of states $\mathcal{O}(n^3)$ times the cost of a small step $\mathcal{O}(n^2)$ times the cost to put a state into a set plus the cost of a store join $\mathcal{O}(\log n^3 + \log n) = \mathcal{O}(\log n)$. Overall the cost is $\mathcal{O}(n^8 \log n)$. **Yikes.**

For the frontier-based semantics, the number of iterations is irrelevant. Instead, the number of times a state can be stepped is relevant because the frontier is only extended with states that need stepping.

Thus, the cost is the number of states $\mathcal{O}(n^3)$ times the number of possible updates $\mathcal{O}(n^3)$ times the cost of a step. The cost to step includes store comparison. We will treat store comparison as logarithmic time since the store age is a number logarithmic in size of the number of store updates [7]. Each state step then takes $\mathcal{O}(|\mathit{Storable}| * \log n)$ time. Therefore overall the complexity of frontier-based semantics is again $\mathcal{O}(n^8 \log n)$.

The original 0CFA uses CPS, so the continuations are part of the program itself. It is also frontier-based. The number of states and storables is $\mathcal{O}(n)$, there is a $\mathcal{O}(n \log n)$ cost per state step, and $\mathcal{O}(n^2)$ many possible store updates, so overall is $\mathcal{O}(n^4 \log n)$. Shivers did not have a complexity result in his dissertation, though folklore claims 0CFA is cubic. The established $\mathcal{O}(n^3)$ bound is for a different, less precise,[8] formulation [71] that monotonically increases each $n$ states at most $n$ times, where each increase possibly affects $n$ more states. Worst-case running times are almost meaningless analyses for CFA, however, since the bounds are so pessimistic on what possible program behaviors there are.

[7] *Often the logarithmic factor involved in store comparisons is ignored because the number of store updates rarely exceeds a machine word.*

[8] *flow-insensitive,* i.e., *order of evaluation does not matter to outcome*

ADDRESSING THE "YIKES" There is a constant battle between precision and performance for analyses. Higher precision often means a theoretically larger state space, but the better precision also can cull execution paths, meaning less work. The contention is that extra precision may not be enough to win back the added cost. Also possible is that the "theoretically larger" state space is met in practice when a "pathological" case that leads to state explosion turns out to be more common than originally thought. A compiler-writer looking for a seriously fast analysis with "good-enough" precision can use this document for abstraction principles, but should look elsewhere [6, 1] for precision/performance tradeoffs. I focus on generally and systematically applicable techniques to "off-the-shelf" languages with effort towards higher precision for verification purposes. In the next chapter we will look at a derivation approach to curtailing the high complexity of this chapter in asymptotics and significantly smaller constant factors.

# 3 ENGINEERING ENGINEERED SEMANTICS

We have seen that AAM provides sound predictive models of program behavior, but in order for such models to be effective, they must be efficiently computable and correct.

Since these analyses so closely resemble a language's interpreter (a) implementing an analysis requires little more than implementing an interpreter, (b) a single implementation can serve as both an interpreter and analyzer, and (c) verifying the correctness of the implementation is straightforward.

Unfortunately, the AAM approach yields analyzers with poor performance relative to hand-optimized analyzers. This chapter takes aim squarely at this "efficiency gap," and narrows it in an equally systematic way through a number of simple steps, many of which are inspired by run-time implementation techniques such as laziness and compilation to avoid interpretative overhead. Each of these steps is proven correct, so the end result is an implementation that is trustworthy and efficient.

The intention is to develop a systematic approach to deriving a practical implementation of an abstract-machine-based analyzer using mostly semantic means rather than tricky and unreliable engineering.

## 3.1 OVERVIEW

This chapter starts with improving the complexity of the widened $CESK_t^*$ semantics of the previous chapter by making reasonable approximations. We then apply our step-by-step optimization techniques in the simplified setting of a core-but-more-realistic[9] functional language. This allows us to explicate the optimizations with a minimal amount of inessential technical overhead. Finally, I give an evaluation of the approach scaled up to an analyzer for a realistic untyped, higher-order imperative language with a number of interesting features and then measure improvements across a suite of benchmarks.

[9] *More realistic than the pure lambda calculus*

At each step during the initial presentation and development, we evaluated the implementation on a set of benchmarks. The highlighted benchmark in figure 2 is from Vardoulakis and Shivers [103] that tests distributivity of multiplication over addition on Church numerals. For the step-by-step development, this benchmark is particularly informative:

1. it can be written in most modern programming languages,

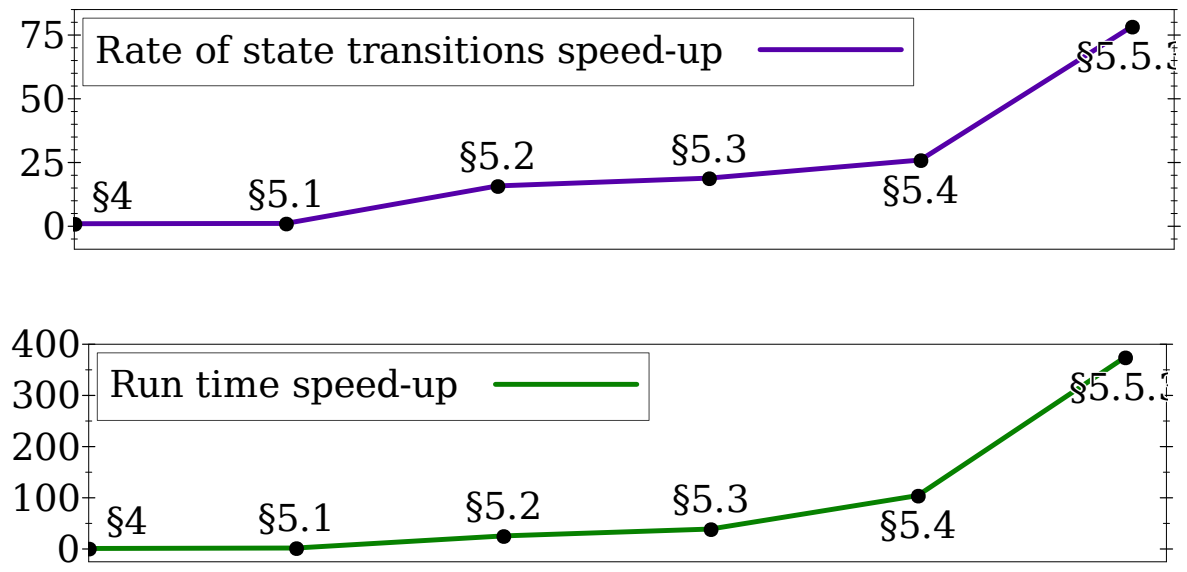

Figure 2: Factor improvements over the baseline analyzer for the Vardoulakis and Shivers benchmark in terms of the rate of state transitions and total analysis time. (Bigger is better.) Each point is marked with the section that introduces the optimization.

2. it was designed to stress an analyzer's ability to deal with complicated environment and control structure arising from the use of higher-order functions to encode arithmetic, and

3. its improvement is about median in the benchmark suite considered in section 3.5, and thus it serves as a good sanity check for each of the optimization techniques considered.

We start, in section 3.2, by developing an abstract interpreter according to the AAM approach of the last chapter. In section 3.3, we perform a further abstraction by store-allocating values originally in continuation frames. The resulting analyzer sacrifices precision for speed and is able to analyze the example in about 1 minute. We therefore take this widened interpreter as the baseline for our evaluation.

Section 3.4 gives a series of simple abstractions and implementation techniques that, in total, speed up the analysis by nearly a factor of 500, dropping the analysis time to a fraction of a second. Figure 2 shows the step-wise improvement of the analysis time for this example.

The techniques we propose for optimizing analysis fall into the following categories:

1. generate fewer states by avoiding the eager exploration of nondeterministic choices that will later collapse into a single join point. We accomplish this by applying lazy evaluation techniques so that nondeterminism is evaluated *by need*.

2. generate fewer states by avoiding unnecessary, intermediate states of a computation. We accomplish this by applying compilation techniques from functional languages to avoid interpretive overhead in the machine transition system.

3. generate states faster. We accomplish this by better algorithm design in the fixed-point computation we use to generate state graphs.

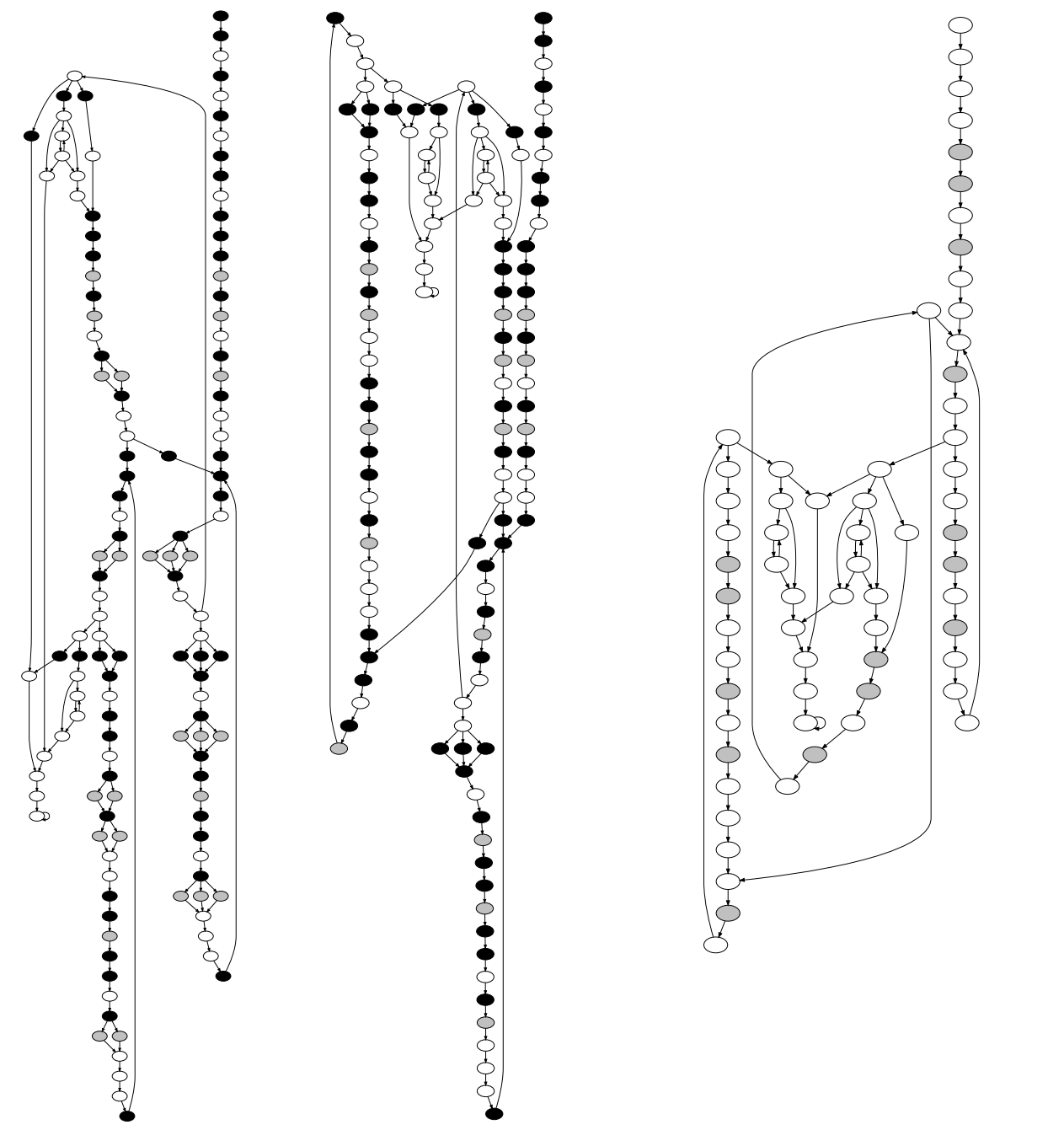

(a) Baseline (b) Lazy (c) Compiled (& lazy)

Figure 3: Example state graphs for Earl et. al. program. Gray states follow variable references, `ev` states are black, and all others are white. Part (a) shows the baseline analyzer result. It has long "corridor" transitions and "diamond" subgraphs that fan-out from nondeterminism and fan-in from joins. Part (b) shows the result of performing nondeterminism lazily and thus avoids many of the diamond subgraphs. Part (c) shows the result of abstract compilation that removes interpretive overhead in the form of intermediate states, thus minimizing the corridor transitions. The end result is a more compact abstraction of the program that can be generated faster.

Figure 3 shows the effect of (1) and (2) for the small motivating example in Earl, et al. [29]. By generating significantly fewer states at a significantly faster rate, we are able to achieve large performance improvements in terms of both time and space.

## 3.2 ABSTRACT INTERPRETATION OF λIF

In this section, we use the AAM approach to define a sound analytic framework for a core higher-order functional language: lambda calculus with conditionals and base type operations. We will call this language λ**IF**. In the subsequent sections, we will explore optimizations for the analyzer in this simplified setting, but scaling these techniques to realistic languages is straightforward and has been done for the analyzer evaluated in section 3.5.

λ**IF** is a family of programming languages parameterized by a set of base values and operations. To make things concrete, we consider a member of the λ**IF** family with integers, booleans, and a few opera-

$$
\begin{array}{llcl}
\text{Expressions} & e & = & x^{\ell} \\
 & & | & \mathtt{lit}^{\ell}(l) \\
 & & | & \lambda^{\ell}\ x.\ e \\
 & & | & (e\ e)^{\ell} \\
 & & | & \mathtt{if}^{\ell}(e, e, e) \\
\text{Variables} & x & = & \mathtt{x} \mid \mathtt{y} \mid \mathtt{...} \\
\text{Literals} & l & = & z \mid b \mid o \\
\text{Integers} & z & = & \mathtt{0} \mid \mathtt{1} \mid \mathtt{-1} \mid \mathtt{...} \\
\text{Booleans} & b & = & \mathtt{tt} \mid \mathtt{ff} \\
\text{Operations} & o & = & \mathtt{zero?} \mid \mathtt{add1} \mid \mathtt{sub1} \mid \mathtt{...}
\end{array}
$$

Figure 4: Syntax of λ**IF**

$$
\begin{array}{lrcl}
\text{Values} & v, u & = & \mathtt{clos}\ (x, e, \rho) \mid l \mid \kappa \\
\text{States} & \varsigma & = & \mathtt{ev}^{t}(e, \rho, \sigma, \kappa) \\
 & & | & \mathtt{co}\ (\kappa, v, \sigma) \\
 & & | & \mathtt{ap}^{t}(v, v, \sigma, \kappa) \\
\text{Continuations} & \kappa & = & \mathtt{halt} \\
 & & | & \mathbf{fun}(v, a_{\kappa}) \\
 & & | & \mathbf{arg}(e, \rho, a_{\kappa}) \\
 & & | & \mathtt{ifk}\ (e, e, \rho, a_{\kappa}) \\
\text{Addresses} & a & \in & \mathit{Addr} \\
\text{Times} & t & \in & \mathit{Time} \\
\text{Environments} & \rho & \in & \mathit{Var} \rightharpoonup \mathit{Addr} \\
\text{Stores} & \sigma & \in & \mathit{Addr} \rightharpoonup \wp(\mathit{Value})
\end{array}
$$

Figure 5: Abstract machine components

tions. Figure 4 defines the syntax of λ**IF**. It includes variables, literals (either integers, booleans, or operations), λ-expressions for defining procedures, procedure applications, and conditionals. Expressions carry a label, $\ell$, which is drawn from an unspecified set and denotes the source location of the expression; labels are used to disambiguate distinct, but syntactically identical pieces of syntax. We omit the label annotation in contexts where it is irrelevant.

The semantics is defined in terms of a machine model. The machine components are defined in figure 5; figure 6 defines the transition relation (unmentioned components stay the same). The evaluation of a program is defined as the set of traces that arise from iterating the machine transition relation. The *traces* function produces the set of all proofs of reachability for any state $\varsigma$ from the injection of program $e$ (from which one could extract a string of states). The machine is a very slight variation on a standard abstract machine for λ**IF** in "eval, continue, apply" form [22]. It can be systematically derived from a

$$\mathit{traces}(e) = \{\mathtt{ev}^{t_0}(e, \varnothing, \varnothing, \mathtt{halt}) \longmapsto\!\!\!\!\!\rightarrow \varsigma\} \text{ where}$$

$$\begin{aligned}
\varsigma &\longmapsto \varsigma' \text{ defined to be the following} \\
& \quad \text{let } t' = \mathit{tick}(\varsigma) \\
\mathtt{ev}^{t}(x\ , \rho, \sigma, \kappa) &\longmapsto \mathtt{co}^{t'}(\kappa, v, \sigma) \text{ if } v \in \sigma(\rho(x)) \\
\mathtt{ev}^{t}(\mathtt{lit}\ (l), \rho, \sigma, \kappa) &\longmapsto \mathtt{co}^{t'}(\kappa, l, \sigma) \\
\mathtt{ev}^{t}(\lambda\ \ x.\ e, \rho, \sigma, \kappa) &\longmapsto \mathtt{co}^{t'}(\kappa, \mathtt{clos}\ (x, e, \rho), \sigma) \\
\mathtt{ev}^{t}((e_0\ e_1)^{\ell}, \rho, \sigma, \kappa) &\longmapsto \mathtt{ev}^{t'}(e_0, \rho, \sigma', \mathbf{arg}^{t}_{\ell}(e_1, \rho, a_\kappa)) \\
& \quad \text{where } a_\kappa = \mathit{allockont}^{t}_{\ell}(\sigma, \kappa) \\
& \qquad \sigma' = \sigma \sqcup [a_\kappa \mapsto \{\kappa\}] \\
\mathtt{ev}^{t}(\mathtt{if}^{\ell}(e_0, e_1, e_2), \rho, \sigma, \kappa) &\longmapsto \mathtt{ev}^{t'}(e_0, \rho, \sigma', \mathtt{ifk}^{t}(e_1, e_2, \rho, a_\kappa)) \\
& \quad \text{where } a_\kappa = \mathit{allockont}^{t}_{\ell}(\sigma, \kappa) \\
& \qquad \sigma' = \sigma \sqcup [a_\kappa \mapsto \{\kappa\}] \\
\mathtt{co}\ (\mathbf{arg}^{t}_{\ell}(e, \rho, a_\kappa), v, \sigma) &\longmapsto \mathtt{ev}^{t}(e, \rho, \sigma, \mathbf{fun}^{t}_{\ell}(v, a_\kappa)) \\
\mathtt{co}\ (\mathbf{fun}^{t}_{\ell}(u, a_\kappa), v, \sigma) &\longmapsto \mathtt{ap}^{t}_{\ell}(u, v, \kappa, \sigma) \text{ if } \kappa \in \sigma(a_\kappa) \\
\mathtt{co}\ (\mathtt{ifk}^{t}(e_0, e_1, \rho, a_\kappa), \mathtt{tt}, \sigma) &\longmapsto \mathtt{ev}^{t'}(e_0, \rho, \sigma, \kappa) \text{ if } \kappa \in \sigma(a_\kappa) \\
\mathtt{co}\ (\mathtt{ifk}^{t}(e_0, e_1, \rho, a_\kappa), \mathtt{ff}, \sigma) &\longmapsto \mathtt{ev}^{t'}(e_1, \rho, \sigma, \kappa) \text{ if } \kappa \in \sigma(a_\kappa) \\
\mathtt{ap}^{t}_{\ell}(\mathtt{clos}\ (x, e, \rho), v, \sigma, \kappa) &\longmapsto \mathtt{ev}^{t'}(e, \rho', \sigma', \kappa) \\
& \quad \text{where } a = \mathit{alloc}(\varsigma) \\
& \qquad \rho' = \rho[x \mapsto a] \\
& \qquad \sigma' = \sigma \sqcup [a \mapsto \{v\}] \\
\mathtt{ap}^{t}_{\ell}(o, v, \sigma, \kappa) &\longmapsto \mathtt{co}\ (\kappa, v', \sigma) \text{ if } v' \in \Delta(o, v)
\end{aligned}$$

Figure 6: Abstract abstract machine for λ**IF**

definitional interpreter through a continuation-passing style transformation and defunctionalization, or from a structural operational semantics using the refocusing construction of Danvy and Nielsen [24].

CONCRETE INTERPRETATION To characterize concrete interpretation, set the implicit parameters of the relation given in figure 6 as follows:

$$\begin{aligned}
\mathit{alloc}(\varsigma) &= a \text{ where } a \notin \mathbf{dom} \text{ of the } \sigma \text{ within } \varsigma \\
\mathit{allockont}^{t}_{\ell}(\sigma, \kappa) &= a_\kappa \text{ where } a_\kappa \notin \mathbf{dom}(\sigma)
\end{aligned}$$

These functions appear to ignore $\ell$ and $t$, but they can be used to determinize the choice of fresh addresses. The $\sqcup$ on stores in the figure is a point-wise lifting of $\cup$: $\sigma \sqcup \sigma' = \lambda a.\sigma(a) \cup \sigma'(a)$. The resulting relation is non-deterministic in its choice of addresses, however it must always choose a fresh address when allocating a continuation or variable binding. If we consider machine states equivalent up to consistent renaming and fix an allocation scheme, this relation defines a deterministic machine (the relation is really a function).

The interpretation of primitive operations is defined by setting $\Delta$ as follows:

$$z+1 \in \Delta(\mathtt{add1}, z) \qquad z-1 \in \Delta(\mathtt{sub1}, z)$$
$$\mathtt{tt} \in \Delta(\mathtt{zero?}, \mathtt{0}) \qquad \mathtt{ff} \in \Delta(\mathtt{zero?}, z) \text{ if } z \neq \mathtt{0}$$

ABSTRACT INTERPRETATION To characterize abstract interpretation, set the implicit parameters just as above, but drop the $a \notin \sigma$ condition. The $\Delta$ relation takes some care to not make the analysis run forever; a simple instantiation is a flat abstraction where arithmetic operations return an abstract top element Z, and zero? returns both tt and ff on Z. This family of interpreters is also non-deterministic in choices of addresses, but it is free to choose addresses that are already in use. Consequently, the machines may be non-deterministic when multiple values reside in a store location.

It is important to recognize from this definition that *any* allocation strategy is a sound abstract interpretation [65]. In particular, concrete interpretation is a kind of abstract interpretation. So is an interpretation that allocates a single cell into which all bindings and continuations are stored. The former is an abstract interpretation with uncomputable reachability and gives only the ground truth of a program's behavior; the latter is an abstract interpretation that is easy to compute but gives little information. Useful program analyses lay somewhere in between and can be characterized by their choice of address representation and allocation strategy. Uniform kCFA [70], presented next, is one such analysis.

UNIFORM kCFA To characterize uniform kCFA, set the allocation strategy as follows, for a fixed constant k:

$$\begin{aligned}
\mathit{Time} &= \mathit{Label}^* \\
t_0 &= \epsilon \\
\mathit{alloc}(\mathtt{ap}^t_\ell(\mathtt{clos}\ (x, e, \rho), v, \sigma, \kappa)) &= x\lfloor \ell t \rfloor_k \\
\mathit{allockont}^t_\ell(\sigma, \kappa) &= \ell t \\
\mathit{tick}(\mathtt{ev}^t(e, \rho, \sigma, \kappa)) &= t \\
\mathit{tick}(\mathtt{co}\ (\mathbf{arg}^t(e, \rho, a_\kappa), v, \sigma)) &= t \\
\mathit{tick}(\mathtt{ap}^t_\ell(u, v, \kappa)) &= \lfloor \ell t \rfloor_k \\
\lfloor t \rfloor_0 = \lfloor \epsilon \rfloor_k &= t_0 \\
\lfloor \ell t \rfloor_{k+1} &= \ell \lfloor t \rfloor_k
\end{aligned}$$

The $\lfloor \cdot \rfloor_k$ notation denotes the truncation of a list of symbols to the leftmost k symbols.

All that remains is the interpretation of primitives. For abstract interpretation, we set $\Delta$ to the function that returns Z on all inputs—a symbolic value we interpret as denoting the set of all integers.

At this point, we have abstracted the original machine to one which has a finite state space for any given program, and thus forms the basis of a sound, computable program analyzer for $\lambda$**IF**.

## 3.3 FROM MACHINE SEMANTICS TO BASELINE ANALYZER

The uniform kCFA allocation strategy would make *traces* in figure 6 a computable abstraction of possible executions, but one that is too inefficient to run, even on small examples. Through this section, we explain a succession of approximations to reach a more appropriate baseline analysis. We ground this path by first formulating the analysis in terms of a classic fixed-point computation.

### 3.3.1 *Static analysis as fixed-point computation*

Conceptually, the AAM approach calls for computing an analysis as a graph exploration: (1) start with an initial state, and (2) compute the transitive closure of the transition relation from that state. All visited states are potentially reachable in the concrete, and all paths through the graph are possible traces of execution.

We can cast this exploration process in terms of a fixed-point calculation. Given the initial state $\varsigma_0$ and the transition relation $\longmapsto$, we define the global transfer function:

$$F_{\varsigma_0} : \wp(\mathit{State}) \times \wp(\mathit{State} \times \mathit{State}) \to \wp(\mathit{State}) \times \wp(\mathit{State} \times \mathit{State}).$$

Internally, this global transfer function computes the successors of all supplied states, and then includes the initial state:

$$\begin{aligned} F_{\varsigma_0}(V, E) &= (\{\varsigma_0\} \cup V', E') \\ E' &= \{(\varsigma, \varsigma') \mid \varsigma \in V \text{ and } \varsigma \longmapsto \varsigma'\} \\ V' &= \{\varsigma' \mid (\varsigma, \varsigma') \in E'\} \end{aligned}$$

Then, the evaluator for the analysis computes the least fixed-point of the global transfer function: $\mathit{eval}(e) = \mathrm{lfp}(F_{\varsigma_0})$, where $\varsigma_0 = \mathtt{ev}^{t_0}(e, \varnothing, \varnothing, \mathtt{halt})$.

The possible traces of execution tell us the most about a program, so we take $\mathit{traces}(e)$ to be the (regular) set of paths through the computed graph. I will elide the construction of the set of edges.

In the next subsection, we fix this with store widening to reach polynomial (albeit of high degree) complexity. This widening effectively lifts the store out of individual states to create a single, global shared store for all.

### 3.3.2 *Store widening*

A common technique to accelerate convergence in flow analyses is to share a common, global store. Formally, we can cast this optimization as a second abstraction or as the application of a widening operator [10] during the fixed-point iteration. The precision is greatly reduced post-widening, but a widening is necessary in order to escape the exponential state space.

[10] *Technically, we would have to copy the value of the global store to all states being stepped to fit the formal definition of a widening, but this representation is order-isomorphic to that.*

Since we can cast this optimization as a widening, there is no need to change the transition relation itself. Rather, what changes is the structure of the fixed-point iteration. In each pass, the algorithm will collect all newly produced stores and join them together. Then, before each transition, it installs this joined store into the current state.

To describe this process, AAM defined a transformation of the reduction relation so that it operates on a pair of a set of contexts (C) and a store (σ). A context includes all non-store components, *e.g.*, the expression, the environment and the stack. The transformed relation, $\widehat{\longmapsto}$, is

$$\begin{aligned}
(C, \sigma) &\mathrel{\widehat{\longmapsto}} (C', \sigma'), \\
\text{where } C' &= \{c' \ :\ \exists c \in C, c', \sigma^c . \mathit{wn}(c, \sigma) \longmapsto \mathit{wn}(c', \sigma^c)\} \\
\sigma' &= \bigsqcup \{\sigma^c \ :\ \exists c \in C, c' . \mathit{wn}(c, \sigma) \longmapsto \mathit{wn}(c', \sigma^c)\} \\
\mathit{wn} &: \mathit{Context} \times \mathit{Store} \to \mathit{State} \\
\mathit{wn}(\mathsf{ev}\ (e, \rho, \kappa), \sigma) &= \mathsf{ev}\ (e, \rho, \sigma, \kappa) \\
\mathit{wn}(\mathsf{co}\ (v, \kappa), \sigma) &= \mathsf{co}\ (v, \kappa, \sigma) \\
\mathit{wn}(\mathsf{ap}\ (u, v, \kappa), \sigma) &= \mathsf{ap}\ (u, v, \sigma, \kappa)
\end{aligned}$$

To retain soundness, this store grows monotonically as the least upper bound of all occurring stores.

### 3.3.3 *Store-allocate all values*

The final approximation we make to get to our baseline is to store-allocate all values that appear, so that any non-machine state that contains a value instead contains an address to a value. The AAM approach stops at the previous optimization. However, the `fun` continuation stores a value, and this makes the space of continuations quadratic rather than linear in the size of the program, for a monovariant analysis like 0CFA. Having the space of continuations grow linearly with the size of the program will drop the overall complexity to cubic (as expected). We also need to allocate an address for the argument position in an `ap` state.

To achieve this linearity for continuations, we allocate an address for the value position when we create the continuation. This address and the tail address are both determined by the label of the application point, so the space becomes linear and the overall complexity drops by a factor of $n$. This is a critical abstraction in languages with $n$-ary functions, since otherwise the continuation space grows super-exponentially ($\mathcal{O}(n^n)$). We extend the semantics to additionally allocate an address for the function value when creating the `fun` continuation. The continuation has to contain this address to remember where to retrieve values from in the store.

The new evaluation rules follow, where $t' = \mathit{tick}(\varsigma)$:

$$\begin{aligned}\mathsf{co}^t(\mathbf{arg}(e,\rho,a_\kappa),v,\sigma) &\longmapsto \mathsf{ev}^{t'}(e,\rho,\sigma',\mathbf{fun}(a,a_\kappa))\\ \text{where } a &= \mathit{alloc}(\varsigma)\\ \sigma' &= \sigma \sqcup [a \mapsto \{v\}]\end{aligned}$$

Now instead of storing the evaluated function in the continuation frame itself, we indirect it through the store for further control on complexity and precision:

$$\begin{aligned}\mathsf{co}^t(\mathbf{fun}(a,a_\kappa),v,\sigma) &\longmapsto \mathsf{ap}_\ell^{t'}(u,a,\kappa,\sigma')\\ &\quad \text{if } \kappa \in \sigma(a_\kappa), u \in \sigma(a)\\ \text{where } a &= \mathit{alloc}(\varsigma)\\ \sigma' &= \sigma \sqcup [a \mapsto \{v\}]\end{aligned}$$

Associated with this indirection, we now apply all functions stored in the address. This nondeterminism is necessary in order to continue with evaluation.

## 3.4 IMPLEMENTATION TECHNIQUES

In this section, we discuss the optimizations for abstract interpreters that yield our ultimate performance gains. We have two broad categories of these optimizations: (1) pragmatic improvement, (2) transition elimination. The pragmatic improvements reduce overhead and trade space for time by utilizing:

1. timestamped stores;
2. store deltas; and
3. imperative, pre-allocated data structures.

The transition-elimination optimizations reduce the overall number of transitions made by the analyzer by performing:

4. frontier-based semantics;
5. lazy nondeterminism; and
6. abstract compilation.

All pragmatic improvements are precision preserving (form complete abstractions), but the "optimizations" are not in some cases[11], for reasons we will describe. We did not observe the precision differences in our evaluation.

[11] *I use scare-quotes since the term "optimization" is usually reserved to semantics-preserving transformations. Sometimes we approximate.*

We apply the frontier-based semantics combined with timestamped stores as our first step. The move to the imperative will be made last in order to show the effectiveness of these techniques in the purely functional realm.

### 3.4.1 *Timestamped frontier*

The semantics given for store widening in section 3.3.2, while simple, is wasteful. It also does not model what typical implementations do. It causes all states found so far to step each iteration, even if they are not revisited. This has negative performance *and* precision consequences (changes to the store can travel back in time in straight-line code). We instead use a frontier-based semantics that corresponds to the classic worklist algorithms for analysis. The difference is that the store is not modified in-place, but updated after all frontier states have been processed. This has implications for the analysis' precision and determinism. Specifically, higher precision, and it is deterministic even if set iteration is not.

The state space changes from a store and set of contexts to a set of seen abstract states (context plus store), S, a set of contexts to step (the frontier), F, and a store to step those contexts with, σ:

$$(S, F, \sigma) \widehat{\longmapsto} (S \cup S', F', \sigma')$$

We constantly see more states, so S is always growing. The frontier, which is what remains to be done, changes. Let's start with the result of stepping all the contexts in F paired with the current store (call it I for intermediate):

$$I = \{(c', \sigma') \mid wn(c, \sigma) \longmapsto wn(c', \sigma'), c \in F\}$$

The next store is the least upper bound of all the stores in I:

$$\sigma' = \bigsqcup \{\sigma \mid (_, \sigma) \in I\}$$

The next frontier is exactly the states that we found from stepping the last frontier, but have not seen before. They must be *states*, so we pair the contexts with the next store:

$$F' = \{c \mid (c, _) \in I, (c, \sigma') \notin S\}$$

Finally, we add what we know we had not yet seen to the seen set:

$$S' = \{(c, \sigma') \mid c \in F'\}$$

To inject a program $e$ into this machine, we start off knowing we have seen the first state, and that we need to process the first state:

$$\mathit{inject}(e) = (\{(c_0, \bot)\}, \{c_0\}, \bot)$$
$$\text{where } c_0 = \mathtt{ev}\ (e, \bot, \mathtt{halt})$$

Notice that now S has several copies of the abstract store in it. As it is, this semantics is much less efficient (but still more precise) than the previously proposed semantics because membership checks have to compare entire stores. Checking equality is expensive because the stores within each state are large, and nearly every entry must be checked against every other due to high similarities amongst stores.

And, there is a better way. Shivers' original work on kCFA was susceptible to the same problem, and he suggested three complementary optimizations: (1) make the store global; (2) update the store imperatively; and (3) associate every change in the store with a version number – its timestamp. Then, put timestamps in states where previously there were stores. Given two states, the analysis can now compare their stores just by comparing their timestamps – a constant-time operation.

There are two subtle losses of precision in Shivers' original timestamp technique that we can fix.

1. In our semantics, the store does not change until the entire frontier has been explored. This avoids cross-branch pollution which would otherwise happen in Shivers' semantics, e.g., when one branch writes to address $a$ and another branch reads from address $a$.

2. The common implementation strategy for timestamps destructively updates each state's timestamp. This loses *temporal* information about the contexts a state is visited in, and in what order. Our semantics has a drop-in replacement of timestamps for stores in the seen set ($\hat{S}$), so we do not experience precision loss.

$$\Sigma \in \mathit{Store}^* \qquad \hat{S} \subseteq \mathbb{N} \times \mathit{Context} \qquad F \subseteq \mathit{Context}$$

$$\begin{aligned}
(\hat{S}, F, \sigma, \Sigma, t) &\mathrel{\widehat{\longmapsto}}^{T} (\hat{S} \cup \hat{S}', F', \sigma', \Sigma', t') \\
\text{where } I &= \{(c', \sigma^c) \mid \mathit{wn}(c, \sigma) \longmapsto \mathit{wn}(c', \sigma^c), c \in F\} \\
\sigma' &= \bigsqcup \{\sigma^c \mid (_, \sigma^c) \in I\} \\
(t', \Sigma') &= \begin{cases} (t+1, \sigma'\Sigma') & \text{if } \sigma' \neq \sigma \\ (t, \Sigma) & \text{otherwise} \end{cases} \\
F' &= \{c \mid (c, _) \in I, (c, t') \notin \hat{S}\} \\
\hat{S}' &= \{(c, t') \mid c \in F'\} \\
\mathit{inject}(e) &= (\{(c_0, 0)\}, \{c_0\}, \bot, \bot{:}\epsilon, 0) \\
\text{where } c_0 &= \mathtt{ev}\ (e, \bot, \mathtt{halt})
\end{aligned}$$

The observation Shivers made was that the store is increasing monotonically, so all stores throughout execution will be totally ordered (form a chain). This observation allows you to replace stores with pointers into this chain. We keep the stores around in $\Sigma$ to achieve a complete abstraction. This corresponds to the temporal information about the execution's effect on the store.

Note also that $F$ is only populated with states that have not been seen at the resulting store. This is what produces the more precise abstraction than the baseline widening.

The general fixed-point combinator we showed above can be specialized to this semantics, as well. In fact, $\widehat{\longmapsto}^{T}$ is a functional relation, so we can get the least fixed-point of it directly.

**Lemma 1.** *$\widehat{\longmapsto}$ maintains the invariant that all stores in $S$ are totally ordered and $\sigma$ is an upper bound of the stores in $S$.*

**Lemma 2.** *$\widehat{\longmapsto}^{T}$ maintains the invariant that $\Sigma$ is in order with respect to $\sqsupset$ and $\sigma = hd(\Sigma)$.*

**Theorem 3.** *$\widehat{\longmapsto}^{T}$ is a complete abstraction of $\widehat{\longmapsto}$.*

The proof follows from the order isomorphism that, in one direction, sorts all the stores in $S$ to form $\Sigma$, and translates stores in $S$ to their distance from the end of $\Sigma$ (their timestamp). In the other direction, timestamps in $\hat{S}$ are replaced by the stores they point to in $\Sigma$.

### 3.4.2 *Locally log-based store deltas*

The above technique requires joining entire (large) stores together. Additionally, there is still a comparison of stores, which we established is expensive. Not every step will modify all addresses of the store, so joining entire stores is wasteful in terms of memory and time. We can instead log store changes and replay the change log on the full store after all steps have completed, noting when there is an actual change. This uses far fewer join and comparison operations, leading to less overhead, and is precision-preserving.

We represent change logs as $\xi \in \mathit{Store}' = (\mathit{Addr} \times \wp(\mathit{Storable}))^*$. Each $\sigma \sqcup [a \mapsto vs]$ becomes a log addition $(a, vs){:}\xi$, where $\xi$ begins empty ($\epsilon$) for each step. Applying the changes to the full store is straightforward:

$$\mathit{replay} : (\mathit{Store}' \times \mathit{Store}) \to (\mathit{Store} \times \mathit{Boolean})$$

$$\begin{aligned} \mathit{replay}([(a_i, vs_i), \ldots], \sigma) &= (\sigma', \delta?(vs_i, \sigma(a_i)) \vee \ldots) \\ \text{where } \sigma' &= \sigma \sqcup [a_i \mapsto vs_i] \sqcup \ldots \\ \delta?(vs, vs') &= vs' \stackrel{?}{=} vs \sqcup vs' \end{aligned}$$

We change the semantics slightly to add to the change log rather than produce an entire modified store. The transition relation is identical except for the addition of this change log. We maintain the invariant that lookups will never rely on the change log, so we can use the originally supplied store unmodified.

A taste of the changes to the reduction relation is as follows:

$$\longmapsto_{\sigma\xi} \subseteq (\mathit{Context} \times \mathit{Store}) \times (\mathit{Context} \times \mathit{Store}')$$

$$\begin{aligned}(\mathtt{ap}^t_\ell(\mathtt{clos}\ (x, e, \rho), a, \kappa), \sigma) &\longmapsto_{\sigma\xi} (\mathtt{ev}^{t'}(e, \rho', \kappa), (a', \sigma(a)){:}\epsilon) \\ \text{where } a' &= \mathit{alloc}(\varsigma) \\ \rho' &= \rho[x \mapsto a']\end{aligned}$$

We lift $\longmapsto_{\sigma\xi}$ to accommodate for the asymmetry in the input and output, and change the frontier-based semantics in the following way:

$$\begin{aligned}(\hat{S}, F, \sigma, \Sigma, t) &\widehat{\longmapsto}_{\sigma\xi} (\hat{S} \cup \hat{S}', F', \sigma', \Sigma', t') \\ \text{where } I &= \{(c', \xi) \mid (c, \sigma) \longmapsto_{\sigma\xi} (c', \xi)\} \\ (\sigma', \Delta?) &= \mathit{replay}(\mathit{appendall}(\{\xi \mid (_, \xi) \in I\}), \sigma) \\ (t', \Sigma') &= \begin{cases} (t+1, \sigma\Sigma) & \text{if } \Delta? \\ (t, \Sigma) & \text{otherwise} \end{cases} \\ F' &= \{c \mid (c, _) \in I, (c, t') \notin \hat{S}\} \\ \hat{S}' &= \{(c, t') \mid c \in F'\} \\ \mathit{appendall}(\varnothing) &= \epsilon \\ \mathit{appendall}(\{\xi\} \cup \Xi) &= \xi {+\!+} \mathit{appendall}(\Xi)\end{aligned}$$

Here *appendall* combines change logs across all non-deterministic steps for a state to later be replayed. The order the combination happens in doesn't matter, because join is associative and commutative.

**Lemma 4.** $(c, \sigma) \longmapsto_{\sigma\xi} (c', \xi)$ *iff* $\mathit{wn}(c, \sigma) \longmapsto \mathit{wn}(c', \mathit{replay}(\xi, \sigma))$

By cases on $\longmapsto_{\sigma\xi}$ and $\longmapsto$.

**Lemma 5** ($\Delta?$ means change)**.** *Let* $\mathit{replay}(\xi, \sigma) = (\sigma', \Delta?)$*.* $\sigma' \neq \sigma$ *iff* $\Delta?$.

By induction on $\xi$.

**Theorem 6.** $\widehat{\longmapsto}_{\sigma\xi}$ *is a complete abstraction of* $\widehat{\longmapsto}^T$.

Follows from previous lemma and that join is associative and commutative.

### 3.4.3 *Lazy nondeterminism*

Tracing the execution of the analysis reveals an immediate shortcoming: there is a high degree of branching and merging in the exploration. Surveying this branching has no benefit for precision. For example, in a function application, `(f x y)`, where `f`, `x` and `y` each have several values each argument evaluation induces $n$-way branching, only to be ultimately joined back together in their respective application positions. Transition patterns of this shape litter the state-graph:

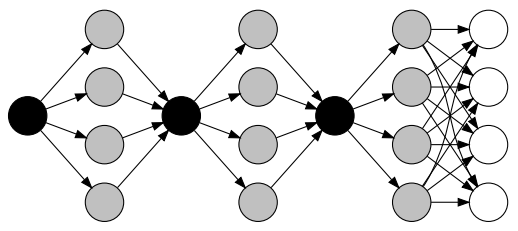

To avoid the spurious forking and joining, we *delay* the nondeterminism until and unless it is needed in *strict contexts* (such as the guard of an `if`, a called procedure, or a numerical primitive application). Doing so collapses these forks and joins into a linear sequence of states:

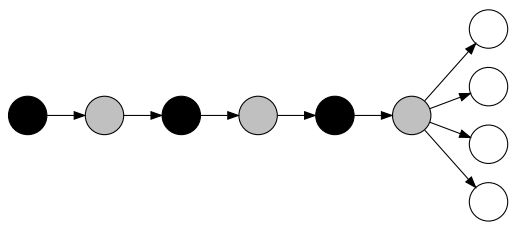

This shift does not change the concrete semantics of the language to be lazy. Rather, it abstracts over transitions that the original nondeterministic semantics steps through. We say the abstraction is *lazy* because it delays splitting on the values in an address until they are *needed* in the semantics. It does not change the execution order that leads to the values that are stored in the address.

We introduce a new kind of value, $\mathtt{addr}\ (a)$, that represents a delayed non-deterministic choice of a value from $\sigma(a)$. The following rules highlight the changes to the semantics:

$$
\begin{aligned}
&\mathit{force} : \mathit{Store} \times \mathit{Value} \to \wp(\mathit{Value}) \\
\mathit{force}(\sigma, \mathtt{addr}\ (a)) &= \sigma(a) \\
\mathit{force}(\sigma, v) &= \{v\} \\
\mathtt{ev}\ (x\ , \rho, \kappa, \sigma) &\longmapsto_{\mathcal{L}} \mathtt{co}\ (\kappa, \mathtt{addr}\ (\rho(x)), \sigma) \\
\mathtt{co}\ (\mathbf{arg}^t_\ell(e, \rho, a_\kappa), v, \sigma) &\longmapsto_{\mathcal{L}} \mathtt{ev}^{t'}(e, \rho, \sigma', \mathbf{fun}^t_\ell(a_f, a_\kappa)) \\
\text{where } a_f &= \mathit{alloc}(\varsigma) \\
\sigma' &= \sigma \sqcup [a \mapsto \mathit{force}(\sigma, v)] \\
\mathtt{co}\ (\mathtt{ifk}^t(e_0, e_1, \rho, a_\kappa), v, \sigma) &\longmapsto_{\mathcal{L}} \mathtt{ev}^{t'}(e_0, \rho, \sigma, \kappa) \\
&\text{if } \kappa \in \sigma(a_\kappa), \mathtt{tt} \in \mathit{force}(\sigma, v)
\end{aligned}
$$

Since `if` guards are in strict position, we must force the value to determine which branch to to take. The middle rule uses *force* only to combine with values in the store - it does not introduce needless nondeterminism.

We have two choices for how to implement lazy nondeterminism.

**option 1: lose precision; simplify implementation** This semantics introduces a subtle precision difference over the baseline. Consider a configuration where a reference to a variable and a binding of a variable will happen in one step, since store widening leads to stepping several states in one big "step." With laziness, the reference will mean the original binding(s) of the variable *or* the new binding, because the actual store lookup is delayed one step (i.e. laziness is administrative).

**option 2: regain precision; complicate implementation** The administrative nature of laziness means that we could remove the loss in precision by storing the result of the lookup in a value representing a delayed nondeterministic choice. This is a more common choice in 0CFA implementations we have seen, but it interferes with the next optimization due to the invariant from store deltas we have that lookups must not depend on the change log.

**Theorem 7** (Soundness)**.** *If* $\varsigma \longmapsto \varsigma'$ *and* $\varsigma \sqsubseteq \hat{\varsigma}$ *then there exists a* $\hat{\varsigma}'$ *such that* $\hat{\varsigma} \longmapsto_{\mathcal{L}} \hat{\varsigma}'$ *and* $\varsigma' \sqsubseteq \hat{\varsigma}'$

Here $\sqsubseteq$ is straightforward — the left-hand side store must be contained in the right-hand-side store, and if values occur in the states, the left-hand-side value must be in the forced corresponding right-hand-side value. The proof is by cases on $\varsigma \longmapsto \varsigma'$.

### 3.4.4 *Abstract compilation*

The prior optimization saved time by doing the same amount of reasoning as before but in fewer transitions. We can exploit the same idea—same reasoning, fewer transitions—with abstract compilation. Abstract compilation transforms complex expressions whose *abstract* evaluation is deterministic into "abstract bytecodes." The abstract interpreter then does in one transition what previously took many. Refer back to figure 3 to see the effect of abstract compilation. In short, abstract compilation eliminates unnecessary allocation, deallocation and branching. The technique is precision preserving without store widening. We discuss the precision differences with store widening at the end of the section.

The compilation step converts expressions into functions that expect the other components of the ev state. Its definition in figure 7 shows close similarity to the rules for interpreting ev states. The next step is to change reduction rules that create ev states to instead call these functions. Figure 8 shows the modified reduction relation. The only change from the previous semantics is that ev state construction is replaced by calling the compiled expression. For notational coherence, we write $\lambda^t(\mathit{args}\ldots)$ for $\lambda(\mathit{args}\ldots, t)$ and $k^t(\mathit{args}\ldots)$ for $k(\mathit{args}\ldots, t)$.

$$
\begin{aligned}
[\![_]\!] &: \mathit{Expr} \to \mathit{Store} \to \mathit{Env} \times \mathit{Store}' \times \mathit{Kont} \times \mathit{Time} \\
&\quad \to \mathit{State} \\
t' &= \mathit{tick}(\ell, \rho, \sigma, t) \\
[\![x\ ]\!]_\sigma &= \lambda^t(\rho, \xi, \kappa).\mathtt{co}\ (\kappa, \mathtt{addr}\ (\rho(x))), \xi \\
[\![\mathtt{lit}\ (l)]\!]_\sigma &= \lambda^t(\rho, \xi, \kappa).\mathtt{co}\ (\kappa, l), \xi \\
[\![\lambda\ \ x.\ e]\!]_\sigma &= \lambda^t(\rho, \xi, \kappa).\mathtt{co}\ (\kappa, \mathtt{clos}\ (x, [\![e]\!], \rho)), \xi \\
[\![(e_0\ e_1)^\ell]\!]_\sigma &= \lambda^t(\rho, \xi, \kappa).[\![e_0]\!]_\sigma^{t'}(\rho, \xi', \mathbf{arg}^t_\ell([\![e_1]\!], \rho, a_\kappa)) \\
&\text{where} \quad a_\kappa = \mathit{allockont}^t_\ell(\sigma, \kappa) \\
&\qquad\qquad \xi' = (a_\kappa, \{\kappa\}){:}\xi \\
[\![\mathtt{if}^\ell(e_0, e_1, e_2)]\!]_\sigma &= \lambda^t(\rho, \xi, \kappa).[\![e_0]\!]_\sigma^{t'}(\rho, \xi', \mathtt{ifk}^t([\![e_1]\!], [\![e_2]\!], \rho, a_\kappa)) \\
&\text{where } a_\kappa = \mathit{allockont}^t_\ell(\sigma, \kappa) \\
&\qquad \xi' = (a_\kappa, \{\kappa\}){:}\xi
\end{aligned}
$$

Figure 7: Abstract compilation

$$
\begin{aligned}
\mathit{traces}(e) &= \{\mathit{inject}([\![e]\!]_\bot^{t_0}(\bot, \epsilon, \mathtt{halt})) \longmapsto\!\!\!\!\!\rightarrow \varsigma\} \text{ where} \\
\mathit{inject}(c, \xi) &= \mathit{wn}(c, \mathit{replay}(\xi, \bot)) \\
\mathit{wn}(c, \sigma) \longmapsto \mathit{wn}(c', \sigma') &\iff c\ [\![\longmapsto]\!]_\sigma\ c', \xi \\
&\xi \text{ is such that } \mathit{replay}(\xi, \sigma) = \sigma'
\end{aligned}
$$

$$
\begin{aligned}
\mathtt{co}\ (\mathbf{arg}^t_\ell(k, \rho, a_\kappa), v)\ [\![\longmapsto]\!]_\sigma\ & k^t(\sigma)(\rho, \xi, \mathbf{fun}^t_\ell(a_f, a_\kappa)) \\
&\text{where } a_f = \mathit{alloc}(\varsigma) \\
&\qquad \xi = (a_f, \mathit{force}(\sigma, v)){:}\epsilon \\
\mathtt{co}\ (\mathbf{fun}^t_\ell(a_f, a_\kappa), v)\ [\![\longmapsto]\!]_\sigma\ & \mathtt{ap}^t_\ell(u, a, \kappa), (a, \mathit{force}(\sigma, v)){:}\epsilon \\
&\text{if } u \in \sigma(a_f), \kappa \in \sigma(a_\kappa) \\
\mathtt{co}\ (\mathtt{ifk}^t(k_0, k_1, \rho, a_\kappa), \mathtt{tt})\ [\![\longmapsto]\!]_\sigma\ & k^t_0(\sigma)(\rho, \epsilon, \kappa) \text{ if } \kappa \in \sigma(a_\kappa) \\
\mathtt{co}\ (\mathtt{ifk}^t(k_0, k_1, \rho, a_\kappa), \mathtt{ff})\ [\![\longmapsto]\!]_\sigma\ & k^t_1(\sigma)(\rho, \epsilon, \kappa) \text{ if } \kappa \in \sigma(a_\kappa) \\
\mathtt{ap}^t_\ell(\mathtt{clos}\ (x, k, \rho), a, \kappa)\ [\![\longmapsto]\!]_\sigma\ & k^{t'}(\sigma)(\rho', \xi, \kappa) \\
&\text{where } \rho' = \rho[x \mapsto a] \\
&\qquad \xi = (a, \sigma(a)){:}\epsilon \\
\mathtt{ap}\ (o, a, \kappa)\ [\![\longmapsto]\!]_\sigma\ & \mathtt{co}\ (\kappa, u), \epsilon \\
&\text{where } v \in \sigma(a), u \in \Delta(o, v)
\end{aligned}
$$

Figure 8: Abstract abstract machine for compiled λ**IF**

CORRECTNESS The correctness of abstract compilation seems obvious, but it has never before been rigorously proved. What constitutes correctness in the case of dropped states, anyway? Applying an abstract bytecode's function does many "steps" in one go, at the end of which, the two semantics line up again (modulo representation of expressions). This constitutes the use of a notion of stuttering. We provide a formal analysis of abstract compilation *without* store widening with a proof of a stuttering bisimulation [13] between this semantics and lazy nondeterminism without widening to show precision preservation.

The number of transitions that can occur in succession from an abstract bytecode is roughly bounded by the amount of expression nesting in the program. We can use the expression containment order to prove stuttering bisimulation with a well-founded equivalence bisimulation (WEB) [59]. WEBs are equivalent to the notion of a stuttering bisimulation, but are more amenable to mechanization since they also only require reasoning over one step of the reduction relation. The trick is in defining a well-founded ordering that determines when the two semantics will match up again, what Manolios calls the pair of functions *erankt* and *erankl* (but we don't need *erankl* since the uncompiled semantics doesn't stutter).

We define a refinement, $r$, from non-compiled to compiled states (built structurally) by "committing" all the actions of an $\mathsf{ev}$ state (defined similarly to $[\![_]\!]$, but immediately applies the functions), and subsequently changing all expressions with their compiled variants. Since WEBs are for single transition systems, a WEB refinement is over the disjoint union of our two semantics, and the equivalence relation we use is just that a state is related to its refined state (and itself). Call this relation $B$.

Before we prove this setup is indeed a WEB, we need one lemma that applying an abstract bytecode's function is equal to refining the corresponding $\mathsf{ev}$ state:

**Lemma 8** (Compile/commit). *Let* $c, \xi' = [\![e]\!]^t_{r(\sigma)}(\rho, \xi, r(\kappa))$. *Let* $wn(c', \sigma') = r(\mathsf{ev}^t(e, \rho, \sigma, \kappa))$. $wn(c, replay(\xi', \sigma)) = wn(c', replay(\xi, \sigma'))$.

The proof is by induction on $e$.

**Theorem 9** (Precision preservation). $B$ *is a WEB on* $\longmapsto_{\mathcal{L}} \uplus [\![\longmapsto]\!]$

The proof follows by cases on $\longmapsto_{\mathcal{L}} \uplus [\![\longmapsto]\!]$ with the WEB *witness* being the well-order on expressions (with a $\bot$ element), and the following *erankt*, *erankl* functions:

$$\begin{aligned} erankt(\mathsf{ev}^t(e, \rho, \sigma, \kappa)) &= e \\ erankt(\varsigma) &= \bot \quad \text{otherwise} \\ erankl(s, s') &= 0 \end{aligned}$$

All cases are either simple steps or appeals to the well-order on *erankt*'s range. The other rank function, *erankl* is unnecessary, so we just make it the constant 0 function. The $[\![\longmapsto]\!]$ cases are trivial.

WIDE STORE AND ABSTRACT COMPILATION It is possible for different stores to occur between the different semantics because abstract compilation can change the order in which the store is changed (across steps). This is the case because some "corridor" expressions may compile down to change the store before some others, meaning there is no stuttering relationship with the wide lazy semantics. Although there is a difference pre- and post- abstract compilation, the result is still deterministic in contrast to Shivers' technique. The soundness is intact since we can add store-widening to the correct unwidened semantics with an easy correctness proof. Call $[\![\widehat{\longmapsto}]\!]$ the result of the widening operator from the previous section on $[\![\longmapsto]\!]$.

### 3.4.5 *Imperative, pre-allocated data structures*

Thus far, we have made our optimizations in a purely functional manner. For the final push for performance, we need to dip into the imperative. In this section, we show an alternative representation of the store and seen set that are more space-efficient and are amenable to destructive updates by adhering to a history for each address.

The following transfer function has several components that can be destructively updated, and intermediate sets can be elided by adding to global sets. In fact, the log of store deltas can be removed as well, by updating the store in-place, and on lookup, using the first value whose timestamp is less than or equal to the current timestamp. We start with the purely functional view.

*Pure setup for imperative implementation*

The store maps to a stack of timestamped sets of abstract values. Throughout this section, we will be taking the parameter $n$ to be the "current time," or the length of the store chain at the beginning of the step.

$$\sigma \in \mathit{Store} = \mathit{Addr} \to \mathit{ValStack}$$
$$V \in \mathit{ValStack} = (\mathbb{N} \times \wp(\mathit{Storable}))^*$$

To allow imperative store updates, we maintain an invariant that we never look up values tagged at a time in the future:

$$\mathcal{L}(V, n) = \begin{cases} vs & \text{if } V = (n', vs){:}V', n' \leqslant n \\ vs' & \text{if } V = (n', vs){:}(n'', vs'){:}V', n' > n \end{cases}$$

To construct this value stack, we have a time-parameterized join operation that also tracks changes to the store. If joining with a time in

$$
\begin{aligned}
[\![_]\!] &: \mathit{Expr} \to \mathbb{N} \to \mathit{Env} \times \mathit{Store} \times \mathit{Kont} \times \mathit{Time} \times \mathit{Boolean} \\
&\quad \to (\wp(\mathit{State}) \times \mathit{Store} \times \mathit{Boolean}) \\
t' &= \mathit{tick}(\ell, \rho, \sigma, t) \\
[\![x\ ]\!]_n &= \lambda^t(\rho, \sigma, \kappa, \Delta?).\{\mathtt{co}\ (\kappa, \mathtt{addr}\ (\rho(x)))\}, \sigma, \Delta? \\
[\![\mathtt{lit}\ (l)]\!]_n &= \lambda^t(\rho, \sigma, \kappa, \Delta?).\{\mathtt{co}\ (\kappa, l)\}, \sigma, \Delta? \\
[\![\lambda\ \ x.\ e]\!]_n &= \lambda^t(\rho, \sigma, \kappa, \Delta?).\{\mathtt{co}\ (\kappa, \mathtt{clos}\ (x, [\![e]\!], \rho))\}, \sigma, \Delta? \\
[\![(e_0\ e_1)^\ell]\!]_n &= \lambda^t(\rho, \sigma, \kappa, \Delta?).[\![e_0]\!]_n^{t'}(\rho, \sigma', \mathbf{arg}_\ell^t([\![e_1]\!], \rho, a_\kappa), \Delta? \vee \Delta?') \\
&\quad \text{where} \qquad a_\kappa = \mathit{allockont}_\ell^t(\sigma, \kappa) \\
&\quad\quad \sigma', \Delta?' = \sigma \sqcup_n [a_\kappa \mapsto \{\kappa\}] \\
[\![\mathtt{if}^\ell(e_0, e_1, e_2)]\!]_\sigma &= \lambda^t(\rho, \xi, \kappa).[\![e_0]\!]_n^{t'}(\rho, \sigma', \mathtt{ifk}^t([\![e_1]\!], [\![e_2]\!], \rho, a_\kappa), \Delta? \vee \Delta?') \\
&\quad \text{where } a_\kappa = \mathit{allockont}_\ell^t(\sigma, \kappa) \\
&\quad\quad \sigma', \Delta?' = \sigma \sqcup_n [a_\kappa \mapsto \{\kappa\}]
\end{aligned}
$$

Figure 9: Abstract compilation for single-threading

the future, we just add to it. Otherwise, we're making a change for the future $(t+1)$, but only if there is an actual change.

$$
\begin{aligned}
\sigma \sqcup_n [a \mapsto vs] &= \sigma[a \mapsto V], \Delta? \\
\text{where } (V, \Delta?) &= \sigma(a) \sqcup_n vs \\
\epsilon \sqcup_n vs &= (n, vs), \mathtt{tt} \\
(n', vs){:}V \sqcup_n vs' &= (n', vs \sqcup vs'){:}V, \delta?(vs, vs') \text{ if } n' > n \\
V \sqcup_n vs &= (n+1, vs^*){:}V, \mathtt{tt} \text{ if } vs_n \neq vs^* \\
\text{where } vs_n &= \mathcal{L}(\sigma(a), n) \\
vs^* &= vs \sqcup vs_n \\
V \sqcup_n vs &= V, \mathtt{ff} \text{ otherwise}
\end{aligned}
$$

The abstract step is better suited for a function interpretation now since there can be multiple output states, but only one store and $\Delta?$.

For the purposes of space, we reuse the $[\![\longmapsto]\!]$ semantics, although the *replay* of the produced $\xi$ objects should be in-place, and the $\mathcal{L}$ function should be using this single-threaded store. Because the store has all the temporal information baked into it, we rephrase the core semantics in terms of a transfer function. The least fixed-point of this function gives a more compact representation of the reduction relation of the previous section.

$$
\begin{aligned}
\mathit{force} &: \mathit{Store} \times \mathit{Value} \times \mathbb{N} \to \wp(\mathit{Value}) \\
\mathit{force}(\sigma, \mathtt{addr}\ (a), n) &= \mathcal{L}(\sigma(a), n) \\
\mathit{force}(\sigma, v, n) &= \{v\}
\end{aligned}
$$

$$
\begin{aligned}
[\![\longmapsto!]\!]^{n}_{\sigma,\Delta?}(\mathtt{co}\ (\mathbf{arg}^{t}_{\ell}(k, \rho, a_\kappa), v)) &= k^{t}(n)(\rho, \sigma', \mathbf{fun}^{t}_{\ell}(a_f, a_\kappa), \Delta? \vee \Delta?') \\
\text{where } a_f &= \mathit{alloc}(\varsigma) \\
\sigma', \Delta?' &= \sigma \sqcup_n [a_f \mapsto \mathit{force}(\sigma, v, n)] \\
[\![\longmapsto!]\!]^{n}_{\sigma,\Delta?}(\mathtt{co}\ (\mathbf{fun}^{t}_{\ell}(a_f, a_\kappa), v)) &= \mathtt{ap}^{t}_{\ell}(u, a, \kappa), \sigma', \Delta? \vee \Delta?' \\
\text{if } u &\in \mathcal{L}(\sigma(a_f), n), \kappa \in \mathcal{L}(\sigma(a_\kappa), n) \\
\text{where } \sigma', \Delta?' &= \sigma \sqcup_n [a \mapsto \mathit{force}(\sigma, v, n)]
\end{aligned}
$$

$$
\begin{aligned}
[\![\longmapsto!]\!]^{n}_{\sigma,\Delta?}(\mathtt{co}\ (\mathtt{ifk}^{t}(k_0, k_1, \rho, a_\kappa), \mathtt{tt})) &= k^{t}_{0}(n)(\rho, \sigma, \kappa, \Delta?) \text{ if } \kappa \in \mathcal{L}(\sigma(a_\kappa), n) \\
[\![\longmapsto!]\!]^{n}_{\sigma,\Delta?}(\mathtt{co}\ (\mathtt{ifk}^{t}(k_0, k_1, \rho, a_\kappa), \mathtt{ff})) &= k^{t}_{1}(n)(\rho, \sigma, \kappa, \Delta?) \text{ if } \kappa \in \mathcal{L}(\sigma(a_\kappa), n)
\end{aligned}
$$

$$
\begin{aligned}
[\![\longmapsto!]\!]^{n}_{\sigma,\Delta?}(\mathtt{ap}^{t}_{\ell}(\mathtt{clos}\ (x, k, \rho), a, \kappa)) &= k^{t'}(n)(\rho', \sigma', \kappa, \Delta? \vee \Delta?') \\
\text{where } \rho' &= \rho[x \mapsto a] \\
\sigma', \Delta?' &= \sigma \sqcup_n [a \mapsto \mathcal{L}(\sigma(a), n)] \\
[\![\longmapsto!]\!]^{n}_{\sigma,\Delta?}(\mathtt{ap}\ (o, a, \kappa)) &= S, \sigma, \Delta? \\
\text{where } S &= \{\mathtt{co}\ (\kappa, u) \ : \ v \in \mathcal{L}(\sigma(a), n), u \in \Delta(o, v)\}
\end{aligned}
$$

Figure 10: Abstract abstract machine for compiled single-threaded λ**IF**

$$
\begin{aligned}
\textit{System} &= (\widehat{\textit{State}} \to \mathbb{N}^*) \times \wp(\widehat{\textit{State}}) \times \textit{Store} \times \mathbb{N} \\
\mathcal{F} &: \textit{System} \to \textit{System} \\
\mathcal{F}(\hat{S}, F, \sigma, t) &= (\hat{S}', F', \sigma', t') \\
\text{where } I &= \{(c', \xi) \mid c \in F, c \; [\![\longmapsto]\!]_{\sigma^*} \; c', \xi\} \\
\sigma^* &= \lambda a.\mathcal{L}(\sigma(a), t) \\
(\sigma', \Delta?) &= \textit{replay}(\textit{appendall}(\{\xi \mid (_, \xi) \in I\}), \sigma) \\
t' &= \begin{cases} t+1 & \text{if } \Delta? \\ t & \text{otherwise} \end{cases} \\
F' &= \{c \mid (c, _) \in I, \Delta? \vee \hat{S}(c) \neq t{:}_\} \\
\hat{S}' &= \lambda c. \begin{cases} t'{:}\hat{S}(c) & \text{if } c \in F' \\ \hat{S}(c) & \text{otherwise} \end{cases}
\end{aligned}
$$

We prove semantic equivalence with the previous semantics with a lock-step bisimulation with the stack of stores abstraction, which follow by equational reasoning from the following lemmas:

**Lemma 10.** *Stores of value stacks completely abstract stacks of stores.*

This depends on some well-formedness conditions about the order of the stacks. The store of value stacks can be translated to a stack of stores by taking successive "snapshots" of the store at different timestamps from the max timestamp it holds down to 0. Vice versa, we replay the changes across adjacent stores in the stack.

We apply a similar construction to the different representation of seen states in order to get the final result:

**Theorem 11.** *$\mathcal{F}$ is a complete abstraction of $[\![\widehat{\longmapsto}]\!]$.*

*Pure to imperative*

The intermediate data structures of the above transfer function can all be streamlined into globals that are destructively updated. In particular, there are 5 globals:

1. $\hat{S}$: the *seen* set, though made a map for faster membership tests and updates.

2. F: the *frontier* set, which must be persisent or copied for the iteration through the set to be correct.

3. $\sigma$: the store, which represents all stores that occur in the machine semantics.

4. $n$: the timestamp, or length of the store chain.

5. $\Delta?$: whether the store changed when stepping states in F.

The reduction relation would then instead of building store deltas, update the global store. We would also not view it as a general relation, but a function that adds all next states to $F$ if they have not already been seen. At the end of iterating through $F$, $\hat{S}$ is updated with the new states at the next timestamp. There is no cross-step store poisoning since the lookup is restricted to the current step's time, which points to the same value throughout the step.

*Pre-allocating the store*

Internally, the algorithm at this stage uses hash tables to model the store to allow arbitrary address representations. But, such a dynamic structure isn't necessary when we know the structure of the store in advance.

In a monovariant allocation strategy, the domain of the store is bounded by the number of expressions in the program. If we label each expression with a unique natural, the analysis can index directly into the store without a hash or a collision. Even for polyvariant analyses, it is possible to compute the maximum number of addresses and similarly pre-allocate either the spine of the store or (if memory is no concern) the entire store.

## 3.5 EVALUATION

[12] *Precision evaluation credit goes to Nicholas Labich*

I implemented, optimized, and evaluated[12] an analysis framework supporting higher-order functions, state, first-class control, compound data, and a large number of primitive kinds of data and operations such as floating point, complex, and exact rational arithmetic. The analysis is evaluated against a suite of Scheme benchmarks drawn from the literature[1]. For each benchmark, I collect analysis times, peak memory usage (as determined by Racket's GC statistics), and the rate of states-per-second explored by the analysis for each of the optimizations discussed in section 3.4, cumulatively applied. The analysis is stopped after consuming 30 minutes of time or 1 gigabyte of space [2]. When presenting *relative* numbers, we use the timeout limits as a lower bound on the actual time required (i.e., one minute versus timeout is at least 30 times faster), thus giving a conservative estimate of improvements.

For those benchmarks that did complete on the baseline, the optimized analyzer outperformed the baseline by a factor of two to three orders of magnitude.

I used the following set of benchmarks:

---

1 Source code of the implementation and benchmark suite available at `https://github.com/dvanhorn/oaam`

2 All benchmarks are calculated as an arithmetic mean of 5 runs on a Linux 3.16 machine with Intel Core i7-3770K CPU / 32GB of memory

Figure 11: Overview performance comparison between baseline and optimized analyzer (entries of t mean timeout, and m mean out of memory). Error is standard deviation rounded to 2 significant figures.

1. **nucleic**: a floating-point intensive application taken from molecular biology that has been used widely in benchmarking functional language implementations [37] and analyses (e.g. [105, 39]). It is a constraint satisfaction algorithm used to determine the three-dimensional structure of nucleic acids.
2. **matrix** tests whether a matrix is maximal among all matrices of the same dimension obtainable by simple reordering of rows and columns and negation of any subset of rows and columns. It is written in continuation-passing style (used in [105, 39]).
3. **nbody**: implementation [106] of the Greengard multipole algorithm for computing gravitational forces on point masses distributed uniformly in a cube (used in [105, 39]).
4. **earley**: Earley's parsing algorithm, applied to a 15-symbol input according to a simple ambiguous grammar. A real program, applied to small data whose exponential behavior leads to a peak heap size of half a gigabyte or more during concrete execution.
5. **maze**: generates a random maze using Scheme's `call/cc` operation and finds a path solving the maze (used in [105, 39]).
6. **church**: tests distributivity of multiplication over addition for Church numerals (introduced by [103]).
7. **lattice**: enumerates the order-preserving maps between two finite lattices (used in [105, 39]).
8. **boyer**: a term-rewriting theorem prover (used in [105, 39]).
9. **mbrotZ**: generates Mandelbrot fractal using complex numbers.
10. **graphs**: counts the number of directed graphs with a distinguished root and $k$ vertices, each having out-degree at most 2. It is written in a continuation-passing style and makes extensive use of higher-order procedures—it creates closures almost as often as it performs non-tail procedure calls (used by [105, 39]).

Figure 11 gives an overview of the benchmark results in terms of absolute time, space, and speed between the baseline and most optimized analyzer. Figure 12 plots the factors of improvement over the baseline for each optimization step. The error bars are the normalized mean errors of the respective benchmark. For example, if comparing baseline over current, say with respective means and standard deviations of $\mu_b$, $\sigma_b$, $\mu_c$, $\sigma_c$, then the error is

$$\mu_{\frac{b}{c}} = \frac{\mu_b}{\mu_c} \cdot \left(\frac{\sigma_b}{\mu_b} + \frac{\sigma_c}{\mu_c}\right).$$

If $b$ times out (respectively runs out of memory), then its fraction in the error is 0, and $\mu_b$ is the timeout length (respectively memory limit).

To determine the impact of each section's technique on precision, we evaluated a singleton variable analysis to find opportunities to inline constants and closed functions. We found no change in the results across all implementations, including Shivers' timestamp approximation – from an empirical point of view, these techniques are precision preserving despite the theoterical loss of precision.

Our step-wise optimizations strictly produce better analysis times with no observed loss of precision. The final result is a systematically derived and verified implementation that operates within a small factor performance loss compared to a hand-optimized, unverified implementation. Moreover, much of the performance gains are achieved with purely functional methods, which allow the use of these methods in rewriting tools and others with restricted input languages. Peak memory usage is considerably improved by the end of the optimization steps.

#### Comparison with other flow analysis implementations

The analysis considered here computes results similar to Earl, et al.'s 0CFA implementation [29], which times out on the Vardoulakis and Shivers benchmark because it does not widen the store as described for our baseline evaluator. So even though it offers a fair point of comparison, a more thorough evaluation is probably uninformative as the other benchmarks are likely to timeout as well (and it would require significant effort to extend their implementation with the features needed to analyze our benchmark suite). That implementation is evaluated against much smaller benchmarks: the largest program is 30 lines.

Vardoulakis and Shivers evaluate their CFA2 analyzer [103] against a variant of 0CFA defined in their framework and the example we draw on is the largest benchmark Vardoulakis and Shivers consider. More work would be required to scale the analyzer to the set of features required by our benchmarks.

The only analyzer we were able to find that proved capable of analyzing the full suite of benchmarks considered here was the Polymorphic splitting system of Wright and Jagannathan [105] [3]. Unfortunately, these analyses compute an inherently different and incom-

3 This is not a coincidence; these papers set a high standard for evaluation, which we consciously aimed to approach.

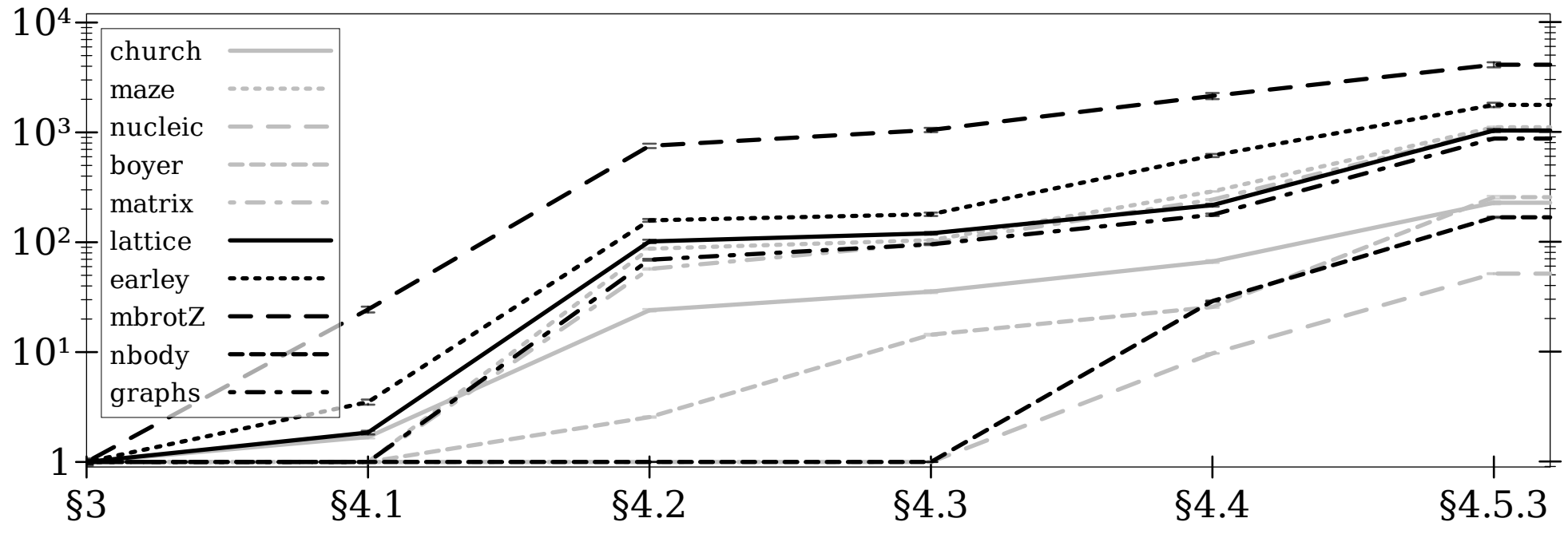

(a) Total analysis time speed-up (baseline / optimized)

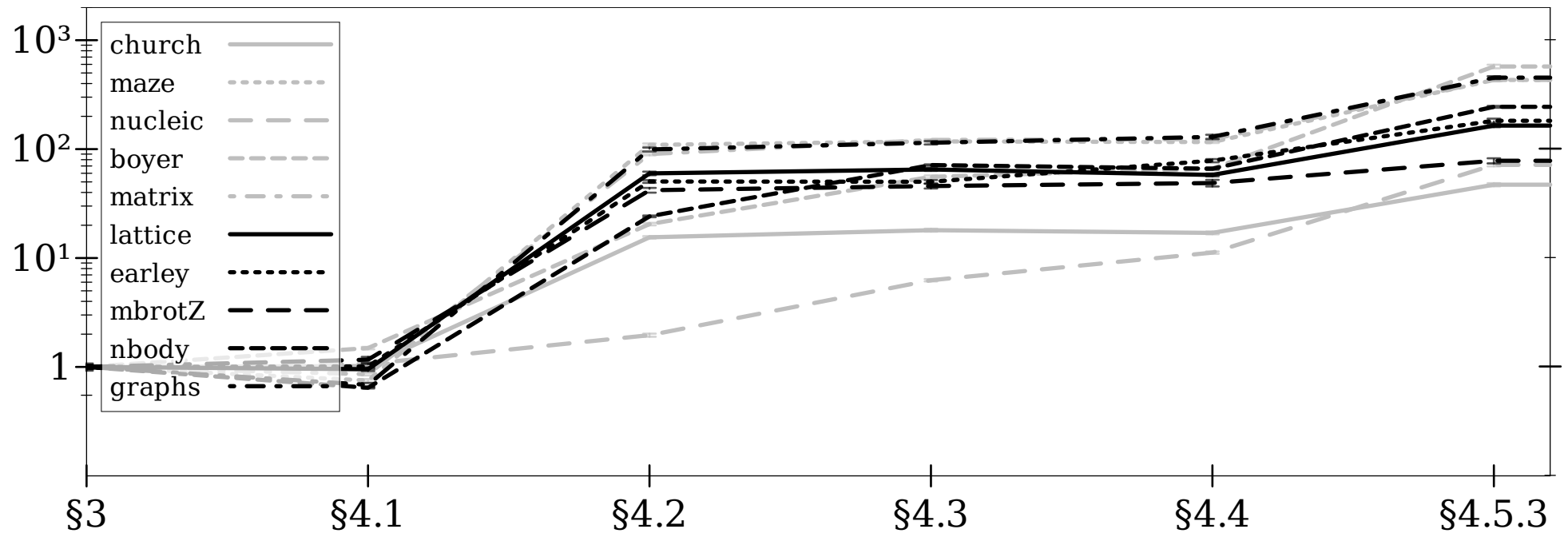

(b) Rate of state transitions speed-up (optimized / baseline)

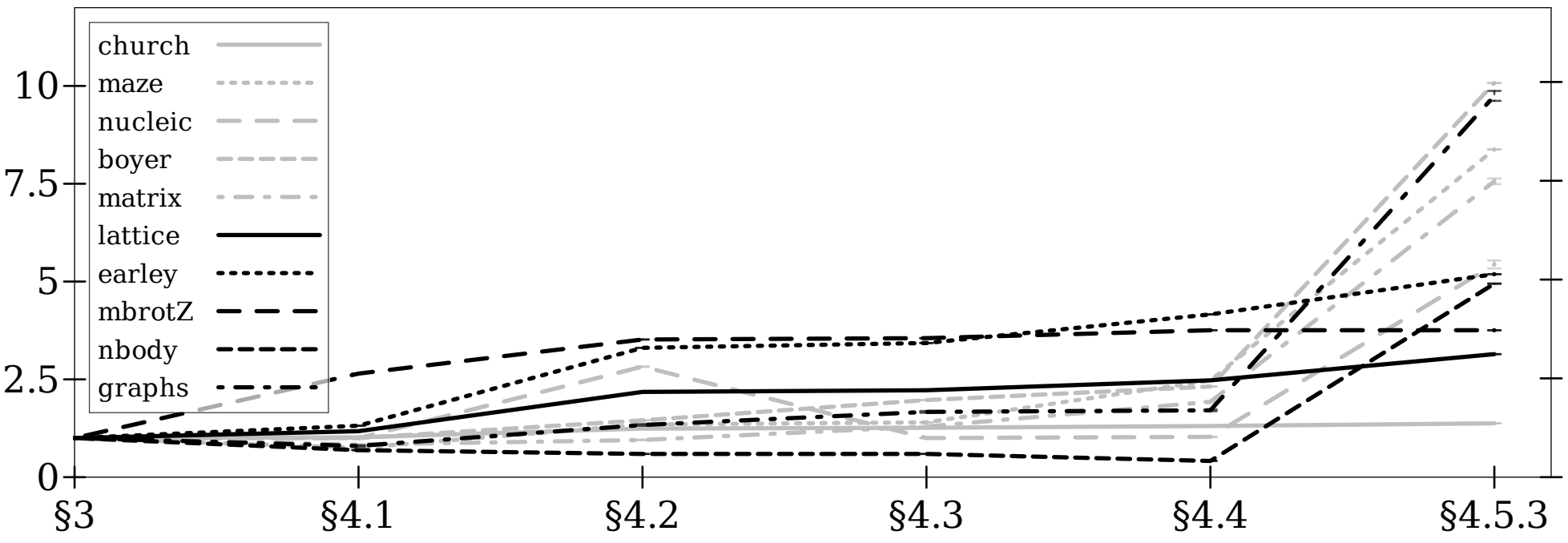

(c) Peak memory usage inverse factor (peak baseline / peak optimized)

Figure 12: Factors of improvement over baseline for each step of optimization (bigger is better).

parable form of analysis via a global acceptability judgment. Consequently, we have omitted a complete comparison with these implementations. The AAM approach provides more precision in terms of temporal-ordering of program states, which comes at a cost that can be avoided in constraint-based approaches. Consequently implementation techniques cannot be "ported" between these two approaches. However, our optimized implementation is within an order of magnitude of the performance of Wright and Jaganathan's analyzer. The optimized AAM approach of this chapter still has many strengths to recommend it in terms of precision, ease of implementation and verification, and rapid design. We can get closer to their performance by relying on the representation of addresses and the behavior of *alloc* to pre-allocate most data structures and split the abstract store out into parts that are more quickly accessed and updated. Our semantic optimizations can still be applied to an analysis that does abstract garbage collection [66], whereas the polymorphic splitting implementation is tied strongly to a single-threaded store.

# 4

# PUSHDOWN ANALYSIS VIA RELEVANT ALLOCATION

Programs in higher-order languages heavily use function calls and method dispatch for control flow. Standard flow analyses' imprecise handling of returns damages all specific analyses' precision. Recent techniques match calls and returns precisely [103, 28] and build smaller models more quickly than a standard 0CFA(evaluation predicted 2-5 times more constant bindings). These works, called CFA2 and PDCFA respectively, use pushdown automata as their approximation's target model of computation. They are hence called "pushdown analyses."[1] CFA2 and PDCFA have difficult details to easily apply to an off-the-shelf semantics—especially if they feature non-local control transfer that breaks the pushdown model.

The AAM method we discussed in Chapter 2 and Chapter 3 is a process to construct *regular* analyses. This chapter describes a systematic process to construct *pushdown* analyses of programming languages, due to the precision benefits.

## 4.1 TRADEOFFS OF APPROXIMATION STRENGTH

Static analysis is the process of soundly predicting properties of programs. It necessarily involves a tradeoff between the precision of those predictions and the computational complexity of producing them. At one end of the spectrum, an analysis may predict nothing, using no resources. At the other end, an analysis may predict everything, at the cost of computability.

Abstract interpretation [20] is a form of static analysis that involves the *approximate* running of a program by interpreting a program over an abstraction of the program's values, e.g. by using intervals in place of integers [19], or types instead of values [56]. By considering the sound abstract interpretation of a program, it is possible to predict the behavior of concretely running the program. For example, if abstract running a program never causes a buffer-overflow, run-time type error, or null-pointer dereference, we can conclude actually running the program can never cause any of these errors either. If a fragment of code is not executed during the abstract running, it can safely be deemed dead-code and removed. More fine-grained properties can be predicted too; to enable inlining, the abstract running of a program can identify all of the functions that are called exactly once

1 I refer to finite model analyses as "regular analyses" after the regular languages of traces they realize.

and the corresponding call-site. Temporal properties can be discovered as well: perhaps we want to determine if one function is always called before another, or if reads from a file occur within the opening and closing of it.

In general, we can model the abstract running of a program by considering each program state as a node in a graph, and track evolution steps as edges, where each node and path through the graph is an *approximation* of concrete program behavior. The art and science of static analysis design is the way we represent this graph of states; how little or how much detail we choose to represent in each state determines the precision and, often, the *cost* of such an analysis. First-order data-structures, numbers, arrays all have an abundance of literature for precise and effective approximations, so this paper focuses on higher-order data: closures and continuations, and their interaction with state evolution.

A major issue with designing a higher-order abstract interpreter is approximating closures and continuations in such a way that the interpreter always terminates while still producing sound and precise approximations. Traditionally, both have been approximated by finite sets, but in the case of continuations, this means the control stack of the abstract interpreter is modeled as a finite graph and therefore cannot be precise with regards to function calls and returns.

WHY PUSHDOWN RETURN FLOW MATTERS: AN EXAMPLE Higher-order programs often create proxies, or monitors, to ensure an object or function interacts with another object or function in a sanitized way. One example of this is behavioral contracts [34]. Simplified, here is how one might write an ad-hoc contract monitor for a given function and predicates for its inputs and outputs:

```
(define (monitor f in? out?)
  (λ (x)
   (if (in? x)
       (let ([r (f x)])
         (if (out? r)
             r
             (blame 'function)))
       (blame 'user))))
```

It is well known that wrapping functions like this thwarts the precision of regular 0CFA and higher kCFA as more wrappings are introduced. In the case of this innocent program

```
(cons (map (monitor render-int int? str?) ℓ)
      (map (monitor fact int? int?) ℓ))
```

according to 0CFA the call to the wrapped `factorial` function within the second `map` may return to within the first `map`. Hence 0CFA is not sufficiently precise to prove `factorial` cannot be blamed. Using more a context-sensitive analysis such as 1CFA, 2CFA, etc., would solve the problem for this example, but would fail for nested proxies. In general, for any k, kCFA will confuse the return flow of some programs as in this example. Yet, a pushdown abstraction that properly matches calls and returns has no trouble with this example, regardless of proxy-nesting depth.

A SYSTEMATIC APPROACH TO PUSHDOWN ANALYSIS At this point, several pushdown analyses for higher-order languages have been developed [100, 28], and the basic idea is simple: instead of approximating a program with a finite state machine, use a pushdown automata. The control stack of the automata models the control stack of the concrete interpreter, while stack frames, which contain closures, are subject to the same abstraction as values in the program.

This approach works well for simple languages which obey the stack discipline of a PDA. But most languages provide features that transgress that discipline, such as garbage collection, first-class control operators, stack inspection, and so on. Some of these features have been successfully combined with pushdown analysis, but required technical innovation and effort [101, 43, 29]. To avoid further one-off efforts, we develop a general technique for creating pushdown analyses for languages with control operators and reflective mechanisms.

## 4.2 REFINEMENT OF AAM FOR EXACT STACKS

We can exactly represent the stack in the $CESK_t^*$ machine with a modified allocation scheme for stacks. The key idea is that if the address is "precise enough," then every path that leads to the allocation will proceed exactly the same way until the address is dereferenced.

"PRECISE ENOUGH": For the $CESK_t^*$ machine, every function evaluates the same way, regardless of the stack. We should then represent the stack addresses as the components of a function call. The one place in the $CESK_t^*$ machine that continuations are allocated is at $(e_0\ e_1)$ evaluation. The expression itself, the environment, the store and the timestamp are necessary components for evaluating $(e_0\ e_1)$, so then we just represent the stack address as those four things. The stack is not relevant for its evaluation, so we do not want to store the stack addresses in the same store – that would also lead to a recursive store structure. I call this new table $\Xi$, because it looks like a stack.

By not storing the continuations in the value store, we separate "relevant" components from "irrelevant" components. We split the stack store from the value store and use only the value store in stack addresses. Stack addresses generally describe the relevant context that lead to their allocation, so we will refer to them henceforth as *contexts*. The resulting state space is updated here:

$$\widehat{State} = \widehat{CESK}_t \times KStore$$

$$\kappa \in Kont ::= \epsilon \mid \phi{:}\tau \qquad \text{overloads } K \text{ in } \widehat{CESK}_t$$

$$\tau \in Context ::= \langle e, \rho, \sigma\rangle_t$$

$$\Xi \in KStore = Context \underset{\text{fin}}{\rightharpoonup} \wp(Kont)$$

| $\hat{\varsigma}, \Xi \longmapsto \hat{\varsigma}', \Xi' \quad a = alloc(\hat{\varsigma}, \Xi) \quad u = tick(\hat{\varsigma}, \Xi)$ | |
|---|---|
| $\langle x, \rho, \sigma, \hat{\kappa}\rangle_t, \Xi$ | $\langle v, \sigma, \hat{\kappa}\rangle_u, \Xi$ if $v \in \sigma(\rho(x))$ |
| $\langle (e_0\ e_1), \rho, \sigma, \hat{\kappa}\rangle_t, \Xi$ | $\langle e_0, \rho, \sigma, \mathbf{appL}(e_1, \rho){:}\tau\rangle_u, \Xi'$ |
| where | $\tau = \langle (e_0\ e_1), \rho, \sigma\rangle_t$ |
| | $\Xi' = \Xi \sqcup [\tau \mapsto \hat{\kappa}]$ |
| $\langle v, \sigma, \mathbf{appL}(e, \rho'){:}\tau\rangle_t, \Xi$ | $\langle e, \rho', \sigma, \mathbf{appR}(v){:}\tau\rangle_u, \Xi$ |
| $\langle v, \rho, \sigma, \mathbf{appR}(\lambda x.\ e, \rho'){:}\tau\rangle_t, \Xi$ | $\langle e, \rho'', \sigma', \hat{\kappa}\rangle_u, \Xi$ if $\hat{\kappa} \in \Xi(\tau)$ |
| where | $\rho'' = \rho'[x \mapsto a]$ |
| | $\sigma' = \sigma \sqcup [a \mapsto v]$ |

Figure 13: $CESK_t^*\Xi$ semantics

The semantics is modified slightly in Figure 13 to use $\Xi$ instead of $\sigma$ for continuation allocation and lookup. Given finite allocation, contexts are drawn from a finite space, but are still precise enough to describe an unbounded stack: they hold all the relevant components to find which stacks are possible. The computed $\longmapsto$ relation thus represents the full description of a pushdown system of reachable states (and the set of paths). Of course this semantics does not always define a pushdown system since *alloc* can have an unbounded codomain. The correctness claim is therefore a correspondence between the same machine but with an unbounded stack, no $\Xi$, and *alloc*, *tick* functions that behave the same disregarding the different representations (a reasonable assumption).

### 4.2.1 *Correctness*

The high level argument for correctness exploits properties of both machines. Where the stack is unbounded (call this $CESK_t$), if every state in a trace shares a common tail in their continuations, that tail is *irrelevant*. This means the tail can be replaced with anything and still produce a valid trace. This property is more generally, "context irrelevance." The $CESK_t^*\Xi$ machine maintains an invariant on $\Xi$ that says that $\hat{\kappa} \in \Xi(\tau)$ represents a trace in $CESK_t$ that starts at the base of $\hat{\kappa}$ and reaches $\tau$ with $\hat{\kappa}$ on top. We can use this invariant and context irrelevance to translate steps in the $CESK_t^*\Xi$ machine into steps in *CESK$_t$*. The other way around, we use a proposition that a full stack is represented by $\Xi$ via unrolling and follow a simple simulation argument.

The common tail proposition we will call *ht* and the replacement function we will call *rt*; they both have obvious inductive and recursive definitions respectively. The invariant is stated with respect to the entire program, $e_{pgm}$:

$$\frac{}{inv_\Xi(\bot)} \qquad \frac{inv_\Xi(\Xi) \qquad \forall \hat\kappa_c \in K.base(\hat\kappa_c) \longmapsto^*_{CESK_t} \langle e_c, \rho_c, \sigma_c, \hat\kappa_c{+}{+}\epsilon\rangle_{t_c}}{inv_\Xi(\Xi[\langle e_c, \rho_c, \sigma_c\rangle_{t_c} \mapsto K])}$$

$$\frac{base(\hat\kappa) \longmapsto^*_{CESK_t} \langle e, \rho, \sigma, \hat\kappa{+}{+}\epsilon\rangle_t \qquad inv_\Xi(\Xi)}{inv(\langle e, \rho, \sigma, \hat\kappa\rangle_t, \Xi)}$$

where

$$base(\epsilon) = \langle e_{pgm}, \bot, \bot, \epsilon\rangle_{t_0}$$
$$base(\phi{:}\langle e_c, \rho_c, \sigma_c\rangle_{t_c}) = \langle e_c, \rho_c, \sigma_c, \epsilon\rangle_{t_c}$$

We use $\cdot{+}{+}\epsilon$ to treat $\tau$ like $\epsilon$ and construct a continuation in *Kont* rather than $\widehat{Kont}$.

**Lemma 12** (Context irrelevance). *For all traces $\pi \in CESK_t{}^*$ and continuations $\kappa$ such that $ht(\pi, \kappa)$, for any $\kappa'$, $rt(\pi, \kappa, \kappa')$ is a valid trace.*

*Proof.* Simple induction on $\pi$ and cases on $\longmapsto_{CESK_t}$. □

**Lemma 13** ($CESK^*_t\Xi$ Invariant). *For all $\varsigma, \varsigma' \in \widehat{State}$, if $inv(\varsigma)$ and $\varsigma \longmapsto \varsigma'$, then $inv(\varsigma')$*

*Proof.* Routine case analysis. □

Note that the injection of $e_{pgm}$ into $CESK^*_t\Xi$, $(\langle e_{pgm}, \bot, \bot, \epsilon\rangle_{t_0}, \bot)$, trivially satisfies *inv*.

The unrolling proposition is the following

$$\frac{}{\epsilon \in unroll_\Xi(\epsilon)} \qquad \frac{\hat\kappa \in \Xi(\tau), \kappa \in unroll_\Xi(\hat\kappa)}{\phi{:}\kappa \in unroll_\Xi(\phi{:}\tau)}$$

**Theorem 14** (Correctness). *For all expressions $e_{pgm}$,*

- **Soundness:** *if $\varsigma \longmapsto_{CESK_t} \varsigma'$, $inv(\varsigma\{\kappa := \hat\kappa\}, \Xi)$, and $\kappa \in unroll_\Xi(\hat\kappa)$, then there are $\Xi', \hat\kappa'$ such that $\varsigma\{\kappa := \hat\kappa\}, \Xi \longmapsto_{CESK^*_t\Xi} \varsigma'\{\kappa := \hat\kappa'\}, \Xi'$ and $\kappa' \in unroll_{\Xi'}(\hat\kappa')$*

- **Local completeness:** *if $\hat\varsigma, \Xi \longmapsto_{CESK^*_t\Xi} \hat\varsigma', \Xi'$ and $inv(\hat\varsigma, \Xi)$, for all $\kappa$, if $\kappa \in unroll_\Xi(\hat\varsigma.\hat\kappa)$ then there is a $\kappa'$ such that $\hat\varsigma\{\hat\kappa := \kappa\} \longmapsto_{CESK_t} \hat\varsigma'\{\hat\kappa := \kappa'\}$ and $\kappa' \in unroll_\Xi(\hat\varsigma'.\hat\kappa)$.*

The completeness result is "local" because it only applies to trace slices, and not entire traces - some starts of traces may not be reachable. As a mode of running, however, there will be no spuriously added states due to the short-circuiting via the memo-use rule. I conjecture full completeness (all traces with $\Xi$ are reachable traces in the stack model) is attainable by adding the calling expression to the representation of a context. By adding the calling expression, there should be an invariant that the range of $\Xi$ is always singleton sets. Thanks to Jens Nicolay for pointing out the incompleteness for traces in the concrete.

REVISITING THE EXAMPLE First we consider what 0CFA gives us, to see where pushdown analysis improves. The important difference is that in kCFA, return points are stored in an address that is linked to the textual location of the function call, plus a $k$-bounded amount of calling history. So, considering the common $k = 0$, the unknown function call within map (either `render-int` or `fact`) returns from the context of the second call to `map` to the context of the first call to `map`. Non-tail calls aren't safe from imprecise return flow: the recursive call to `map` returns directly to both calls in the outer `cons`. All nonsense.

In our presentation, return points are stored in an address that represents the *exact* calling context with respect to the abstract machine's components. This means when there is a "merging" of return points, it really means that two places in the program have requested the exact same thing of a function, even with the same global values. The function *will* return to both places. The predicted control flow in the example is as one would expect, or *hope*, an analysis would predict: the correct flow.

### 4.2.2 *Engineered semantics for efficiency*

I cover three optimizations that may be employed to accelerate the fixed-point computation.

1. Continuations can be "chunked" more coarsely at function boundaries instead of at each frame in order to minimize table lookups.

2. We can globalize $\Xi$ with no loss in precision, unlike a global store; it will not need to be stored in the frontier but will need to be tracked by seen states. The seen states only need comparison, and a global $\Xi$ increases monotonically, so we can use Shivers' timestamp technique [87]. The timestamp technique does not store an entire $\Xi$ in the seen set at each state, but rather how many times the global $\Xi$ has increased.

3. Since evaluation is the same regardless of the stack, we can memoize results to short-circuit to the answer. The irrelevance

$$\hat{\varsigma} \in \widehat{CESIK} = \langle e, \rho, \sigma, \iota, \hat{\kappa} \rangle \quad \iota \in LKont = Frame^*$$
$$\hat{\kappa} \in Kont ::= \epsilon \mid \tau$$

Figure 14: *CESIK*$^*\Xi$ semantic spaces

of the stack then precludes the need for timestamping the global $\Xi$.

This last optimization will be covered in more detail in Section 4.5. From here on, this chapter will not explicitly mention timestamps.

A secondary motivation for the representation change in 1 is that flow analyses commonly split control-flow graphs at function call boundaries to enable the combination of intra- and inter-procedural analyses. In an abstract machine, this split looks like installing a continuation prompt at function calls. We borrow a representation from literature on delimited continuations [8] to split the continuation into two components: the continuation and meta-continuation. Our delimiters are special since each continuation "chunk" until the next prompt has bounded length. The bound is roughly the deepest nesting depth of an expression in functions' bodies. Instead of "continuation" and "meta-continuation" then, I will use terminology from CFA2 and call the top chunk a "local continuation," and the rest the "continuation."[13]

[13] *Since the continuation is either $\epsilon$ or a context, CFA2 calls these "entries" to mean execution entry into the program ($\epsilon$) or a function ($\tau$). One can also understand these as entries in a table ($\Xi$). I stay with the "continuation" nomenclature because they represent full continuations.*

Figure 15 has a visualization of a hypothetical state space. Reduction relations can be thought of as graphs: each state is a node, and if a state $\varsigma$ reduces to $\varsigma'$, then there is an edge $\varsigma \longmapsto \varsigma'$. We can also view our various environments that contain pointers (addresses, contexts) as graphs: each pointer is a node, and if the pointer $\tau$ references an object $\iota$ that contains another pointer $\tau'$, then there is a labeled edge $\tau \xrightarrow{\iota} \tau'$. States' contexts point into $\Xi$ to associate each state with a *regular language* of continuations. The reversed $\Xi$ graph can be read as a collection of finite state machines that accepts all the continuations that are possible at each state that the reversed pointers lead to. The `halt` continuation is this graph's starting state.

The resulting shuffling of the semantics to accommodate this new representation is in Figure 16. The extension to $\Xi$ happens in a different rule – function entry – so the shape of the context changes to hold the function, argument, and store. We have a choice of whether to introduce an administrative step to dereference $\Xi$ once $\iota$ is empty, or to use a helper metafunction to describe a "pop" of both $\iota$ and $\kappa$. Suppose we choose the second because the resulting semantics has a 1-to-1 correspondence with the previous semantics. A first attempt might land us here:

$$pop(\phi{:}\iota, \hat{\kappa}, \Xi) = \{(\phi, \iota, \hat{\kappa})\}$$
$$pop(\epsilon, \tau, \Xi) = \{(\phi, \iota, \hat{\kappa}) \ : \ (\phi{:}\iota, \hat{\kappa}) \in \Xi(\tau)\}$$

However, tail calls make the dereferenced $\tau$ lead to $(\epsilon, \tau')$. Because abstraction makes the store grow monotonically in a finite space, it's possible that $\tau' = \tau$ and a naive recursive definition of *pop* will diverge chasing these contexts. Now *pop* must save all the contexts it dereferences in order to guard against divergence. So $pop(\iota, \hat{\kappa}, \Xi) = pop^*(\iota, \hat{\kappa}, \Xi, \emptyset)$ where

$$\begin{aligned}
pop^*(\epsilon, \epsilon, \Xi, G) &= \emptyset \\
pop^*(\phi{:}\iota, \hat{\kappa}, \Xi, G) &= \{(\phi, \iota, \hat{\kappa})\} \\
pop^*(\epsilon, \tau, \Xi, G) &= \{(\phi, \iota, \hat{\kappa}) \ : \ (\phi{:}\iota, \hat{\kappa}) \in \Xi(\tau)\} \\
&\cup \bigcup_{\tau' \in G'} pop^*(\epsilon, \tau', \Xi, G \cup G') \\
\text{where } G' &= \{\tau' \ : \ (\epsilon, \tau') \in \Xi(\tau)\} \setminus G
\end{aligned}$$

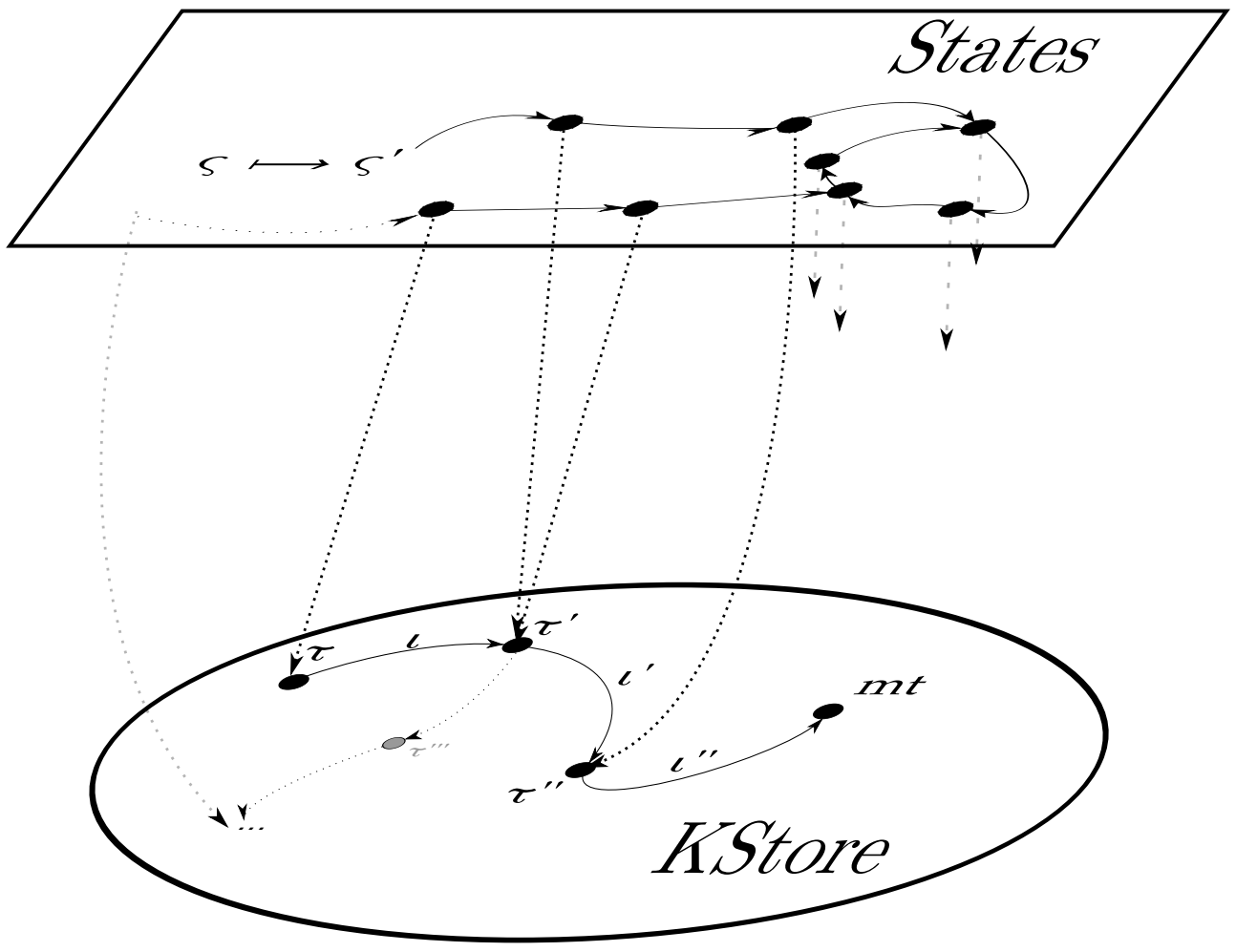

Figure 15: Graph visualization of states and Ξ

In practice, one would not expect G to grow very large. Had we chosen the first strategy, the issue of divergence is delegated to the machinery from the fixed-point computation.[2] However, when adding the administrative state, the "seen" check requires searching a far larger set than we would expect G to be.

We run the the stepping relation along all nondeterministic paths. The continuation table can be global and use the same counting mechanism we used for the global stores in Chapter 3, without loss of precision. For ease of exposition, I will keep a map from state without Ξ to largest Ξ at which it has been seen. The fixed point computation thus computes over the following system:

$$\widehat{System} = (\widehat{CESK}_t \underset{\text{fin}}{\rightharpoonup} KStore) \times \wp(\widehat{CESK}_t^2) \times \wp(\widehat{CESK}_t) \times KStore$$

We compute all next steps from the frontier, combine all changes to Ξ, and continue with a new frontier of states we stepped to that we haven't seen at the current Ξ.

2 CFA2 employs the first strategy and calls it "transitive summaries."

$$\hat{\varsigma}, \Xi \longmapsto \hat{\varsigma}', \Xi' \qquad a = \mathit{alloc}(\hat{\varsigma}, \Xi)$$

| | |
|---|---|
| $\langle x, \rho, \sigma, \iota, \hat{\kappa}\rangle, \Xi$ | $\langle v, \sigma, \iota, \hat{\kappa}\rangle, \Xi$ if $v \in \sigma(\rho(x))$ |
| $\langle (e_0\ e_1), \rho, \sigma, \iota, \hat{\kappa}\rangle, \Xi$ | $\langle e_0, \rho, \sigma, \mathbf{appL}(e_1, \rho){:}\iota, \hat{\kappa}\rangle, \Xi$ |
| $\langle v, \sigma, \iota, \hat{\kappa}\rangle, \Xi$ | $\langle e, \rho', \sigma, \mathbf{appR}(v, \rho){:}\iota', \hat{\kappa}'\rangle, \Xi$ |
| | if $\mathbf{appL}(e, \rho'), \iota', \hat{\kappa}' \in \mathit{pop}(\iota, \hat{\kappa}, \Xi)$ |
| $\langle v, \sigma, \iota, \hat{\kappa}\rangle, \Xi$ | $\langle e, \rho[x \mapsto a], \sigma', \epsilon, \tau\rangle, \Xi'$ |
| | if $\mathbf{appR}(\lambda x.\ e, \rho), \iota', \hat{\kappa}' \in \mathit{pop}(\iota, \hat{\kappa}, \Xi)$ |
| where | $\sigma' = \sigma \sqcup [a \mapsto v]$ |
| | $\tau = (\langle \lambda x.\ e, \rho\rangle, v, \sigma)$ |
| | $\Xi' = \Xi \sqcup [\tau \mapsto (\iota, \hat{\kappa})]$ |

Figure 16: *CESIK**Ξ semantics

$$\begin{aligned}
\mathcal{F}_e(S, R, F, \Xi) &= (S \lhd S', R \cup R', F', \Xi') \\
I &= \bigcup_{\hat{\varsigma} \in F} \{(\langle \hat{\varsigma}, \hat{\varsigma}'\rangle, \Xi') \ :\ \hat{\varsigma}, \Xi \longmapsto \hat{\varsigma}', \Xi'\} \\
R' &= \pi_0 I \qquad \Xi' = \bigsqcup \pi_1 I \\
S' &= [\hat{\varsigma} \mapsto \Xi' \ :\ \hat{\varsigma} \in \pi_1(R')] \\
F' &= \{\hat{\varsigma} \in \mathbf{dom}(S') \ :\ S'(\hat{\varsigma}) \neq S(\hat{\varsigma})\}
\end{aligned}$$

For a program $e$, we will say $(\bot, \emptyset, \{\langle e, \bot, \bot, \epsilon, \epsilon\rangle\}, \bot)$ is the bottom element of $\mathcal{F}_e$'s domain. The "analysis" then is then the pair of the $R$ and $\Xi$ components of $\mathbf{lfp}(\mathcal{F}_e)$.

CORRECTNESS The correctness argument for this semantics is not about single steps but instead about the entire relation that $\mathcal{F}$ computes. The argument is that the $R$ and $\Xi$ components of the system represent a slice of the unbounded relation $\longmapsto_{CESK_t}$ (restricted to reachable states). We will show that traces in any $n \in \mathbb{N}$ times we *unfold* $\longmapsto_{CESK_t}$ from the initial state, there is a corresponding $m$ applications of $\mathcal{F}$ that reify into a relation that exhibit the same trace. Conversely, any trace in the reification of $\mathcal{F}_e^m(\bot)$ has the same trace in some $n$ unfoldings of $\longmapsto_{CESK_t}$. For an arbitrary *alloc* function, we cannot expect $\mathcal{F}$ to have a fixed point, so this property is the best we can get. For a finite *alloc* function, Kleene's fixed point theorem dictates there is a $m$ such that $\mathcal{F}_e^m(\bot)$ is a fixed point, so every trace in the reified relation is also a trace in an unbounded number of unfoldings of $\longmapsto_{CESK_t}$. This is the corresponding local completeness argument for the algorithm.

$$\begin{aligned}
\mathit{unfold}(\varsigma_0, \longmapsto, 0) &= \{(\varsigma_0, \varsigma) : \varsigma_0 \longmapsto \varsigma\} \\
\mathit{unfold}(\varsigma_0, \longmapsto, n+1) &= \mathit{unfold}_1(\mathit{unfold}(\varsigma_0, \longmapsto, n)) \\
\text{where } \mathit{unfold}_1(R) &= R \cup \{(\varsigma, \varsigma') : (_, \varsigma) \in R, \varsigma \longmapsto \varsigma'\}
\end{aligned}$$

The reification simply realizes all possible complete continuations that a state could have, given $\Xi$:

$$\frac{\langle\langle e, \rho, \sigma, \iota, \hat{\kappa}\rangle, \langle e', \rho', \sigma', \iota', \hat{\kappa}'\rangle\rangle \in R \qquad \kappa \in \mathit{unroll}'_\Xi(\hat{\kappa})}{\langle e, \rho, \sigma, \iota{+}{+}\kappa\rangle \longmapsto_{\mathit{reify}(S,R,F,\Xi)} \langle e', \rho', \sigma', \iota'{+}{+}\kappa\rangle}$$

The *unroll′* judgment is like *unroll*, but with prepending of local continuations:

$$\frac{}{\epsilon \in \mathit{unroll}'_\Xi(\epsilon)} \qquad \frac{(\iota, \hat{\kappa}) \in \Xi(\tau) \qquad \kappa \in \mathit{unroll}'_\Xi(\hat{\kappa})}{\iota{+}{+}\kappa \in \mathit{unroll}'_\Xi(\tau)}$$

**Theorem 15** (Correctness)**.** *For all* $e_0$*, let* $\varsigma_0 = \langle e_0, \bot, \bot, \epsilon\rangle$ *in* $\forall n \in \mathbb{N}, \varsigma, \varsigma' \in \mathit{CESK}_t$*:*

- *if* $(\varsigma, \varsigma') \in \mathit{unfold}(\varsigma_0, \longmapsto_{\mathit{CESK}_t}, n)$ *then there is an* $m$ *such that* $\varsigma \longmapsto_{\mathit{reify}(\mathcal{F}^m_{e_0}(\bot))} \varsigma'$
- *if* $\varsigma \longmapsto_{\mathit{reify}(\mathcal{F}^n_{e_0}(\bot))} \varsigma'$ *then there is an* $m$ *such that* $(\varsigma, \varsigma')$ *is in* $\mathit{unfold}(\varsigma_0, \longmapsto_{\mathit{CESK}_t}, m)$

*Proof.* By induction on $n$. □

### 4.2.3 *Remarks about cost*

The common tradeoff for performance over precision is to use a global store. A representation win originally exploited by Shivers [87] is to represent the seen states' stores by the *age* of the store. A context in this case contains the store age for faster comparison. Old stores are mostly useless since the current one subsumes them, so a useful representation for the seen set is as a map from the *rest of the state* to the store age it was last visited with. We will align with the analysis literature and call these "rest of state" objects *points*. Note that since the store age becomes part of the state representation due to "context," there are considerably more points than in the comparable finite state approach. When we revisit a state because the store age (or $\Xi$ age) is different from the last time we visited it (hence we're visiting a *new state*), we can clobber the old ages. A finite state approach will use less memory because the seen set will have a smaller domain (fewer distinctions made because of the lack of a "context" component).

## 4.3 STACK INSPECTION AND RECURSIVE METAFUNCTIONS

Since we just showed how to produce a pushdown system from an abstract machine, some readers may be concerned that we have lost the ability to reason about the stack as a whole. This is not the case. The semantics may still refer to $\Xi$ to make judgments about the possible stacks that can be realized at each state. A metafunction in the semantics that operates over a whole stack can be recast as a transition system that we overapproximate and run to fixed point using the AAM methodology.

Some semantic features allow a language to inspect some arbitrarily deep part of the stack, or compute a property of the whole stack before continuing. Java's access control security features are an example of the first form of inspection, and garbage collection is an example of the second. I will demonstrate both forms are simple first-order metafunctions that the AAM methodology will soundly interpret. Access control can be modeled with continuation marks, so I demonstrate with the CM machine of Clements and Felleisen.

Semantics that inspect the stack do so with metafunction calls that recur down the stack. Recursive metafunctions have a semantics as well, hence fair game for AAM. And, they should always terminate (otherwise the semantics is hosed). We can think of a simple pattern-matching recursive function as a set of rewrite rules that apply repeatedly until it reaches a result. Interpreted via AAM, non-deterministic metafunction evaluation leads to a set of possible results.

The finite restriction on the state space carries over to metafunction inputs, so we can always detect infinite loops that abstraction may have introduced and bail out of that execution path. Specifically, a metafunction call can be seen as an initial state, $s$, that will evaluate through the metafunction's rewrite rules $\longmapsto$ to compute all terminal states (outputs):

$$\begin{aligned}
&\mathit{terminal} : \forall A.\text{relation } A \times A \to \wp(A) \\
&\mathit{terminal}(\longmapsto, s) = \mathit{terminal}^*(\emptyset, \{s\}, \emptyset) \\
&\quad \text{where } \mathit{terminal}^*(S, \emptyset, T) = T \\
&\quad \mathit{terminal}^*(S, F, T) = \mathit{terminal}^*(S \cup F, F', T \cup T') \\
&\qquad \text{where } T' = \bigcup_{s \in F} \mathit{post}(s) \stackrel{?}{=} \emptyset \to \{s\}, \emptyset \\
&\qquad F' = \bigcup_{s \in F} \mathit{post}(s) \setminus S \\
&\quad \mathit{post}(s) = \{s' \;:\; s \longmapsto s'\}
\end{aligned}$$

This definition is a typical worklist algorithm. It builds the set of terminal terms, $T$, by exploring the frontier (or worklist), $F$, and only adding terms to the frontier that have not been seen, as represented by S. If s has no more steps, *post*(s) will be empty, meaning s should be added to the terminal set T.

We prove a correctness condition that allows us to reason equationally with *terminal* later on:

**Lemma 16** (*terminal*$^*$ correct)**.** *Fix* $\longmapsto$. *Constrain arbitrary* S, F, T *such that* $T \sqsubseteq S$ *and* $\forall s \in S, post(s) = \emptyset \iff s \in T$, $F \cap S = \emptyset$, *and for all* $s \in S$, $post(s) \subseteq S \cup F$.

- ***Soundness:*** *for all* $s \in S \cup F$, *if* $s \longmapsto^* s_t$ *and* $post(s_t) = \emptyset$ *then* $s_t \in terminal^*(S, F, T)$.
- ***Local completeness:*** *for all* $s \in terminal^*(S, F, T)$ *there is an* $s_0 \in S \cup F$ *such that* $s_0 \longmapsto^* s$ *and* $post(s) = \emptyset$.

*Proof.* By induction on *terminal*$^*$'s recursion scheme. □

Note that it is possible for metafunctions' rewrite rules to themselves use metafunctions, but the *seen* set (S) for *terminal* must be bound with a dynamic variable – it cannot restart at $\emptyset$ upon reentry. Without this precaution, the host language will exceed its stack limits when an infinite path is explored, rather than bail out.

### 4.3.1 *Case study for stack traversal: GC*

Garbage collection is an example of a language feature that needs to crawl the stack, specifically to find live addresses. We are interested in garbage collection because it can give massive precision boosts to analyses [66, 29]. Unadulterated, abstract GC inflicts an exponential state space that can destroy performance. The following function will produce the set of live addresses in the stack:

$$
\begin{aligned}
K\mathcal{LL} &: \mathit{Frame}^* \to \wp(\mathit{Addr}) \\
K\mathcal{LL}(\kappa) &= K\mathcal{LL}^*(\kappa, \emptyset) \\
K\mathcal{LL}^*(\epsilon, L) &= L \\
K\mathcal{LL}^*(\phi{:}\kappa, L) &= K\mathcal{LL}^*(\kappa, L \cup \mathcal{T}(\phi)) \\
\text{where } \mathcal{T}(\mathbf{appL}(e, \rho)) = \mathcal{T}(\mathbf{appR}(e, \rho)) &= \mathcal{T}(e, \rho) \\
\mathcal{T}(e, \rho) &= \{\rho(x) \ : \ x \in \mathit{fv}(e)\}
\end{aligned}
$$

When interpreted via AAM, the continuation is indirected through $\Xi$ and leads to multiple results, and possibly loops through $\Xi$. Thus this is more properly understood as

$$
\begin{aligned}
K\mathcal{LL}(\Xi, \hat{\kappa}) &= \mathit{terminal}(\longmapsto, K\mathcal{LL}^*(\Xi, \hat{\kappa}, \emptyset)) \\
K\mathcal{LL}^*(\Xi, \epsilon, L) &\longmapsto L \\
K\mathcal{LL}^*(\Xi, \phi{:}\tau, L) &\longmapsto K\mathcal{LL}^*(\Xi, \hat{\kappa}, L \cup \mathcal{T}(\phi)) \text{ if } \hat{\kappa} \in \Xi(\tau)
\end{aligned}
$$

A garbage collecting semantics can choose to collect the store with respect to each live set (call this $\Gamma^*$), or, soundly, collect with respect to their union (call this $\hat{\Gamma}$).[3] On the one hand we could have tighter collections but more possible states, and on the other hand we can leave some precision behind in the hope that the state space will be smaller. In the general idea of relevance versus irrelevance, the continuation's live addresses are relevant to execution, but are already implicitly represented in contexts because they must be mapped in the store's domain.

A state is "collected" only if live addresses remain in the domain of $\sigma$. We say a value $v \in \sigma(a)$ is live if $a$ is live. If a value is live, any addresses it touches are live; this is captured by the computation in $\mathcal{R}$:

$$\mathcal{R}(\mathit{root}, \sigma) = \{b \ : \ a \in \mathit{root}, a \leadsto_\sigma^* b\}$$

$$\frac{v \in \sigma(a) \qquad b \in \mathcal{T}(v)}{a \leadsto_\sigma b}$$

So the two collection methods are as follows. Exact GC produces different collected states based on the possible stacks' live addresses:[4]

$$\begin{aligned}\Gamma^*(\hat{\varsigma}, \Xi) &= \{\hat{\varsigma}\{\sigma := \hat{\varsigma}.\sigma|_L\} \ : \ L \in \mathit{live}^*(\hat{\varsigma}, \Xi)\} \\ \mathit{live}^*(\langle e, \rho, \sigma, \hat{\kappa}\rangle, \Xi) &= \{\mathcal{R}(\mathcal{T}(e, \rho) \cup L, \sigma) \ : \ L \in K\mathcal{L}\mathcal{L}(\Xi, \hat{\kappa})\}\end{aligned}$$

$$\frac{\hat{\varsigma}, \Xi \longmapsto \hat{\varsigma}', \Xi' \qquad \hat{\varsigma}' \in \Gamma^*(\varsigma', \Xi')}{\hat{\varsigma}, \Xi \longmapsto_{\Gamma^*} \hat{\varsigma}', \Xi}$$

And inexact GC produces a single state that collects based on all (known) stacks' live addresses:

$$\begin{aligned}\hat{\Gamma}(\hat{\varsigma}, \Xi) &= \hat{\varsigma}\{\sigma := \hat{\varsigma}.\sigma|_{\widehat{\mathit{live}}(\hat{\varsigma}, \Xi)}\} \\ \widehat{\mathit{live}}(\langle e, \rho, \sigma, \hat{\kappa}\rangle, \Xi) &= \mathcal{R}(\mathcal{T}(e, \rho) \cup \bigcup K\mathcal{L}\mathcal{L}(\Xi, \hat{\kappa}), \sigma)\end{aligned}$$

$$\frac{\hat{\varsigma}, \Xi \longmapsto \hat{\varsigma}', \Xi'}{\hat{\varsigma}, \Xi \longmapsto_{\hat{\Gamma}} \hat{\Gamma}(\varsigma', \Xi'), \Xi'}$$

Without the continuation store, the baseline GC is

$$\begin{aligned}\Gamma(\varsigma) &= \varsigma\{\sigma := \varsigma.\sigma|_{\mathit{live}(\varsigma)}\} \\ \mathit{live}(e, \rho, \sigma, \kappa) &= \mathcal{R}(\mathcal{T}(e, \rho) \cup K\mathcal{L}\mathcal{L}(\kappa), \sigma)\end{aligned}$$

$$\frac{\varsigma \longmapsto \varsigma'}{\varsigma \longmapsto_\Gamma \Gamma(\varsigma')}$$

3 The garbage collecting version of PDCFA [43] evaluates the $\hat{\Gamma}$ strategy.

4 It is possible and more efficient to build the stack's live addresses piecemeal as an additional component of each state, precluding the need for $K\mathcal{L}\mathcal{L}$. Each stack in $\Xi$ would also store the live addresses to restore on pop.

Suppose at arbitrary times we decide to perform garbage collection rather than continue with garbage. So when $\hat{\varsigma} \longmapsto \hat{\varsigma}'$, we instead do $\hat{\varsigma} \longmapsto_{\Gamma} \hat{\varsigma}'$. The times we perform GC do not matter for soundness, since we are not analyzing GC behavior. However, garbage stands in the way of completeness. Mismatches in the GC application for the different semantics lead to mismatches in resulting state spaces, not just up to garbage in stores, but in spurious paths from dereferencing a reallocated address that was not first collected.

The state space compaction that continuation stores give us makes ensuring GC times match up for the completeness proposition tedious. Our statement of local completeness then will assume both semantics perform garbage collection on every step. Call this step relation $\longmapsto_{\Gamma CESK_t}$.

The generalization of "context irrelevance" to stack-relevant computation is "context congruence", where we use an equivalence relation $\equiv_K$ to constrain traces. Define a semantics to be congruent mod $\equiv_K$ the following way:

$$\begin{aligned}
\mathit{ctx\text{-}congruent} &: \forall S.\wp(S \times S) \times \wp(\mathit{Kont} \times \mathit{Kont}) \to \mathit{Prop} \\
\mathit{ctx\text{-}congruent}(\longmapsto, \equiv_K) &= \forall \pi \in S^*, \kappa.\Pi_{wf}(\pi, \longmapsto) \mathit{ht}(\pi, \kappa) \implies \\
&\quad \forall \kappa'.\kappa \equiv_K \kappa' \implies \\
&\quad \Pi_{wf}(\mathit{rt}(\pi, \kappa, \kappa'), \longmapsto)
\end{aligned}$$

In this case, continuations are equivalent if they touch the same addresses:

$$\frac{\mathcal{T}(\kappa) = \mathcal{T}(\kappa')}{\kappa \equiv_\Gamma \kappa'}$$

The following lemma is $\mathit{ctx\text{-}congruent}(\mathit{step}_{\Gamma CESK_t}, \equiv_\Gamma)$ restated with less symbols.

**Lemma 17** (Context congruence). *For all traces $\pi \in \Gamma CESK_t{}^*$ and continuations $\kappa$ such that $ht(\pi, \kappa)$, for any $\kappa'$ such that $\kappa \equiv_K \kappa'$, $rt(\pi, \kappa, \kappa')$ is a valid trace.*

*Proof.* Simple induction on $\pi$ and cases on $\longmapsto_{\Gamma CESK_t}$. □

**Lemma 18** (Correctness of $K\mathcal{L}\mathcal{L}$). *For all $\Xi, \kappa, \hat{\kappa}, L$,*

- ***Soundness:*** *if $\kappa \in unroll_\Xi(\hat{\kappa})$ then $K\mathcal{L}\mathcal{L}^*(\kappa, L) \in terminal(\longmapsto, K\mathcal{L}\mathcal{L}^*(\Xi, \hat{\kappa}, L))$*
- ***Local completeness:*** *for all $L' \in K\mathcal{L}\mathcal{L}^*(\Xi, \hat{\kappa}, L)$ there is a $\kappa \in unroll_\Xi(\hat{\kappa})$ such that $L' = K\mathcal{L}\mathcal{L}^*(\kappa, L)$.*

*Proof.* Soundness follows by induction on the unrolling. Local completeness follows by induction on the trace from local completeness in Lemma 16. □

**Theorem 19** (Correctness of $\Gamma^* CESK_t^*\Xi$). *For all expressions* $e_0$,

- **Soundness:** *if* $\varsigma \longmapsto_{\Gamma CESK_t} \varsigma'$ *and* $\varsigma.\kappa \in unroll_\Xi(\hat\kappa)$, *then there are* $\Xi', \hat\kappa', \sigma'$ *such that* $\varsigma\{\kappa := \hat\kappa\}, \Xi \longmapsto_{\Gamma^* CESK_t^*\Xi} \hat\varsigma', \Xi'$ *where* $\hat\varsigma' = \varsigma'\{\kappa := \hat\kappa', \sigma := \sigma'\}$ *and* $\varsigma'.\kappa \in unroll_{\Xi'}(\hat\kappa')$ *and finally there is an* $L \in live^*(\hat\varsigma', \Xi')$ *such that* $\sigma'|_L = \varsigma'.\sigma|_{live(\varsigma')}$
- **Local completeness:** *if* $\hat\varsigma \equiv \langle e, \rho, \sigma, \hat\kappa\rangle, \Xi \longmapsto_{\Gamma^* CESK_t^*\Xi} \hat\varsigma', \Xi'$ *and there is an* $L_\kappa \in K\mathcal{L}\mathcal{L}(\Xi, \hat\kappa)$ *such that* $\sigma|_L = \sigma$ *(where* $L = \mathcal{R}(\mathcal{T}(e, \rho) \cup L_\kappa, \sigma)$*) and* $inv(\hat\varsigma, \Xi)$, *for all* $\kappa \in unroll_\Xi(\hat\kappa)$ *such that* $K\mathcal{L}\mathcal{L}(\kappa) = L_\kappa$, *there is a* $\kappa'$ *such that* $\hat\varsigma\{\hat\kappa := \kappa\} \longmapsto_{\Gamma CESK_t} \hat\varsigma'\{\hat\kappa := \kappa'\}$ *(a GC step) and* $\kappa' \in unroll_\Xi(\hat\varsigma'.\hat\kappa)$

**Theorem 20** (Soundness of $\hat\Gamma CESK_t^*\Xi$). *For all expressions* $e_0$, *if* $\varsigma \longmapsto_{\Gamma CESK_t} \varsigma'$, $inv(\varsigma\{\kappa := \hat\kappa\}, \Xi)$, *and* $\varsigma.\kappa \in unroll_\Xi(\hat\kappa)$, *then there are* $\Xi', \hat\kappa', \sigma''$ *such that* $\varsigma\{\kappa := \hat\kappa\}, \Xi \longmapsto_{\hat\Gamma CESK_t^*\Xi} \hat\varsigma', \Xi'$ *where* $\hat\varsigma' = \varsigma'\{\kappa := \hat\kappa', \sigma := \sigma''\}$ *and* $\varsigma'.\kappa \in unroll_{\Xi'}(\hat\kappa')$ *and finally* $\varsigma'.\sigma|_{live(\varsigma')} \sqsubseteq \sigma''|_{\widehat{live}(\hat\varsigma', \Xi')}$

The proofs are straightforward, and use the usual lemmas for GC, such as idempotence of $\Gamma$ and $\mathcal{T} \subseteq \mathcal{R}$.

### 4.3.2 *Case study analyzing security features: the CM machine*

The CM machine provides a model of access control: a dynamic branch of execution is given permission to use some resource if a continuation mark for that permission is set to "grant." There are three new forms we add to the lambda calculus to model this feature: `grant`, `frame`, and `test`. The `grant` construct adds a permission to the stack. The concern of unforgeable permissions is orthogonal, so we simplify with a set of permissions that is textually present in the program:

$P \in \mathit{Permissions}$ a set

$$\mathit{Expr} ::= \ldots \mid \texttt{grant}\ P\ e \mid \texttt{frame}\ P\ e \mid \texttt{test}\ P\ e\ e$$

The `frame` construct ensures that only a given set of permissions are allowed, but not necessarily granted. The security is in the semantics of `test`: we can test if all marks in some set P have been granted in the stack without first being denied; this involves crawling the stack:

$$\begin{aligned}
\mathcal{OK}(\emptyset, \kappa) &= \mathit{True} \\
\mathcal{OK}(P, \epsilon^m) &= \mathit{pass?}(P, m) \\
\mathcal{OK}(P, \phi{:}^m\kappa) &= \mathit{pass?}(P, m)\ \textbf{and}\ \mathcal{OK}(P \setminus m^{-1}(\textbf{Grant}), \kappa) \\
\text{where } \mathit{pass?}(P, m) &= P \cap m^{-1}(\textbf{Deny}) \stackrel{?}{=} \emptyset
\end{aligned}$$

The set subtraction is to say that granted permissions do not need to be checked farther down the stack.

| | |
|---|---|
| $\langle \texttt{grant P } e, \rho, \sigma, \kappa\rangle$ | $\langle e, \rho, \sigma, \kappa[P \mapsto \textbf{Grant}]\rangle$ |
| $\langle \texttt{frame P } e, \rho, \sigma, \kappa\rangle$ | $\langle e, \rho, \sigma, \kappa[\overline{P} \mapsto \textbf{Deny}]\rangle$ |
| $\langle \texttt{test P } e_0\ e_1, \rho, \sigma, \kappa\rangle$ | $\langle e_0, \rho, \sigma, \kappa\rangle$ if $\mathit{True} = \mathcal{OK}(P, \kappa)$ |
| | $\langle e_1, \rho, \sigma, \kappa\rangle$ if $\mathit{False} = \mathcal{OK}(P, \kappa)$ |

Figure 17: CM machine semantics

Continuation marks respect tail calls and have an interface that abstracts over the stack implementation. Each stack frame added to the continuation carries the permission map. The empty continuation also carries a permission map. Crucially, the added constructs do not add frames to the stack; instead, they update the permission map in the top frame, or if empty, the empty continuation's permission map.

$$\begin{aligned} m \in \mathit{PermissionMap} &= \mathit{Permissions} \underset{\text{fin}}{\rightharpoonup} \mathit{GD} \\ \mathit{gd} \in \mathit{GD} &::= \textbf{Grant} \mid \textbf{Deny} \\ \kappa \in \mathit{Kont} &::= \epsilon^m \mid \phi{:}^m\kappa \end{aligned}$$

Update for continuation marks:

$$\begin{aligned} m[P \mapsto \mathit{gd}] &= \lambda x. x \overset{?}{\in} P \to \mathit{gd}, m(x) \\ m[\overline{P} \mapsto \mathit{gd}] &= \lambda x. x \overset{?}{\in} P \to m(x), \mathit{gd} \end{aligned}$$

The abstract version of the semantics has one change on top of the usual continuation store. The `test` rules are now

$$\begin{aligned} \langle \texttt{test P } e_0\ e_1, \rho, \sigma, \hat{\kappa}\rangle, \Xi &\longmapsto \langle e_0, \rho, \sigma, \hat{\kappa}\rangle, \Xi \text{ if } \mathit{True} \in \widehat{\mathcal{OK}}(\Xi, P, \hat{\kappa}) \\ &\longmapsto \langle e_1, \rho, \sigma, \hat{\kappa}\rangle, \Xi \text{ if } \mathit{False} \in \widehat{\mathcal{OK}}(\Xi, P, \hat{\kappa}) \end{aligned}$$

where the a new $\widehat{\mathcal{OK}}$ function uses *terminal* and rewrite rules:

$$\begin{aligned} \widehat{\mathcal{OK}}(\Xi, P, \hat{\kappa}) &= \mathit{terminal}(\longmapsto, \widehat{\mathcal{OK}}^*(\Xi, P, \hat{\kappa})) \\ \widehat{\mathcal{OK}}^*(\Xi, \emptyset, \hat{\kappa}) &\longmapsto \mathit{True} \\ \widehat{\mathcal{OK}}^*(\Xi, P, \epsilon^m) &\longmapsto \mathit{pass?}(P, m) \\ \widehat{\mathcal{OK}}^*(\Xi, P, \phi{:}^m\tau) &\longmapsto \begin{cases} \widehat{\mathcal{OK}}(\Xi, P \setminus m^{-1}(\textbf{Grant}), \hat{\kappa}) & \text{if } \mathit{pass?}(P, m) \\ \mathit{False} & \text{otherwise} \end{cases} \\ &\quad \text{where } \hat{\kappa} \in \Xi(\tau) \end{aligned}$$

**Lemma 21** (Correctness of $\widehat{\mathcal{OK}}$). *For all* $\Xi, P, \kappa, \hat{\kappa}$,

- ***Soundness:*** *if* $\kappa \in \mathit{unroll}_\Xi(\hat{\kappa})$ *then* $\mathcal{OK}(P, \kappa) \in \widehat{\mathcal{OK}}(\Xi, P, \hat{\kappa})$.
- ***Local completeness:*** *if* $b \in \widehat{\mathcal{OK}}(\Xi, P, \hat{\kappa})$ *then there is a* $\kappa \in \mathit{unroll}_\Xi(\hat{\kappa})$ *such that* $b = \mathcal{OK}(P, \kappa)$.

*Proof.* Soundness follows by induction on the unrolling. Local completeness follows by induction on the trace from local completeness in Lemma 16. □

With this lemma in hand, the correctness proof is almost identical to the core proof of correctness.

**Theorem 22** (Correctness)**.** *The abstract semantics is sound and locally complete in the same sense as Theorem 14.*

## 4.4 RELAXING CONTEXTS FOR DELIMITED CONTINUATIONS

In Section 4.2 we showed how to get a pushdown abstraction by separating continuations from the value store. This separation breaks down when continuations themselves become values via first-class control operators. The glaring issue is that continuations become "storable" and relevant to the execution of functions. But, it was precisely the *irrelevance* that allowed the separation of $\sigma$ and $\Xi$. Specifically, the store components of continuations become elements of the store's codomain — a recursion that can lead to an unbounded state space and therefore a non-terminating analysis. We apply the AAM methodology to cut out the recursion; whenever a continuation is captured to go into the store, we allocate an address to approximate the store component of the continuation.

We introduce a new environment, $\chi$, that maps these addresses to the stores they represent. The stores that contain addresses in $\chi$ are then *open*, and must be paired with $\chi$ to be *closed*. This poses the same problem as before with contexts in storable continuations. Therefore, we give up some precision to regain termination by *flattening* these environments when we capture continuations. Fresh allocation still maintains the concrete semantics, but we necessarily lose some ability to distinguish contexts in the abstract.

### 4.4.1 *Case study of first-class control: shift and reset*

I choose to study `shift` and `reset` [23] because delimited continuations have proven useful for implementing web servers [75, 61], providing processes isolation in operating systems [50], representing computational effects [33], modularly implementing error-correcting parsers [86], and finally undelimited continuations are *passé* for good reason [49]. Even with all their uses, however, their semantics can yield control-flow possibilities that surprise their users. A *precise* static analysis that illuminates their behavior is then a valuable tool.

Our concrete test subject is the abstract machine for shift and reset adapted from Biernacki et al. [8] in the "**eval**, **continue**" style in Figure 18. The figure elides the rules for standard function calls. The new additions to the state space are a new kind of value, $\mathbf{comp}(\kappa)$, and a *meta-continuation*, $C \in \mathit{MKont} = \mathit{Kont}^*$ for separating continuations by their different prompts. Composable continuations are indistinguishable from functions, so even though the meta-continuation is concretely a list of continuations, its conses are notated as function composition: $\kappa \circ C$.

| $\varsigma \longmapsto_{SR} \varsigma'$ | |
|---|---|
| $\mathtt{ev}\ (\mathtt{reset}\ e, \rho, \sigma, \kappa, C)$ | $\mathtt{ev}\ (e, \rho, \sigma, \epsilon, \kappa \circ C)$ |
| $\mathtt{co}\ (\epsilon, \kappa \circ C, v, \sigma)$ | $\mathtt{co}\ (\kappa, C, v, \sigma)$ |
| $\mathtt{ev}\ (\mathtt{shift}\ x.e, \rho, \sigma, \kappa, C)$ | $\mathtt{ev}\ (e, \rho[x \mapsto a], \sigma', \epsilon, C)$ |
| where | $\sigma' = \sigma \sqcup [a \mapsto \mathbf{comp}(\kappa)]$ |
| $\mathtt{co}\ (\mathbf{fun}(\mathbf{comp}(\kappa'))\mathord{:}\kappa, C, v, \sigma)$ | $\mathtt{co}\ (\kappa', \kappa \circ C, v, \sigma)$ |

Figure 18: Machine semantics for shift/reset

### 4.4.2 *Reformulated with continuation stores*

The machine in Figure 18 is transformed now to have three new tables: one for continuations ($\Xi_\kappa$), one to close stored continuations ($\chi$), and one for meta-continuations ($\Xi_C$). The first is like previous sections, albeit continuations may now have the approximate form that is storable. The meta-continuation table is more like previous sections because meta-contexts are not storable. Meta-continuations do not have simple syntactic strategies for bounding their size, so I choose to bound them to size 0. They could be paired with lists of $\widehat{\mathit{Kont}}$ bounded at an arbitrary $n \in \mathbb{N}$, but I simplify for presentation.

Contexts for continuations are still at function application, but now contain the $\chi$. Contexts for meta-continuations are in two places: manual prompt introduction via `reset`, or via continuation invocation. At continuation capture time, continuation contexts are approximated to remove $\hat{\sigma}$ and $\chi$ components. The different context spaces are thus:

$$\begin{aligned}
\dot{\tau} \in \mathit{ExactContext} &::= \langle e, \rho, \hat{\sigma}, \chi \rangle \\
\hat{\tau} \in \widehat{\mathit{Context}} &::= \langle e, \rho, a \rangle \\
\tau \in \mathit{Context} &::= \hat{\tau} \mid \dot{\tau} \\
\gamma \in \mathit{MContext} &::= \langle e, \rho, \hat{\sigma}, \chi \rangle \mid \langle \tilde{\kappa}, \hat{v}, \hat{\sigma}, \chi \rangle
\end{aligned}$$

Revisiting the graphical intuitions of the state space, we have now $\tilde{\kappa}$ in states' stores, which represent an *overapproximation* of a set of continuations. We augment the illustration from Figure 15 in Figure 19 to include the new *CStore* and the overapproximating behavior of $\tilde{\kappa}$. The informal notation $\sigma \rightsquigarrow \tilde{\kappa}$ suggests that the state's store *contains*, or *refers to* some $\tilde{\kappa}$.

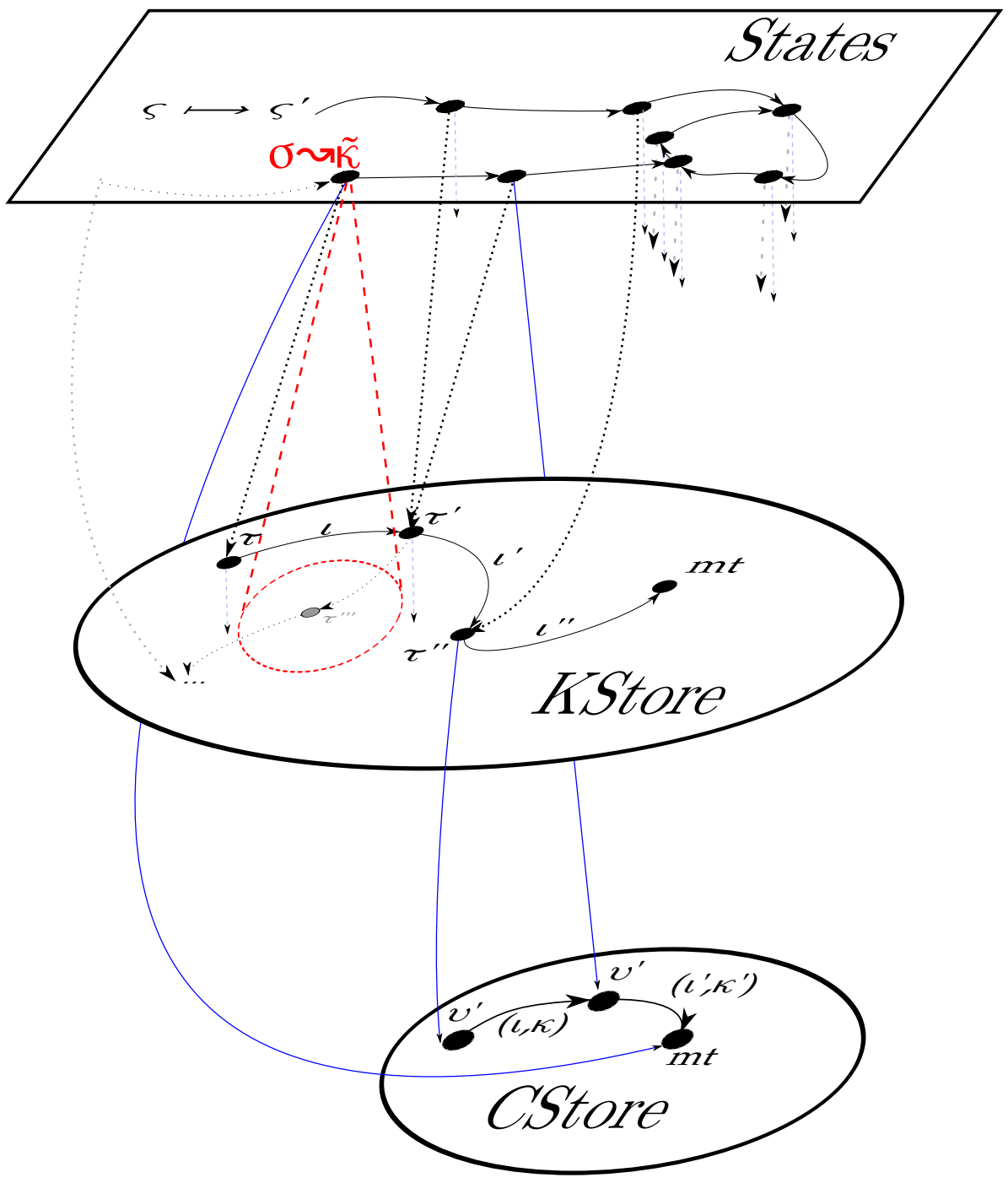

Figure 19: Graphical visualization of states, $\Xi_{\hat{\kappa}}$ and $\Xi_{\hat{C}}$.

$$
\begin{aligned}
\hat{\varsigma} \in \widehat{SR} &::= \mathsf{ev}\ (e, \rho, \hat{\sigma}, \chi, \hat{\kappa}, \hat{C}) \mid \mathsf{co}\ (\hat{\kappa}, \hat{C}, \hat{v}, \hat{\sigma}, \chi) \\
\mathit{State} &::= \hat{\varsigma}, \Xi_{\hat{\kappa}}, \Xi_{\hat{C}} \\
\chi \in \mathit{KClosure} &= \mathit{Addr} \underset{\text{fin}}{\rightharpoonup} \wp(\mathit{Store}) \\
\Xi_{\hat{\kappa}} \in \mathit{KStore} &= \mathit{ExactContext} \underset{\text{fin}}{\rightharpoonup} \wp(\widehat{\mathit{Kont}}) \\
\Xi_{\hat{C}} \in \mathit{CStore} &= \mathit{MContext} \underset{\text{fin}}{\rightharpoonup} \wp(\widehat{\mathit{Kont}} \times \widehat{\mathit{MKont}}) \\
\hat{\kappa} \in \widehat{\mathit{Kont}} &::= \epsilon \mid \phi{:}\tau \mid \tau \qquad \tilde{\kappa} \in \widetilde{\mathit{Kont}} ::= \epsilon \mid \hat{\tau} \\
\hat{C} \in \widehat{\mathit{MKont}} &::= \epsilon \mid \gamma \qquad\qquad \hat{v} \in \widehat{\mathit{Value}} ::= \tilde{\kappa} \mid (\ell, \rho)
\end{aligned}
$$

Figure 20: Shift/reset abstract semantic spaces

The approximation and flattening happens in $\mathbb{A}$:

$$\mathbb{A} : \mathit{KClosure} \times \mathit{Addr} \times \widehat{\mathit{Kont}} \to \mathit{KClosure} \times \widetilde{\mathit{Kont}}$$

$$\begin{aligned}
\mathbb{A}(\chi, a, \epsilon) &= \chi, \epsilon \\
\mathbb{A}(\chi, a, \phi{:}\tau) &= \chi', \phi{:}\hat{\tau} \text{ where } (\chi', \hat{\tau}) = \mathbb{A}(\chi, a, \tau) \\
\mathbb{A}(\chi, a, \langle e, \rho, \hat{\sigma}, \chi' \rangle) &= \chi \sqcup \chi' \sqcup [a \mapsto \hat{\sigma}], \phi{:}\langle e, \rho, a \rangle \\
\mathbb{A}(\chi, a, \langle e, \rho, b \rangle) &= \chi \sqcup [a \mapsto \chi(b)], \phi{:}\langle e, \rho, a \rangle
\end{aligned}$$

The third case is where continuation closures get flattened together. The fourth case is when an already approximate continuation is approximated: the approximation is inherited. Approximating the context and allocating the continuation in the store require two addresses, so we relax the specification of *alloc* to allow multiple address allocations in this case.

Each of the four rules of the original shift/reset machine has a corresponding rule that we explain piecemeal. I will use $\to$ for steps that do not modify the continuation stores for notational brevity. We use the above $\mathbb{A}$ function in the rule for continuation capture, as modified here.

$$\mathtt{ev}\ (\mathtt{shift}\ x.e, \rho, \hat{\sigma}, \chi, \hat{\kappa}, \hat{C}) \to \mathtt{ev}\ (e, \rho', \hat{\sigma}', \chi', \epsilon, \hat{C})$$

where

$$\begin{aligned}
(a, a') &= \mathit{alloc}(\hat{\varsigma}, \Xi_{\hat{\kappa}}, \Xi_{\hat{C}}) & \rho' &= \rho[x \mapsto a] \\
(\chi', \tilde{\kappa}) &= \mathbb{A}(\chi, a', \hat{\kappa}) & \hat{\sigma}' &= \hat{\sigma} \sqcup [a \mapsto \tilde{\kappa}]
\end{aligned}$$

The rule for `reset` stores the continuation and meta-continuation in $\Xi_{\hat{C}}$:

$$\begin{aligned}
\mathtt{ev}\ (\mathtt{reset}\ e, \rho, \hat{\sigma}, \chi, \hat{\kappa}, \hat{C}), \Xi_{\hat{\kappa}}, \Xi_{\hat{C}} &\longmapsto \mathtt{ev}\ (e, \rho, \hat{\sigma}, \chi, \epsilon, \gamma), \Xi_{\hat{\kappa}}, \Xi'_{\hat{C}} \\
\text{where } \gamma &= \langle e, \rho, \hat{\sigma}, \chi \rangle \\
\Xi_{\hat{C}} &= \Xi_{\hat{C}} \sqcup [\gamma \mapsto (\hat{\kappa}, \hat{C})]
\end{aligned}$$

The prompt-popping rule simply dereferences $\Xi_{\hat{C}}$:

$$\mathtt{co}\ (\epsilon, \gamma, \hat{v}, \hat{\sigma}, \chi) \to \mathtt{co}\ (\hat{\kappa}, \hat{C}, \hat{v}, \hat{\sigma}, \chi) \text{ if } (\hat{\kappa}, \hat{C}) \in \Xi_{\hat{C}}(\gamma)$$

The continuation installation rule extends $\Xi_{\hat{C}}$ at the different context:

$$\begin{aligned}
\mathtt{co}\ (\hat{\kappa}, \hat{C}, \hat{v}, \hat{\sigma}, \chi), \Xi_{\hat{\kappa}}, \Xi_{\hat{C}} &\longmapsto \mathtt{co}\ (\tilde{\kappa}, \gamma, \hat{v}, \hat{\sigma}, \chi), \Xi_{\hat{\kappa}}, \Xi'_{\hat{C}} \\
&\text{if } (\mathbf{appR}(\tilde{\kappa}), \hat{\kappa}') \in \mathit{pop}(\Xi_{\hat{\kappa}}, \chi, \hat{\kappa}) \\
\text{where } \gamma &= \langle \tilde{\kappa}, \hat{v}, \hat{\sigma}, \chi \rangle \\
\Xi_{\hat{C}} &= \Xi_{\hat{C}} \sqcup [\gamma \mapsto (\hat{\kappa}', \hat{C})]
\end{aligned}$$

Again we have a metafunction *pop*, but this time to interpret approximated continuations:

$$
\begin{aligned}
pop(\Xi_{\hat\kappa}, \chi, \hat\kappa) &= pop^*(\hat\kappa, \emptyset)\\
\text{where } pop^*(\epsilon, G) &= \emptyset\\
pop^*(\phi{:}\tau, G) &= \{(\phi, \tau)\}\\
pop^*(\tau, G) &= \bigcup_{\hat\kappa \in G'} (pop^*(\hat\kappa, G \cup G'))\\
\text{where } G' &= \bigcup_{\dot\tau \in I(\tau,\chi)} \Xi_{\hat\kappa}(\dot\tau) \setminus G\\
I(\dot\tau, \chi) &= \{\dot\tau\}\\
I(\langle e, \rho, a\rangle, \chi) &= \{\langle e, \rho, \hat\sigma, \chi'\rangle \in \mathbf{dom}(\Xi_{\hat\kappa}) \;:\; \hat\sigma \in \chi(a), \chi' \sqsubseteq \chi\}
\end{aligned}
$$

Notice that since we flatten $\chi$s together, we need to compare for containment rather than for equality (in I). A variant of this semantics with GC is available in the PLT redex models.

COMPARISON TO CPS TRANSFORM TO REMOVE `shift` AND `reset`: We lose precision if we use a CPS transform to compile away `shift` and `reset` forms, because variables are treated less precisely than continuations. Consider the following program and its CPS transform for comparison:

```
(let* ([id (λ (x) x)]
       [f (λ (y) (shift k (k (k y))))]
       [g (λ (z) (reset (id (f z))))])
  (⩽ (g 0) (g 1)))
```

```
(let* ([id (λ (x k) (k x))]
       [f (λ (y j) (j (j y)))]
       [g (λ (z h)
           (h (f z (λ (fv)
                     (id fv (λ (i) i))))))])
  (g 0 (λ (g0v) (g 1 (λ (g1v) (⩽ g0v g1v))))))
```

The $CESK_t^*\Xi$ machine with a monovariant allocation strategy will predict the CPS'd version returns true or false. In analysis literature, "monovariant" means variables get one address, namely themselves. Our specialized analysis for delimited control will predict the non-CPS'd version returns true.

$$\frac{}{\mathbf{appL}(e,\rho) \sqsubseteq_{\Xi,\chi} \mathbf{appL}(e,\rho)} \qquad \frac{v \sqsubseteq_{\Xi,\chi} \hat{v}}{\mathbf{appR}(v) \sqsubseteq_{\Xi,\chi} \mathbf{appR}(\hat{v})}$$

$$\frac{}{\epsilon \sqsubseteq \mathit{unroll}_{\Xi,\chi}(\epsilon)} \qquad \frac{\phi \sqsubseteq_{\Xi,\chi} \hat{\phi} \qquad \kappa \sqsubseteq \mathit{unroll}_{\Xi,\chi}(\tau)}{\phi{:}\kappa \sqsubseteq \mathit{unroll}_{\Xi,\chi}(\hat{\phi}{:}\tau)}$$

$$\frac{\hat{\kappa} \in \Xi(\dot{\tau}) \quad \kappa \sqsubseteq \mathit{unroll}_{\Xi,\chi}(\hat{\kappa})}{\kappa \sqsubseteq \mathit{unroll}_{\Xi,\chi}(\dot{\tau})} \qquad \frac{\dot{\tau} \in I(\Xi,\chi,\hat{\tau}) \quad \kappa \sqsubseteq \mathit{unroll}_{\Xi,\chi}(\dot{\tau})}{\kappa \sqsubseteq \mathit{unroll}_{\Xi,\chi}(\hat{\tau})}$$

$$\frac{}{\epsilon \sqsubseteq \mathit{unrollC}_{\Xi_{\hat{\kappa}},\Xi_{\hat{C}},\chi}(\epsilon)}$$

$$\frac{(\hat{\kappa},\hat{C}) \in \Xi_{\hat{C}}(\gamma) \qquad \kappa \sqsubseteq \mathit{unroll}_{\Xi_{\hat{\kappa}},\chi}(\hat{\kappa}) \qquad C \sqsubseteq \mathit{unrollC}_{\Xi_{\hat{\kappa}},\Xi_{\hat{C}},\chi}(\hat{C})}{\kappa \circ C \sqsubseteq \mathit{unrollC}_{\Xi_{\hat{\kappa}},\Xi_{\hat{C}},\chi}(\gamma)}$$

Figure 21: Order on (meta-)continuations

### 4.4.3 *Correctness*

We impose an order on values since stored continuations are more approximate in the analysis than in *SR*:

$$\frac{}{v \sqsubseteq_{\Xi,\chi} v} \qquad \frac{\kappa \sqsubseteq \mathit{unroll}_{\Xi,\chi}(\tilde{\kappa})}{\mathbf{comp}(\kappa) \sqsubseteq_{\Xi,\chi} \tilde{\kappa}} \qquad \frac{\forall v \in \sigma(a).\exists \hat{v} \in \hat{\sigma}(a).v \sqsubseteq_{\Xi,\chi} \hat{v}}{\sigma \sqsubseteq_{\Xi,\chi} \hat{\sigma}}$$

$$\frac{\kappa \sqsubseteq \mathit{unroll}_{\Xi_{\hat{\kappa}},\chi}(\hat{\kappa}) \qquad C \sqsubseteq \mathit{unrollC}_{\Xi_{\hat{\kappa}},\Xi_{\hat{C}},\chi}(\hat{C}) \qquad \sigma \sqsubseteq_{\Xi_{\hat{\kappa}},\chi} \hat{\sigma}}{\mathsf{ev}\ (e,\rho,\sigma,\kappa,C) \sqsubseteq \mathsf{ev}\ (e,\rho,\hat{\sigma},\chi,\hat{\kappa},\hat{C}),\Xi_{\hat{\kappa}},\Xi_{\hat{C}}}$$

$$\frac{v \sqsubseteq_{\Xi_{\hat{\kappa}},\chi} \hat{v} \qquad \kappa \sqsubseteq \mathit{unroll}_{\Xi_{\hat{\kappa}},\chi}(\hat{\kappa}) \qquad C \sqsubseteq \mathit{unrollC}_{\Xi_{\hat{\kappa}},\Xi_{\hat{C}},\chi}(\hat{C}) \qquad \sigma \sqsubseteq_{\Xi_{\hat{\kappa}},\chi} \hat{\sigma}}{\mathsf{co}\ (\kappa,C,v,\sigma) \sqsubseteq \mathsf{co}\ (\hat{\kappa},\hat{C},\hat{v},\hat{\sigma},\chi),\Xi_{\hat{\kappa}},\Xi_{\hat{C}}}$$

Unrolling differs from the previous sections because the values in frames can be approximate. Thus, instead of expecting the exact continuation to be in the unrolling, we have a judgment that an unrolling approximates a given continuation in Figure 21 (note we reuse $I$ from *pop**'s definition).

**Theorem 23** (Soundness)**.** *If* $\lozenge \longmapsto_{SR} \blacklozenge$*, and* $\lozenge \sqsubseteq \square$ *then there is* $\blacksquare$ *such that* $\square \longmapsto_{SRS\chi_t} \blacksquare$ *and* $\blacklozenge \sqsubseteq \blacksquare$.

FRESHNESS IMPLIES COMPLETENESS The high level proof idea is that fresh allocation separates evaluation into a sequence of bounded length paths that have the same store, but the store only grows and distinguishes contexts such that each continuation and metacontinuation have a unique unrolling. It is an open question whether the addition of garbage collection preserves completeness. Each context with the same store will have different expressions in them since expressions can only get smaller until a function call, at which point the store grows. This forms an order on contexts: smaller store means smaller context, and same store but smaller expression (indeed a subexpression) means a smaller context. Every entry in each enviroment $(\hat{\sigma}, \chi, \Xi_{\hat{\kappa}}, \Xi_{\hat{C}})$ will map to a unique element, and the continuation stores will have no circular references (the context in the tail of a continuation is strictly smaller than the context that maps to the continuation). There can only be one context that I maps to for approximate contexts because of the property of stores in contexts.

We distill these intuitions into an invariant about states that we will then use to prove completeness.

$$\frac{\begin{array}{c}\forall a \in \mathbf{dom}(\hat{\sigma}).\exists \hat{v}.\hat{\sigma}(a) = \{\hat{v}\} \wedge \hat{v} \preceq_{\chi} \Xi_{\hat{\kappa}} \\ \forall a \in \mathbf{dom}(\chi).\exists \hat{\sigma}'.\chi(a) = \{\hat{\sigma}'\} \wedge \hat{\sigma}' \in \pi_3(\mathbf{dom}(\Xi_{\hat{\kappa}})) \\ \forall \dot{\tau} \in \mathbf{dom}(\Xi_{\hat{\kappa}}).\exists \hat{\kappa}.\Xi_{\hat{\kappa}}(\dot{\tau}) = \{\hat{\kappa}\} \wedge \hat{\kappa} \sqsubset_{\chi}^{\Xi_{\hat{\kappa}}} \dot{\tau} \\ \forall \gamma \in \mathbf{dom}(\Xi_{\hat{C}}).\exists \hat{C}.\Xi_{\hat{C}}(\gamma) = \{\hat{C}\} \wedge \hat{C} \sqsubset \gamma\end{array}}{\mathit{inv}^*(\hat{\sigma}, \chi, \Xi_{\hat{\kappa}}, \Xi_{\hat{C}})}$$

$$\frac{\begin{array}{c}\mathit{inv}^*(\hat{\sigma}, \chi, \Xi_{\hat{\kappa}}, \Xi_{\hat{C}}) \qquad \langle e, \rho, \hat{\sigma}, \chi\rangle \sqsubset \mathbf{dom}(\Xi_{\hat{\kappa}}) \cup \mathbf{dom}(\Xi_{\hat{C}}) \\ (\exists \langle e_c, \rho, \hat{\sigma}, \chi\rangle \in \mathbf{dom}(\Xi_{\hat{\kappa}})) \implies e \in \mathit{subexpressions}(e_c) \\ \hat{\kappa} \preceq_{\chi} \Xi_{\hat{\kappa}} \qquad \hat{C} \preceq \Xi_{\hat{C}}\end{array}}{\mathit{inv}_{\mathit{fresh}}(\mathsf{ev}\ (e, \rho, \hat{\sigma}, \chi, \hat{\kappa}, \hat{C}), \Xi_{\hat{\kappa}}, \Xi_{\hat{C}})}$$

$$\frac{\mathit{inv}^*(\hat{\sigma}, \chi, \Xi_{\hat{\kappa}}, \Xi_{\hat{C}}) \qquad \hat{v} \preceq_{\chi} \Xi_{\hat{\kappa}} \qquad \hat{\kappa} \preceq_{\chi} \Xi_{\hat{\kappa}} \qquad \hat{C} \preceq \Xi_{\hat{C}}}{\mathit{inv}_{\mathit{fresh}}(\mathsf{co}\ (\hat{\kappa}, \hat{C}, \hat{v}, \hat{\sigma}, \chi), \Xi_{\hat{\kappa}}, \Xi_{\hat{C}})}$$

Where the order $\preceq$ states that any contexts in the (meta-)continuation are mapped in the given table.

$$\frac{}{(\ell, \rho) \preceq_{\chi} \Xi_{\hat{\kappa}}} \qquad \frac{}{\epsilon \preceq_{\chi} \Xi_{\hat{\kappa}}} \qquad \frac{}{\epsilon \preceq \Xi_{\hat{C}}} \qquad \frac{\dot{\tau} \in \mathbf{dom}(\Xi_{\hat{\kappa}})}{\dot{\tau} \preceq_{\chi} \Xi_{\hat{\kappa}}} \qquad \frac{\gamma \in \mathbf{dom}(\Xi_{\hat{C}})}{\gamma \preceq \Xi_{\hat{C}}}$$

$$\frac{\exists \hat{\sigma}.\chi(a) = \{\hat{\sigma}\} \qquad \exists! \chi'.\langle e, \rho, \hat{\sigma}, \chi'\rangle \in \mathbf{dom}(\Xi_{\hat{\kappa}}) \wedge \chi' \sqsubseteq \chi}{\langle e, \rho, a\rangle \preceq_{\chi} \Xi_{\hat{\kappa}}}$$

And the order $\sqsubset$ states that the contexts in the (meta-)continuation are strictly smaller than the given context.

$$\frac{}{\epsilon \sqsubset_{\chi}^{\Xi_{\hat{\kappa}}}} \quad \frac{}{\epsilon \sqsubset \gamma} \quad \frac{\tau \sqsubset_{\chi}^{\Xi_{\hat{\kappa}}} \dot{\tau}}{\phi{:}\tau \sqsubset_{\chi}^{\Xi_{\hat{\kappa}}} \dot{\tau}}$$

$$\frac{e' \in \mathit{subexpressions}(e)}{\langle e', \rho, \hat{\sigma}, \chi\rangle \sqsubset_{\chi}^{\Xi_{\hat{\kappa}}} \langle e, \rho, \hat{\sigma}, \chi\rangle} \quad \frac{e' \in \mathit{subexpressions}(e)}{\langle e', \rho, \hat{\sigma}, \chi\rangle \sqsubset \langle e, \rho, \hat{\sigma}, \chi\rangle}$$

$$\frac{\mathbf{dom}(\hat{\sigma}) \sqsubset \mathbf{dom}(\hat{\sigma}')}{\langle _, _, \hat{\sigma}, _\rangle \sqsubset_{\chi}^{\Xi_{\hat{\kappa}}} \langle _, _, \hat{\sigma}', _\rangle} \quad \frac{\mathbf{dom}(\hat{\sigma}) \sqsubset \mathbf{dom}(\hat{\sigma}')}{\langle _, _, \hat{\sigma}, _\rangle \sqsubset \langle _, _, \hat{\sigma}', _\rangle}$$

$$\frac{\forall \dot{\tau}' \in I(\Xi_{\hat{\kappa}}, \chi, \hat{\tau}) \qquad \dot{\tau}' \sqsubset_{\chi}^{\Xi_{\kappa}} \dot{\tau}}{\hat{\tau} \sqsubset_{\chi}^{\Xi_{\hat{\kappa}}} \dot{\tau}}$$

**Lemma 24** (Freshness invariant)**.** *If alloc produces fresh addresses, $inv_{fresh}(\hat{\varsigma}, \Xi_{\hat{\kappa}}, \Xi_{\hat{C}})$ and $\hat{\varsigma}, \Xi_{\hat{\kappa}}, \Xi_{\hat{C}} \longmapsto \hat{\varsigma}', \Xi'_{\hat{\kappa}}, \Xi'_{\hat{C}}$ then $inv_{fresh}(\hat{\varsigma}', \Xi'_{\hat{\kappa}}, \Xi'_{\hat{C}})$.*

*Proof.* By case analysis on the step. □

**Theorem 25** (Complete for fresh allocation)**.** *If alloc produces fresh addresses then the resulting semantics is complete with respect to states satisfying the invariant.*

*Proof sketch.* By case analysis and use of the invariant to exploit the fact the unrollings are unique and the singleton codomains pigeonhole the possible steps to only concrete ones. □

## 4.5 SHORT-CIRCUITING VIA "SUMMARIZATION"

All the semantics of previous sections have a performance weakness that many analyses share: unnecessary propagation. Consider two portions of a program that do not affect one another's behavior. Both can change the store, and the semantics will be unaware that the changes will not interfere with the other's execution. The more possible stores there are in execution, the more possible contexts in which a function will be evaluated. Multiple independent portions of a program may be reused with the same arguments and store contents they depend on, but changes to irrelevant parts of the store lead to redundant computation. The idea of skipping from a change past several otherwise unchanged states to uses of the change is called "sparseness" in the literature [76, 104, 73].

Memoization is a specialized instance of sparseness; the base stack may change, but the evaluation of the function does not, so given an already computed result we can jump straight to the answer. I use the vocabulary of "relevance" and "irrelevance" so that future work can adopt the ideas of sparseness to reuse contexts in more ways.

Recall the core notion of irrelevance: if we have seen the results of a computation before from a different context, we can reuse them. The semantic counterpart to this idea is a memo table that we extend when popping and appeal to when about to push. This simple idea works well with a deterministic semantics, but the nondeterminism of abstraction requires care. In particular, memo table entries can end up mapping to multiple results, but not all results will be found at the same time. Note the memo table space:

$$M \in \mathit{Memo} = \mathit{Context} \underset{\text{fin}}{\rightharpoonup} \wp(\mathit{Relevant})$$
$$\mathit{Relevant} ::= \langle e, \rho, \sigma \rangle$$

There are a few ways to deal with multiple results:

1. rerun the analysis with the last memo table until the table doesn't change (expensive),

2. short-circuit to the answer but also continue evaluating anyway (negates most benefit of short-circuiting), or

3. use a frontier-based semantics like in Section 4.2.2 with global $\Xi$ and $M$, taking care to

   a) at memo-use time, still extend $\Xi$ so later memo table extensions will "flow" to previous memo table uses, and

   b) when $\Xi$ and $M$ are extended at the same context at the same time, also create states that act like the $M$ extension point also returned to the new continuations stored in $\Xi$.

I will only discuss the final approach. The same result can be achieved with a one-state-at-a-time frontier semantics, but I believe this is cleaner and more parallelizable. Its second sub-point I will call the "push/pop rendezvous." The rendezvous is necessary because there may be no later push or pop steps that would regularly appeal to either (then extended) table at the same context. The frontier-based semantics then makes sure these pushes and pops find each other to continue on evaluating. In pushdown and nested word automata literature, the push to pop short-circuiting step is called a "summary edge" or with respect to the entire technique, "summarization." I find the memoization analogy appeals to programmers' and semanticists' operational intuitions.

A second concern for using memo tables is soundness. Without the completeness property of the semantics, memoized results in, *e.g.*, an inexactly GC'd machine, can have dangling addresses since the possible stacks may have grown to include addresses that were previously garbage. These addresses would not be garbage at first, since they must be mapped in the store for the contexts to coincide, but during the function evaluation the addresses can become garbage. If they are supposed to then be live, and are used (presumably they are reallocated post-collection), the analysis will miss paths it must explore for soundness. Thus we generalized context irrelevance to context congruence.

Context congruence is a property of the semantics *without* continuation stores, so there is an additional invariant to that of Section 4.2 for the semantics with $\Xi$ and $M$: $M$ respects context congruence. Contexts must carry enough information to define an *acceptability* proposition to apply context congruence. A context abstracts over a set of continuations, so all continuations in this set must be congruent to each other.

Let's abstract a bit from our specific representations with some named concepts. A context is extendable to a state in the following way:

$$\begin{aligned} &\mathit{extend} : \mathit{Context} \times \mathit{Kont} \to \mathit{CESK}_t \\ &\mathit{extend}(\langle e_c, \rho_c, \sigma_c \rangle, \kappa) = \langle e_c, \rho_c, \sigma_c, \kappa \rangle \end{aligned}$$

A result is plugged into a context to create a state in the following way:

$$\begin{aligned} &\mathit{plug} : \mathit{Relevant} \times \mathit{Kont} \to \mathit{CESK}_t \\ &\mathit{plug}(\langle e_r, \rho_r, \sigma_r \rangle, \kappa) = \langle e_r, \rho_r, \sigma_r, \kappa \rangle \end{aligned}$$

So if for $\longmapsto_M \subseteq \mathit{CESK}_t \times \mathit{CESK}_t$ the notion of acceptability is well-behaved,

$$\begin{aligned} &\mathit{ctx\text{-}congruent}(\longmapsto_M, \equiv_K), \text{ and} \\ &\forall \tau, \kappa, \kappa'. A(\tau, \kappa) \wedge A(\tau, \kappa') \implies \kappa \equiv_K \kappa' \end{aligned}$$

then we state the invariant on $M$ as follows:

$$\frac{}{\mathit{inv}_M(\bot)}$$

$$\frac{\begin{array}{c}\mathit{inv}_M(M) \\ \forall r \in R, \kappa. A(\tau, \kappa) \implies \exists \pi \equiv \mathit{extend}(\tau, \kappa) \longmapsto_M^* \mathit{plug}(r, \kappa). \mathit{ht}(\pi, \kappa)\end{array}}{\mathit{inv}_M(M[\tau \mapsto R])}$$

We can prove this invariant with appeals to the context congruence lemma and the $\Xi$ invariant to stitch together the trace.

Inexact GC does *not* respect context congruence for the same reasons it is not complete: some states are spurious due to inequivalent continuations' effect on GC. This means that some memo table entries will be spurious, and the expected path in the invariant will not exist. The reason we use unrolled continuations instead of simply $\epsilon$ for this (balanced) path is precisely for stack inspection reasons.

The rules in Figure 22 are the importantly changed rules from Section 4.2 that short-circuit to memoized results. The technique looks more like memoization with a $CESIK^*_t\Xi$ machine, since the memoization points are truly at function call and return boundaries. The *pop* function would need to also update M if it dereferences through a context, but otherwise the semantics are updated *mutatis mutandis*.

| $\hat{\varsigma}, \Xi, M \longmapsto \hat{\varsigma}', \Xi', M'$ | |
|---|---|
| $\langle (e_0\ e_1), \rho, \sigma, \hat{\kappa} \rangle, \Xi, M$ | $\langle e_0, \rho, \sigma, \mathbf{appL}(e_1, \rho){:}\tau \rangle, \Xi, M$ |
| | if $\tau \notin \mathbf{dom}(M)$, or |
| | $\langle e', \rho', \sigma', \hat{\kappa} \rangle, \Xi', M$ |
| | if $\langle e', \rho', \sigma' \rangle \in M(\tau)$ |
| where | $\tau = \langle (e_0\ e_1), \rho, \sigma \rangle$ |
| | $\Xi' = \Xi \sqcup [\tau \mapsto \hat{\kappa}]$ |
| $\langle v, \sigma, \mathbf{appR}(\lambda x.\ e, \rho){:}\tau \rangle, \Xi, M$ | $\langle e, \rho', \sigma', \hat{\kappa} \rangle, \Xi, M'$ if $\hat{\kappa} \in \Xi(\tau)$ |
| where | $\rho' = \rho[x \mapsto a]$ |
| | $\sigma' = \sigma \sqcup [a \mapsto v]$ |
| | $M' = M \sqcup [\tau \mapsto \langle e, \rho', \sigma' \rangle]$ |

Figure 22: Important memoization rules

$$\mathcal{F}_e(S, R, F, \Xi, M) = (S \cup F, R \cup R', F' \setminus S, \Xi', M')$$

where

$$\begin{aligned}
I &= \bigcup_{\varsigma \in F} \{(\langle \varsigma, \varsigma' \rangle, \Xi', M') \ : \ \varsigma, \Xi, M \longmapsto \varsigma', \Xi', M'\} \\
R' &= \pi_0 I \qquad \Xi' = \bigsqcup \pi_1 I \qquad M' = \bigsqcup \pi_2 I \\
\Delta\Xi &= \Xi' \setminus \Xi \quad \Delta M = M' \setminus M \\
F' &= \pi_1 R' \cup \{\langle e, \rho, \sigma, \hat{\kappa} \rangle : \tau \in \mathbf{dom}(\Delta\Xi) \cap \mathbf{dom}(\Delta M). \\
&\qquad \hat{\kappa} \in \Delta\Xi(\tau), \langle e, \rho, \sigma \rangle \in \Delta M(\tau)\}
\end{aligned}$$

The $\pi_i$ notation is for projecting out pairs, lifted over sets. This worklist algorithm describes unambiguously what is meant by "rendezvous." After stepping each state in the frontier, the differences to the $\Xi$ and $M$ tables are collected and then combined in $F'$ as calling contexts' continuations matched with their memoized results.

$$\text{REIFY}\epsilon \quad \frac{\langle\langle e,\rho,\sigma,\hat{\kappa}\rangle,\langle e',\rho',\sigma',\hat{\kappa}\rangle\rangle \in R \qquad \kappa \in \mathit{unroll}_{\Xi}(\hat{\kappa})}{\langle e,\rho,\sigma,\kappa\rangle \longmapsto_{\mathit{reifyM}(S,R,F,\Xi,M)} \langle e',\rho',\sigma',\kappa\rangle}$$

$$\text{REIFY}\pm \quad \frac{\langle\langle e,\rho,\sigma,\phi{:}\tau\rangle,\langle e',\rho',\sigma',\phi'{:}\tau\rangle\rangle \in R \qquad \hat{\kappa} \in \Xi(\tau) \qquad \kappa \in \mathit{unroll}_{\Xi}(\hat{\kappa})}{\langle e,\rho,\sigma,\phi{:}\kappa\rangle \longmapsto_{\mathit{reifyM}(S,R,F,\Xi,M)} \langle e',\rho',\sigma',\phi'{:}\kappa\rangle}$$

$$\text{REIFY+} \quad \frac{\langle\langle e,\rho,\sigma,\hat{\kappa}\rangle,\langle e',\rho',\sigma',\phi{:}\langle e,\rho,\sigma\rangle\rangle\rangle \in R \qquad \kappa \in \mathit{unroll}_{\Xi}(\hat{\kappa})}{\langle e,\rho,\sigma,\kappa\rangle \longmapsto_{\mathit{reifyM}(S,R,F,\Xi,M)} \langle e',\rho',\sigma',\phi{:}\kappa\rangle}$$

$$\text{REIFY-} \quad \frac{\begin{array}{c}\langle\langle e,\rho,\sigma,\mathbf{appR}(v){:}\tau\rangle,\langle e',\rho',\sigma',\hat{\kappa}'\rangle\rangle \in R \\ \hat{\kappa} \in \Xi(\tau) \qquad \kappa \in \mathit{unroll}_{\Xi}(\hat{\kappa})\end{array}}{\langle e,\rho,\sigma,\kappa\rangle \longmapsto_{\mathit{reifyM}(S,R,F,\Xi,M)} \langle e',\rho',\sigma',\kappa\rangle}$$

**Theorem 26** (Correctness)**.** *For all* $e_0$*, let* $\varsigma_0 = \langle e_0, \bot, \bot, \epsilon\rangle$ *in* $\forall n \in \mathbb{N}, \varsigma, \varsigma' \in \mathit{CESK}_t$*:*

- *if* $(\varsigma, \varsigma') \in \mathit{unfold}(\varsigma_0, \longmapsto_{\mathit{CESK}_t}, n)$ *then there is an* $m$ *such that* $\varsigma \longmapsto_{\mathit{reifyM}(\mathcal{F}^m_{e_0}(\bot))} \varsigma'$
- *if* $\varsigma \longmapsto_{\mathit{reifyM}(\mathcal{F}^n_{e_0}(\bot))} \varsigma'$ *then there is an* $m$ *such that* $(\varsigma, \varsigma')$ *is in* $\mathit{unfold}(\varsigma_0, \longmapsto_{\mathit{CESK}_t}, m)$

The proof appeals to the invariant on M whose proof involves an additional argument for the short-circuiting step that reconstructs the path from a memoized result using both context congruence and the table invariants.

Part II

# ALGORITHMIC CONSTRUCTIONS

# INTRODUCTION TO PART II: ALGORITHMIC CONSTRUCTIONS

*"Besides black art, there is only automation and mechanization."*
*~Federico Garcia Lorca*

The previous part of this dissertation set up a methodology for approaching analysis design. This part formalizes and automates the methodology with a language of abstractable semantics. Existing tools for writing semantics focus on executing the provided semantics in a more explicit model that offers no built-in support for abstraction [80, 32]. They are fine platforms on which to build AAM-style constructions, type-checkers, and program verifiers as in Nguyen et al. [69], Tobin-Hochstadt and Van Horn [92], Van Horn and Might [95], Johnson and Van Horn [44], Kuan et al. [55], Rosu [79]. Indeed the PLT Redex tool for semantics engineering inspired the language I describe in this part of the dissertation.

The process of translating a concrete semantics in Redex to its "AAM-ified" abstract semantics again in Redex is turn-the-crank, mindless, yet error prone grunt work. An "abstract semantics" in the AAM sense is still executable, and can be seen as concrete on a different level. Because the semantics is executable, Redex and the K framework are still valid tools to use, if unwieldy at times. The further transformations for more efficient interpretation add insult to injury, contorting one's prototype into the brittle mess they probably hoped to avoid.

A goal for our semantics is to be expressive and flexible enough to support standard abstract machines for concrete execution, and *abstract* abstract machines for abstract execution (analysis). The boundary between concrete and abstract is mitigated by user-provided allocation functions, for implicit and explicit resource allocation.

The first chapter in this part details the syntax and concrete semantics of a core Redex-like language. The second chapter adds the machinery necessary to soundly abstract allocation while still staying faithful to the concrete semantics if allocation happens to be fresh. The language natively supports inserting abstraction at all allocation points in order to guard against an unbounded state space. Further, supporting both concrete and abstract execution in one meta-semantics necessitates precise equality and object identity judgements. The core language supports both equality and object identity via a native cardinality analysis.

The final technical chapter is a case study that demonstrates the feasibility of automatic analysis construction for complicated semantics. The case study details a novel semantics for temporal higher-order contracts, written in the language previously described. The analysis my language produces is itself a novel research contribution.

# 5 A LANGUAGE FOR ABSTRACT MACHINES

As simple as the AAM method is to apply in many cases, the fact that a human is "the compiler" from a standard to non-standard machine makes the process error-prone. Translating large machines means a large opportunity for error, even though the translation is "simple." Worse, when a semantics needs a notion of object identity or structural equality, the naive AAM translation is unsound or uselessly imprecise. A better solution is to have a language of semantics that can interpret the meaning of an object language on programs both in the concrete and in the abstract.

The abstract machines we have seen in the previous chapters have all had a similar shape - a list of reduction rules like

$$\mathit{Pattern} \longmapsto \mathit{Template}\ [\mathit{optional\ side\text{-}condition}]$$

A meta-language for expressing a language's semantics with reduction rules of this shape must then have a semantics of pattern matching, side-condition evaluation and template evaluation. This chapter details a small language of reduction rules that can appeal to first-order metafunctions. I give the concrete semantics only, to help the reader understand the linguistic constructs before we dive into abstractions, Galois connections, and dragons.

The concrete semantics relies on external parameters for address allocation, data structure allocation, interpretation of "external values," etc. If any assumptions required for "concrete execution" are violated by these parameters, the semantics is undefined. These parameters are the knobs for altering the power for our eventual abstraction, so I introduce them in this chapter to get the reader comfortable with their placement.

## 5.1 REPRESENTING AN ABSTRACT MACHINE

An abstract machine in our sense is a collection of reduction rules for transforming machine states. A machine state ($\varsigma$) is

$$\varsigma \in \mathit{State} = \mathit{Store} \times \mathit{Term} \times \mathit{Time}$$

Each of these components has a role in the overall evaluation model.

TERM:

$t \in \mathit{Term} = \mathbf{Variant}(n, \bar{t}) \mid \mathbf{External}(E, v) \mid \mathbf{EAddr}(a) \mid \mathbf{IAddr}(a)$

where $E$ is a description of an external value space, like *Number*, and $v$ is a meta-meta-language value that $E$ uses to represent elements of the space.

A term is what a pattern manipulates and what an expression constructs. An object language's representation of a "state" and everything in it is a term. For example, since the store is already provided, the CESK machine's term will be a CEK tuple: $(\mathsf{ev}\ e\ \rho\ \kappa)$.

A *Term* is one of four (4) kinds:

1. **Variant:** a named $n$-tuple of terms;
2. **External:** a meta-meta-language value paired with an external space descriptor of the value space's operations (in the concrete, only equality);
3. **Explicit address:** an address (which can be anything) with an identity;
4. **Indirect address:** an address that stands for the term it maps in the store.

The language in which we express an abstract machine's reduction relation does not allow direct access to terms. Instead the language offers a language of expressions. A variant must be constructed out of subexpressions' evaluations, an external value may be written, an explicit address must be allocated, and an indirect address can only be created with the variant construction external parameter.

STORE:

$$\sigma \in \mathit{Store} = \mathit{Addr} \xrightarrow[\text{fin}]{} \mathit{Term}$$

The core ideas of AAM are those of address allocation and store-allocating data. In order to perform "the AAM transform," we need linguistic support to allocate addresses, update and read from a store.

TIME:

$\tau \in \mathit{Time}$ a user-provided set with no restrictions.

$\mathit{tick} : \mathit{State} \to \mathit{Time}$ a user-provided update function

The *Time* component is inherited from core AAM: it is user-provided to help guide allocation.

For example, in the concrete semantics of kCFA, the *Time* component is a list of all functions called, in the order that they were called. The list of functions called provides a distinguishing feature for storing binders in the store. As a result, no two function calls create the same $\langle \text{variable}, \tau \rangle$ pair, so allocation is "fresh."

## 5.2 DISCUSSION OF THE DESIGN SPACE

The language we develop in this and the following chapter is distinct from other semantics modeling systems in important ways. The first is that the language reuses its language of "rules" to allow intermediate computation via calls to metafunctions defined as a list of rules. The second is that we explicitly support a single-threaded store within the language. The third is that we have two flavors of address into the store: explicit and implicit.

Store-passing is by no means pleasant or impervious to error, but the main motivation to linguistically support a store is to help automatically abstract it à la AAM. With stores come addresses. Explicit addresses are for direct use by rules; they are what one usually thinks of an address. Implicit addresses are a tool that we use to hide that some nested data structures are actually threaded through the store. For now, our established familiarity with AAM's strategy to "store-allocate recursion" should be enough foreshadowing to motivate implicit addresses. We will see a longer discussion of the utility of implicit addresses in Section 6.5.

Abstract machines have a "small-step" semantics, meaning we represent intermediate computation with explicit state transitions as governed by "rules." But, unlike in term reduction systems [52], some intermediate computation can be hidden with function calls. Therefore we allow rules to call functions in order to compute the machine's overall "next step," even if the metafunction takes several steps of computation. We call such functions "metafunctions" because of their role in expressing a semantics. For example, the CM machine in Section 4.3.2 has a rule that uses the metafunction $\mathcal{OK}$ to recursively crawl the store for access control information.

Recursive metafunctions are written as ordered conditional reduction rules on terms that represent the call to the function, *e.g.* factorial looks like[14]

$$\begin{aligned}
\texttt{factorial} &: \texttt{Nat} \to \texttt{Nat} \\
\texttt{(factorial (Zero))} &\longmapsto \texttt{(S (Zero))} \\
\texttt{(factorial } n\texttt{)} &\longmapsto \textbf{Call}(\texttt{mult}, \langle n, \textbf{Call}(\texttt{factorial}, \langle m \rangle)\rangle) \\
&\quad \textbf{where } \texttt{(S } m\texttt{)} = n \\
&\quad \textbf{and } n \in \texttt{Nat ::= (Zero) | (S } n\texttt{)}
\end{aligned}$$

[14] *An abbreviated form for illustrative purposes.*

**Call** is the metalanguage's built-in metafunction for interpreting user-defined metafunctions. A **Call** uses the named function's associated reduction rules to rewrite the **Call** to the function's result. Note the big-step flavor of this - the reason is that these are metafunctions for a semantics description and are therefore expected to be total. We can understand functions as out-of-band rewriting rules for intermediate computation. A reduction rule can thus refer to the output of a recursive metafunction without the metafunction evaluation contributing to any of the machine's trace history. Metafunctions' reduction rules are additionally, as a nicety, viewed as a top-level pattern match: rules are applied in order, and stop when a rule applies.

CALL SYNTAX The asymmetry between the function pattern and function call serves to syntactically distinguish variants (*e.g.*, `Zero` and `S`) and functions. When we create a variant **Variant**$(n, \langle t \ldots \rangle)$, the meta-semantics invokes an external parameter to create an alternative representation of the variant that is equivalent to **Variant**$(n, \langle t \ldots \rangle)$. When we introduce approximation, the external parameter may choose to abstract its subterms to curtail any unbounded nesting.

When we call a function, the arguments are packaged into a variant sharing the same name as the function and are then immediately destructured by the rewrite rules. I use this strategy to reuse the matching machinery for function evaluation. Function calls themselves are not data structures, so we don't need to worry about allocating space for them. The syntax distinguishes calls and variant construction to draw attention to their different allocation behavior.

## 5.3 THE GRAMMAR OF PATTERNS AND RULES

In the previous part of this dissertation, we've seen some examples of abstract machine rules. For instance, an abstract machine rule can be unconditional:

CESK variable lookup:

$$\texttt{(ev}\ x\ \rho\ \sigma\ \kappa\texttt{)} \longmapsto \texttt{(co}\ \kappa\ \texttt{(lookup}\ \sigma\ \rho(x)\texttt{)}\ \sigma\texttt{)}$$

conditional:

Vector reference:

$$\texttt{(ap vector-ref}\ \langle \texttt{vec}(s, vs), i, v\rangle\ \sigma\ \kappa\texttt{)} \longmapsto \texttt{(co}\ \kappa\ \sigma(vs(i))\ \sigma\texttt{)}$$
$$\text{where } 0 \leqslant i \leqslant |vs|$$

side-effecting:

Box update:

$$\texttt{(ap set-box!}\ \langle \texttt{boxv}(a), v\rangle\ \sigma\ \kappa\texttt{)} \longmapsto \texttt{(co}\ \kappa\ \texttt{void}\ \sigma[a \mapsto v]\texttt{)}$$

and can even appeal to metafunctions:

Stack-based security:

$$(\texttt{ev test P } e_0\ e_1\ \rho\ \sigma\ \kappa) \longmapsto (\texttt{ev } e_0\ \rho\ \sigma\ \kappa)$$
$$\text{where } \texttt{tt} = \texttt{Call}(\mathcal{OK}, \texttt{P}, \kappa)$$

Our metalanguage must therefore support side-conditions, store updates, and calls to metafunctions. To evaluate a rule on a machine state, we first *match* the left hand side, *apply* the side-conditions, and if the side-conditions don't fail, we further *evaluate* the right hand side. The informal *Pattern* $\longmapsto$ *Template*[*optional side-conditions*] mental model of rules is replaced with a generalized form where the *Template* can perform more computation than simply fill in the holes of a templated term. The notion of a *Template* is generalized to a simple language of *Expression* that includes metafunction calls. Side-conditions are written using the same expression language.

A PATTERN is like a term with named holes and simple predicates to match shape. There are five (5) pattern kinds that fit into three categories: predicate, binding, and structure.

There are three (3) predicate patterns. They are "predicate" patterns because they check some property of a term in order to match:

1. **Wild**: matches anything (also written _);
2. **Is-Addr**: matches any explicit address;
3. **Is-External**$(E)$ matches any **External**$(E, v)$ term.

There is one (1) binding pattern, which puns as both reference and binding. For simplicity, shadowing is not allowed.

4. **Name**$(x, p)$: binds a metavariable to a term matching the given pattern. If the variable is already bound, the matched term must be equal to the term already bound.

A pattern that contains the an already bound variable, or the same variable more than once, is called a *non-linear* pattern.

There is one (1) structure pattern that match terms structurally:

5. **Variant**$(n, \langle p \ldots \rangle)$ matches a term **Variant**$(n, \langle t \ldots \rangle)$ where $p \ldots$ and $t \ldots$ are the same length and match pairwise, where any metavariables bound in $p_i$ are in scope for following patterns.

The *scope* of a metavariable is the set of patterns and expressions in which that metavariable may be referenced. In this language, scope extends in a tree postorder traversal of a pattern to the following expressions and side conditions. For example, in the following rule,

```
(X (Y n) n) ⟼ (Z n m)
              [where m (W n 0)]
              [where tt (Call test m n)]
```

$$
\begin{aligned}
p \in \mathit{Pattern} &::= \mathbf{Name}(x, p) \mid \mathbf{Variant}(n, \overline{p}) \mid \mathbf{Is\text{-}External}(E) \\
&\quad \mid \mathbf{Is\text{-}Addr} \mid \mathbf{Wild} \\
e \in \mathit{Expr} &::= \mathbf{Ref}(x) \mid \mathbf{Variant}(n, \mathit{tag}, \overline{e}) \mid \mathbf{Let}(\overline{\mathit{bu}}, e) \mid \mathbf{Call}(f, \overline{e}) \\
&\quad \mid \mathbf{Alloc}(\mathit{tag}) \mid \mathbf{Deref}(e) \\
\mathit{rule} \in \mathit{Rule} &::= \mathbf{Rule}(p, e, \overline{\mathit{bu}}) \\
\mathit{bu} \in \mathit{BU} &::= \mathbf{Where}(p, e) \mid \mathbf{Update}(e, e) \\
\mathit{MF} &::= \mathbf{User}(\overline{\mathit{rule}}) \mid \mathbf{ExtMF}(\mathit{emf}) \\
\mathit{emf} &\in \mathit{State} \times \mathit{Term}^* \to \mathit{EvRes}[\mathit{Term}] \text{ in meta-meta-language} \\
x &\in \mathit{Metavariable} \text{ some set of names} \\
n &\in \mathit{Variant\text{-}Name} \text{ some set of names} \\
f &\in \mathit{Metafunction\text{-}Name} \text{ some set of names} \\
\mathit{tag} &\in \mathit{Tag} \text{ some set}
\end{aligned}
$$

Figure 23: Patterns, expressions and rules

The $n$ metavariable is in scope for the side conditions' expressions and the right-hand-side, and the $m$ metavariable is in scope for following side conditions and the right-hand-side expression. The `tt` pattern is a variant representing "true" and does not bind any metavariables.

AN EXPRESSION is a control string that ultimately creates a term. An expression can call a metafunction, allocate an address, or construct a variant, or dereference the store. With the **Let** form, an expression may perform pattern matching and update the store before evaluating the **Let** body expression. Both pattern matching and store update do not result in a term. We therefore classify them as a separate type called a "binding/update" (*BU*). A rule's side conditions are also expressed using binding/updates.

Address allocation and variant construction are guided by external parameters. These two forms thus carry an arbitrary *tag* to distinguish the forms for the external parameters to recognize[15].

[15] *A default tag is the tree address of the expression through the description of the entire semantics.*

A RULE is like a pattern match "clause" in a language with pattern matching. A rule consists of a left-hand-side pattern, a list of side-conditions, and the right-hand-side expression. We say a rule "fires" on a term if both the rule's pattern match succeeds, and each of the side-conditions' pattern matches succeed. The result of a rule is the evaluation of the right-hand-side expression in the environment of both the left-hand-side pattern match, and all the pattern matches of the side conditions. The entire grammar of (the abstract syntax of) the language is shown in Figure 23.

## 5.4 TERM EQUALITY

Concrete equality defines structural equality of two concrete terms in a store. Equality results in a yes or no answer.

The difficulty with equality is that terms can be cyclic due to addresses. Equality of cyclic terms is logically equivalent to the equality of infinite terms. Thus, equality in our setting is a coinductive proposition.

The usual trick to deciding coinductive propositions is to build a set of "guarded truths." In a sense, this is a set of hedges: if the original proposition is true (which we don't know yet), then all guarded truths are true; if the original proposition is false, then the "guarded truths" imply a falsehood. Operationally, we attempt to derive more guarded truths until there is nothing more to derive (indicating consistency), or we derive a falsehood (indicating our hedges don't actually hold). If we ever derive a falsehood, we know the original proposition (concrete equality of two terms) must have been false.

The magic of coinduction is that if we ever need to prove what we are trying to prove while we're proving it, then we've proved it[16]. To visualize coinductive term equality graphically, consider that when we've seen the same two terms while deciding equality, we've found a cycle back to earlier in the term graph. The presence of this cycle means that (for the path we followed) the equality of two terms does not imply a falsehood. If we find that all structural comparisons we make lead to either a bottomed out recursion or such cycle detection, then the set of "guarded truths" is justified for later use.

[16] *Terms and conditions may apply.*

The result type for the workhorse of concrete equality is the following:

$$\begin{aligned} \mathit{EqRes} &= \text{option } \mathit{Pairs} \\ \mathit{ps} \in \mathit{Pairs} &= \wp(\mathit{Term} \times \mathit{Term}) \end{aligned}$$

The set of term pairs is our set of "guarded propositions of concrete term equality." We thread the set through subsequent equality tests for possible cycle detection. If we find reason to contradict our set of truths, for example we see $0 = 1$, then we throw the set away and return **None**. Consequently, if the overall result of equality is **None**, then the two terms are not equal. If the result is some set of truths, the two terms are equal.

We use $\mathit{guard}(\sigma)(t_0, t_1, \mathit{ps})$ to guard against cycles when comparing $t_0$ and $t_1$ for equality, where $\mathit{ps}$ is a set of term pairs. If $\langle t_0, t_1 \rangle \in \mathit{ps}$, then the two terms are coinductively equal. If a pair of terms is previously unseen, we add it to the set and continue the structural comparison in a helper function, $\mathit{tequal}^*$.

Figure 24 shows the definition of concrete term equality. The definition sometimes uses `do` notation for easily manipulating the *option* (AKA *Maybe*) monad. The *ps* set is threaded through all successful

$$\begin{aligned}
tequal(\sigma)(t_0, t_1) &= \texttt{case}\ guard(\sigma)(t_0, t_1, \emptyset)\ \texttt{of} \\
&\qquad \mathbf{Some}(ps) : \texttt{tt} \\
&\qquad \mathbf{None} : \texttt{ff}
\end{aligned}$$

$$\text{where } guard : Store \to Term \times Term \times Pairs \to EqRes$$

$$\begin{aligned}
guard(\sigma)(t_0, t_1, ps) &= \texttt{if}\ (t_0, t_1) \overset{?}{\in} ps\ \texttt{then} \\
&\qquad \mathbf{Some}(ps) \\
&\qquad \texttt{else}\ tequal^*(\sigma)(t_0, t_1, ps \cup \{(t_0, t_1)\})
\end{aligned}$$

Let **Ex** abbreviate **External** and $\mathcal{V}abs$ abbreviate **Variant**.

$$tequal^* : Store \to Term \times Term \times Pairs \to EqRes$$

$$\begin{aligned}
tequal^*(\sigma)(\mathbf{EAddr}(a), \mathbf{EAddr}(a), ps) &= \mathbf{Some}(ps) \\
tequal^*(\sigma)(\mathbf{IAddr}(a), t_1, ps) &= guard(\sigma)(\sigma(a), t_1, ps) \\
tequal^*(\sigma)(t_0, \mathbf{IAddr}(a), ps) &= guard(\sigma)(t_0, \sigma(a), ps) \\
tequal^*(\sigma)(\mathbf{Ex}(E, v_0), \mathbf{Ex}(E, v_1), ps) &= E.\equiv (\sigma, v_0, v_1, ps) \\
tequal^*(\sigma)(\mathbf{V}(n, \mathbf{t}), \mathbf{V}(n, \mathbf{t}'), ps) &= V_{=}(\sigma)(\mathbf{t}, \mathbf{t}')(ps) \\
tequal^*(\sigma)(t_0, t_1, ps) &= \mathbf{None} \text{ otherwise}
\end{aligned}$$

$$\begin{aligned}
\text{where } V_{=} &: Store \to Term^* \times Term^* \to Pairs \to EqRes \\
V_{=}(\sigma)(\langle\rangle, \langle\rangle)(ps) &= \mathbf{Some}(ps) \\
V_{=}(\sigma)(t_0\mathbf{t}, t_0'\mathbf{t}')(ps) &= \texttt{do}\ ps' \leftarrow guard(\sigma)(t_0, t_0', ps) \\
&\qquad V_{=}(\sigma)(\mathbf{t}, \mathbf{t}')(ps') \\
V_{=}(\sigma)(_, _)(_) &= \mathbf{None} \text{ otherwise}
\end{aligned}$$

Figure 24: Concrete term equality

equality checks because the whole judgment of equality is not yet finished. A visual to keep in mind is a more traditional judgment derivation; the higher up we are in the derivation, the larger the set of term pairs is. An answer of **None** means that no judgment derivation of equality exists.

Equality of address terms depends on their equality modalities. Identity compares for the same actual address.

External equality is trusted to do The Right Thing. Variants compare pointwise with the helper $V_{=}$, carefully threading through the term pairs; mismatched lengths are caught by failure to match both empty or both non-empty lists of terms. The $V_{=}$ function is curried for a cleaner correctness proof.

## 5.5 PATTERN MATCHING

A pattern can match a term in at most one way. If a pattern matches, then its result is the extended environment of bindings. An indirect address is automatically dereferenced if it is either bound via **Name**,

$$
\begin{aligned}
M &: \mathit{Store} \to \mathit{Pattern} \times \mathit{Term} \times \mathit{MEnv} \to \mathit{MRes} \\
M(\sigma)(\mathbf{Name}(x, p), t, \rho) &= \texttt{if } x \overset{?}{\in} \mathbf{dom}(\rho) \texttt{ then} \\
&\qquad \texttt{if } \mathit{tequal}(\sigma)(\rho(x), t) \texttt{ then} \\
&\qquad\quad M(\sigma)(p, t, \rho) \\
&\qquad \texttt{else } \mathbf{None} \\
&\quad \texttt{else } M(\sigma)(p, t, \rho[x \mapsto \mathit{demand}(\sigma, t)]) \\
M(\sigma)(\mathbf{Wild}, t, \rho) &= \mathbf{Some}(\rho) \\
M(\sigma)(\mathbf{Is\text{-}Addr}, \mathbf{EAddr}(_), \rho) &= \mathbf{Some}(\rho) \\
M(\sigma)(\mathbf{Is\text{-}External}(E), \mathbf{External}(E, _), \rho) &= \mathbf{Some}(\rho) \\
M(\sigma)(\mathbf{Variant}(n, \overline{p}), \mathbf{Variant}(n, \overline{t}), \rho) &= V_M(\sigma)(\overline{p}, \overline{t}, \rho) \\
M(\sigma)(p, \mathbf{IAddr}(a), \rho) &= M(\sigma)(p, \sigma(a), \rho) \\
M(\sigma)(p, _, \rho) &= \mathbf{None} \\
\text{where } V_M &: \mathit{Store} \to \mathit{Pattern}^* \times \mathit{Term}^* \times \mathit{MEnv} \to \mathit{MRes} \\
V_M(\sigma)(\epsilon, \epsilon, \rho) &= \mathbf{Some}(\rho) \\
V_M(\sigma)(p_0\overline{p}, t_0\overline{t}, \rho) &= \texttt{do } \rho' \leftarrow M(\sigma)(p_0, t_0, \rho) \\
&\qquad V_M(\sigma)(\overline{p}, \overline{t}, \rho') \\
V_M(\sigma)(_, _, \rho) &= \mathbf{None}
\end{aligned}
$$

Figure 25: Pattern matching

or a pattern needs to inspect it. We call inspecting a term *demanding* the term. We have a helper function to that effect:

$$
\begin{aligned}
\mathit{demand} &: \mathit{Store} \times \mathit{Term} \to \mathit{Term} \\
\mathit{demand}(\sigma, \mathbf{IAddr}(a)) &= \sigma(a) \\
\mathit{demand}(\sigma, t) &= t
\end{aligned}
$$

The pattern matcher is defined in Figure 25. The result type is either **Some** extended metalanguage binding environment, or **None** to signify no match exists:

$$
\begin{aligned}
\mathit{MRes} &= \text{option } \mathit{MEnv} \\
\rho \in \mathit{MEnv} &= \mathit{Metavariable} \to \mathit{Term}
\end{aligned}
$$

## 5.6 EXPRESSION EVALUATION

Once a rule's left hand side matches, the right hand side can perform some computation to produce the following term. The computation language is our small grammar of expressions.

We evaluate an expression with *Ev* defined in Figure 26. If evaluation gets stuck, say if a **Let** binding has a failed match, then *Ev*

returns **None**. Otherwise, evaluation completes with both a term and an updated store.

For reader clarity, we hide the store-passing and failure (stuckness) in the *MaybeState* monad. An evaluation result is therefore

$$\mathit{EvRes}[T] = \mathit{Store} \to \mathit{MaybeState}(\mathit{Store}, T)$$

where the underlying data structure is

$$\mathit{MaybeState}(\mathit{Store}, T) = \textbf{None} \mid \textbf{Some}\langle\sigma, T\rangle$$

with operations

$$\begin{aligned}
\mathit{return}(a) &= \lambda\sigma.\textbf{Some}(\langle\sigma, a\rangle) \\
\mathit{fail}() &= \lambda\sigma.\textbf{None} \\
\mathit{bind}(\textbf{Some}(\langle\sigma, a\rangle), f) &= \lambda_.f(\sigma)(a) \\
\mathit{bind}(\textbf{None}, f) &= \lambda_.\textbf{None} \\
\textbf{MVariant}(\varsigma, n, \mathit{tag}, \bar{t}, \rho) &= \lambda\sigma.\textbf{Some}(\mathit{mkV}(\sigma, n, \mathit{tag}, \bar{t}, \rho))
\end{aligned}$$

We see in the definition of expression evaluation (Figure 26) that address allocation takes all of the store, the tag, and the current environment in order to compute the address. Allocation cannot side-effect the store, as a result. However, variant construction with the external parameter, *mkV*, can. The *mkV* parameter, as we will see in the next chapter, is a critical component to provide control over the state space abstraction.

A **Let** expression can both locally bind metavariables by using pattern-matching, and globally update the store.

The binding/update forms in **Let** are also used for side-conditions, so we have three result kinds rather than expression evaluation's two. If any expression evalution in a binding/update form gets stuck, then the whole form is stuck. A stuck side-condition is an error in the semantics description, and brings evaluation to a grinding halt: the rule application itself is considered stuck. If a **Where** form's pattern doesn't match, then the side-condition should signal a rule is "unapplicable," and we should try the next rule. Stuckness is distinct from applicability. Rule- and side-condition evaluation therefore have three possible outcomes:

$$\mathit{Rule\text{-}result}(a) ::= \textbf{Stuck} \mid \textbf{Unapplicable} \mid \textbf{Fires}(\sigma, a)$$

Another way to think about this trichotomy is that it encodes a behavior that is emergent in a compilation of "rules in order" to a "set of rules." In the set interpretation of the list of rules, say $\bar{r}, r_0, \overline{r'}$, the side conditions on $r_0$ have as a precondition the negation of all of $\bar{r}$'s left-hand-side patterns and side-conditions. In the compiled form, the order we try the rules doesn't matter, but any stuck side condition means that the rest of the rule won't evaluate.

$$
\begin{aligned}
\mathit{Ev} &: \mathit{State} \to \mathit{Expr} \times \mathit{MEnv} \to \mathit{EvRes}[\mathit{Term}] \\
\mathit{Ev}(\varsigma)(\mathbf{Ref}(x), \rho) &= \mathit{return}(\rho(x)) \\
\mathit{Ev}(\varsigma)(\mathbf{Alloc}(\mathit{tag}), \rho) &= \texttt{do}\ \ \sigma \leftarrow \texttt{get} \\
&\qquad \mathit{return}(\mathbf{EAddr}(\mathit{alloc}(\varsigma, \sigma, \mathit{tag}, \rho))) \\
\mathit{Ev}(\varsigma)(\mathbf{Variant}(n, \mathit{tag}, \overline{e}), \rho) &= \texttt{do}\ \ \overline{t} \leftarrow \overline{\mathit{Ev}}(\varsigma)(\overline{e}, \rho) \\
&\qquad \mathbf{MVariant}(\varsigma, n, \mathit{tag}, \overline{t}, \rho) \\
\mathit{Ev}(\varsigma)(\mathbf{Let}(\overline{\mathit{bu}}, e), \rho) &= \texttt{do}\ \ \rho' \leftarrow \overline{\mathit{Ev}_{\mathit{bu}}}(\varsigma)(\overline{\mathit{bu}})(\rho) \\
&\qquad \mathit{Ev}(\varsigma)(e, \rho') \\
\mathit{Ev}(\varsigma)(\mathbf{Call}(f, \overline{e}), \rho) &= \texttt{do}\ \ \overline{t} \leftarrow \overline{\mathit{Ev}}(\varsigma)(\overline{e}, \rho) \\
&\qquad \mathit{Ev}_{\mathit{mf}}(\varsigma)(f, \overline{t}) \\
\mathit{Ev}(\varsigma)(\mathbf{Deref}(e), \rho) &= \texttt{do}\ \ t \leftarrow \mathit{Ev}(\varsigma)(e, \rho) \\
&\qquad \texttt{case t of} \\
&\qquad\quad \mathbf{EAddr}(a) : \texttt{do}\ \ \sigma \leftarrow \texttt{get} \\
&\qquad\qquad\qquad \mathit{return}(\sigma(a)) \\
&\qquad\quad _ : \mathit{fail}() \\
\overline{\mathit{Ev}} &: \mathit{State} \to \mathit{Expr}^* \times \mathit{MEnv} \to \mathit{EvRes}[\mathit{Term}^*] \\
\overline{\mathit{Ev}}(\varsigma)(\epsilon, \rho) &= \mathit{return}(\epsilon) \\
\overline{\mathit{Ev}}(\varsigma)(e_0\overline{e}, \rho) &= \texttt{do}\ \ t_0 \leftarrow \mathit{Ev}(\varsigma)(e_0, \rho) \\
&\qquad\ \ \overline{t} \leftarrow \overline{\mathit{Ev}}(\varsigma)(\overline{e}, \rho) \\
&\qquad\ \ \mathit{return}(t_0\overline{t})
\end{aligned}
$$

Figure 26: Expression evaluation

A **Let** expression treats **Unapplicable** and **Stuck** as synonymous, since it is not "top-level." The *return*, *fail*, **stuck**, and *bind* operations for *Rule-result* are the following:

$$\begin{aligned}
\mathit{bind}(\mathbf{Stuck}, f) &= \lambda_.\mathbf{Stuck} \\
\mathit{bind}(\mathbf{Unapplicable}, f) &= \lambda_.\mathbf{Unapplicable} \\
\mathit{bind}(\mathbf{Fires}(\sigma, a), f) &= \lambda_.f(\sigma)(a) \\
\mathit{return}(a) &= \lambda\sigma.\mathbf{Fires}(\sigma, a) \\
\mathit{fail}() &= \lambda\sigma.\mathbf{Unapplicable} \\
\mathbf{stuck}() &= \lambda\sigma.\mathbf{Stuck}
\end{aligned}$$

When in the *Maybe* monad, we implicitly treat non-**Fires** as **None** to avoid notational bloat. Likewise, in the *Rule-result* monad, we implicitly treat **None** as **Stuck**, and **Some** as **Fires**.

METAFUNCTION EVALUATION $\mathit{Ev}_{\mathit{mf}}$, applies its user-provided rules in order until it reaches a result or gets stuck. When we try apply a list of rules in order (until one fires), we are only concerned with stuckness or firedness. Therefore, $\mathit{Ev}_{\overline{\mathit{rule}}}$ returns an $\mathit{EvRes}[\mathit{Term}]$. Rule evaluation and metafunction evaluation are defined in Figure 28. All metafunction calls depend on the meta-meta-language's runtime stack to match calls with returns. In other words, a metafunction returns when $\mathit{Ev}_{\mathit{mf}}$ returns.

Each metafunction must be named. The semantics takes as a parameter a map $\Xi : \mathit{Metafunction\text{-}Name} \underset{\text{fin}}{\rightharpoonup} \mathit{MF}$. An *MF* is the meaning of a metafunction, which is either a list of rules (within the language), or a meta-meta-language function that consumes the current machine state and outputs a "result." All metafunctions are in each other's scope, so general recursion is possible within expression evaluation. The language makes no restrictions to force totality, but does distinguish divergence from stuckness.

A call to metafunction $f$ with arguments $\bar{t}$ creates a variant $\mathbf{Variant}(f, \bar{t})$ for the rules to interpret.

$$
\begin{aligned}
\mathit{Ev}_{bu} &: \mathit{State} \to \mathit{BU} \times \mathit{MEnv} \to \mathit{Rule\text{-}Result}(\mathit{MEnv}) \\
\mathit{Ev}_{bu}(\varsigma)(\epsilon, \rho) &= \mathit{return}(\rho) \\
\mathit{Ev}_{bu}(\varsigma)(\mathbf{Where}(p, e), \rho) &= \texttt{do}\ \ \sigma \leftarrow \texttt{get} \\
&\qquad \texttt{case run}(\mathit{Ev}(\varsigma)(e, \rho), \sigma)\ \texttt{of} \\
&\qquad\quad \mathbf{Fires}(\sigma', t) : \texttt{case}\ M(\sigma')(p, t, \rho)\ \texttt{of} \\
&\qquad\qquad\qquad\qquad \mathbf{Some}(\rho') : \mathit{return}(\rho') \\
&\qquad\qquad\qquad\qquad \mathbf{None} : \mathit{fail}() \\
&\qquad\quad _ : \mathbf{stuck}() \\
\mathit{Ev}_{bu}(\varsigma)(\mathbf{Update}(e_a, e_v), \rho) &= \texttt{do}\ \ t_a \leftarrow \mathit{Ev}(\varsigma)(e_a, \rho) \\
&\qquad t_v \leftarrow \mathit{Ev}(\varsigma)(e_v, \rho) \\
&\qquad \sigma \leftarrow \texttt{get} \\
&\qquad \texttt{case}\ t_a\ \texttt{of} \\
&\qquad\quad \mathbf{EAddr}(a) : \texttt{do put}\ \sigma[a \mapsto t_v] \\
&\qquad\qquad\qquad\qquad \mathit{return}(\rho) \\
&\qquad\quad _ : \mathbf{stuck}() \\
\overline{\mathit{Ev}_{bu}} &: \mathit{State} \to \mathit{BU}^* \to \mathit{MEnv} \to \mathit{Rule\text{-}Result}(\mathit{MEnv}) \\
\overline{\mathit{Ev}_{bu}}(\varsigma)(\epsilon)(\rho) &= \mathit{return}(\rho) \\
\overline{\mathit{Ev}_{bu}}(\varsigma)(\mathit{bu}_0\overline{\mathit{bu}})(\rho) &= \mathit{bind}(\mathit{Ev}_{bu}(\varsigma)(\mathit{bu}_0, \rho), \overline{\mathit{Ev}_{bu}}(\overline{\mathit{bu}}))
\end{aligned}
$$

Figure 27: Side-condition evaluation

$$
\begin{aligned}
&\mathit{Ev}_{\mathit{rule}} : \mathit{State} \to \mathit{Rule} \times \mathit{Term} \times \mathit{MEnv} \to \mathit{Rule\text{-}Result}(\mathit{Term})\\
&\mathit{Ev}_{\mathit{rule}}(\varsigma)(\textbf{Rule}(p, e, \overline{bu}), t, \rho) = \texttt{do}\ t \leftarrow \mathit{Ev}(\varsigma)(e, \rho)\\
&\qquad \sigma \leftarrow \texttt{get}\\
&\qquad \texttt{case}\ M(\sigma)(p, t, \rho)\ \texttt{of}\\
&\qquad\quad \textbf{Some}(\rho') : \texttt{do}\ \texttt{put}\ \sigma\\
&\qquad\qquad \rho'' \leftarrow \overline{\mathit{Ev}_{bu}}(\varsigma)(\overline{bu})(\rho')\\
&\qquad\qquad \mathit{Ev}(\varsigma)(e, \rho'')\\
&\qquad\quad \textbf{None} : \mathit{fail}()\\
&\mathit{Ev}_{\overline{\mathit{rule}}} : \mathit{State} \to \overline{\mathit{Rule}} \times \mathit{Term} \times \mathit{MEnv} \to \mathit{EvRes}[\mathit{Term}]\\
&\mathit{Ev}_{\overline{\mathit{rule}}}(\varsigma)(\epsilon, t, \rho) = \textbf{None}\\
&\mathit{Ev}_{\overline{\mathit{rule}}}(\varsigma)(\mathit{rule}_0\overline{\mathit{rule}}, t, \rho) = \texttt{case}\ \mathit{Ev}_{\mathit{rule}}(\varsigma)(\mathit{rule}_0, t, \rho)\ \texttt{of}\\
&\qquad \textbf{Fires}(t') : \mathit{return}(t')\\
&\qquad \textbf{Stuck} : \textbf{None}\\
&\qquad \textbf{Unapplicable} : \mathit{Ev}_{\overline{\mathit{rule}}}(\varsigma)(\bar{r}, t, \rho)\\
&\mathit{Ev}_{\mathit{mf}} : \mathit{State} \to \mathit{Metafunction\text{-}Name} \times \mathit{Term}^* \to \mathit{EvRes}[\mathit{Term}]\\
&\mathit{Ev}_{\mathit{mf}}(\varsigma)(f, \bar{t}) = \texttt{case}\ \mathcal{M}(f)\ \texttt{of}\\
&\qquad \textbf{User}(\overline{\mathit{rule}}) : \mathit{Ev}_{\overline{\mathit{rule}}}(\varsigma)(\overline{\mathit{rule}}, \textbf{Variant}(f, \bar{t}), \bot)\\
&\qquad \textbf{ExtMF}(\mathit{mf}) : \mathit{mf}(\varsigma, \bar{t})
\end{aligned}
$$

Figure 28: Rule and metafunction evaluation

## 5.7 RUNNING A MACHINE

AN ABSTRACT MACHINE is

- $\mathcal{S} : \wp_{\text{fin}}(\mathit{Rule})$: a set of rules (its "reduction relation");
- $\mathcal{M} : \mathit{Metafunction\text{-}Name} \underset{\text{fin}}{\rightharpoonup} \mathit{MF}$: the definitions of metafunctions;
- $\mathit{alloc} : \mathit{State} \times \mathit{Store} \times \mathit{Tag} \times \mathit{Env} \to \mathit{Addr}$: an address allocation function;
- $\mathit{mkV} : \mathit{State} \times \mathit{Store} \times \mathit{Variant\text{-}Name} \times \mathit{Tag} \times \mathit{Term}^* \times \mathit{Env} \to (\mathit{Store} \times \mathit{Term})$: a variant construction function;
- $\mathit{tick} : \mathit{State} \to \mathit{Time}$: a *Time* update function.

A machine "runs" by applying its reduction relation until stuck. If a state has no next step, then we call the state "final." If at most one rule applies at any one time, then the set of rules is deterministic. If not, the semantics is non-deterministic. For full generality, we define partial functions *step* and *run* that do not assume the rules are deterministic.

The step function is like $\mathit{Ev}_{\overline{\mathit{rule}}}$, except it returns either **Some** set of next states, or **None** because the input state is final.

$$
\begin{aligned}
\mathit{Step\text{-}result} &= \text{option } \wp(\mathit{State}) \\
\mathit{step} &: \mathit{State} \to \mathit{Step\text{-}result} \\
\mathit{step}(\overbrace{\sigma, t, \tau}^{\varsigma}) &= \mathit{step\text{-}all}(\varsigma, \mathcal{S}, t, \sigma, \emptyset, \emptyset) \\
\mathit{step\text{-}all} &: \mathit{State} \times \wp(\mathit{Rule}) \times \mathit{Term} \times \mathit{Store} \times \wp(\mathit{Term}) \\
&\quad \to \mathit{Step\text{-}result} \\
\mathit{step\text{-}all}(\varsigma, \emptyset, t, \sigma, \emptyset) &= \textbf{None} \\
\mathit{step\text{-}all}(\varsigma, \emptyset, t, \sigma, \mathit{next}) &= \textbf{Some}(\mathit{next}) \\
\mathit{step\text{-}all}(\varsigma, \{r\} \cup \mathcal{S}, t, \sigma, \mathit{next}) &= \texttt{case run}(\mathit{Ev}_{\mathit{rule}}(\varsigma)(r, t, \bot), \sigma) \texttt{ of} \\
&\qquad \textbf{Fires}(\sigma', t') : \mathit{step\text{-}all}(\varsigma, \mathcal{S}, t, \sigma, \{\langle \sigma', t', \mathit{tick}(\varsigma) \rangle\} \cup \mathit{next}) \\
&\qquad _ : \mathit{step\text{-}all}(\varsigma, \mathcal{S}, t, \sigma, \mathit{next})
\end{aligned}
$$

With the ability to step according to all the semantic rules, we can repeatedly apply *step* until all states are final. We do this by stepping each state in a set individually, to find both the next states and the final states, until the set of states to step is empty:

$$\begin{aligned}
\mathit{run} &: \mathit{State} \rightharpoonup \wp(\mathit{State})\\
\mathit{run}(\varsigma) &= \mathit{find\text{-}final}(\{\varsigma\}, \emptyset, \emptyset)\\
\mathit{find\text{-}final} &: \wp(\mathit{State}) \times \wp(\mathit{State}) \times \wp(\mathit{State}) \to \wp(\mathit{State})\\
\mathit{find\text{-}final}(\emptyset, \emptyset, \mathit{final}) &= \mathit{final}\\
\mathit{find\text{-}final}(\emptyset, \mathit{next}, \mathit{final}) &= \mathit{find\text{-}final}(\mathit{next}, \emptyset, \mathit{final})\\
\mathit{find\text{-}final}(\{\varsigma\} \cup \mathit{todo}, \mathit{next}, \mathit{final}) &= \texttt{case}\ \mathit{step}(\varsigma)\ \texttt{of}\\
&\qquad \mathbf{None} : \mathit{find\text{-}final}(\mathit{todo}, \mathit{next}, \{\varsigma\} \cup \mathit{final})\\
&\qquad \mathbf{Some}(\Sigma) : \mathit{find\text{-}final}(\mathit{todo}, \Sigma \cup \mathit{next}, \mathit{final})
\end{aligned}$$

The assumptions required of external parameters in this meta-semantics are that

1. allocation is fresh:

$$\mathit{alloc}(\sigma, \mathit{tag}, \tau, \rho) \notin \mathbf{dom}(\sigma)$$

2. variant construction both conservatively extends the store and creates an equivalent variant:

$$\begin{aligned}
\forall \sigma, \rho. \exists \sigma', t_v. \mathit{mkV}(\sigma, n, \mathit{tag}, \bar{t}, \rho) = \mathbf{Some}(\sigma', t_v)\\
\wedge \forall a \in \mathbf{dom}(\sigma). \mathit{tequal}(\sigma')(\sigma(a), \sigma'(a)) = \texttt{tt}\\
\wedge\, \mathit{tequal}(\sigma')(t_v, \mathbf{Variant}(n, \bar{t})) = \texttt{tt}
\end{aligned}$$

3. external metafunctions maintain state well-formedness. A state is well-formed if all the addresses it mentions are in the domain of the store. Formalizing this requires reifying the runtime stack to ensure that all live addresses within intermediate computation are kept live. I leave this informally stated and just say, "be reasonable."

This simple little language is our platform for introducing abstraction. We want to relax the conditions on our external parameters such that the resulting semantics is a sound simulation of this concrete semantics. The abstract semantics we define in the next chapter straddles the boundary of concrete and abstract interpretation: we will have the ability to strengthen the external parameters to recover the concrete semantics. The guarantee on top of soundness is that, if the above conditions hold of the abstract semantics' parameters, then the resulting semantics has a bisimulation with the concrete semantics. The next chapter thus strictly generalizes this one by giving a semantics for concrete, abstract, and anywhere in between, abstract machines.

# 6 A LANGUAGE FOR AAM

This chapter reconstructs the previous chapter while wearing an approximation hat. We develop an abstract semantics of reduction that natively supports the AAM abstraction tool: resource allocation.

## 6.1 INTRODUCTION

We judge correctness of the semantics by guaranteeing that approximate allocation functions lead to approximate rule applications. For example, if *alloc* freshly allocates, and $\widehat{alloc}$ is 0CFA-like with an appropriate structural abstraction of the freshly allocated addresses, then we can expect that the simulation property holds like in AAM:

$$\frac{\alpha(\square) \sqsubseteq \lozenge \qquad \square \longmapsto_{alloc} \blacksquare}{\exists \blacklozenge . \lozenge \longmapsto_{\widehat{alloc}} \blacklozenge \text{ and } \alpha(\blacksquare) \sqsubseteq \blacklozenge}$$

In English, this states, "if $\lozenge$ approximates $\square$ and $\square$ concretely steps to $\blacksquare$, then $\lozenge$ abstractly steps to a $\blacklozenge$ that approximates $\blacksquare$." Informally, all concrete steps are overapproximated in the abstract. Therefore, if a step does not exist in the abstract, it absolutely does not exist in the concrete. This soundness guarantee means that we can prove the absence of bad program behavior with a computable approximation. The metalanguage semantics developed in this chapter is designed with simulation in mind.

I will refer to rules that the metalanguage interprets as either "an object semantics" or "user-provided rules." Anything the metalanguage semantics uses but is left undefined is an *external parameter*, or "user-provided X" where X is the parameter's role (for example, *alloc* is both an external parameter and a user-provided allocation function). I sometimes refer to a user as an analysis designer.

CONCEPT OVERVIEW Four concepts we cover in this chapter are the following:

1. *weak equality* of data structures and finite functions (terms). An equality is "weak" if it is uncertain (due to approximation);
2. *weak matching* of non-linear patterns against terms. A match is "weak" if
   a) a non-linear pattern's equality is weak;
   b) an approximate term (which represents multiple terms) has both a successful match and a failing match, or one has a weak match;

3. *weak evaluation* of a simple expression language. An evaluation is "weak" if
   a) progress (conversely, stuckness) is uncertain due to weak matching in side-conditions;
   b) a store update uses an abstract address (a "weak update").
4. *worthwhile splitting* of a state into multiple refined states, only when it benefits the precision of a specific task.

The preferable alternative to a weak update is a "strong update," which can replace contents of the store instead of merging contents. Merging is the enemy of precision. If a concrete store is $\mathit{Addr} \rightharpoonup T$ (for some T) then an abstract store is $\widehat{\mathit{Addr}} \rightharpoonup \hat{T}$, where they are adjoined with a Galois connection $\langle \wp(T), \subseteq \rangle \underset{\alpha}{\overset{\gamma}{\leftrightarrows}} \langle \hat{T}, \sqsubseteq \rangle$. A strong update is only justified if the address is fresh, but abstract addresses are not necessarily fresh.

Equality, matching, and evaluation each have three parts:

1. a concrete semantics (no merging, and thus requires fresh allocation), which we've already seen;
2. an abstract semantics (freshness not required, so merging may happen) that exactly approximates [17] the concrete through structural abstraction; and
3. a splitting abstract semantics that splits the state space based on *store refinements*.

[17] *This is abstract interpretation vernacular. If* $f$ *is a "concrete" function and* $f^\sharp$ *is an "approximate" function, and* $f \circ \gamma = \gamma \circ f^\sharp$*, then* $f^\sharp$ *is called an exact approximation. Notice that left-composition with* $\alpha$ *cancels on the right hand side if* $\langle \gamma, \alpha \rangle$ *form a Galois insertion.*

CHAPTER OVERVIEW In Section 6.2 I explain the components of an abstract abstract machine state and motivate their inclusion with respect to the overall design space. I then show that we can "run" machines in many different ways, with varying tradeoffs. I then explain the role of store refinements in Section 6.4. Before we jump into all the technical details of how the semantics works, I motivate by example the high level ideas behind the design choices for the available tuning knobs in Section 6.5. Section 6.6 defines exteral descriptors for external terms, and the abstract term Galois connection. Weak equality's abstract components are fully developed as $\widehat{\mathit{tequal}}$, and $\widehat{\mathit{tequal}}_S$ (S for splitting) in Section 6.7. Both weak matching (Section 6.8) and expression evaluation (Section 6.9) are presented using only the splitting version for brevity; the non-splitting versions should be easily recoverable by the reader.

The different ways of "running" are discussed in more detail in Section 6.10. That section additionally specifies the external components that can be plugged in to make an object semantics. We wrap up in Section 6.11 with a discussion of candidates for the external components.

**N.B.** Any missing proofs are long and are moved to the appendix.

## 6.2 REPRESENTING AN ABSTRACT ABSTRACT MACHINE

An abstract abstract machine is still a collection of reduction rules for transforming machine states. An abstract abstract machine state ($\hat{\varsigma}$) is

$$\hat{\varsigma} \in \widehat{State} = \widehat{Store} \times \widehat{Term} \times \widehat{Time}$$

The different components resemble their concrete counterparts, but with extra support for approximation.

TERM:

$$\hat{t} \in \widehat{Term} = PreTerm \cup \{\mathbf{Delay}(\hat{a}) \ : \ \hat{a} \in \widehat{Addr}\} \cup NDTerm$$

An abstract term has more to it than a concrete term. A *PreTerm* resembles a concrete term, and an *NDTerm* (nondeterministic term) is an approximation of a set of *PreTerm*:

$$\begin{aligned}
PreTerm &::= \widehat{st} \mid \mathbf{External}(E, v) \text{ where } v \in E.ty \\
\widehat{st} \in STerm &::= \mathbf{Variant}(n, \bar{\hat{t}}) \mid \mathbf{EAddr}(\hat{a}) \mid \mathbf{IAddr}(\hat{a}, lm) \\
NDTerm &::= \mathbf{NDT}(\widehat{ts}, Es)
\end{aligned}$$

$$\text{where } \widehat{ts} \in \wp(STerm)$$

$$Es \in External\text{-}map = External\text{-}Descriptor \underset{\text{fin}}{\rightharpoonup} Meta\text{-}meta\text{-}value$$

$$\text{where } Es(E) \in E.ty$$

$$lm \in Lookup\text{-}modality$$

We will discuss *External-Descriptor* and *NDTerm* in further detail in Section 6.6.

The indirect address in *STerm* has an additional component on top of the concrete semantics' **IAddr**: *lm*, for *lookup modality*. The *lm* flag determines the subtly different store dereference semantics for that address. We will see more of *lm* later in Section 6.5, but one modality is to delay the lookup. We represent delayed lookup with the $\mathbf{Delay}(\hat{a})$ form, like in Section 3.4.3.

I will sometimes use physics terminology to refer to an *NDTerm* as a term in *superposition*, and choosing an abstract term from an *NDTerm* is *collapsing* it.

Not every subset of *PreTerms* is representable, so we will have a Galois connection $\langle \wp(PreTerm), \subseteq \rangle \underset{\alpha}{\overset{Choose}{\leftrightarrows}} \langle NDTerm, \sqsubseteq \rangle$. We will see this Galois connection defined formally in Section 6.6.

For the sake of notational brevity, I will write $\mathbf{NDT}(\{\widehat{st}\ldots\})$ to mean $\mathbf{NDT}(\{\widehat{st}\ldots\}, \bot)$, and write variants in prefix notation. I also optionally parenthesize nullary variants and hypothetical external descriptors, so `(Cons 1 Nil)` is an abbreviation for

$$\mathbf{Variant}(\texttt{'Cons}, \langle \mathbf{External}(Number, 1), \mathbf{Variant}(\texttt{'Nil}, \langle\rangle)\rangle)$$

We will write $\bot$ to mean the empty map when the context expects a map, or more generally the bottom element of a lattice.

STORE: An abstract store is a pair of a *Heap* and a *Count*:

$$\hat{\sigma} \in \widehat{Store} ::= \langle h, \mu \rangle$$
$$h \in Heap = \widehat{Addr} \underset{\text{fin}}{\rightharpoonup} NDTerm$$
$$\mu \in Count = \widehat{Addr} \underset{\text{fin}}{\rightharpoonup} \hat{\mathbb{N}}$$
$$\hat{\mathbb{N}} = \{0, 1, \omega\}$$
$$\hat{\mathbb{N}} \text{ linearly ordered } 0 < 1 < \omega \text{ with } \sqcup = \max$$

An abstract store is in two pieces in order to straddle the boundary between concrete and abstract. A concretely allocated address is fresh; it has a unique identity. In the abstract, an address may be allocated multiple times, which $\mu$ tracks. If the abstract address $\hat{a}$ is previously unallocated then we know that it is fresh ($\mu(\hat{a}) = 1$). If not, then the address denotes more than one concrete address and we can't say for certain that a self-comparison in the abstract is always true in the concrete. Since "more than one" is all it takes to make imprecise predictions, we overapproximate "more than one" as $\omega$. An address that is not fresh is called "used." To model semantics that use object identity or strong updates, or just run in "concrete mode," the freshness of an address is necessary information.

TIME:

$$\widehat{Time} \text{ a user-provided set with no restrictions.}$$
$$\widehat{tick} : \widehat{State} \to \widehat{Time} \text{ a user-provided update function}$$

The $\widehat{Time}$ component can be anything, but is often some representation of the trace history in order to inform allocation of the current execution context. As a formality, $\widehat{Time}$ is required to be in Galois connection with $\wp(Time)$, but I will gloss over this unimportant detail.

We say that the $\widehat{Time}$ component *distinguishes* a state because different $\widehat{Time}$ values mean different state representations. The more distinctions are made with the $\widehat{Time}$ space, the more the state space is "split" or "partitioned." The traces of states are correspondingly partioned. Trace partitioning is an important component of high precision, low false-alarm analyses [60], as it better refines the context to understand the execution of the current term.

## 6.3 OVERVIEW OF RUNNING

Applying all of a semantics' reduction rules to a state is called "stepping" the state. Let's call the function that does this *step*, which is defined in Section 6.10. The ways in which we step states gives us a few notions of "running" a term in a given semantics:

- Nondeterministic run: repeatedly apply *step* on an arbitrarily chosen output state until stuck; report the final state as the result:

$$\begin{aligned} run(t) &= run^*(inject(t)) \\ run^*(\hat{\varsigma}) &= \hat{\varsigma} \text{ if } step(\hat{\varsigma}) = \emptyset \\ run^*(\hat{\varsigma}) &= run^*(\mathit{Choice\text{-}function}(step(\hat{\varsigma})) \text{ otherwise} \end{aligned}$$

  Pro: depending on the choice function, we can pin-point bad states without much overhead. Con: likely to diverge.

- All runs: treat the initial state as a singleton set "frontier" to repeatedly step:

$$\begin{aligned} run(t) &= run^*(\{t\}) \\ run^*(F) &= \texttt{case} \bigcup_{\hat{\varsigma} \in F} step(\hat{\varsigma}) \texttt{ of} \\ &\qquad \emptyset : F \\ &\qquad F' : run^*(F') \end{aligned}$$

  Pro: explores the whole state space. Con: diverges if the abstracted program diverges.

- Loop-detecting: run like the previous mode, but don't re-step already seen states:

$$\begin{aligned} run(t) &= run^*(\emptyset, \{t\}) \\ run^*(S, F) &= \texttt{case} \bigcup_{\hat{\varsigma} \in F} step(\hat{\varsigma}) \texttt{ of} \\ &\qquad \emptyset : S \\ &\qquad F' : run^*(S \cup F', F' \setminus S) \end{aligned}$$

  Pro: finite allocation and finite externals implies this will always terminate with the reachable states. Con: does not represent control flow for post-processing.

- Reduction relation-grounding: create a concrete representation of the reduction relation as used to evaluate the given term:

$$\begin{aligned} run(t) &= run^*(\emptyset, \{t\}, \emptyset) \\ run^*(S, F, R) &= \texttt{case } \{(\hat{\varsigma}, \hat{\varsigma}') \ : \ \hat{\varsigma} \in F, \hat{\varsigma}' \in step(\hat{\varsigma})\} \texttt{ of} \\ &\qquad \emptyset : R \\ &\qquad R' : run^*(S \cup \pi_1(R'), \pi_1(R') \setminus S, R \cup R') \end{aligned}$$

  Pro: allows other tools to consume the reduction relation as a model. Con: larger memory footprint.

Once we define equality, matching, and expression evaluation, we get all these notions of "running" the machine. The notions that are likely to diverge can always be given "fuel" to stop after the fuel runs out. Semantics engineering tools like PLT Redex, the K framework, and Maude, all provide multiple modes of running for user convenience.

## 6.4 STORE REFINEMENTS

User-provided trace partitioning is not the only way to split execution traces. The semantics additionally splits the state space based on fresh address contents via *store refinements*. If an abstract state ($\hat{\varsigma}$)'s store maps a fresh abstract address ($\hat{a}$) to some set of terms $\{t_0, \ldots, t_n\}$, then we can refine that state into multiple states with more specific stores:

$$\hat{\varsigma}[\hat{\sigma}.h := \hat{\varsigma}.\hat{\sigma}.h[\hat{a} \mapsto t_0]],$$
$$\vdots$$
$$\hat{\varsigma}[\hat{\sigma}.h := \hat{\varsigma}.\hat{\sigma}.h[\hat{a} \mapsto t_n]]$$

Each state can be stepped individually with this more specific information about the store. Refinements enable the semantics to be more precise when comparing terms for equality, or in the presence of a template creating a tuple like $(x, x)$. The choice made for the first $x$ determines the choice for the second $x$.

For example, suppose we run the JavaScript program in Figure 29 with a collecting semantics that does no trace partitioning. The state-

```
function foo(b,x,y) {
  var z;
  if (b) { z = new x }
  else { z = new y }
  return z; }
var n = foo(unknown,
            function () { this.bar = 0 },
            function () { this.bar = 1 });
if ((n.bar + n.bar) % 2 !== 0) { launch_the_missiles() }
```

Let's say the `unknown` variable comes from an arbitrary context so that the abstract semantics must explore each branch.

Figure 29: Example exemplifying the benefit of store refinements

ment after the `if` constitutes a *join point*, where the store and count of the incoming states are joined together. The boolean condition is too abstract to determine if it is always truthy or always falsy, so both branches of the `if` must be analyzed. The contents of `z` are stored in a fresh address, but at the join point its contents are either the object `{ bar: 0 }` or the object `{ bar: 1 }`. The object in `n` has a `bar` field that is either 0 or 1, so the addition will be 0, 1 or 2; the 1 means that `launch_the_missiles` is called[18].

Store refinements allow us to leverage the knowledge that `z`'s address is fresh. We can refine the store after the join point to split the state space once we access `n.bar`. In one case, `n.bar` will always mean 0, and in the other always 1. Both refinements lead to an even sum, so we can show that `launch_the_missiles` is never called.

[18] *Missile launch protocol is not written in JavaScript[citation needed], but bad things can nevertheless happen if contextual assumptions are invalidated by overapproximation.*

THE THEORY If an address has only been allocated once (is "fresh"), then it can be treated as a concrete entity. Allocations are counted, so an address $\hat{a}$ is fresh if $\mu(\hat{a}) = 1$. Freshness information is necessary for strong equality judgments. Strong equality judgments are crucial to support concrete execution with the same semantic framework we use for abstract execution. In addition to strong equality, fresh addresses allow the semantics to perform case splitting on a fresh address's contents. If an address is fresh but the store maps it to a non-singleton set, $\{b_0, b_1, \ldots\}$, then the choice of $b_i$ on lookup can be written back as a singleton, $\{b_i\}$ (and all choices are explored). This write-back is called "refining" the store.

A fresh address can map to a non-singleton set when control flow merges at one point, say after an `if` statement: If we need to read the address's contents, we can split the state space based on *which* of the terms is chosen. Going forward, the different states' stores will map the fresh address to the respective choice.

This might seem odd or wrong; how can we still say `z`'s address is fresh when a concrete allocator is free to assign different addresses at the different `new` expressions? Fresh addresses' physical identities are unimportant in the same way as binders are in syntax. Therefore they can be renamed to match, and thus the abstract address still identifies one concrete address, but the address can map to multiple values.

A map from fresh addresses to their choices is called a *Refinement*.

$$\delta \in \mathit{Refinement} = \widehat{\mathit{Addr}} \underset{\text{fin}}{\rightharpoonup} \mathit{PreTerm}$$

Refinements are only valid on fresh addresses and actual store contents, so we use the following definitions for well-formedness ("$\delta$ refines $\hat{\sigma}$"), and for the family of sets of all well-formed refinements:

$$\begin{aligned} \mathit{refines}(\delta, \langle h, \mu \rangle) &= \forall \hat{a} \in \mathbf{dom}(\delta).\mu(\hat{a}) = 1 \wedge \delta(\hat{a}) \in \mathit{Choose}(h(\hat{a})) \\ \mathit{Refinements}(\hat{\sigma}) &= \{\delta \ : \ \mathit{refines}(\delta, \hat{\sigma})\} \end{aligned}$$

## 6.5 DESIGN MOTIVATION BY EXAMPLE

In the abstract world, function evaluation and implicit addresses have important new roles to play. Let's take a look at some rules we want to write, how we want them to be abstracted, and how our semantics' non-standard concepts get us there.

### 6.5.1 *Overview of explicit versus implicit addresses*

One way to interpret an address is as just a stand-in for what it points to. Under this interpretation, a pattern match implicitly dereferences the address and continues matching on the stored contents. For example, a language implementation will implicitly store-allocate nested data when introducing a `cons`, and implicitly dereference the store when eliminating via `car` or `cdr`.

Alternatively, we can view an address is an object that the semantics can explicitly manipulate with lookups and updates. We need to instruct the pattern matcher to not dereference and instead bind the address itself. For example, consider the CESK machine's function call rule[19], which explicitly allocates and binds an address:

[19] *An abbreviated form for illustrative purposes.*

$$\overbrace{v, \rho, \sigma, \mathbf{appR}(\lambda\, x.\; e, \rho'){:}K}^{\varsigma} \longmapsto e, \rho[x \mapsto a], \sigma[a \mapsto (v, \rho)], K$$
$$\text{where } a = \mathit{alloc}(\varsigma, x)$$

All binding uses pattern matching, so $a$'s binding is the result of matching a trivial pattern. Since $a$ came from an explicit *alloc*, it is an explicit address. If the allocated address were implicit, the pattern matcher would immediately try to dereference the address, causing evaluation to get stuck.

Dual to explicit allocation is explicit dereference; in the CESK machine this is in variable reference:

$$x, \rho, \sigma, K \longmapsto v, \rho', \sigma, K$$
$$\text{where } (v, \rho') = \sigma(\rho(x))$$

Let's discuss the `cons`, `car`, `cdr` example I hinted at for motivating implicit dereferencing. We would like to write the obvious rule for interpreting the `cons` primitive to construct the `consv` value:

$$\mathbf{ap}(\texttt{cons}, \langle v_0, v_1 \rangle, \sigma, \kappa) \longmapsto \mathbf{co}(\kappa, \texttt{consv}(v_0, v_1), \sigma)$$

A rule like this, with a structural `consv` value, can create unboundedly many states and cause the semantics to diverge[20]. For example, a program like the following might diverge in a naive analysis:

[20] *Which is highly undesirable for an analysis.*

```
(define (bad x) (bad (cons 'more x)))
(bad 'start)
```

If we don't introduce some approximation, the bindings for `x` keep growing:

```
'start,
(cons 'more 'start),
(cons 'more (cons 'more 'start)),...
```

The motto of AAM is to "store-allocate recursion," but this is more accurately understood as, "disallow unbounded nesting of data by indirecting through store allocations." Suppose we had a way to interpret the above "obvious" rule as the following rule:

$$\mathbf{ap}(\mathtt{cons}, \langle v_0, v_1 \rangle, \sigma, \kappa) \longmapsto \mathbf{co}(\kappa, \mathtt{consv}(a_A, a_D), \sigma \sqcup [a_A \mapsto v_0, a_D \mapsto v_1])$$
$$\text{where } a_A = \mathit{alloc}(\varsigma, \mathtt{car}), a_D = \mathit{alloc}(\varsigma, \mathtt{cdr})$$

In this case, a finite allocation strategy leads to finitely many representable cons cells. Finite allocation with this rule protects us from the above example of divergence. We can't just rewrite the obvious rule to this rule, because we want to leave the car (and similarly, cdr) rules unchanged as

$$\mathbf{ap}(\mathtt{car}, \langle \mathtt{consv}(v_0, v_1) \rangle, \sigma, \kappa) \longmapsto \mathbf{co}(\kappa, v_0, \sigma)$$

Notice the mismatch between $v_0$ here and $a_A$ above. Since the consv contains addresses, but the car rule's result expects an address, we need the rule to implicitly dereference the address in the consv. I said above that naming an address in a rule is by definition explicit, so the rewritten rule has no way to mark $a_A$ or $a_D$ as implicit. Thus, the only way to introduce an implicit address is within implicit allocation.

When we allocate something like a cons, the semantics calls an external parameter, *mkV*, for allocating variants. When we construct a consv with values $v_0$ and $v_1$, the *mkV* parameter can choose to represent the variant as (consv $\mathbf{IAddr}(a_A, \mathit{lm})$ $\mathbf{IAddr}(a_D, \mathit{lm})$), meaning $a_A$ and $a_D$ will be implicitly dereferenced as guided by *lm*, to be discussed below. Then *mkV* can update the store to map addresses $a_A$ and $a_D$ to $v_0$ and $v_1$ respectively.

Another behavior *mkV* could have is to simply construct (consv $v_0$ $v_1$) structurally. An analysis designer may know an invariant that some structural constructions are safe to perform - they won't introduce divergence. For example, an $n$-ary function application will create a continuation frame that contains the list of evaluated function and arguments. We know *a priori* that the list is bounded by the syntactic form's list of expressions. The program is a one-time input to start the analysis at an initial state, so there are a finite number of function application expressions with finitely many argument expressions in each.

An **IAddr** can be implicitly dereferenced in subtly different ways. One way we might expect the car rule to be interpreted is the following:

$$\mathbf{ap}(\mathtt{car}, \langle \mathtt{consv}(a_A, a_D) \rangle, \sigma, \kappa) \longmapsto \mathbf{co}(\kappa, v_0, \sigma)$$
$$\text{where } v_0 \in \sigma(a_A).$$

One might assume the "where" clause should nondeterministically split the execution on the values stored at $a_A$, resolving its nondeterminism. However, immediately splitting execution on address contents leads to the explosive and usually unproductive fan-outs that we saw before we added lazy nondeterminism in Section 3.4.3.

An implicit address therefore has a *modality* to guide the pattern matcher's action when binding an implicit address to a variable. A *lookup modality* is one of:

- `'resolve`: immediately split the state space based on the choice of term out of the stored *NDTerm*. Refine the store to the chosen term iff the address is fresh.
- `'deref`: dereference the address to get the stored *NDTerm* without splitting the state space. Matching on it later will split in order to resolve the nondeterminism.
- `'delay`: delay dereferencing the address. The address is treated like `'resolve` when a term is matched with a pattern that inspects structure (not wild nor named wild).

The second two lookup modalities are the candidate implementations of lazy nondeterminism discussed in Section 3.4.3 as, respectively, option 1 and option 2. The lookup modalities drive how the semantics should refer to an address's contents, so the expression for store-lookup also has a lookup modality.

### 6.5.2 *Weak matching: rule ordering and prediction strength*

Abstract addresses and abstract terms motivate the notion of *weak matching*. Equality judgments from non-linear patterns can be inexact: rules *may* fire, leading to nondeterministic state exploration. For example say we have a metafunction, `rem`, that removes duplicate adjacent elements of a list:

$$\begin{aligned}
\texttt{(rem '())} &\longmapsto \texttt{'()} \\
\texttt{(rem (cons } x \texttt{ '()))} &\longmapsto \texttt{(cons } x \texttt{ '())} \\
\texttt{(rem (cons } x \texttt{ (cons } x \; \mathit{lst}\texttt{)))} &\longmapsto \texttt{(cons } x \; \mathbf{Call}(\texttt{rem}, \langle \mathit{lst} \rangle)\texttt{)} \\
\texttt{(rem (cons } x \texttt{ (cons } y \; \mathit{lst}\texttt{)))} &\longmapsto \texttt{(cons } x \\
&\qquad \mathbf{Call}(\texttt{rem}, \langle \texttt{(cons } y \; \mathit{lst}\texttt{)} \rangle)\texttt{)}
\end{aligned}$$

But we may not be able to say with certainty in the third rule that the adjacent elements are equal. Say we have a call

$$\mathbf{Call}(\texttt{rem}, \langle \texttt{(cons } a \texttt{ (cons } a \texttt{ (nil)))} \rangle)$$

where the store contains $[a \mapsto \mathbf{NDT}(\{0, 1\})]$ with $a$ used. Since we can't refine $a$ to one of the two numbers it maps to, possible concretizations of the input include

```
(cons 0 (cons 0 (nil))) and
(cons 0 (cons 1 (nil))).
```

Thus both the third and fourth rule *may* fire, leading to the nondeterminism. If $a$ were fresh, then the third rule would *strongly* fire, with two different store refinements mapping $a$ to either 0 or 1.

Now that we have covered the high level concepts, let's talk details. First, let's discuss the rest of the structure of terms.

## 6.6 externals and *NDTerm*

An external value is paired with an *external descriptor*, which contains the operations the semantics needs to interact with external values (*e.g.*, equality, join, ordering). The semantics handles the switch between abstract and concrete seamlessly for non-external terms, but an external term itself might have a different representation for the two.

external descriptors An external descriptor contains its concrete equality operation as well as the following operations and "types":

- "type" *ty*: Racket has one type: Racket value. To make conceptual matters clearer here though, I write *ty* for an intended flavor of Racket value that represents the external's abstract value representation.
- "type" *concrete*: the flavor of Racket value that represents the external's concrete value representation.
- $\sqcup : \widehat{State} \times Refinement \times \widehat{\Delta Store} \to ty \times ty \to ty$:
  takes some context, including evaluation's changes to the store (defined in Section 6.9), and two values to produce a combination of the two that soundly represents both. The function need not be a lattice-theoretic "join" (least upper bound). In fact, to guarantee convergence, $\sqcup$ should not produce any infinitely ascending chains of values[21]. We use $\sqcup$ for notational simplicity.
- $\sqsubseteq \subseteq ty \times ty$: approximation ordering
- $\hat{=} : \widehat{State} \to ty \times ty \to \widehat{EqResM}$:
  takes some context and two values and produces the output type for abstract term equality (definition upcoming, along with why we need the set of term pairs).
- $\hat{=}_S : \widehat{State} \to ty \times ty \to \widehat{EqResM}_S$:
  like the previous, but for *splitting* abstract term equality.
- $\gamma : ty \to \wp(concrete)$: the Galois connection's concretization function, which we use only in proofs.

[21] *This is generally referred to as a "widening" in abstract interpretation literature, and is commonly notated* $\nabla$.

- $\equiv$: *Store* $\times$ *concrete* $\times$ *concrete* $\times$ *Pairs* $\to$ *EqRes*:
  judges the equality of two concrete external values, returning the concrete equality result type from the previous chapter.

NONDETERMINISTIC TERMS An *NDTerm* is intended to be a representation of a set of *PreTerm*s. However, we cannot simply use a set representation because a *PreTerm* may contain external values. External values may be drawn from an unbounded space, so a set of them may grow unbounded. Therefore, for external values $v$ and $v'$ with the same descriptor, $E$, the set $\{\textbf{External}(E, v), \textbf{External}(E, v')\}$ is represented as a safe (overapproximating) combination $\textbf{External}(E, E.\sqcup(v, v'))$[22]. A set of *PreTerm* can include external values from different descriptors, so *NDTerm* represents the set of external values as a map from external descriptor $E$ to value of type $E.ty$.

[22] The "no infinite ascending chains" condition provides that $\sqcup$ has to stop growing the values eventually.

The Galois connection between *NDTerm* and $\wp(\mathit{PreTerm})$ is the following:

$$\langle \wp(\mathit{PreTerm}), \subseteq \rangle \underset{\alpha}{\overset{\mathit{Choose}}{\leftrightarrows}} \langle \mathit{NDTerm}, \sqsubseteq \rangle$$

$$\mathit{Choose}(\textbf{NDT}(\widehat{ts}, Es)) = \widehat{ts} \cup \{\textbf{External}(E, v) \ : \ Es(E) = v\}$$

$$\alpha(S) = \textbf{NDT}(\{\widehat{st} \in S\}, [E \mapsto \bigsqcup_{\textbf{External}(E,v) \in S} v \ : \ \textbf{External}(E, _) \in S])$$

## 6.7 TERM EQUALITY

Concrete terms are either equal or not. In the abstract though, equality can return weak "yes and no" answers. Equality in the abstract can represent both answers because abstract an abstract term can represent multiple concrete terms. One choice from a pair of two terms' concretizations can be equal, and yet another choice can be unequal. The possibilities are thus,

- **strongly equal:** when all concretizations are concretely equal;
- **strongly unequal:** when all concretizations are concretely unequal;
- **weakly equal:** when (exactly) the previous two don't apply, or (soundly) whenever.

An abstract equality function is an *exact* approximation when it outputs a weak result *exactly* when there is no strong result.

OVERVIEW We first define an abstract term equality in Section 6.7.1 that gives the appropriate strong or weak result. Not all exact approximations (defined in Section 6.1) are created equal; in Section 6.7.2 we show that in the context of state space exploration, we can do better than exact. Then in Section 6.7.3 we define and prove useful properties about *worthwhile* refinements. We finish in Section 6.7.4 with another exact approximation of term equality that additionally splits the state space if it is worthwhile to do so.

An executable form of the semantics (in Haskell) in this chapter is available in Appendix 18. I take notational shortcuts in this chapter to not overburden the exposition.

### 6.7.1 *Abstract term equality*

We have a gold standard in hand for structural term equality. If we had the concretization function, could we just use concrete equality for abstract equality? Perhaps the following diagram would work:

$$\mathit{Abstract\text{-}Input} \xrightarrow{\gamma} \wp(\mathit{Concrete\text{-}Input}) \xrightarrow{map(tequal_C)} \wp(\mathit{Boolean}) \xrightarrow{\alpha} \widehat{\mathit{Equality}}$$

where

$$\begin{aligned} \widehat{\mathit{Equality}} &::= \mathbf{Equal} \mid \mathbf{Unequal} \mid \mathbf{May} \\ \alpha(\{\mathtt{tt}\}) &= \mathbf{Equal} \\ \alpha(\{\mathtt{ff}\}) &= \mathbf{Unequal} \\ \alpha(\{\mathtt{tt}, \mathtt{ff}\}) &= \mathbf{May} \end{aligned}$$

An equality on abstract terms is an exact approximation if it performs the same thing as this diagram. The problem with a direct approach like this is that $\gamma$ routinely produces infinite sets. The middle arrow takes a while to give an answer in that case.

We can be more clever, but when we switch over to abstract execution, we raise some difficulties and questions:

1. an abstract address $\hat{a}$ can be used, so address identity is lost; $\hat{a} \stackrel{?}{=} \hat{a}$ can represent both true and false concrete equality comparisons;

2. a structural or delayed address maps to a representation of a *set* of terms, so they all have to be equal in order for a strong result;

3. if we have a fresh address $\hat{a}$ that maps to $\mathbf{NDT}(\{0, 1\})$ appear twice in a term, we have to remember $\hat{a}$'s choice of value within equality;

4. do we remember the choice of value for a fresh address even after checking equality?

Let's look at some examples that illuminate these issues.

EXAMPLE EQUALITIES Abstract term equality returns one of three results: strongly equal (**Equal**), strongly unequal (**Unequal**), or weakly equal (**May**). In the following examples of each kind of result, I use $=$, $\neq$, and $\approx$ to stand for strongly equal, strongly unequal, and weakly equal, respectively:

- strongly equal:
  - no approximate structure: `(unit)` $=$ `(unit)`
  - fresh address identity: $\mathbf{EAddr}(\hat{a}) = \mathbf{EAddr}(\hat{a})$ when $\mu(\hat{a}) = 1$.
  - fresh address structural equality: $\mathbf{IAddr}(\hat{a}, \texttt{'deref}) = \mathbf{IAddr}(\hat{a}, \texttt{'resolve})$ when $\mu(\hat{a}) = 1$ and

    $$\hat{\sigma}.h(\hat{a}) = \mathbf{NDT}(\{\texttt{(unit)}, \texttt{(top)}\}, \bot)$$

    Recall that a fresh address denotes exactly one concrete address, say $a$. During concrete execution, $a$ may hold either `(unit)` or `(top)`, but certainly not both. Say $a$ maps to `(unit)`; since there are no other concretizations of $\hat{a}$ in the concrete, we can forget about `(top)` and keep running with $[a \mapsto \texttt{(unit)}]$. In the abstract, this means we have a finite braching factor to search for better equality predictions given fresh addresses.

    Call this store $\hat{\sigma}_2$ for later examples.
- strongly unequal:
  - structure mismatch: $\texttt{(A }\mathbf{EAddr}(\hat{a})\texttt{)} \neq \texttt{(B }\mathbf{EAddr}(\hat{a})\texttt{)}$
  - address non-identity: $\mathbf{EAddr}(\hat{a}) \neq \mathbf{EAddr}(\hat{b})$
- weakly equal:
  - used address identity: $\mathbf{EAddr}(\hat{a}) \approx \mathbf{EAddr}(\hat{a})$ when $\mu(\hat{a}) = \omega$
  - fresh address structural (in)equality: $\mathbf{IAddr}(\hat{a}, \texttt{'deref}) \approx \texttt{(unit)}$ when $\mu(\hat{a}) = 1$ and we have $\hat{\sigma}_2$
  - used address structural equality: $\mathbf{IAddr}(\hat{a}, \texttt{'deref}) \approx \mathbf{IAddr}(\hat{a}, \texttt{'deref})$ when $\mu(\hat{a}) = \omega$ and we have $\hat{\sigma}_2$

Most of these examples should not be surprising. The last bullets of strongly equal and weakly equal are worth elaborating.

If an address is fresh, but maps to a representation of more than one term, we still have a strong equality. Say $\hat{a}$ represents a single $a$ in the concrete. The only store concretizations (restricted to $\alpha^{-1}(\hat{a})$) are

$$\sigma_0(a) = \texttt{(unit)}, \text{ and}$$
$$\sigma_1(a) = \texttt{(top)}.$$

Therefore the self equality of the structural address must be strongly equal in order to have an exact approximation.

In the same setup, except with $\mu(\hat{a}) = \omega$, there can be unboudedly many concretizations of $\hat{a}$ and its corresponding mappings[23]. For example, if $\alpha^{-1}(\hat{a}) = \{a_0, a_1, \ldots\}$, then we have these concretizations:

$$\begin{aligned}[a_0 \mapsto \texttt{(unit)}]\\ [a_0 \mapsto \texttt{(top)}]\\ [a_0 \mapsto \texttt{(unit)}, a_1 \mapsto \texttt{(top)}]\\ [a_0 \mapsto \texttt{(top)}, a_1 \mapsto \texttt{(unit)}]\\ \vdots\\ [a_{2i} \mapsto \texttt{(unit)}, \ldots, a_{2i+1} \mapsto \texttt{(top)}, \ldots]\\ \vdots\end{aligned}$$

Then the $\hat{a}$ on the left of the equality can be $a_0$, and the $\hat{a}$ on the right can be $a_1$. In the context of some stores, concrete term equality judges the two structural addresses as equal, whereas in other stores, the two are unequal.

[23] *While $\alpha$ must be a surjection, it's not required that each abstract address have an unbounded preimage in $\alpha$. It is the case that both the $k$CFA and $m$CFA addressing schemes have unboundedly many concrete addresses for each abstract address.*

INTERNAL REFINEMENTS Although equality is with respect to a store, an abstract store represents multiple concrete stores. Any inspection of a fresh address's contents must collapse its nondeterminism in order to get the equality behavior for fresh structural addresses that we expected in the above examples. Collapsing nondeterminism splits our abstract store into multiple representations that, all together, have the same concretization. The split representation gives us more power to identify the contents of fresh addresses.

Our function that decides abstract term equality thus internally splits its representation of the store with store refinements. Not all addresses need to have nondeterminism collapsed in order to determine the equality of two abstract terms; the fewer split stores we have to consider, the better for efficiency. Our function that decides abstract term equality therefore works over a set of store refinements, that, once applied to the current store, represent the split space of stores. Similar to concrete equality, we additionally carry a set of term pairs. A term pair represents guarded equality of two terms *with respect to the store refinement which led to the decision*. If we find that both equality and inequality are possible outcomes, we can forget all the store refinements involved, since we've already lost the precision we were trying for.

Thus, strong equalities are witnessed by a map from store refinement to a set of term pairs. Strong inequalities are witnessed only by the original store refinement since we don't need to carry forward any refinements to continue to witness an inequality. Weak equalities are witnessed by a set of term pairs.

The map from store refinement to set of term pairs must adequately represent the entire store. Our notion for "adequate" is formalized by the definition of a *cut* of the set of all store refinements.

CUTTING THE REFINEMENT SPACE A cut, C, of a finite poset $\langle P, \sqsubseteq \rangle$, is a set of elements that separate P into elements either less than or greater than elements of C. In other words, each element of P is comparable to some element of C. Additionally, no element of C is comparable to any other element of C (each chain is "cut" at exactly one element).

$$\textsc{Comparable}\ \frac{c \sqsubseteq p \vee p \sqsubseteq c}{c \circledast p} \qquad \textsc{Cut}\ \frac{\forall p \in P.\exists c \in C.c \circledast p \qquad \forall \delta, \delta' \in C.\delta \circledast \delta' \implies \delta = \delta'}{\mathit{Cut}(C, \langle P, \sqsubseteq \rangle)}$$

A cut of store refinements maintains the same overall concretization, but allow us to split the space in the abstract:

**Theorem 27** (Concretization split)**.** *For all* C *such that* $\mathit{Cut}(C, \mathit{Refinements}(\hat{\sigma}))$, $\gamma_S(\hat{\sigma}) = \bigcup \{\gamma_S(\hat{\sigma} \blacktriangleleft \delta) \ : \ \delta \in C\}$

The notation for "apply refinement $\delta$ to store $\hat{\sigma}$" is $\hat{\sigma} \blacktriangleleft \delta$. The operation casts the *PreTerm* in $\delta$ to an *NDTerm* and strongly updates the store.

$$\begin{aligned}
\langle h, \mu \rangle \blacktriangleleft \delta &\triangleq \langle h \triangleleft \lambda \hat{a}.\lceil \delta(\hat{a}) \rceil_h, \mu \rangle \\
\text{where } f \triangleleft g &\triangleq \lambda x.x \overset{?}{\in} \mathbf{dom}(g) \to g(x), f(x) \\
\lceil \mathbf{External}(E, v) \rceil_h &= \mathbf{NDT}(\emptyset, [E \mapsto v]) \\
\lceil \mathbf{Delay}(\hat{a}) \rceil_h &= h(\hat{a}) \\
\lceil \mathbf{NDT}(\widehat{ts}, \mathit{Es}) \rceil_h &= \mathbf{NDT}(\widehat{ts}, \mathit{Es}) \\
\lceil \widehat{st} \rceil_h &= \mathbf{NDT}(\{\widehat{st}\}, \bot) \text{ otherwise}
\end{aligned}$$

A minimal cut is preferable since it less eagerly splits the state space, but is not necessary.

The example above of "fresh structural address identity" requires that we search both refinements

$[\hat{a} \mapsto$ `(unit)`$]$, and

$[\hat{a} \mapsto$ `(top)`$]$

in which case our equality result would have a map of the first refinement to some set of term pairs, and the second refinement to some other set of term pairs. The sets of term pairs are separated this way to denote, "the equality of these pairs of terms is consistent with a store refined by the given refinement."

$$
\begin{aligned}
\mathbf{Must}(R) \sqcup \mathbf{Fail} &= \mathbf{May}(\mathit{squash}(R)) \\
\mathbf{Must}(R) \sqcup \mathbf{Must}(R') &= \mathbf{Must}(\lambda\delta.R(\delta) \cup R'(\delta)) \\
\mathbf{Must}(R) \sqcup \mathbf{May}(\mathit{ps}) &= \mathbf{May}(\mathit{ps} \cup \mathit{squash}(\mathit{dp})) \\
\mathbf{May}(\mathit{Us}) \sqcup \mathbf{Fail} &= \mathbf{May}(\mathit{Us}) \\
\mathbf{May}(\mathit{Us}) \sqcup \mathbf{May}(\mathit{Us}') &= \mathbf{May}(\mathit{Us} \cup \mathit{Us}') \\
\mathbf{Fail} \sqcup \mathbf{Fail} &= \mathbf{Fail}
\end{aligned}
$$

where

$$\mathit{squash}(R) = \bigcup \mathbf{rng}(R)$$

Figure 30: Join operation for $\mathit{Res}[U]$

We thus have the following intermediate equality result type:

$$
\begin{aligned}
\widehat{eq} \in \widehat{\mathit{EqRes}} &= \mathit{Res}[\widehat{\mathit{Term}} \times \widehat{\mathit{Term}}] \\
\text{where } \mathit{Res}[U] &::= \mathbf{Fail} \mid \mathbf{Must}(R) \mid \mathbf{May}(\mathit{Us}) \\
R \in \mathit{Refmap}[U] &= \mathit{Refinement} \underset{\text{fin}}{\rightharpoonup} \wp(U) \\
\mathit{Us} &\in \wp(U) \\
u &\in U
\end{aligned}
$$

We will use the following metavariables for the instantiated generic forms:

$$\mathit{dp} \in \mathit{Refmap}[\widehat{\mathit{Term}} \times \widehat{\mathit{Term}}] \text{ and } \mathit{ps} \in \widehat{\mathit{Pairs}} = \wp_{\text{fin}}(\widehat{\mathit{Term}} \times \widehat{\mathit{Term}})$$

The result type is embedded in a not-quite-monad type (we get monads in pattern matching and evaluation):

$$\widehat{em} \in \widehat{\mathit{EqResM}} = \mathit{Refinement} \times \widehat{\mathit{Pairs}} \to \widehat{\mathit{EqRes}}$$

We require that the domain of a *Refmap* must cut the set of refinements for the current abstract store.

When we find an inequality, we throw away all the term pairs because they entail a falsehood. The *dp* map splits up the sets of “guarded truths” by the refinement used to justify them.

If we find that both strong equality and strong inequality are possible, then we join the results to jump to a **May** equality. To make $\widehat{\mathit{EqRes}}$ a join semilattice, we pointwise-union the sets of term pairs in **Must** and **May**. The join operation is the symmetric closure of the rules in Figure 30.

For combined equality for terms like **Variant**s, we will want to sequence our operations to thread *ps* through. The equality sequencing operation is defined in Figure 31. If any individual term is strongly

$$
\begin{aligned}
seq &: \widehat{EqResM} \to \widehat{EqResM} \to \widehat{EqResM} \\
seq(\hat{em}, \hat{em}')(\delta, ps) &= \texttt{case}\ \hat{em}(\delta, ps)\ \texttt{of} \\
&\qquad \mathbf{Fail} : \mathbf{Fail} \\
&\qquad \mathbf{May}(ps') : weaken(\hat{em}'(\delta, ps')) \\
&\qquad \mathbf{Must}(dp) : \bigsqcup_{\delta' \in \mathbf{dom}(dp)} \hat{em}'(\delta', dp(\delta')) \\
\text{where } weaken(\mathbf{Must}(dp)) &= \mathbf{May}(squash(dp)) \\
weaken(\mathrm{r}) &= \mathrm{r} \text{ otherwise} \\
success &= \lambda(\delta, ps).\mathbf{Must}([\delta \mapsto ps]) \\
maybe &= \lambda(\delta, ps).\mathbf{May}(ps) \\
fail &= \lambda(\delta, ps).\mathbf{Fail}
\end{aligned}
$$

Figure 31: Operations for $\widehat{EqResM}$

unequal, the entire equality is strongly unequal, so the operation should short-circuit on **Fail**. Otherwise, if any individual term is weakly equal, then regardless of the other terms in the variant, the entire equality is weakly equal. Since **Must** splits the search space up by refinement, we apply $\mathtt{f}$ to each refinement and its corresponding set of term pairs.

NOTE ON NOTATION: In meta-meta-language definitions such as that of $\widehat{tequalaux}$, I use *resolvable* as both a (meta-)pattern synonym for

$$\mathbf{IAddr}(_, _) \vee \mathbf{Delay}(_) \vee \mathbf{NDT}(_, _), \text{ and}$$

in the right-hand-side, *resolvable* refers to the term that (meta-)matches the (meta-)pattern.

A term (meta-)matching *resolvable* has an associated resolution operation, *resolve*:

$$
\begin{aligned}
resolve &: \widehat{Store} \to \widehat{Term} \times (\widehat{Term} \to \widehat{EqResM}) \to \widehat{EqResM} \\
resolve(\hat{\sigma})(\mathbf{IAddr}(\hat{a}, _), \mathrm{f}) &= select(\hat{\sigma}, \hat{a}, \mathrm{f}) \\
resolve(\hat{\sigma})(\mathbf{Delay}(\hat{a}), \mathrm{f}) &= select(\hat{\sigma}, \hat{a}, \mathrm{f}) \\
resolve(\hat{\sigma})(\mathbf{NDT}(\widehat{ts}, Es), \mathrm{f}) &= \lambda(\delta, ps). \bigsqcup_{\hat{t} \in Choose(\mathbf{NDT}(\widehat{ts}, Es))} \mathtt{f}(\hat{t})(\delta, ps)
\end{aligned}
$$

where

$$\mathit{select} : \widehat{\mathit{Store}} \times \widehat{\mathit{Addr}} \times (\widehat{\mathit{Term}} \to \widehat{\mathit{EqResM}}) \to \widehat{\mathit{EqResM}}$$

$$\begin{aligned}
\mathit{select}(\hat{\sigma}, \hat{a})(\delta, \mathit{ps}) = &\ \texttt{if}\ \ \hat{a} \overset{?}{\in} \mathbf{dom}(\delta)\ \ \texttt{then} \\
&\quad \mathtt{f}(\delta(\hat{a}))(\delta, \mathit{ps}) \\
&\ \texttt{else if}\ \hat{\sigma}.\mu(\hat{a}) \overset{?}{=} 1\ \texttt{then} \\
&\quad \bigsqcup_{\hat{t} \in \mathit{Choose}(\hat{\sigma}.h(\hat{a}))} \mathtt{f}(\hat{t})(\delta[\hat{a} \mapsto \hat{t}], \mathit{ps}) \\
&\ \texttt{else} \quad \bigsqcup_{\hat{t} \in \mathit{Choose}(\hat{\sigma}.h(\hat{a}))} \mathtt{f}(\hat{t})(\delta, \mathit{ps})
\end{aligned}$$

is how we interpret addresses in the context of the store refinement. If we already know the value of a fresh address, we use it. If we don't yet know, and the address is fresh, we internally split the search space by creating different store refinements that assign the address its possible values. Otherwise, the address is "used" and can mean any one of its mapped terms without splitting the search.

The *bind* and *resolve* operations allow us to easily define abstract term equality, $\widehat{\mathit{tequal}}$:

$$\widehat{\mathit{tequal}} : \widehat{\mathit{State}} \times \widehat{\mathit{Store}} \to \widehat{\mathit{Term}} \times \widehat{\mathit{Term}} \to \widehat{\mathit{Equality}}$$

$$\text{where } \widehat{\mathit{Equality}} ::= \mathbf{Equal} \mid \mathbf{Unequal} \mid \mathbf{May}.$$

The full definition is in Figure 32.

The internal refinements that equality builds have extra structure that this definition does not leverage. If we determine that the reasons for strong equality and strong inequality *don't overlap*, then we can learn more about the state of the store once we consume an equality. We will see in the next subsection that sometimes it's advantageous to not immediately throw up our hands when we determine that both equality and inequality are possible.

### 6.7.2 *Better than exact: Term equality with splitting*

Equality is not the only role of our meta-semantics; results of equality must be *consumed* to guide further computation. If we can't prove two terms are definitively equal or unequal, in some cases we perform a *case split*, and learn something about the shape of the store in either case[24]. This means that (prehaps unintuitively) the way we consume an equality can affect the precision of our semantics in later steps of computation. Thus, an "exact" abstract term equality can be less precise than another "exact" abstract term equality in the grand scheme of the whole analysis. For example, consider a program

[24] *Readers familiar with Typed Racket can relate this to occurrence typing.*

```
(if (equal? a b)
```

For notational brevity, let $\mathbf{V} = \mathbf{Variant}$, and $\mathbf{Ex} = \mathbf{External}$.

$$\begin{aligned}
\widehat{tequal}(ctx)(t_0, t_1) &= \widehat{equality}(\widehat{guard}(ctx)(t_0, t_1, \bot, \emptyset)) \\
\text{where } \widehat{equality}(\mathbf{Must}(dp)) &= \mathbf{Equal} \\
\widehat{equality}(\mathbf{Fail}) &= \mathbf{Unequal} \\
\widehat{equality}(\mathbf{May}(ps)) &= \mathbf{May} \\
\widehat{guard}(ctx)(t_0, t_1)(\delta, ps) &= \mathtt{if}\ (t_0, t_1) \overset{?}{\in} ps\ \mathtt{then} \\
&\qquad success(\delta, ps) \\
&\quad \mathtt{else}\ \widehat{tequalaux}(ctx)(t_0, t_1)(\delta, ps \cup \{(t_0, t_1)\})
\end{aligned}$$

$$\begin{aligned}
&\widehat{tequalaux}(\hat{\varsigma}, \langle _, \mu \rangle)(\mathbf{EAddr}(\hat{a}), \mathbf{EAddr}(\hat{a})) = identical?(\mu, \hat{a}) \\
&\widehat{tequalaux}(ctx)(resolvable, t_1) = resolve(resolvable, \lambda t_0'.\widehat{guard}(ctx)(t_0', t_1)) \\
&\widehat{tequalaux}(ctx)(t_0, resolvable) = resolve(resolvable, \lambda t_1'.\widehat{guard}(ctx)(t_0, t_1')) \\
&\widehat{tequalaux}(ctx)(\mathbf{Ex}(E, v_0), \mathbf{Ex}(E, v_1)) = E.\widehat{\equiv}(ctx)(v_0, v_1) \\
&\widehat{tequalaux}(ctx)(\mathbf{V}(n, \mathbf{t}), \mathbf{V}(n, \mathbf{t}')) = V_A(ctx)(\mathbf{t}, \mathbf{t}') \\
&\widehat{tequalaux}(ctx)(t_0, t_1) = fail \text{ otherwise}
\end{aligned}$$

where

$$\begin{aligned}
identical?(\mu, \hat{a}) &= \mathtt{if}\ \mu(\hat{a}) \overset{?}{\leqslant} 1\ \mathtt{then} \\
&\qquad success \\
&\quad \mathtt{else}\ maybe \\
V_A &: \widehat{State} \times \widehat{Store} \to \widehat{Term}^* \times \widehat{Term}^* \to \widehat{EqResM} \\
V_A(ctx)(\langle\rangle, \langle\rangle) &= success \\
V_A(ctx)(t_0\mathbf{t}, t_0'\mathbf{t}') &= seq(\widehat{guard}(ctx)(t_0, t_0'), V_A(ctx)(\mathbf{t}, \mathbf{t}')) \\
V_A(ctx)(_, _) &= fail \text{ otherwise}
\end{aligned}$$

Figure 32: Abstract term equality

```
        (car a)
        (not (cdr b)))
```

where

a is bound to (cons F F),

b is bound to (cons F **IAddr**(*addr*, 'delay)), and

*addr* maps to **NDT**({T, F}, ⊥) in the store.

Regardless of the freshness of *addr*, the first equality is true and false, given different concretizations of b.

PRECISE FOR ONE STEP The first equality is a weak **May** result to represent that the concrete equalities were both true and false. The abstract semantics must then explore each of the "then" and "else" branches. The "then" branch evaluates to F. The "else" branch evaluates to either T or F; if *addr* is fresh, then the state space gets split when not inspects the contents of *addr*.

PRECISE FOR MORE STEPS If *addr* is fresh, we could refine the state space to collapse *addr* to either T or F. We test equality in the split state space and determine that in one world, a and b are strongly equal, and in the other world, a and b are strongly unequal. If we keep stepping the computation in these parallel worlds, we find that the "else" branch evaluates only to F. This happens because the "else" branch is only reachable when the store has *addr* mapped to T.

If we could (magically) produce some number of refinements up front, apply them, and query $\widehat{tequal}$ in the split space, that'd be one way to get trace partitioning. What is this magic, and how do we determine that it's not making us do useless work?

### 6.7.3 *Worthwhile splitting*

Suppose we had our hands on some magic partitioning. We might ask ourselves what properties it should have.

For one, it should find refinements that produce *only* strong results when possible. We should only get a **May** result when no partitioning yields only strong results. The full spectrum of equality judgments is what happens when we apply each possible refinement. To first approximation, we will say this is $tequal_S$:

$$tequal_{S?}(\hat{\varsigma}, \hat{\sigma})(\hat{t}_0, \hat{t}_1)(\delta) = tequal^*_{S?}(\hat{\varsigma}, \hat{\sigma})(\hat{t}_0, \hat{t}_1)(\delta, \emptyset)$$
$$\text{where}$$
$$tequal^*_{S?}(\hat{\varsigma}, \hat{\sigma})(\hat{t}_0, \hat{t}_1)(\delta, ps) = [\delta' \mapsto \widehat{tequalaux}(\hat{\varsigma}, \hat{\sigma})(\hat{t}_0, \hat{t}_1)(\delta', ps) \\ : \delta' \in \mathit{Refinements}(\hat{\sigma}), \delta \sqsubseteq \delta']$$

The refinements considered for non-splitting equality must be extensions of the base refinement, $\delta$. Applications of store refinements are strong updates that inject into **NDT**.

This definition is obviously over-eager, and less obviously insufficient. Many refinements will be irrelevant to the results, meaning the state space is split before it *needs* to be. We also don't want to split the state space **at all** if we still have to consider a **May** equality. There can be some refinements that are too small to make a strong prediction (*e.g.*, no refinement at all: $\bot$), so if there are any **May** results, there still might be strong results.

A refined equality P is a function

$$P : \mathit{Refinements}(\hat{\sigma}) \to \widehat{\mathit{Equality}}$$

Let's say that if C is a cut of (the domain of) an equality result P, and all refinements in C map to a non-**May** answer in P, then we say that C is an *almost worthwhile* cut. If P is additionally antitone (less refined means more imprecise), then C is a *worthwhile* cut. We need the antitone property on P to make sure that we don't make some absurd jump from a refinement $\delta$ justifying **Equal** to a larger refinement justifying **Unequal** or **May**. Refinements have the property that once they're precise enough for a strong result, no extra information will refute or weaken it.

$$\textsc{Worthwhile cut}\quad \frac{\mathit{Cut}(C, \mathbf{dom}(P)) \qquad P \text{ antitone} \qquad \forall\delta \in C.P(\delta) \neq \mathbf{May}}{\mathit{worthwhile}(C, P)}$$

If there are no worthwhile cuts, then $\mathit{tequal}_S(\hat{\varsigma}, \hat{\sigma})(\hat{t}_0, \hat{t}_1, \delta) = \mathbf{May}$. There can be several (even minimal) worthwhile cuts, so $\mathit{tequal}_S$ is not yet a function. We need a function to split the state space for the operational semantics written with a *step* function. The definitions here do give us a space of acceptable answers, so that $\mathit{tequal}_S$ need only return *some* worthwhile cut if it exists:

$$\begin{aligned} \mathit{tequal}_S(\hat{\varsigma}, \hat{\sigma})(\hat{t}_0, \hat{t}_1, \delta) \in \tilde{P} &\stackrel{?}{=} \emptyset \to \{\mathbf{May}\}, \tilde{P} \\ \text{where } P &= \mathit{tequal}_{S?}(\hat{\varsigma}, \hat{\sigma})(\hat{t}_0, \hat{t}_1) \\ \tilde{P} &= \{P|_C \;:\; \mathit{worthwhile}(C, P)\} \end{aligned}$$

For inductive reasoning, we do need to have a connection between combinations of $\mathit{tequal}_{S?}$ and combinations of terms. We say that two worthwhile cuts are *conflicting* if elements that map to different polarities are comparable:

$$\frac{\mathit{worthwhile}(C', P') \qquad \begin{array}{c}\mathit{worthwhile}(C, P) \\ \exists\delta \in C, \delta' \in C'.\delta \circledast \delta' \wedge P(\delta) \sqcup P'(\delta') = \mathbf{May}\end{array}}{\mathit{conflicting}(C, P, C', P')}$$

Sets of refinements $C$ and $C'$ are combined by taking the max of all comparable refinements. For example, if $t_0$ and $t_1$ are equal with a $\bot$ refinement, but $t_0'$ and $t_1$ are equal with a $[\hat{a} \mapsto 0]$ refinement, then $\mathbf{NDT}(\{t_0, t_0'\}, \bot)$ is equal to $t_1$ with only a $[\hat{a} \mapsto 0]$ refinement. We want small cuts, but we grow them as needed. Cuts combine by taking the largest of comparable elements:

$$C \sqcup C' = \{\delta \in C \cup C' \ : \ \forall \delta' \in C \cup C'. \delta' \circledast \delta \implies \delta' \sqsubseteq \delta\}$$

**Lemma 28** (Worthwhile composition). *Given total* $P, P'$ : *Refinements*$(\hat{\sigma}) \to$ *Equality, if worthwhile*$(C, P)$*, worthwhile*$(C', P')$ *and* $\neg$*conflicting*$(C, P, C', P')$ *then worthwhile*$(C \sqcup C', P \sqcup P')$.

**Lemma 29** (Conflicting composition never worthwhile). *If conflicting*$(C, P, C', P')$*, then for all* $C''$, $\neg$*worthwhile*$(C'', P \sqcup P')$.

When we have a recursive call that creates a **May** result, we need to know that no matter what, we can't extend it to finagle a strong result.

**Lemma 30** (Fruitless extension). *If (for all* $C$, $\neg$*worthwhile*$(C, P)$*), then for all* $P', C$, $\neg$*worthwhile*$(C, P \sqcup P')$.

*Proof.* Results can only get worse via $\sqcup$, so whichever $\delta \in C$ leads to $P(\delta) = \mathbf{May}$ from the hypothesis, we get $(P \sqcup P')(\delta) = \mathbf{May}$. $\square$

Now we have the metatheory to handle a compositional splitting equality function. Let's move on to define *tequal*$_S$, which will build **a** worthwhile cut, if at least one exists.

### 6.7.4 *Abstract term equality with worthwhile splitting*

Without access to an oracle for a worthwhile cut, we need a way to produce one bottom-up as we check term equality. Fortunately, we need only change our definitions of $\widehat{EqRes}$ and its associated operators, $\sqcup$ and *bind*, and equality's use of **Fail**. The definition of abstract term equality with splitting in Figure 33 almost exactly mirrors equality without splitting, except an **Unequal** result now stores the current refinement in a set, as $\mathbf{Both}(\bot, \{\delta\})$, and $\mathbf{Equal}(dp)$ is written $\mathbf{Both}(dp, \emptyset)$.

With splitting equalities, we must additionally remember which refinements lead to strong *inequality*. If refinements justifying strong equality do not "conflict" with refinements justifying strong inequality, then we have a worthwhile splitting.

We have a notion of refinement overlap that helps us decide if two cuts are conflicting:

$$\textsc{Overlapping refinements}\quad \frac{\delta \in \Delta \qquad \delta' \in D \qquad \delta \circledast \delta'}{\Delta \bowtie D}$$

$$\widehat{tequal}_S : \widehat{State} \times \widehat{Store} \to \widehat{Term} \times \widehat{Term} \times Refinement \to \widehat{Equality}_S$$

$$\begin{aligned}
\widehat{tequal}_S(ctx)(t_0, t_1, \delta) &= \widehat{equality}_S(\widehat{guard}_S(ctx)(t_0, t_1)(\delta, \emptyset)) \\
\widehat{equality}_S(\mathbf{Both}(\bot, \Delta)) &= \mathbf{Unequal} \\
\widehat{equality}_S(\mathbf{Both}(dp, \emptyset)) &= \mathbf{Equal} \\
\widehat{equality}_S(\mathbf{Both}(dp, \Delta)) &= \mathbf{Split}(\mathbf{dom}(dp), \Delta) \\
\widehat{equality}_S(\mathbf{May}(ps)) &= \mathbf{May} \\
\widehat{guard}_S : \widehat{State} \times \widehat{Store} &\to \widehat{Term} \times \widehat{Term} \to \widehat{EqResM}_S \\
\widehat{guard}_S(ctx)(t_0, t_1)(\delta, ps) &= \mathtt{if}\ (t_0, t_1) \overset{?}{\in} ps\ \mathtt{then} \\
&\qquad \mathbf{Both}([\delta \mapsto ps], \emptyset) \\
&\quad \mathtt{else}\ \widehat{tequal}^*_S(ctx)(t_0, t_1)(\delta, ps \cup \{(t_0, t_1)\})
\end{aligned}$$

$$\begin{aligned}
&\widehat{tequal}^*_S : \widehat{State} \times \widehat{Store} \to \widehat{Term} \times \widehat{Term} \to \widehat{EqResM}_S \\
&\widehat{tequal}^*_S(\hat{\varsigma}, \hat{\sigma})(\mathbf{EAddr}(\hat{a}), \mathbf{EAddr}(\hat{a})) = identical?_S(\hat{\sigma}.\mu, \hat{a}) \\
&\widehat{tequal}^*_S(ctx)(resolvable, t_1) = eq\text{-}resolve_S(ctx)(resolvable, \lambda t_0'.\widehat{guard}_S(ctx)(t_0', t_1)) \\
&\widehat{tequal}^*_S(ctx)(t_0, resolvable) = eq\text{-}resolve_S(ctx)(resolvable, \lambda t_1'.\widehat{guard}_S(ctx)(t_0, t_1')) \\
&\widehat{tequal}^*_S(ctx)(\mathbf{V}(n, \mathbf{t}), \mathbf{V}(n, \mathbf{t}')) = V_S(ctx)(\mathbf{t}, \mathbf{t}') \\
&\widehat{tequal}^*_S(ctx)(\mathbf{Ex}(E, v_0), \mathbf{Ex}(E, v_1)) = E.\hat{\equiv}_S(ctx)(v_0, v_1) \\
&\widehat{tequal}^*_S(ctx)(t_0, t_1) = fail_S \text{ otherwise}
\end{aligned}$$

where

$$\begin{aligned}
identical?_S(\mu, \hat{a}) &= \mathtt{if}\ \mu(\hat{a}) \overset{?}{\leqslant} 1\ \mathtt{then} \\
&\qquad success_S \\
&\quad \mathtt{else}\ maybe_S \\
V_S : \widehat{State} \times \widehat{Store} &\to \widehat{Term}^* \times \widehat{Term}^* \to \widehat{EqResM}_S \\
V_S(ctx)(\langle\rangle, \langle\rangle) &= success_S \\
V_S(ctx)(t_0\mathbf{t}, t_0'\mathbf{t}') &= seq_S(guard_S(ctx)(t_0, t_0'), V_S(ctx)(\mathbf{t}, \mathbf{t}')) \\
V_S(ctx)(_, _) &= fail_S \text{ otherwise}
\end{aligned}$$

Figure 33: Splitting term equality

The bind and join operations on a new equality result type are changed under the hood. The intermediate result type combines strong equality and inequality into a single variant, since non-overlapping refinements can justify different outcomes.

$$\begin{aligned} \widehat{EqRes}_S &= Res_S[\widehat{Term} \times \widehat{Term}] \\ Res_S[U] &::= \mathbf{Both}(R, \Delta) \mid \mathbf{May}(Us) \\ \text{where } R \in \mathit{Refmap}[U] &\quad Us \in \wp(U) \\ \Delta &\subseteq_{\text{fin}} \mathit{Refinement} \end{aligned}$$

The result type for the entire equality only suggests a case split if doing so is worthwhile. If all refinements we chased ended up proving strong equality, then we don't (yet) need to know why. We only split if we have a **Both** result with both non-empty equality judgments, and non-empty inequality judgments:

$$\widehat{Equality}_S ::= \mathbf{Equal} \mid \mathbf{Unequal} \mid \mathbf{May} \mid \mathbf{Split}(\Delta, \Delta)$$

Our equality type is again wrapped in a not-quite-monad type:

$$\widehat{em}_S \in \widehat{EqResM}_S = \widehat{Pairs} \to ResM_S[\widehat{Term} \times \widehat{Term}]$$

but we do prepare ourselves for upcoming sections with the $ResM_S[U]$ monad type:

$$ResM_S[U] = \mathit{Refinement} \to Res_S[U]$$

The $Res_S[U]$ type forms a join semilattice with $\mathbf{Both}(\bot, \emptyset)$ as bottom; the join operation is defined in Figure 34. The join operation is the symmetric closure of the rules in Figure 34. We use the $\bowtie$ relation defined above to determine if two instances of **Both**, as interpreted as both a cut and a refined equality function, satisfy the *conflicting* proposition of the previous subsection. A **Both** result is interpreted as a cut and refined equality function by *as-W*:

$$\begin{aligned} \mathit{as\text{-}W} &: \widehat{EqRes}_S \to (\wp(\mathit{Refinement}) \times (\mathit{Refinement} \to \widehat{Equality})) \\ \mathit{as\text{-}W}(\mathbf{Both}(dp, \Delta)) &= \left\langle \mathbf{dom}(dp) \cup \Delta, \lambda\delta. \begin{cases} \mathbf{Equal} & \text{if } \exists \delta' \in \mathbf{dom}(dp). \delta \circledast \delta' \\ \mathbf{Unequal} & \text{if } \exists \delta' \in \mathbf{dom}(\Delta). \delta \circledast \delta' \end{cases} \right\rangle \end{aligned}$$

Note that the refined equality function *as-W* returns is total.

We have a the same few operations on $\widehat{EqResM}_S$ as we did on $\widehat{EqResM}$: *seq*, *eq-resolve*, *success*, *fail*, and *maybe*. The definitions are in

$$
\begin{aligned}
\mathbf{May}(\mathit{Us}) \sqcup \mathbf{May}(\mathit{Us}') &= \mathbf{May}(\mathit{Us} \cup \mathit{Us}') \\
\mathbf{Both}(R, _) \sqcup \mathbf{May}(\mathit{Us}) &= \mathbf{May}(\mathit{Us} \cup \mathit{squash}(R)) \\
\mathbf{Both}(R, \Delta) \sqcup \mathbf{Both}(R', \Delta') &= \texttt{if}\ \ \Delta \overset{?}{\bowtie} \mathbf{dom}(R')\ \mathbf{orelse}\ \Delta' \overset{?}{\bowtie} \mathbf{dom}(R)\ \ \texttt{then} \\
&\qquad \mathbf{May}(\mathit{squash}(R) \cup \mathit{squash}(R')), \\
&\qquad \mathbf{Both}(R \sqcup R', \Delta \sqcup \Delta')
\end{aligned}
$$

where

$$
R_0 \sqcup R_1 = [\delta \mapsto \bigcup_{\delta' \sqsubseteq \delta} R_0(\delta') \cup R_1(\delta')\ :\ \delta \in \mathbf{dom}(R_0) \sqcup \mathbf{dom}(R_1)]
$$

Figure 34: $\mathit{Res}_S[U]$ join rules

Figure 35. The *eq-resolve* function depends on the lower level $\mathit{resolve}_S$ of the *ResM* monad:

$$
\begin{aligned}
\mathit{resolve}_S &: \widehat{\mathit{Store}} \to \widehat{\mathit{Term}} \times (\widehat{\mathit{Term}} \to \mathit{ResM}[U]) \to \mathit{ResM}[U] \\
\mathit{resolve}_S(\hat{\sigma})(\mathbf{IAddr}(\hat{a}, _), f) &= \mathit{select}_S(\hat{\sigma}, \hat{a}, f) \\
\mathit{resolve}_S(\hat{\sigma})(\mathbf{Delay}(\hat{a}), f) &= \mathit{select}_S(\hat{\sigma}, \hat{a}, f) \\
\mathit{resolve}_S(\hat{\sigma})(\mathbf{NDT}(\widehat{\mathit{ts}}, \mathit{Es}), f) &= \lambda(\delta). \bigsqcup_{\hat{t} \in \mathit{Choose}(\mathbf{NDT}(\widehat{\mathit{ts}}, \mathit{Es}))} f(\hat{t})(\delta)
\end{aligned}
$$

where

$$
\begin{aligned}
\mathit{select}_S &: \widehat{\mathit{Store}} \times \widehat{\mathit{Addr}} \times (\widehat{\mathit{Term}} \to \mathit{ResM}[U]) \to \mathit{ResM}[U] \\
\mathit{select}_S(\hat{\sigma}, \hat{a})(\delta) &= \texttt{if}\ \ \hat{a} \overset{?}{\in} \mathbf{dom}(\delta)\ \ \texttt{then} \\
&\qquad f(\delta(\hat{a}))(\delta) \\
&\quad \texttt{else if}\ \ \hat{\sigma}.\mu(\hat{a}) \overset{?}{=} 1\ \ \texttt{then} \\
&\qquad \bigsqcup_{\hat{t} \in \mathit{Choose}(\hat{\sigma}(\hat{a}))} f(\hat{t})(\delta[\hat{a} \mapsto \hat{t}]) \\
&\quad \texttt{else} \bigsqcup_{\hat{t} \in \mathit{Choose}(\hat{\sigma}.h(\hat{a}))} f(\hat{t})(\delta)
\end{aligned}
$$

Refmaps are not joined pointwise since their domains are part of a cut (the $\Delta$ set is the other part). Cuts are joined with the max operation of the previous section. We treat unmapped refinements in a *Refmap* as mapping to $\emptyset$ here.

$$
\begin{aligned}
seq_S &: \widehat{EqResM}_S \times \widehat{EqResM}_S \to \widehat{EqResM}_S \\
seq_S(\hat{em}_S, \hat{em}'_S)(ps)(\delta) &= \texttt{case } \hat{em}_S(ps)(\delta) \texttt{ of} \\
&\qquad \mathbf{Both}(dp, \Delta) : \mathbf{Both}(\bot, \Delta) \sqcup \bigsqcup_{\delta' \in \mathbf{dom}(dp)} \hat{em}'_S(\delta', dp(\delta')) \\
&\qquad \mathbf{May}(ps) : weaken_S(\hat{em}'_S(\delta, ps)) \\
weaken_S &: Res_S[\mathsf{U}] \to Res_S[\mathsf{U}] \\
weaken_S(\mathbf{Both}(\bot, \Delta)) &= \mathbf{Both}(\bot, \Delta) \\
weaken_S(\mathbf{May}(Us)) &= \mathbf{May}(Us) \\
weaken_S(\mathbf{Both}(\mathsf{R}, _)) &= \mathbf{May}(squash(\mathsf{R})) \text{ otherwise}
\end{aligned}
$$

with the varying success operations

$$
\begin{aligned}
success_S(ps)(\delta) &= \mathbf{Both}([\delta \mapsto ps], \emptyset) \\
maybe_S(ps)(\delta) &= \mathbf{May}(ps) \\
fail_S(ps)(\delta) &= \mathbf{Both}(\bot, \{\delta\})
\end{aligned}
$$

and the resolution operation

$$
\begin{aligned}
&\textit{eq-resolve} : \widehat{State} \times \widehat{Store} \to \widehat{Term} \times (\widehat{Term} \to \widehat{EqResM}_S) \to \widehat{EqResM}_S \\
&\textit{eq-resolve}(\hat{\varsigma}, \hat{\sigma})(\hat{t}, f)(\delta, ps) = resolve_S(\hat{\sigma})(\hat{t}, f)(\delta)
\end{aligned}
$$

Figure 35: $\widehat{EqResM}_S$ sequencing

In Figure 35 we define a bind operator for sequencing equality judgments through variants and maps as a form of "this *and* that are equal." The behavior should short-circuit when unequal, jump to top if we pass through a **May** without further inequalities, and combine possibilities on equalities.

## 6.8 PATTERN MATCHING

The semantics of patterns is defined by matching, which has a few pieces to consider. The fixed inputs are the current state and address cardinalities, $\hat{\varsigma}$ and $\mu$. The variable inputs are the following:

- $p \in \mathit{Pattern}$: the pattern;
- $t \in \widehat{\mathit{Term}}$: the term to match;
- $\rho \in \widehat{\mathit{MEnv}}$: the metavariable binding environment mapping pattern names to terms;
- $\delta \in \mathit{Refinement}$: the currently pursued refinement.

The inexactness of terms means that a pattern can match in multiple different ways. Matching must therefore return a *set* of metalanguage environments when it successfully matches. Similar to term equality, matching builds up store refinements as it either resolves indirect addresses to further match, or uses splitting term equality in non-linear patterns. The match function uses the same $\mathit{Res}_S$ return container as splitting abstract term equality, but stores output binding environments instead of term pairs. Additionally, the return type $\widehat{\mathit{MatchM}}$ is an actual monad:

$$\begin{aligned}
\widehat{\mathit{MatchM}} &= \mathit{ResM}_S[\widehat{\mathit{MEnv}}] \\
\text{where } \mathit{ResM}_S[U] &= \mathit{Refinement} \to \mathit{Res}_S[U] \\
\rho \in \widehat{\mathit{MEnv}} &= \mathit{Name} \underset{\text{fin}}{\rightharpoonup} \widehat{\mathit{Term}}
\end{aligned}$$

$$\begin{aligned}
\mathit{de} &\in \mathit{Refmap}[\widehat{\mathit{MEnv}}] \\
\mathit{Rs} &\in \wp(\widehat{\mathit{MEnv}})
\end{aligned}$$

Similarly, the top level return type has four variants:

$$\widehat{\mathit{Match}} ::= \mathbf{Success}(\mathit{Rs}) \mid \mathbf{Fail} \mid \mathbf{May}(\mathit{Rs}) \mid \mathbf{Split}(\mathit{de}, \Delta)$$

The monad operations in Figure 36 manage the nondeterminism.

Matching is driven structurally by well-founded patterns, except insofar as term resolution eventually results in terms with more structure. Matching and resolution are extended with a guard set in order to catch unproductive recursion. This detail distracts from the overall presentation, so it appears only in the Haskell implementation in Appendix 18

$$\begin{aligned}
&\qquad\qquad \mathit{bind}_S : \mathit{ResM}_S[U] \times (U \to \mathit{ResM}_S[V]) \to \mathit{ResM}_S[V] \\
&\mathit{bind}_S(\hat{m}, f)(\delta) = \texttt{case}\ \hat{m}(\delta)\ \texttt{of} \\
&\qquad\qquad \mathbf{Both}(R,\Delta) : \mathbf{Both}(\bot,\Delta) \sqcup \bigsqcup_{\delta' \in \mathbf{dom}(R), u \in R(\delta')} f(u)(\delta') \\
&\qquad\qquad \mathbf{May}(\mathit{Us}) : \mathit{weaken}_S(\bigsqcup_{u \in \mathit{Us}} f(u)(\delta)) \\
&\mathit{return}_S(u)(\delta) = \mathbf{Both}([\delta \mapsto \{u\}], \emptyset)
\end{aligned}$$

Figure 36: Monad operations on $\mathit{ResM}_S$

When we match a term with the pattern $\mathbf{Name}(x, p)$, the term is demanded and bound to $x$ in the $\widehat{\mathit{MEnv}}$, and the demanded the term is further inspected by $p$. If the term is an **IAddr**, then its lookup modality is consulted to drive the matching semantics. Additionally if the term is *resolvable*, then its nondeterminism is resolved unless the pattern is **Wild**[25].

[25] *It is sound to resolve anyway, but wasteful to unnecessarily split the binding environments.*

Let's take a look at how the different lookup modalities drive pattern matching in a small example. The example term we'll match on is

$$\mathbf{Variant}(\texttt{pair}, \langle \mathbf{IAddr}(\hat{a}, \mathit{lm}), \mathbf{NDT}(\{0, 1\})\rangle).$$

Our example pattern is "bind the first subterm to $x$, and the second subterm to $y$, insofar as $y$ is a number." The pattern for this is

$$\mathbf{Variant}(\texttt{pair}, \langle \mathbf{Name}(x, \mathbf{Wild}), \mathbf{Name}(y, \mathbf{Is\text{-}External}(\mathit{Number}))\rangle).$$

The context we have is that $\hat{a}$ is fresh, and the store maps $\hat{a}$ to $\mathbf{NDT}(\{\mathsf{T}, \mathsf{F}\})$. Let's vary $\hat{a}$'s lookup modality and see what we get for a match result:

- $\mathit{lm} =$ `'resolve`: the match produces two possible store refinements, each mapped to a metalanguage environment. The $\hat{a}$ refinement determines the term to which $x$ is bound, and $y$ is bound to either 0 or 1 (the **Is-External** pattern ensures the **NDT** is demanded).

$$\begin{aligned}
\mathbf{Must}([[\hat{a} \mapsto \mathsf{T}] &\mapsto \{[x \mapsto \mathsf{T}, y \mapsto 0], \\
&\qquad [x \mapsto \mathsf{T}, y \mapsto 1]\}, \\
[\hat{a} \mapsto \mathsf{F}] &\mapsto \{[x \mapsto \mathsf{F}, y \mapsto 0], \\
&\qquad [x \mapsto \mathsf{F}, y \mapsto 1]\}]).
\end{aligned}$$

- $\mathit{lm} =$ `'delay`: produces two environments with no refinement

$$\begin{aligned}
\mathbf{Must}([\bot \mapsto \{&[x \mapsto \mathbf{Delay}(\hat{a}), y \mapsto 0], \\
&[x \mapsto \mathbf{Delay}(\hat{a}), y \mapsto 1]\}]).
\end{aligned}$$

When **Delay**$(\hat{a})$ is inspected, $\hat{a}$ will be *at least* be in $\{\mathsf{T}, \mathsf{F}\}$ since the store currently maps $\hat{a}$ to **NDT**$(\{\mathsf{T}, \mathsf{F}\})$. Further, when inspected, $\hat{a}$ might have gone from fresh to used. Both of these possibilities exist because between delay and demand time, $\hat{a}$ can be reallocated and/or updated with additional terms.

- $lm =$ `'deref`: produces two environments with no refinement

  $$\mathbf{Must}([\bot \mapsto \{[x \mapsto \mathbf{NDT}(\{\mathsf{T}, \mathsf{F}\}), y \mapsto 0], \\ [x \mapsto \mathbf{NDT}(\{\mathsf{T}, \mathsf{F}\}), y \mapsto 1]\}])$$

  so $x$ denotes the *NDTerm* that is *either* $\mathsf{T}$ or $\mathsf{F}$.

Weak matching is defined in Figure 37. The simplest cases depend on identity in the metalanguage. Modalities, variant names and external space descriptors should be unique values, so this is not problematic.

Abstract pattern matching has four differences from concrete pattern matching:

1. the monad is changed to support refinement and nondeterminism;
2. non-linear patterns use abstract term equality, splitting and weakening the match result as the equality result dictates;
3. our notion of demand can split the match space by store refinements if the demanded term has a `'resolve` lookup modality;
4. there are other terms than **IAddr** to resolve before continuing matching: **NDT** and **Delay**.

The abstract notion of demand is

$$\begin{aligned}
\widehat{demand} : \widehat{Term} \times \widehat{Store} \times \mathit{Refinement} \times \mathit{Pattern} &\to \wp(\mathit{PreTerm} \times \mathit{Refinement}) \\
\widehat{demand}(\mathbf{IAddr}(\hat{a}, lm), \hat{\sigma}, \delta, p) &= \mathtt{case}\ lm\ \mathtt{of} \\
&\qquad \mathtt{'delay} : \{\langle \mathbf{Delay}(\hat{a}), \delta\rangle\} \\
&\qquad \mathtt{'deref} : \{\langle \mathit{deref}(\hat{\sigma}, \delta, \hat{a}), \delta\rangle\} \\
&\qquad \mathtt{'resolve} : \mathit{select}(\hat{\sigma}, \delta, \hat{a}) \\
\widehat{demand}(\mathbf{NDT}(\widehat{ts}, Es), \hat{\sigma}, \delta, \mathbf{Wild}) &= \{\langle \mathbf{NDT}(\widehat{ts}, Es), \delta\rangle\} \\
\widehat{demand}(\mathbf{NDT}(\widehat{ts}, Es), \hat{\sigma}, \delta, p) &= \{\langle \hat{t}, \delta\rangle \ :\ \hat{t} \in \mathit{Choose}(\mathbf{NDT}(\widehat{ts}, Es))\} \\
\widehat{demand}(\hat{t}, \hat{\sigma}, \delta, p) &= \{\langle \hat{t}, \delta\rangle\}
\end{aligned}$$

Where *deref* defines dereferencing an address without resolving it:

$$\mathit{deref} : \widehat{Store} \times \mathit{Refinement} \times \widehat{Addr} \to \widehat{Term}$$

$$\begin{aligned}
\mathit{deref}(\hat{\sigma}, \delta, \hat{a}) = \ &\mathtt{if}\ \hat{a} \overset{?}{\in} \mathbf{dom}(\delta)\ \mathtt{then} \\
&\quad \delta(\hat{a}) \\
&\mathtt{else}\ \hat{\sigma}.h(\hat{a})
\end{aligned}$$

$$\hat{M}_S : \widehat{State} \times \widehat{Store} \to Pattern \times \widehat{Term} \times \widehat{MEnv} \times Refinement \times \to \widehat{Match}$$

$$\begin{aligned}
\hat{M}_S(ctx)(p, t, \rho, \delta) &= match(\hat{M}^*_S(ctx)(p, t, \rho)(\delta)) \\
match(\mathbf{Both}(\bot, \Delta)) &= \mathbf{Fail} \\
match(\mathbf{Both}(de, \emptyset)) &= \mathbf{Success}(squash(de)) \\
match(\mathbf{Both}(de, \Delta)) &= \mathbf{Split}(de, \Delta) \\
match(\mathbf{May}(Us)) &= \mathbf{May}(Us)
\end{aligned}$$

$$\hat{M}^*_S \quad : \quad \widehat{State} \times \widehat{Store} \quad \to \quad Pattern \times \widehat{Term} \times \widehat{MEnv} \quad \to \quad \widehat{MatchM}$$

$$\hat{M}^*_S(ctx)(\mathbf{Name}(x, p), t, \rho) = \lambda\delta.$$

$\texttt{if } x \overset{?}{\in} \mathbf{dom}(\rho) \texttt{ then}$

$\quad \texttt{case } \widehat{tequal}_S(ctx)(\rho(x), t, \delta) \texttt{ of}$

$\quad\quad \mathbf{Equal} : \hat{M}^*_S(ctx)(p, t, \rho)(\delta)$

$\quad\quad \mathbf{Unequal} : fail_S(\delta)$

$\quad\quad \mathbf{Split}(\Delta_=, \Delta_{\neq}) : \mathbf{Both}(\bot, \Delta_{\neq}) \sqcup \bigsqcup_{\delta' \in \Delta_=} \hat{M}^*_S(ctx)(p, t, \rho)(\delta')$

$\quad\quad \mathbf{May} : weaken_S(\hat{M}^*_S(ctx)(p, t, \rho)(\delta))$

$\texttt{else} \quad \bigsqcup_{\langle t', \delta' \rangle \in \widehat{demand}(t, \hat{\sigma}, \delta, p)} \hat{M}^*_S(p, t', \rho[x \mapsto t'])(\delta')$

$$\begin{aligned}
\hat{M}^*_S(ctx)(\mathbf{Wild}, t, \rho) &= return_S(\rho) \\
\hat{M}^*_S(ctx)(\mathbf{Is\text{-}Addr}, \mathbf{EAddr}(_), \rho) &= return_S(\rho) \\
\hat{M}^*_S(ctx)(\mathbf{Is\text{-}External}(E), \mathbf{Ex}(E, _), \rho) &= return_S(\rho) \\
\hat{M}^*_S(ctx)(\mathbf{V}(n, \overline{p}), \mathbf{V}(n, \overline{t}), \rho) &= V_{\hat{M}}(ctx)(\overline{p}, \overline{t}, \rho) \\
\hat{M}^*_S(ctx)(p, resolvable, \rho) &= \\
& resolves_S(ctx.\hat{\sigma})(resolvable, \lambda t'.\hat{M}^*_S(ctx)(p, t', \rho)) \\
\hat{M}^*_S(ctx)(p, t, \rho) &= fail_S \text{ otherwise}
\end{aligned}$$

where

$$\begin{aligned}
V_{\hat{M}}(ctx)(\langle\rangle, \langle\rangle, \rho) &= return_S(\rho) \\
V_{\hat{M}}(ctx)(p_0\overline{p}, t_0\overline{t}, \rho) &= bind_S(\hat{M}^*_S(ctx)(p_0, t_0, \rho), V_{\hat{M}}(ctx)(\overline{p}, \overline{t}, \rho)) \\
V_{\hat{M}}(ctx)(_, _, _) &= fail_S \text{ otherwise}
\end{aligned}$$

Figure 37: Weak pattern matching

If p is not **Wild**, then we must *match through* the term, resolving any nondeterminism in t since it is now in a *strict* position.

The version of pattern matching without worthwhile splitting ($\hat{\mathcal{M}}$) is a minor change to this definition, which I won't fully reconstruct. The result type uses the non-splitting container, $\mathit{Res}[\widehat{\mathit{MEnv}}]$. The monad operations change so that $\mathit{return}_S(\rho) = \mathbf{Must}(\{\rho\})$, $\mathit{fail} = \mathbf{Fail}$, $\mathcal{B}$ uses the join operation on $\mathit{Res}[U]$, and we use the non-splitting equality function.

The correctness criteria are then the exact approximation and worthwhile splitting requirements that we proved for the abstract term equalities.

**Theorem 31** (Non-splitting match is an exact approximation)**.** $\gamma' \circ \hat{\mathcal{M}} = \mathcal{M} \circ \gamma$ *where* $\gamma$ *is the structural concretization of* $\hat{\mathcal{M}}$*'s inputs, and* $\gamma'$ *is the concretization of* $\mathit{Res}[\widehat{\mathit{MEnv}}]$.

We generalize *worthwhile* to support the different $\mathit{Res}_S[U]$ result type (any **May** result is bad):

$$\textsc{Worthwhile cut}\quad \frac{\mathit{Cut}(C, \mathbf{dom}(P)) \qquad P \text{ antitone} \qquad \forall \delta \in C, \mathit{Us}.P(\delta) \neq \mathbf{May}(\mathit{Us})}{\mathit{worthwhile}'(C, P)}$$

If a result is worthwhile, we can use a refined enough input that produces a single strong result.

**Theorem 32** (Matching worthwhile)**.** $\hat{\mathcal{M}}^*_S(\hat{\varsigma}, \hat{\sigma})(p, t, \delta, \rho)$ *is in*

$$\begin{aligned}
&\texttt{if } \tilde{P} \stackrel{?}{=} \emptyset \texttt{ then} \\
&\quad \{\hat{\mathcal{M}}(\hat{\varsigma}, \hat{\sigma})(p, t, \rho)(\delta)\} \\
&\texttt{else} \\
&\quad \{\mathbf{Both}([\delta \mapsto U : P(\delta) = \mathit{return}(\delta, U)], P^{-1}(\mathbf{Fail})) : P \in \tilde{P}\} \\
\mathit{where}\ & P = [\delta \mapsto \hat{\mathcal{M}}(\hat{\varsigma}, \hat{\sigma})(p, t, \rho)(\delta') : \delta' \in \mathit{Refinements}(\hat{\sigma}), \delta \sqsubseteq \delta'] \\
& \tilde{P} = \{P|_C : \mathit{worthwhile}'(C, P)\}
\end{aligned}$$

Now that we have equality and matching defined, we've covered the semantics for the left hand side of rules. We now need to handle the right hand side: expressions.

## 6.9 EXPRESSION EVALUATION

The expression language is modified slightly from the previous chapter on its concrete semantics. A store lookup expression has an additional lookup modality, which is functionally a no-op in the concrete. The whole expression grammar is in Figure 38.

$$
\begin{aligned}
e \in \mathit{Expr} &::= \mathbf{Ref}(x) \mid \mathbf{Variant}(n, \mathit{tag}, \overline{e}) \mid \mathbf{Call}(f, \overline{e}) \mid \mathbf{Let}(\overline{bu}, e) \\
&\mid \mathbf{Deref}(e, \mathit{lm}) \mid \mathbf{Alloc}(\mathit{tag}) \\
\mathit{lm} \in \mathit{Lookup\text{-}Modality} &::= \texttt{'resolve} \mid \texttt{'deref} \mid \texttt{'delay} \\
\mathit{bu} \in \mathit{BU} &::= \mathbf{Update}(e, e) \mid \mathbf{Where}(p, e) \\
f &\in \mathit{Metafunction\text{-}Names}
\end{aligned}
$$

Figure 38: Grammar of expressions

### 6.9.1 *Representation of evaluation results*

An expression can introduce changes to the store, so its evaluation result type includes both the term it evaluates to, and any store *changes*. In the concrete, we simply updated the store in-place and passed it along. In the abstract, that strategy introduces too much unnecessary overhead. Instead, expressions evaluate to

$$
\begin{aligned}
&\widehat{\mathit{EvResult}}_S[T] = \widehat{\mathit{State}} \to \mathit{Refinement} \to \widehat{\Delta\mathit{Store}} \to \mathit{Res}_S[T \times \widehat{\Delta\mathit{Store}}] \\
&\text{where } \widehat{\Delta\mathit{Store}} = \widehat{\mathit{Addr}} \underset{\text{fin}}{\rightharpoonup} \mathit{Change} \\
&\quad \mathit{ct} \in \mathit{Change} = \mathbf{Strong}(\mathit{PreTerm}) \mid \mathbf{Weak}(\mathit{AbsTerm}) \mid \mathbf{Reset}(\mathit{AbsTerm})
\end{aligned}
$$

Each possible result can change the store in different ways, in different store refinements, so an output $T$ is wrapped as such. The arguments to the left are for us to interpret expressions in *the abstract interpretation monad*. We will see the monad operations in the next subsection.

A STORE CHANGE object, *Change*, represents the ways we can update the store. A fresh address can be strongly updated to something entirely different. A used address can only have some updates joined in. Finally, a fresh address can be first strongly updated *and then reallocated* and updated again; the first strong update means that we *reset* the contents to a now monotonically growing *NDTerm*. We understand a $\widehat{\Delta\mathit{Store}}$ as its effect on an abstract store:

$$
\begin{aligned}
\mathit{apply}\Delta &: \widehat{\mathit{Store}} \times \widehat{\Delta\mathit{Store}} \to \widehat{\mathit{Store}} \\
\mathit{apply}\Delta(\hat{\sigma}, \bot) &= \hat{\sigma} \\
\mathit{apply}\Delta(\langle h, \mu\rangle, \partial\hat{\sigma}[\hat{a} \mapsto \mathbf{Strong}(\hat{t})]) &= \mathit{apply}\Delta(\langle h \blacktriangleleft [\hat{a} \mapsto \hat{t}], \mu[\hat{a} \mapsto 1]\rangle, \partial\hat{\sigma}) \\
\mathit{apply}\Delta(\langle h, \mu\rangle, \partial\hat{\sigma}[\hat{a} \mapsto \mathbf{Weak}(\hat{t})]) &= \mathit{apply}\Delta(\langle h \sqcup [\hat{a} \mapsto \hat{t}], \mu\rangle, \partial\hat{\sigma}) \\
\mathit{apply}\Delta(\langle h, \mu\rangle, \partial\hat{\sigma}[\hat{a} \mapsto \mathbf{Reset}(\hat{t})]) &= \mathit{apply}\Delta(\langle h[\hat{a} \mapsto \hat{t}], \mu[\hat{a} \mapsto \omega]\rangle, \partial\hat{\sigma})
\end{aligned}
$$

FAILED EVALUATION or "stuckness" is a possible evaluation result. We use a $\mathit{Res}_S$ container because expression evaluation can be strongly or weakly *progressing* (conversely, stuck). A strongly progressing expression uses a *Refmap* to map refinements to possible result payloads. The refinements in a *Refmap* cannot represent the sequence of operations "resolve $\hat{a}$ to $v$, then strongly update $\hat{a}$ to $v'$," so we only use the *Refmap* domain to represent resolutions and not updates. For example, without separating strong update from store refinement, we can can confuse the evaluation of the following:

```
(match (lookup t)
  [F F [Update t T]])
```

where the metalanguage enviroment maps t to $\mathbf{EAddr}(\hat{a})$ where

$$\hat{\sigma} = \langle[\hat{a} \mapsto \mathbf{NDT}(\{\mathsf{T}, \mathsf{F}\}, \bot)], [\hat{a} \mapsto 1]\rangle.$$

The match is not total, so when the match resolves $\hat{a}$ to T, the lack of a rule means the match is stuck. If we reuse the store refinement for strong updates, then at the end of evaluating the match there is only the one refinement that represents both an answer and stuckness. We lose the information that states, "*if* we evaluate expression $e$ under refinement $\delta$, then we get result $r$." Instead we only have, "at the end of evaluating $e$, the possible writes to the store are $X$, with result terms $Y$." So, the types of store refinement and strong update may be the same, but their interpretation is different.

Both the binding/update forms and rules have a different result type, just like in concrete evaluation. Now instead of strong success and strong failure, we have the extra third mode: strong stuckness. Recall that if an expression get stuck during the evalution of a *BU* form, then the form is considered stuck. A stuck rule is considered to have "applied" and just done nothing. A failed match in a **Where** form just means that the form is unapplicable, and we can move on to try another rule.

$$\widehat{\mathit{Rule\text{-}result}}_S[T] ::= \mathbf{FireStuckUnapplicable}(\mathit{ets}, \Delta, \Delta) \mid \mathbf{May}(E)$$
$$\mathit{ets} \in \mathit{Refmap}[T \times \widehat{\Delta\mathit{Store}}] \text{ and } E \in \wp(T \times \widehat{\Delta\mathit{Store}})$$

I will abbreviate **FireStuckUnapplicable** as **FSU**

Expression evaluation and rule evaluation can be converted back and forth. A stuck expression is a stuck rule, and a progressing expression is a firing rule. A firing rule is a progressing expression, and a non-firing rule is a stuck expression. I leave the conversions' definitions to the appendix.

The expression evaluation functions therefore have the following type signatures:

$$\mathit{Ev} : \mathit{Expr} \times \widehat{\mathit{MEnv}} \to \widehat{\mathit{EvResult}}_S[\widehat{\mathit{Term}}]$$
$$\overline{\mathit{Ev}} : \mathit{Expr}^* \times \widehat{\mathit{MEnv}} \to \widehat{\mathit{EvResult}}_S[\widehat{\mathit{Term}}^*]$$
$$\mathit{Ev}_{mf} : \mathit{Metafunction\text{-}Name} \times \widehat{\mathit{Term}}^* \to \widehat{\mathit{EvResult}}_S[\widehat{\mathit{Term}}]$$
$$\mathit{Ev}_{bu} : \mathit{BU} \times \widehat{\mathit{MEnv}} \to \widehat{\mathit{Rule\text{-}result}}_S[\mathit{MEnv}]$$
$$\mathit{Ev}_{\overline{bu}} : \mathit{BU}^* \times \widehat{\mathit{MEnv}} \to \widehat{\mathit{Rule\text{-}result}}_S[\mathit{MEnv}]$$

A metafunction's meaning allows user-defined rules and external implementations. The output of an external metafunction in this abstract semantics has a different output type than in the concrete:

$$\begin{aligned} r \in \mathit{Rule} &= \mathbf{Rule}(p, e, \overline{bu}) \\ \mathit{Metafunction\text{-}meaning} &= \mathbf{User}(\overline{r}) \mid \mathbf{ExtMF}(\widehat{mf}) \\ \widehat{mf} &: \widehat{Term}^* \to \widehat{EvResult}_S[\widehat{Term}] \end{aligned}$$

Metafunctions' treatment is discussed in the evaluation of **Call** expressions.

When an expression's evaluation depends on a **Let** binding succeeding, but it strongly fails, the evaluation is stuck (though we still use **Fail**). Stuckness can happen after some store refinements, updates, address allocations, and nondeterministic matching. If evaluation follows two paths, one that is **Must** and the other that is **Fail**, with overlapping refinements, the evaluation is considered weak. Any changes to the store are rolled back between a nondeterministic choice and stuckness. The reason why is motivated by the following example.

EXAMPLE Consider we evaluate $\mathbf{Let}(\mathbf{Where}(0, \mathbf{Term}(\mathbf{NDT}(\{0,1\}, \bot))), \texttt{'body})$. The match in the **Let** is weak due to the *NDTerm*, so we should be in agreement that we only weakly evaluate to `'body`. Suppose we separate the nondeterministic binding from the structural match. Consider then we first bind $x$ to the resolution of used address $a$, so $\hat{\sigma} = \langle [a \mapsto \mathbf{NDT}(\{0,1\}, \bot)] \rangle [a \mapsto \omega]$. The binding is

$$bu = \mathbf{Where}(\mathit{Name}(x, \mathbf{Wild}), \mathbf{IAddr}(a, \texttt{'resolve})).$$

Each value in $a$ gets bound to $x$, and the match is *strong*. The full expression is then $\mathbf{Let}(bu\ \mathbf{Where}(0, \mathbf{Ref}(x)), \texttt{'body})$. One binding of $x$ will be 0, so the second match will be *strong* and evaluation will *strongly* get to the body. However, the other binding will be 1 and the second **Where** match fails and evaluation is stuck. The two evaluations have the same justifications, so the strengths should weaken to **May** as we expect.

### 6.9.2 *The abstract interpretation monad*

Expression evaluation requires and builds state. The whole venture is laborious without the right abstraction. We work within a monad that has built-in commands for interacting with the external parameters and running through non-determinism. First we'll catologue the operations for $\widehat{EvResult}_S[T]$ and their purpose before we define them.

- $\mathit{return}(\hat{t})$: in the current context, we've evaluated to $\hat{t}$ (translates to $\mathbf{Must}([\delta \mapsto \{\langle \hat{t}, \partial\hat{\sigma} \rangle\}]))$;

- $ev \gg= f$: (also written **bind**) evaluate *ev* to some $t$, then evaluate $f(t)$;
- *fail*(): evaluation is stuck (translates to $\mathbf{Fail}(\{\delta\})$);
- $mkV(n, tag, \bar{\hat{t}}, \rho)$: the *mkV* external itself plugs into the monad to create a variant named $n$ with allocation site tagged *tag* out of terms $\bar{\hat{t}}$ after matching has resolved to the environment $\rho$;
- $\mathbf{alloc}(tag, \rho)$: create an explicit address, weakening any store changes to that address;
- $\mathbf{with\text{-}lookup}(\hat{a}, lm, f)$: look up $\hat{a}$ in the appropriate lookup mode and pass the resulting term to $f$ in the updated context;
- $\mathbf{choose}(\Delta, ev)$: combine the evaluations of *ev* under the different refinements from $\Delta$;
- $\mathbf{resolve}(\hat{t})$: get a grounded form of $\hat{t}$;
- $\mathbf{update\text{-}res}(\hat{a}, \hat{t}, ev)$: set $\hat{\sigma}.h(\hat{a})$ to $\hat{t}$ with the appropriate strength, then run *ev* in the new context.

We give *mkV* and *alloc* access to not only the current state and allocation site tag, but also to the binding environment. This choice captures the way that the original AAM paper resolves some non-determinism and then passes the choice to later parameters[26].

[26] See Van Horn and Might [95] figure 5, where both alloc and tick are passed a continuation that was previously selected from the store.

Running an evalution is just applying it to the current state, refinement, and store changes:

$$run\text{-}ev(ev, \hat{\varsigma}, \delta, \partial\hat{\sigma}) = ev(\hat{\varsigma}, \delta, \partial\hat{\sigma})$$

The first is as straightforward as described:

*return*:

$$return(a) = \lambda\hat{\varsigma}, \delta, \partial\hat{\sigma}.\mathbf{Both}([\delta \mapsto \{\langle a, \partial\hat{\sigma}\rangle\}], \bot)$$

Bind requires multiple runs in different contexts and joining the results:

**bind**:

$$\begin{aligned} ev \gg= f \triangleq \lambda\hat{\varsigma}, \delta, \partial\hat{\sigma}.\mathtt{case}\ & run\text{-}ev(ev, \hat{\varsigma}, \delta, \partial\hat{\sigma})\ \mathtt{of} \\ & \mathbf{Both}(R, \Delta) : \mathbf{Fail}(\Delta) \sqcup \bigsqcup_{\delta' \in \mathbf{dom}(R), \langle t, \partial\hat{\sigma}'\rangle \in R(\delta')} f(t, \hat{\varsigma}, \delta', \partial\hat{\sigma}') \\ & \mathbf{May}(Us) : weaken_S(\bigsqcup_{\langle t, \partial\hat{\sigma}'\rangle \in Us} f(t, \hat{\varsigma}, \delta, \partial\hat{\sigma}')) \end{aligned}$$

Failure captures the current refinement:

*fail*:

$$\mathit{fail}() = \lambda\hat{\varsigma}, \delta, \partial\hat{\sigma}.\mathbf{Fail}(\{\delta\})$$

The allocation external gets read-only access to the monad context. Reallocating an address must signal the weakening of any current use of the address already Say we allocate some address $\hat{a}$. If we know the $\partial\hat{\sigma}$ maps $\hat{a}$ to a strong update, then the address is now used and the strong update must be demoted to a **Reset**. Any further updates to $\hat{a}$ will grow the *NDTerm* in the mapped **Reset** object.

Allocating an address further extends the store to map an "uninitialized" term – the bottom element of the term lattice: $\mathbf{NDT}(\emptyset, \bot)$ (we just write $\bot$).

**alloc**:

$$\begin{aligned}\mathbf{alloc}(\mathit{tag}, \rho) = \lambda\hat{\varsigma}, \delta, \partial\hat{\sigma}.\mathit{return}\ \hat{a}\ \hat{\varsigma}\ \delta\ (&\mathtt{case}\ \partial\hat{\sigma}(\hat{a})\ \mathtt{of}\\ &\mathbf{Strong}(\hat{t}) : \partial\hat{\sigma}[\hat{a} \mapsto \mathbf{Reset}(\hat{t})]\\ &\mathit{ct} : \partial\hat{\sigma}\\ &\bot : \partial\hat{\sigma}[\hat{a} \mapsto \mathtt{if}\ \hat{\varsigma}.\hat{\sigma}.\mu(\hat{a}) = 0\ \mathtt{then}\\ &\qquad\qquad \mathbf{Strong}(\bot)\\ &\qquad\quad \mathtt{else}\ \mathbf{Weak}(\bot)])\end{aligned}$$

where $\hat{a} = \mathit{alloc}(\mathit{tag}, \rho)\ \hat{\varsigma}\ \delta\ \partial\hat{\sigma}$

Store lookup resolves as necessary. First, if an address is locally modified, take any strong contents and run with them; weak contents must be reconciled with the current store first. If no modifications, then the address might be refined; run with that if it exists. If no modifications or refinements, then look up from the store and, given the lookup mode, refine then run.

**with-lookup**:

$$\begin{aligned}\mathbf{with\text{-}lookup}(\hat{a}, \mathtt{'delay}, \mathsf{f}) &= \mathsf{f}(\mathbf{Delay}(\hat{a}))\\ \mathbf{with\text{-}lookup}(\hat{a}, \mathit{lm}, \mathsf{f}) &= \lambda\hat{\varsigma}, \delta, \partial\hat{\sigma}.\mathtt{case}\ \partial\hat{\sigma}(\hat{a})\ \mathtt{of}\end{aligned}$$

$$\begin{aligned}&\mathbf{Strong}(\hat{t}) : \mathsf{f}\ \hat{t}\ \hat{\varsigma}\ \delta\ \partial\hat{\sigma}\\ &\mathbf{Reset}(\hat{t}) : \mathsf{f}\ \hat{t}\ \hat{\varsigma}\ \delta\ \partial\hat{\sigma}\\ &\mathbf{Weak}(\hat{t}) : \mathsf{f}(\hat{t} \sqcup \hat{\varsigma}.\hat{\sigma}.\mathsf{h}(\hat{a}))\ \hat{\varsigma}\ \delta\ \partial\hat{\sigma}\\ &\bot : \mathtt{case}\ \delta(\hat{a})\ \mathtt{of}\\ &\quad \hat{t} : \mathsf{f}\ \hat{t}\ \hat{\varsigma}\ \delta\ \partial\hat{\sigma}\\ &\quad \bot : \mathtt{case}\ \mathit{lm}\ \mathtt{of}\\ &\qquad \mathtt{'resolve} : \bigsqcup_{\hat{t} \in \mathit{Choose}(\hat{\varsigma}.\hat{\sigma}.\mathsf{h}(\hat{a}))} \mathsf{f}\ \hat{t}\ \hat{\varsigma}\ \delta[\hat{a} \mapsto \hat{t}]\ \partial\hat{\sigma}\\ &\qquad \mathtt{'deref} : \mathsf{f}(\hat{\varsigma}.\hat{\sigma}.\mathsf{h}(\hat{a}))(\hat{\varsigma})(\delta)(\partial\hat{\sigma})\end{aligned}$$

**choose**:

$$\textbf{choose}(\Delta, ev) = \lambda\hat{\varsigma}, \delta, \partial\hat{\sigma}. \bigsqcup_{\delta' \in \Delta} ev\ \hat{\varsigma}\ \delta'\ \partial\hat{\sigma}$$

Term resolution only gets stuck if there are no terms to resolve. Otherwise, each term is packaged up in evaluation object. We call **resolve** when a term is demanded, so we do not respect the lookup modality for an implicit address - we simply resolve.

**resolve**:

$$\begin{aligned}
\textbf{resolve}(\textbf{NDT}(\emptyset, \bot)) &= \mathit{fail}() \\
\textbf{resolve}(\textbf{NDT}(ts, es)) &= \lambda\hat{\varsigma}, \delta, \partial\hat{\sigma}.\textbf{Both}([\delta \mapsto \{\langle \hat{t}, \partial\hat{\sigma}\rangle : \\
&\qquad \hat{t} \in \mathit{Choose}(\textbf{NDT}(ts, es))\}], \bot) \\
\textbf{resolve}(\textbf{Delay}(\hat{a})) &= \textbf{with-lookup}(\hat{a}, \texttt{'resolve}, \mathit{return}) \\
\textbf{resolve}(\textbf{IAddr}(\hat{a}, _)) &= \textbf{with-lookup}(\hat{a}, \texttt{'resolve}, \mathit{return}) \\
\textbf{resolve}(\hat{t}) &= \mathit{return}(\hat{t})
\end{aligned}$$

Finally, we have the store update. A danger to soundness is a strong update to an address that somewhere in the machine is lying dormant in a **Delay**. The delayed lookup should refer to the current value, and not the value post-update. There are a few ways to approach this: disallow strong updates, never delay and always deref, or find and replace $\textbf{Delay}(\hat{a})$ with the $\textbf{NDT}(\hat{\varsigma}.\hat{\sigma}.h(\hat{a}))$ before doing the update. I punt on handling delays with deus ex machina (which can also be used before allocation):

$$\mathit{undelay} : \widehat{Addr} \to \widehat{EvResult}_S[\mathsf{T}] \to \widehat{EvResult}_S[\mathsf{T}].$$

**update-res**:

$$\begin{aligned}
\textbf{update-res}(\hat{a}, \hat{t}, ev) = \mathit{undelay}\ \hat{a}\ \lambda\hat{\varsigma}, \delta, \partial\hat{\sigma}.& \\
\texttt{case}\ \partial\hat{\sigma}(\hat{a})\ \texttt{of}& \\
\textbf{Strong}(_) : ev\ \hat{\varsigma}\ \delta\ \partial\hat{\sigma}[\hat{a} \mapsto \textbf{Strong}(\hat{t})]& \\
\textbf{Weak}(\hat{t}') : ev\ \hat{\varsigma}\ \delta\ \partial\hat{\sigma}[\hat{a} \mapsto \textbf{Weak}(\hat{t} \sqcup \hat{t}')]& \\
\textbf{Reset}(\hat{t}') : ev\ \hat{\varsigma}\ \delta\ \partial\hat{\sigma}[\hat{a} \mapsto \textbf{Reset}(\hat{t} \sqcup \hat{t}')]& \\
\bot : ev\ \hat{\varsigma}\ \delta\ \partial\hat{\sigma}[\hat{a} \mapsto ct]&
\end{aligned}$$

where $ct = \texttt{if}\ \hat{\varsigma}.\hat{\sigma}.\mu(\hat{a}) \stackrel{?}{=} 1\ \texttt{then}\ \textbf{Strong}(\hat{t})\ \texttt{else}\ \textbf{Weak}(\hat{t})$

The **Weak** and **Reset** forms both join terms, but recall that they have different semantics when updating the store. A **Reset** will perform a strong update with the joined contents and that **Weak** further joins its contents with what the store currently holds for the address.

We've built our hammer. Let's find some nails.

$$Ev : Expr \times \widehat{MEnv} \to EvResult_S[\widehat{Term}]$$

$$
\begin{aligned}
Ev(\mathbf{Ref}(x), \rho) &= return(\rho(x)) \\
Ev(\mathbf{Alloc}(tag), \rho) &= \mathbf{alloc}(tag, \rho) \\
Ev(\mathbf{Variant}(n, tag, \overline{e}), \rho) &= \texttt{do}\ \ \overline{t} \leftarrow \overline{Ev}(\overline{e}) \\
&\qquad mkV(n, tag, \overline{t}, \rho) \\
Ev(\mathbf{Let}(\overline{bu}, e), \rho) &= \texttt{do}\ \ \rho' \leftarrow \overline{B}(\overline{bu}, \rho) \\
&\qquad Ev(e, \rho') \\
Ev(\mathbf{Deref}(e, lm), \rho) &= \texttt{do}\ \ \hat{t} \leftarrow Ev(e, \rho) \\
&\qquad \hat{t}' \leftarrow \mathbf{resolve}(\hat{t}) \\
&\qquad \texttt{case}\ \hat{t}'\ \texttt{of} \\
&\qquad\quad \mathbf{EAddr}(\hat{a}) : \mathbf{with\text{-}lookup}(\hat{a}, lm, return) \\
&\qquad\quad _ : fail() \\
Ev(\mathbf{Call}(f, \overline{e}), \rho) &= \texttt{do}\ \ \overline{t} \leftarrow \overline{Ev}(\overline{e}) \\
&\qquad Ev_{mf}(f, \overline{t})
\end{aligned}
$$

Figure 39: Expression evaluation (scaffolding)

### 6.9.3 *Finishing the semantics of expression evaluation*

Given the monad language we've built up, evaluation's definition is strongly reminiscent of the previous chapter's concrete semantics.

LET expression evaluation bounces between evaluating bindings/updates and evaluating expressions. First we see what evaluating a single *bu* looks like, then their sequencing, and then their combination with expressions in **Let**'s evaluation rule. We expect a successful match to output its extended environments at the appropriate strength, splitting if necessary. A failing match should populate the "unapplicable" set of refinements.

$$B : BU \times \widehat{MEnv} \to \widehat{Rule\text{-}results}_S[\widehat{MEnv}]$$

$$
\begin{aligned}
B(\mathbf{Where}(p, e), \rho) &= \texttt{do}\ \ \hat{t} \leftarrow Ev(e, \rho) \\
&\qquad \hat{M}_S(p, \hat{t}, \rho) \\
B(\mathbf{Update}(e_a, e_v), \rho) &= \texttt{do}\ \ \hat{t}_a \leftarrow Ev(e_a, \rho) \\
&\qquad \hat{t}'_a \leftarrow \mathbf{resolve}(\hat{t}_a) \\
&\qquad \hat{t}_v \leftarrow Ev(e_v, \rho) \\
&\qquad \texttt{case}\ \hat{t}'_a\ \texttt{of} \\
&\qquad\quad \mathbf{EAddr}(\hat{a}) : \mathbf{update\text{-}res}(\hat{a}, \hat{t}_v, return(\rho)) \\
&\qquad\quad _ : fail()
\end{aligned}
$$

We see here that B co-opts matching for **Where** and (via an unshown mundate coercion) injects its results into *EvResult*. For **Update**, we update defined for allocation so that updates to fresh addresses are refinements, and updates to used addresses are not. The store can only be updated with addresses, so non-addresses make evaluation stuck.

The $\overline{B}$ function folds *bind* down the list.

$$\begin{aligned}\overline{B} &: BU^* \times \widehat{MEnv} \to \widehat{Rule\text{-}result}_S[\widehat{MEnv}] \\ \overline{B}(\epsilon, \rho) &= return(\rho) \\ \overline{B}(bu : \overline{\mathtt{bu}}, \rho) &= \mathtt{do}\ \ \rho' \leftarrow \mathrm{B}(bu, \rho) \\ &\qquad\quad \overline{B}(\overline{\mathtt{bu}}, \rho')\end{aligned}$$

METAFUNCTION CALL evaluation looks like the **Variant** case, except at the end it calls out to the $Ev_{mf}$ function.

Metafunctions are supposed to be total in the concrete, but abstract inputs can lead to divergence. Consider the call

**Call**(`rem`, ⟨(`cons b b`)⟩)

where the store contains [`b` ↦ **NDT**({(`nil`), (`cons b b`)})]. This term represents the circular data structure

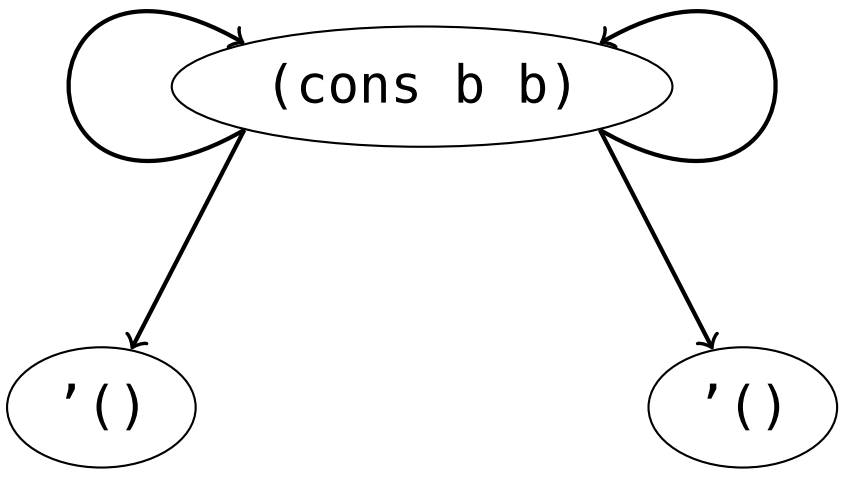

The circular structure is an abstraction of infinitely many concrete terms that unroll the self-reference arbitrarily many times before bottoming out at `'()`. Nondeterministic inputs can thus lead to nondeterministic outputs; metafunctions can evaluate to multiple possible answers. Recursion on a circular structure like this does not terminate unless one tracks the already seen inputs to stop on any revisited input.

Metafunction evaluation tries to apply rules before evaluating the right-hand-sides. The self-reference allowed by metafunction rules can lead us into non-terminating evaluations. We can catch all the non-terminating cases when the state space is finitized by tracking whether we have seen the same combination of inputs before. I leave this detail out of the following formalism, but the implementation is straightforward. An implementation sketch: I chose to use dynamic binding to create a memo-table if there wasn't one already bound. A typical memo table has an indefinite lifetime, but with dynamic binding, leaving the top-level context of a metafunction call frees up precious memory.

$$\begin{aligned} Ev_{mf}(mf, \bar{t}) = \texttt{case}\ \mathcal{M}(mf)\ \texttt{of} \\ &\mathbf{ExtMF}(\widehat{mf}) : \widehat{mf}(\bar{t}) \\ &\mathbf{User}(\bar{r}) : oapp(\bar{r}, \mathbf{Variant}(mf, \bar{t})) \end{aligned}$$

External metafunction evaluation punts to the given function. User metafunction evaluation applies rules in order until it finds a strongly firing rule (or keeps going with weakly firing rules).

$$\begin{aligned} oapp &: Rule^* \times \widehat{Term} \to EvResult_S[\widehat{Term}] \\ oapp(\epsilon, \hat{t}) &= fail() \\ oapp(r : \bar{r}, \hat{t}) &= maybefire(Ev_{rule}(r, t), oapp(\bar{r}, \hat{t})) \end{aligned}$$

The *maybefire* function runs the first rule evaluation to see if it strongly fires, and if there are left over obligations, continues (and combines) with the ordered evaluation.

Notice that *oapp* does not take an environment: metafunction calls are all top level. A richer language would allow locally defined metafunctions and pattern matching in expressions, where both use this same machinery.

$$\begin{aligned} maybefire &: \widehat{Rule\text{-}result}_S[T] \times EvResult_S[T] \to EvResult[T] \\ maybefire(er, ev) &= \lambda\hat{\varsigma}, \delta, \partial\hat{\sigma}.\texttt{case}\ er\ \hat{\varsigma}\ \delta\ \partial\hat{\sigma}\ \texttt{of} \\ &\quad \mathbf{FSU}(R, \Delta_S, \Delta_U) : \mathbf{Both}(R, \Delta_S) \sqcup (\mathbf{choose}(\Delta_U, ev)\ \hat{\varsigma}\ \delta\ \partial\hat{\sigma}) \\ &\quad \mathbf{May}(E) : \mathbf{May}(E) \sqcup (ev\ \hat{\varsigma}\ \delta\ \partial\hat{\sigma}) \end{aligned}$$

Finally, we need to know the definition of rule evalution, $Ev_{rule}$. It's what we expect: match, run $\overline{bu}$, then run the right hand side. The monad handles failure and stuckness.

$$\begin{aligned} Ev_{rule} &: Rule \times \widehat{Term} \to \widehat{Rule\text{-}result}_S[\widehat{Term}] \\ Ev_{rule}(\mathbf{Rule}(p, e, \overline{bu}), \hat{t}) &= \texttt{do}\ \rho \leftarrow \hat{M}_S(p, t, \bot) \\ &\qquad \rho' \leftarrow \overline{B}(\overline{bu}, \rho) \\ &\qquad Ev(e, \rho') \end{aligned}$$

So, the overall meaning of $Ev_{mf}$ is that if a rule strongly applies, we evaluate its right-hand-side and return that as the result. If a rule weakly applies, then we both evaluate the right-hand-side (but weaken its strength) and keep trying to apply rules (also weakening).

Every interpreted metafunction call translates to a call to $Ev_{mf}$, a function in the meta-meta-language: calls and returns are properly matched by construction. The metafunction evaluation strategy uses the metalanguage's call stack and thus enjoys proper call/return matching in the same way as Vardoulakis' Big CFA2 [98] and Glück's context-free language parser [35].

## 6.10 COMBINING IT ALL

A (conditional) reduction rule for an abstract machine takes the form of

$$\mathit{Pattern} \longmapsto \mathit{Expr}[\mathit{bindings}/\mathit{updates}]$$

Patterns match the machine state with the previous section's matching semantics, and the resulting binding environment(s) drive the evaluation of the right-hand-side. The rule's bindings can further rule out whether the rule actually *fires*, since binding is introduced by may-fail pattern-matching.

The meaning of a reduction rule is induced by matching ($M$), bindings/updates evaluation ($B^*$), and expression evaluation (*Ev*), all of which have results that are strong or weak. We refer to a result's quality of strong or weak as its *strength*. The strength of the evaluation of the rule's right-hand-side expression is irrelevant to the strength of the rule *firing*, so the strength of whether that expression entirely evaluates is discarded and replaced with the combined strength of the initial match and the rule's bindings. Binding (matching) can fail, which is one point of strength, but before matching even happens, the right-hand-side expression of the **Where** form must be evaluated. Expressions themselves can have bindings and updates, so the strength of an expression evaluation is inherited from the strengths of its internal points of failure.

Let's recall that the semantics is parameterized by

- $\mathcal{S} \subset_{\text{fin}} \mathit{Rule}$: a collection of rewrite rules on the terms carried in a state;
- $\mathcal{M} : \mathit{Metafunction\text{-}Name} \underset{\text{fin}}{\rightharpoonup} \mathit{Metafunction\text{-}Meaning}$: the metafunction environment;
- $\mathit{alloc} : \mathit{Tag} \times \widehat{\mathit{MEnv}} \to \widehat{\mathit{State}} \to \mathit{Refinement} \to \widehat{\Delta\mathit{Store}} \to \widehat{\mathit{Addr}}$: for allocating addresses given the state being stepped, the store and binding environment at the allocation point, as indicated by a *Tag*;
- $\mathit{mkV} : \mathit{Name} \times \mathit{Tag} \times \widehat{\mathit{Term}}^* \times \widehat{\mathit{MEnv}} \to \mathit{EvResult}_{\mathcal{S}}[\widehat{\mathit{Term}}]$: for optionally creating an abstracted version of a **Variant** that is about to be constructed;
- $\tau_0 : \widehat{\mathit{Time}}$: the initial "additional element";
- $\mathit{tick} : \widehat{\mathit{MEnv}} \to \widehat{\mathit{State}} \to \mathit{Refinement} \to \widehat{\Delta\mathit{Store}} \to \widehat{\mathit{Time}}$: combines the stepped state with the components of what is about to become a state to produce the *Time* component of this state.

With these components in place, we can create and step abstract machine states:

$$\begin{aligned}
\mathit{inject} &: \widehat{\mathit{Term}} \to \widehat{\mathit{State}} \\
\mathit{inject}(t) &= \mathbf{State}(t, \bot, \bot, \tau_0) \\
\mathit{step} &: \widehat{\mathit{State}} \to \wp(\widehat{\mathit{State}}) \\
\mathit{step}(\hat{\varsigma}) &= \mathit{finalize}(\hat{\varsigma}.\hat{\sigma}, \bigsqcup_{r \in \mathcal{S}} \mathit{apply}(r)\ \hat{\varsigma}\ \bot\ \bot)
\end{aligned}$$

where we define a different rule evaluation function that returns the output term as well as the next $\widehat{\mathit{Time}}$ element.

$$\begin{aligned}
\mathit{apply} &: \mathit{Rule} \times \widehat{\mathit{Term}} \to \mathit{EvResult}_S[\widehat{\mathit{Term}} \times \widehat{\mathit{Term}}] \\
\mathit{apply}(\mathbf{Rule}(p, e, \mathbf{bu}), \hat{t}) &= \texttt{do}\ \ \rho \leftarrow \hat{M}_S(p, t, \bot) \\
&\qquad\quad \rho' \leftarrow \overline{B}(\overline{bu}, \rho) \\
&\qquad\quad \hat{t}' \leftarrow \mathit{Ev}(e, \rho') \\
&\qquad\quad \mathit{return\text{-}tick}(\hat{t}', \rho')
\end{aligned}$$

The *return-tick* monad operation uses the current state to call the *tick* external and tuple it with the given term (not shown).

The output to state transformation is defined as *finalize*. The strength of the reduction is forgotten, but it could be remembered as a label on the “edge” between states, for diagnostic purposes. The store changes and refinements are applied to the previous state’s store to ultimately create the next set of states. We only refine addresses that have not been modified.

$$\begin{aligned}
\mathit{finalize}(\hat{\sigma}, \mathbf{Both}(R, _)) &= \{\mathbf{State}(\hat{t}, \mathit{apply}\Delta(\hat{\sigma}, \partial\hat{\sigma}) \blacktriangleleft \delta|_{\overline{\mathbf{dom}(\partial\hat{\sigma})}}, \tau')\ :\ ((\hat{t}, \tau), \partial\hat{\sigma}) \in R(\delta)\} \\
\mathit{finalize}(\hat{\sigma}, \mathbf{May}(E)) &= \{\mathbf{State}(t, \mathit{apply}\Delta(\hat{\sigma}, \partial\hat{\sigma}), \tau)\ :\ ((\hat{t}, \tau), \partial\hat{\sigma}) \in E\}
\end{aligned}$$

An additional step we might add is garbage collection, which we’ve covered before, but we define the $\mathcal{T}$ function here.

$$\begin{aligned}
\mathcal{T}(\mathbf{Variant}(n, \langle t \ldots \rangle)) &= \bigcup \{\mathcal{T}(t) \ldots\} \\
\mathcal{T}(\mathbf{IAddr}(a, _) \vee \mathbf{Delay}(a)) &= \{a\} \\
\mathcal{T}(\mathbf{External}(E, v)) &= E.\mathcal{T}(v) \\
\mathcal{T}(\mathbf{NDT}(\mathit{ts}, \mathit{Es})) &= \bigcup_{t \in \mathit{Choose}(\mathbf{NDT}(\mathit{ts}, \mathit{Es}))} \mathcal{T}(t)
\end{aligned}$$

All of this now defined, we can give a few notions of “running” a term in a given semantics:

- Nondeterministic run: repeatedly apply *step* on an arbitrarily chosen output state until stuck; report the final state as the result:

$$
\begin{aligned}
\mathit{run}(\mathsf{t}) &= \mathit{run}^*(\mathit{inject}(\mathsf{t})) \\
\mathit{run}^*(\hat{\varsigma}) &= \hat{\varsigma} \text{ if } \mathit{step}(\hat{\varsigma}) = \emptyset \\
\mathit{run}^*(\hat{\varsigma}) &= \mathit{run}^*(\mathit{Choice\text{-}function}(\mathit{step}(\hat{\varsigma})) \text{ otherwise}
\end{aligned}
$$

- All runs: treat the initial state as a singleton set "frontier" to repeatedly step:

$$
\begin{aligned}
\mathit{run}(\mathsf{t}) &= \mathit{run}^*(\{\mathsf{t}\}) \\
\mathit{run}^*(\mathsf{F}) &= \mathtt{case}\ \bigcup_{\hat{\varsigma} \in \mathsf{F}} \mathit{step}(\hat{\varsigma})\ \mathtt{of} \\
&\qquad \emptyset : \mathsf{F} \\
&\qquad \mathsf{F}' : \mathit{run}^*(\mathsf{F}')
\end{aligned}
$$

- Loop-detecting: run like the previous mode, but don't re-step already seen states:

$$
\begin{aligned}
\mathit{run}(\mathsf{t}) &= \mathit{run}^*(\emptyset, \{\mathsf{t}\}) \\
\mathit{run}^*(\mathsf{S}, \mathsf{F}) &= \mathtt{case}\ \bigcup_{\hat{\varsigma} \in \mathsf{F}} \mathit{step}(\hat{\varsigma})\ \mathtt{of} \\
&\qquad \emptyset : \mathsf{S} \\
&\qquad \mathsf{F}' : \mathit{run}^*(\mathsf{S} \cup \mathsf{F}', \mathsf{F}' \setminus \mathsf{S})
\end{aligned}
$$

- Reduction relation-grounding: create a concrete representation of the reduction relation as used to evaluate the given term:

$$
\begin{aligned}
\mathit{run}(\mathsf{t}) &= \mathit{run}^*(\emptyset, \{\mathsf{t}\}, \emptyset) \\
\mathit{run}^*(\mathsf{S}, \mathsf{F}, \mathsf{R}) &= \mathtt{case}\ \{(\hat{\varsigma}, \hat{\varsigma}') \ :\ \hat{\varsigma} \in \mathsf{F}, \hat{\varsigma}' \in \mathit{step}(\hat{\varsigma})\}\ \mathtt{of} \\
&\qquad \emptyset : \mathsf{R} \\
&\qquad \mathsf{R}' : \mathit{run}^*(\mathsf{S} \cup \pi_1(\mathsf{R}'), \pi_1(\mathsf{R}') \setminus \mathsf{S}, \mathsf{R} \cup \mathsf{R}')
\end{aligned}
$$

The full generality of refinements, state splitting, and per-state stores is not the most practical to use as a static analysis. Another way we might run a term is in a *collecting semantics*, that is one that treats one part of the state as an anchor for the other parts to monotonically grow on. A *collecting semantics* represents the state space not as a set of states, but as a monotonic map from *some parts* of a state to *the other parts* of the state. For our purposes, say

$$
\begin{aligned}
\mathit{State} &\cong \mathsf{C} \times \mathsf{D} \text{ for some spaces C and D where D a join-semilattice} \\
\mathit{Collecting} &: \mathsf{C} \xrightarrow{\text{mono}} \mathsf{D}
\end{aligned}
$$

The collecting semantics' representation is more conservative than the set of states representation. Every state that has the same C component has its D component merged together. The merged components are "collected" at the shared C.

In our case, $C = \widehat{Term} \times \widehat{Time}$ and $D = \widehat{Store}$. The $\widehat{Term}$ component of C is determined by the object language's reduction rules, but *Time* and its updater, *tick*, are external parameters.

For the collecting semantics, we don't finalize each state, since we instead apply the changes to what is known at the anchor points. The "seen set" becomes the "anchor map" for our $\langle \mathrm{term}, \tau\rangle$ pairs to map their anchored $\hat{\sigma}$.

If an anchor is updated in a step, then it is added to the frontier, since the new value needs to propagate through the semantics.

$$
\begin{aligned}
\mathit{run}(t) &= \mathit{run}^*([\langle t, \tau_0\rangle \mapsto \langle \bot, \bot\rangle], \{\langle t, \tau_0\rangle\}, \emptyset) \\
\mathit{run}^* &: (\widehat{Term} \times \widehat{Time} \overset{\text{mono}}{\to} \widehat{Store}) \times \wp(\widehat{Term} \times \widehat{Time}) \times \wp((\widehat{Term} \times \widehat{Time})^2) \\
\mathit{run}^*(S, F, R) &= F' \overset{?}{=} \emptyset \to \langle S, R\rangle, \mathit{run}^*(S \triangleleft S', F', R \cup R') \\
&\text{where } I : \wp(\widehat{Term} \times \widehat{Time} \times \wp(\widehat{State})) \\
&\qquad I = \bigcup_{\langle t, \tau\rangle \in F} \langle t, \tau, \mathit{step}(\mathbf{State}(t, S(t, \tau), \tau))\rangle \\
&\qquad S' = [\langle t, \tau\rangle \mapsto S(t, \tau) \sqcup \hat{\sigma} \ : \ (_, _, N) \in I, \mathbf{State}(t, \hat{\sigma}, \tau) \in N] \\
&\qquad R' = \{(t, \tau, t', \tau') \ : \ (t, \tau, N) \in I, \mathbf{State}(t', _, \tau') \in N\} \\
&\qquad F' = \{t\tau \in \mathbf{dom}(S') \ : \ S(t\tau) \neq S'(t\tau)\}
\end{aligned}
$$

The large store joins that are required at each state step (see $S'$) can be prohibitively expensive. Two different implementation strategies are promising to generalize to this semantics. One is from the TAJS project due to Jensen et al. [40] that lazily performs joins at each demanded address, propagating backwards from the constructed anchor graph. The other is due to Staiger-Stöhr [89], which separates the construction of anchor connections from store joins to use fast graph algorithms on a novel on-the-fly SSA data structure. The use of SSA for sparse analysis allows one to flow information directly from change to use points, and not worry about irrelevant re-analysis.

## 6.11 PATHS TO ABSTRACTION

When we have a semantics, S, written in the core language developed in this chapter, there are many paths one can take to get a static analysis of programs in S. The primary concern we have is bounding the state space. The moving parts are the semantics' parameters: *alloc*, *tick*, and $\mathcal{M}$'s external meanings must have finite ranges, external descriptors' $\sqcup$ must produce finite chains, and *mkV* must ensure some bound on nesting depth. In AAM, the transformation focuses on *mkV*, as *alloc*, and *tick* are already parameters (and external stuff is trusted).

A heavy-handed abstraction for *mkV* is to heap-allocate all subterms in all constructed variants and maps. This means a larger, more varied store that will be referenced and updated more – it is slow. In addition, there's still the problem of addressing. A sound but terribly imprecise *mkV* would map everything to a single address. We could synthesize the addresses as (a hash of) the tree address through the rules to the **Variant** construction that calls *mkV*, paired with the state's $\widehat{Time}$ component. This is also not great, since it doesn't capture the common addressing idiom of additionally varying the addresses by the *program expression* that is driving the semantics to the **Variant**. At this point, we need deeper insight into the semantics itself and ask the user for assistance in constructing *mkV* (by, say, producing the code for a skeleton function to fill in). If *mkV* allocates everything, that is a lot of boilerplate user assistance.

The problematic points of a variant are where the nesting can be unbounded. A soft type analysis of the user's semantics itself can reconstruct the recursive structure that our *mkV* generator could use to cull the requests of the user. At this point, the machinery might be too smart without linguistic support for a conversation with the *mkV* generator. By "too smart," I mean that an analysis's results can be surprising, and expectations are not checked or even expressible.

The final consideration is to remove the need for *mkV* entirely. The AAM transformation removes the external dependency on *mkV*. With *mkV* removed and its functionality apparent in the rules themselves, we have more information to optimize the analysis behind the scenes. The fewer implicit state splits, address allocations and store lookups, the better. This call for additional support is motivation for a proper language (not a calculus) to enable abstraction. This language is a topic of future work.

# 7 CASE STUDY: TEMPORAL HIGHER-ORDER CONTRACTS

Software systems are large, consist of many modules, and have invariants that are either outright inexpressible or too costly to express (and prove) in the language's static type system—if it has one. When this is the case, one might hope to rely on software contracts, a concept first introduced in Eiffel by Bertrand Meyer [63], to give dynamic guarantees about the behavior of a system. In modern higher-order languages, the question of "who violated the contract?" becomes non-trivial as pre- and post-conditions on behavioral values must be delayed. Findler and Felleisen [34] gave the first semantic account of blame in a higher-order language with contracts, spawning a large body of research on behavioral contracts. More recently, Disney, Flanagan, and McCarthy (DFM) proposed a system of *temporal higher-order contracts* to provide a linguistic mechanism for describing temporal properties of behavioral values flowing through a program. Example temporal properties are, "a file can only be closed if it has been opened" and, in the higher-order setting, "if function A is given a function B, then B may not be called once A returns." Such invariants are important for interfaces that have set-up and tear-down protocols to follow, or even pure interfaces that have particular compositions of calls needed to construct some object.

Despite the significant engineering benefits of contracts, there are downsides; contracts offer no static guarantees and can impose prohibitive runtime overhead, particularly in the temporal case. Since contracts are runtime monitors, they do not themselves ensure correctness—though their blame reporting helps the process of constructing correct programs. Verification technology provides an additional level of confidence in correctness or pinpoints means for failure; its early use can even accelerate development with its bug-reporting capabilities. A sound model-checker can justify safely removing contract checking and have a performance-and-correctness return on the initial investment in contract design.

Statically verifying temporal contracts poses additional challenges over that of behavioral contracts, as they monitor the progression of interactions with a module over time, and not just localized interactions at module boundaries. This chapter uses the AAM framework through the *Limp* analysis modeling tool to to check for reachability of a temporal contract blame. I propose a semantics of temporal higher-order contracts that has an operational interpretation by means of regular expression derivatives [14]. The formal semantics differs from DFM's published formalism, but captures the spirit of its implementation. The interpretation via regular expression derivatives allows us to add the temporal property checking as part of the language's semantics itself. The sound and computable abstractions *Limp* offers make verification "fall out" of the language specification. Model checking a program with respect to its temporal contracts just amounts to running it—post abstraction—to either witness blame (perhaps spuriously) or confirm its absence (absolutely).

I provide an intuition for how temporal contracts work (§ 7.1), and present their syntax with examples. The semantics of DFM and a discussion of its problem points follow (§ 7.2), with my new semantics along with a proven-correct notion of derivative for regular expressions with back-references in the scope of our non-standard semantics of negation. Next, I show the semantics of a miniature Scheme-like language with temporal higher-order contracts (§7.3) in *Limp*, where the derivative function to give an operational interpretation of "stepping" temporal contracts is a simple metafunction. Finally in § 7.4, I show how the abstract semantics *Limp* produces is computable and is able to verify some programs' blame freedom.

## 7.1 OVERVIEW OF TEMPORAL HIGHER-ORDER CONTRACTS

Temporal contracts provide a declarative language for monitoring the temporal ordering of actions that pass through module boundaries. We begin with a simple example that exhibits the expressiveness of temporal higher-order contracts to frame the discussion. The following example comes from DFM, and its behavior led us to explore an alternative semantics.

### 7.1.1 *DFM's sort example, revised*

This example presents the contract for a hypothetical *sort* function which takes two arguments: a comparator and a list (of positive numbers), is non-reentrant, and furthermore cannot make its given comparator available to be called after it's done sorting. We express the structural component of the contract (with labels given to function components) like so:

$$sort : (cmp : Pos \to Pos \to Bool)\ (\texttt{List Pos}) \to (\texttt{List Pos})$$

We can express the temporal component of the contract in a natural way with the following:

$$\neg(\ldots\texttt{call}(sort,_,_)\ !\texttt{ret}(sort,_)^{*}\ \texttt{call}(sort,_,_))$$
$$\cap\neg(\ldots\ \langle\texttt{call}(sort,?cmp,_)\rangle\ \ldots\ \texttt{ret}(sort,_)\ \ldots\ \texttt{call}(cmp,_,_))$$

The first clause expresses non-re-entrance, phrased as a negation of a trace reentering the function: after a call to *sort* and some actions that aren't returns from *sort*, there is another call to *sort*.

The second clause of the temporal component specifies a higher-order property: given a call to *sort*, its associated *cmp* argument cannot be called after *sort* returns. We use angle brackets around actions that we want to bind values from, using the ? binding form. Since *cmp* will be wrapped with its higher-order contract at each call, which creates new values, the bindings for *cmp* will be distinct across execution. The intention of the regular-expression notation is to say, "as long as the trace is a prefix of these strings of actions, the temporal contract is satisfied." For example, $A$ satisfies the contract $ABC$, but $ABD$ doesn't.

Since the semantics is prefix closed, we can take the state of a regular expression parser as an indicator for whether the contract system should blame. I chose regular expression derivatives for this purpose due to their simplicity, though we extended them to allow for back-referencing values captured in binding forms. Consider the following faulty trace for an interaction with *sort* that violates the higher-order component of the temporal contract:

$$\texttt{call}(\mathit{sort}, \leqslant, \text{'}(2\ 1))\ \texttt{call}(\mathit{wrap}, 2, 1)\ \texttt{ret}(\mathit{sort}, \text{'}(1\ 2)))\ \texttt{call}(\mathit{wrap}, 0, 1)$$
$$\text{where } \mathit{wrap} = \lambda xy.(\texttt{if}\ (\texttt{and}\ (\texttt{pos?}\ x)\ (\texttt{pos?}\ y))\ (\leqslant\ x\ y)\ \texttt{blame}^{\mathit{client}}_{\mathit{contract}})$$

The contract system applies the regular expression derivative for each action in the trace as it arrives, and blames as soon as derivation fails. Recall that a regular expression derivative is with respect to some character $c$ such that $w \in \partial_c R$ iff $cw \in R$. In our system, we have structured characters, *actions*, that carry values that we can reference later via binding. The derivatives for this faulty trace, cumulatively (the $T$ for the previous equality has one less prime), are

$$\begin{aligned}
\partial_{\texttt{call}(\mathit{sort}, \leqslant, '(2\ 1))} T &= \neg \cup \{T_0, !\texttt{ret}(\mathit{sort}, _)^*\ \texttt{call}(\mathit{sort}, _, _)\}, \\
&\quad \cap \neg \cup \{T_1, (\ldots\ \texttt{ret}(\mathit{sort}, _)\ \ldots\ \texttt{call}(\mathit{cmp}, _, _), \rho)\} \\
\partial_{\texttt{call}(\leqslant, \mathit{wrap}, 2, 1)} T' &= T' \\
\partial_{\texttt{ret}(\mathit{sort}, '(1\ 2))} T'' &= \neg T_0 \cap \neg \cup \{T_1, T_2\} \\
\partial_{\texttt{call}(\mathit{wrap}, 0, 1)} T''' &= \neg T_0 \cap \neg \cup \{T_1, T_2, \epsilon\} = \neg T_0 \cap \bot = \bot \\
\text{where } T_0 &= \ldots \texttt{call}(\mathit{sort}, _, _)\ !\texttt{ret}(\mathit{sort}, _)^*\ \texttt{call}(\mathit{sort}, _, _) \\
T_1 &= \ldots\ \langle \texttt{call}(\mathit{sort}, ?\mathit{cmp}, _) \rangle\ \ldots\ \texttt{ret}(\mathit{sort}, _)\ \ldots\ \texttt{call}(\mathit{cmp}, _, _) \\
T_2 &= \ldots\ \texttt{call}(\mathit{cmp}, _, _), \rho \\
\rho &= [\mathit{cmp} \mapsto \mathit{wrap}]
\end{aligned}$$

The final state has a negated nullable contract, which we regard as a failing state. This is because a regular expression derives to $\epsilon$ through a string $w$ iff $w$ is accepted by the regular expression. We

$$S \in \mathit{Structural} ::= \texttt{flat}(e) \mid \ell : S \underset{e}{\mapsto} S \mid \langle S, S \rangle$$
$$e \in \mathit{Expr} ::= \texttt{tmon}\ (e, T) \mid \texttt{smon}_j^{k,l}(S, e) \mid \text{other forms}$$
$$\ell, k, l, j \in \mathit{Label} \text{ an infinite set}$$
$$\eta \in \mathit{Timeline} \text{ an infinite set}$$

Figure 40: Syntax of structural contracts with labels

interpret non-empty regular expressions as contracts that are following an allowed path, and have not yet violated it. This interpretation implies prefix closure of our partial trace semantics in Section 7.2.

### 7.1.2 *Syntax of contracts*

The general forms for expressing structural contracts and monitoring values are in Figure 40. The temporal arrow contract $\ell : S \underset{e}{\mapsto} S$ has two additional components on top of a standard arrow contract: a timeline which $e$ must evaluate to, and a name $\ell$. A *timeline* ($\eta$) is generated by evaluating the `(new-timeline)` expression. Each timeline $\eta$ is associated with a runtime monitor $\tau(\eta)$ that tracks contract adherence of calls and returns of functions temporally contracted on the timeline. A *name* on a temporal arrow contract is inherited from DFM: it allows a contract to refer to any wrapping of a function instead of a specific wrapping.

The other forms of the language are orthogonal, but we of course assume the existence of $\lambda$ expressions. The labels $k, l, j$ in an `smon`[27] expression are the identities of parties engaged in a software contract. There are three parties: the provider of a value ($k$, the server), the consumer of a value ($l$, the client), and the provider of the contract ($j$, the contract).

[27] *Short for "structural contract monitor."*

A structural monitor $\texttt{smon}_j^{k,l}(S, e)$ is a behavioral monitor [25]:the structural contract is given by $S$, where a temporally contracted value's actions are sent to the runtime monitor associated with the contract's timeline. A temporal contract can be attached to a timeline with the imperative `tmon` command. Each addition $T$ to a timeline $\eta$ sets the timeline's monitor to the extended $\cap\{T, \tau(\eta)\}$.

The syntax is presented in Figure 41.

Structural contracts include subsets of first-order data that satisfy a particular predicate ($\texttt{flat}(e)$), functions with associated structural contracts on their domain and range in addition to a label to reference within the temporal contract ($\ell : S_D \underset{e}{\mapsto} S_R$), and `cons` pairs whose components are contracted ($\langle S_A, S_D \rangle$). Temporal contracts include actions ($A$), negated actions ($!A$), action matching scoped over a following contract ($\langle A \rangle\ T$), concatenation ($T \cdot T$) (often written using juxtaposition), negated contracts ($\neg T$), Kleene closure ($T^*$), union

$$
\begin{aligned}
T^\circ \in \mathit{Temporal}^\circ &::= A \mid !A \mid \epsilon \mid \neg T^\circ \mid T^\circ \cdot T^\circ \mid T^{\circ *} \\
&\quad\mid \cup\tilde{T^\circ} \mid \cap\tilde{T^\circ} \mid \langle A \rangle\, T^\circ \\
T \in \mathit{Temporal} &= \text{same rules as } \mathit{Temporal}^\circ \text{ for T plus } \mid T^\circ, \rho \\
\rho \in \mathit{Env} &= \mathit{Var} \to \wp(\mathit{Value}) \\
A \in \mathit{Action} &::= \mathit{ac}(n, \mathit{pat}) \mid \texttt{Any} \\
\mathit{ac} \in \mathit{FAction} &::= \texttt{call} \mid \texttt{ret} \\
\pi \in \mathit{Trace} &= \mathit{Action}^* \\
\mathit{pat} \in \mathit{TPattern} &::= v \mid ?x \mid n \mid (\texttt{cons}\ \mathit{pat}\ \mathit{pat}) \mid !\mathit{pat} \mid \texttt{Any} \mid \texttt{None} \\
x \in \mathit{Var} &\quad \text{an infinite set} \\
n \in \mathit{Name} &::= x \mid \ell
\end{aligned}
$$

Figure 41: Syntax of temporal contracts

($\cup\tilde{T}$), intersection ($\cap\tilde{T}$), the empty temporal contract ($\epsilon$), and an open temporal contract closed by an environment ($T^\circ, \rho$). We consider the fail contract $\perp$ as a macro for $\cup\{\}$, and DFM's universal contract ... a macro for $\cap\{\}$. The difference between $!A$ and $\neg T$ is that the first must be an action and force time to step forward once, whereas the second may match arbitrarily many actions.

Actions themselves are expressed as patterns denoting calls ($\texttt{call}(n, \mathit{pat})$) or returns ($\texttt{ret}(n, \mathit{pat})$), with respect to a particular function named $n$ and with its argument or result matching a pattern *pat*. If $n$ is a label ($\ell$), we simply check that the monitor wrapping the function has the same label. Arrow contract monitors impose their label on the contracted function. However, if $n$ is a variable ($x$), then we consult a binding environment that the monitoring system builds as we pass binding actions to determine if the function is exactly equal to the value bound. Patterns can match values ($v$), variable bindings and references ($?x$, $x$), labeled functions ($\ell$), structured data (($\texttt{cons}$ *pat pat*)), negated patterns that do not bind (!*pat*), anything or nothing (`Any`, `None`).

### 7.1.3 *File example*

$$
\begin{aligned}
&\mathit{FileSystemContract} = \mathit{open} : \mathit{String} \to \mathit{FileContract} \\
&\mathit{FileContract} = \mathit{Record} \\
&\quad \mathit{read} : \mathit{Unit} \to \mathit{String} \\
&\quad \mathit{write} : \mathit{String} \to \mathit{Unit} \\
&\quad \mathit{close} : \mathit{Unit} \to \mathit{Unit} \\
&\text{where} \quad ...\ \texttt{ret}(\mathit{close}, _)
\end{aligned}
$$

This example gives the contract for a hypothetical file system, which can be used to open files by giving the *open* function a filename (a *String*); the client is then given a file handle contracted by *FileContract*. A file handle, in turn, is a record of functions which interact with the file: *read*, *write*, and *close*, all which perform the expected behaviors. The temporal contract is what is interesting: it is not phrased in terms of a negation, but rather an affirmation. Its goal is to state that a user of the file is forbidden from making use of the file handle (through the use of its component functions) after the user has *close*d the file. It is phrased such that there is no "..." at the end of its trace; this means that the last reference one can make to such a monitored record is returning from *close*; after that, it cannot be used. Note that this is not a *liveness property*; this does not mean that a return from *close must* happen, as traces are *prefix-closed*.[28] Instead, the property is a *safety property*, though expressed in the affirmative.

[28] *A set of strings* $S$ *is called "prefix-closed" if* $s \in S$ *and* $s' \sqsubseteq s$ *implies* $s' \in S$ *where* $\sqsubseteq$ *is "is a prefix of" ordering.*

## 7.2 SEMANTICS

I present and analyze a slightly different formulation than DFM's temporal contracts that allows for more precise specification of how functions and values in general are used. But first, we must discuss why we do not import DFM's semantics directly.

### 7.2.1 *DFM's semantics*

DFM give a denotational semantics to their temporal contracts looks almost identical to a textbook definition of the denotation of regular expressions, with the key difference being the inclusion of binding forms. The details of the full definition are unimportant, and look similar to our denotation of full traces ($\mathsf{F}[\![_]\!]$, next subsection), but with two crucial differences. The first is their denotation of negation:

$$[\![\neg T^\circ]\!]_E = \mathit{Trace} \setminus [\![T^\circ]\!]_E$$

They use a module semantics based on an *EF* machine that tracks the bindings shared across module boundaries, $E$, and a stack of module boundaries to return to, $F$. Regardless of how this machine works, the denotation of a temporal contract attached to a structural contract, $[\![S\ \mathtt{where}\ T^\circ]\!]$, is generated by traces of *EF* that are driven by sent or received calls and return actions (roughly):

$$\left\{ \begin{array}{l} \mathtt{ret}(\mathit{start}, h)\pi \in \mathit{prefixes}([\![T^\circ]\!]_E) : \\ \qquad \langle E_0, \mathit{start}\rangle \Rightarrow^{\pi} \langle E, F\rangle \wedge E_0 = \epsilon, h : S \end{array} \right\}$$

The use of *prefixes* in this definition is problematic, and negation is the culprit. Contracts that state anything about how a trace may not end would allow just such traces since *extensions* to such "bad traces" are acceptable, and prefix closure will throw the "bad traces" back into what is acceptable. For example, the denotation of temporal contracts from DFM allows $AA \in \mathit{prefixes}(\llbracket \neg A \rrbracket)$, and because of prefix closure, $A \in \mathit{prefixes}(\llbracket \neg A \rrbracket)$.

The second difference is in the semantics for referring to functions; we give a different account that captures the spirit of their prose describing their system, and more closely reflects their implementation. The temporal component of the example discussed in Section 7.1.1 was originally the following:

$$\neg(\ldots\ \texttt{call}(\mathit{sort}, _, _)\ !\texttt{ret}(\mathit{sort}, _)^{*}\ \texttt{call}(\mathit{sort}, _, _))$$
$$\cap \neg(\ldots\ \texttt{ret}(\mathit{sort}, _, _) \ldots\ \texttt{call}(\mathit{cmp}, _, _))$$

In contrast to our restatement, the flat use of *labels* instead of *bindings* would cause a second call to a supposedly-correct *sort* to fail, since it internally calls the comparator of the same label, but of a different monitor construction. Their implementation works around this by additionally adding a monitor-wrapping action, that generates a new label to pair with the function label to uniquely identify it. Our semantics is functionally the same, since we use pointer equality on wrapped values. Wrapping a value constructs a new container, so the fresh monitor label plays the same role as the monitor's pointer identity.

### 7.2.2 *My semantics*

I change three aspects of DFM's semantics:

1. temporal contracts' denotational semantics is split into both *full trace* and *partial trace* interpretations, with a non-standard interpretation of negation;
2. matching includes *uncertainty*, to allow for sound approximations (inherited from *Limp*);
3. we use *temporal monitors as modules* to interpret module interactions.

Our semantics (Figure 42) alternates between full traces ($F\llbracket _ \rrbracket$) and partial traces ($P\llbracket _ \rrbracket$) to combat the problems introduced by DFM's original use of *prefixes* on top of a semantics of full traces. Our interpretation of negation disallows any future observation to redeem a trace: a negated temporal contract will reject all non-empty full traces of the given contract, as well as any extension of such traces[1]. By Theorem 33, no rejected trace can be extended to an accepted one. I claim that this semantics is what DFM intended their system to mean, as it matches up with the expectations of their prose, the

1 This semantics of negation does not satisfy double-negation elimination (DNE).

$$
\begin{aligned}
B[\![\cdot]\!]_\bullet &: (\mathit{Temporal}^\circ \times \mathit{Env}) \cup \mathit{Temporal} \to \mathit{Trace} \\
B[\![\cup \tilde{T^\circ}]\!]_\rho &= \bigcup B[\![T^\circ]\!]_\rho \ldots \\
B[\![\cap \tilde{T^\circ}]\!]_\rho &= \bigcap B[\![T^\circ]\!]_\rho \ldots \\
B[\![\epsilon]\!]_\rho &= \{\epsilon\} \\
B[\![\neg T^\circ]\!]_\rho &= \neg F[\![T^\circ]\!]_\rho \\
P[\![\cdot]\!]_\bullet &: (\mathit{Temporal}^\circ \times \mathit{Env}) \cup \mathit{Temporal} \to \mathit{Trace} \\
P[\![T^{\circ *}]\!]_\rho &= F[\![T^{\circ *}]\!]_\rho \cdot P[\![T^\circ]\!]_\rho \\
P[\![T^\circ_0 \cdot T^\circ_1]\!]_\rho &= P[\![T^\circ_0]\!]_\rho \cup F[\![T^\circ_0]\!]_\rho \cdot P[\![T^\circ_1]\!]_\rho \\
P[\![\langle A \rangle\ T^\circ]\!]_\rho &= \{\epsilon\} \cup \{d\pi : d, \rho' \in (\!|A|\!)_\rho, \\
&\qquad\qquad \pi \in P[\![T^\circ]\!]_{\rho'}\} \\
P[\![A]\!]_\rho &= \{\epsilon\} \cup F[\![A]\!]_\rho \\
F[\![\cdot]\!]_\bullet &: (\mathit{Temporal}^\circ \times \mathit{Env}) \cup \mathit{Temporal} \to \mathit{Trace} \\
F[\![T^\circ_0 \cdot T^\circ_1]\!]_\rho &= F[\![T^\circ_0]\!]_\rho \cdot F[\![T^\circ_1]\!]_\rho \\
F[\![T^{\circ *}]\!]_\rho &= F[\![T^\circ]\!]^*_\rho \\
F[\![\langle A \rangle\ T^\circ]\!]_\rho &= \{d\pi \ :\ d, \rho' \in (\!|A|\!)_\rho, \pi \in F[\![T^\circ]\!]_{\rho'}\} \\
F[\![A]\!]_\rho &= \{d \ :\ d, \rho' \in (\!|A|\!)_\rho\} \\
(\!|A|\!)_\rho &= \{d, \rho' \ :\ \rho' = m(A, d, \rho)\} \\
\neg \Pi &= \{\epsilon\} \cup \{\pi \ :\ \forall \pi' \in \Pi \setminus \{\epsilon\}. \pi' \not\leq \pi\}
\end{aligned}
$$

Figure 42: Denotational Semantics of Temporal Contracts (B means both P and F)

test cases in their implementation, and additionally raises blame on programs that DFM were surprised their implementation accepted — in particular, a program that produces the faulty trace discussed in Section 7.1.1.

**Theorem 33** (Prefix closure)**.** $\mathit{prefixes}(P[\![T^\circ]\!]_\rho) = P[\![T^\circ]\!]_\rho$

This theorem is provable with structural induction on $T^\circ$ and the following lemma:

**Lemma 34** (Full in prefix)**.** $F[\![T^\circ]\!]_\rho \subseteq P[\![T^\circ]\!]_\rho$

The semantics and derivatives here are simplified to the concrete case since *Limp* automatically abstracts for us.

Matching allows binding arbitrary values from the language for later comparison, so the space of temporal contract derivatives is unbounded. After abstraction the value space becomes finite, but comparison for (concrete) equality is not decidable, so we fall back on *Limp*'s built-in support for managing equality judgments.

Matching against sets of values makes it possible that we have several possible matches. Thus $m$ returns a set of environment, possible

$$
\begin{aligned}
\partial^{\cdot}_{\cdot}\bullet &: \mathit{Action} \times \mathit{Temporal}^{\circ} \times \mathit{Env} \to \mathit{Temporal} \\
\partial^{\rho}_{d}\epsilon &= \bot \\
\partial^{\rho}_{d}A &= \begin{cases} \epsilon & \text{if } \rho = m(A, d, \rho) \\ \bot & \text{if } \#\mathtt{f} = m(A, d, \rho) \end{cases} \\
\partial^{\rho}_{d}\langle A \rangle\ T^{\circ} &= \begin{cases} T^{\circ}, \rho' & \text{if } \rho' = m(A, d, \rho) \\ \bot & \text{if } \#\mathtt{f} = m(A, d, \rho) \end{cases} \\
\partial^{\rho}_{d}T^{\circ}_{0} \cdot T^{\circ}_{1} &= \cup\{\partial^{\rho}_{d}T^{\circ}_{0} \cdot (T^{\circ}_{1}, \rho),\ \nu(T^{\circ}_{0}) \cdot \partial^{\rho}_{d}T^{\circ}_{1}\} \\
\partial^{\rho}_{d}\cup\tilde{T^{\circ}} &= \cup\partial^{\rho}_{d}T^{\circ}\ldots \\
\partial^{\rho}_{d}\cap\tilde{T^{\circ}} &= \cap\partial^{\rho}_{d}T^{\circ}\ldots \\
\partial^{\rho}_{d}T^{\circ *} &= \partial^{\rho}_{d}T^{\circ} \cdot (T^{\circ *}, \rho) \\
\partial^{\rho}_{d}\neg T^{\circ} &= \mathtt{if}\ \nu(\partial^{\rho}_{d}T^{\circ}) \stackrel{?}{=} \epsilon\ \mathtt{then}\ \bot\ \mathtt{else}\ \neg\partial^{\rho}_{d}T^{\circ}
\end{aligned}
$$

Figure 43: Derivatives of Temporal Contracts

$$
\begin{aligned}
\nu(\epsilon) &= \nu(T^{\circ *}) = \nu(\neg T^{\circ}) = \epsilon \\
\nu(\langle A \rangle\ T^{\circ}) &= \nu(A) = \bot \\
\nu(\cup\tilde{T^{\circ}}) &= \bigvee \nu(T^{\circ})\ldots \\
\nu(\cap\tilde{T^{\circ}}) &= \bigwedge \nu(T^{\circ})\ldots \\
\nu(T^{\circ}_{0} \cdot T^{\circ}_{1}) &= \nu(T^{\circ}_{0}) \wedge \nu(T^{\circ}_{1}) \\
\nu(T^{\circ}, \rho) &= \nu(T^{\circ})
\end{aligned}
$$

Figure 44: Nullability of Temporal Contracts

$$m : \mathit{Pattern} \times \mathit{Data} \times \mathit{Env} \to \wp(\mathit{MR})$$

$$\begin{aligned}
p \in \mathit{Pattern} &::= \mathit{TPattern} \text{ rules plus } \mid c(\overline{p}) \\
\mathit{mr} \in \mathit{MR} &::= \rho \mid \texttt{\#f} \\
S &\subset \mathit{MR} \\
\text{Let } \rho_\bot &= \texttt{\#f} \\
d \in \mathit{Data} &= \tilde{d} \mid v \mid c(\overline{d}) \\
c \in \mathit{Constructors} &= \{\texttt{call}, \texttt{ret}, \texttt{cons}\} \\
S \bowtie S' &= \{\mathit{mr} \triangleleft \mathit{mr}' \ : \ \mathit{mr} \in S, \mathit{mr}' \in S'\} \\
\rho \triangleleft \rho' &= (\lambda x.(\texttt{if } x \in \mathbf{dom}(\rho') \\
&\qquad\qquad \rho'(x) \\
&\qquad\qquad \rho(x)))
\end{aligned}$$

Figure 45: Spaces and functions for matching

$$\begin{aligned}
m(\texttt{Any}, _, \rho) &= \rho \\
m(\texttt{None}, _, \rho) &= \texttt{\#f} \\
m(\ell, \lambda^\ell x.\ e, \rho) &= \rho \\
m(!\mathit{pat}, d, \rho) &= \mathit{case}\ m(\mathit{pat}, d, \rho) \\
&\quad \mid \texttt{\#f} \Rightarrow \rho \\
&\quad \mid \rho' \Rightarrow \texttt{\#f} \\
m(?x, d, \rho) &= \rho[x \mapsto d] \\
m(x, d, \rho) &= m(\rho(x), d, \rho) \\
m(c(\overline{p}), c(\overline{d}), \rho) &= (\bowtie S \ldots) \\
&\text{where } S \ldots = m(\mathit{pat}, d, \rho) \ldots \\
m(p, \tilde{d}, \rho) &= \bigcup m(p, d, \rho) \ldots \\
m(d, d', \rho) &= \rho_{d \simeq d'} \\
m(p, d, \rho) &= \texttt{\#f} \quad \text{otherwise}
\end{aligned}$$

Figure 46: Semantics of matching

from matching a given pattern against some data, and if failure is possible (`#f`). The interesting case is for constructed data, where we must combine results for each tree element. We simply left-associate $\bowtie$ over results to get a cross-product of the different match combinations. The $\triangleleft$ operator extends the left environment with the bindings of the right, though the order doesn't matter considering that binding patterns may not bind the same variable twice.

The denotational semantics is not executable, so the correctness of derivatives with respect to the denotational semantics is crucial to the correctness of our monitoring system. We must show the correctness of both the full and partial interpretations of both open and closed temporal contracts, though all have similar proofs. We say $e$ satisfies a temporal contract $T$ its trace of actions sent to the temporal monitor is in the denotation of the temporal contract ($P[\![T]\!]$). The proof that our monitoring system ensures an expression either satisfies its contract or blames is mostly technical. This is because monitors are generated during reduction, but the proof hinges mainly on the correctness of derivatives:

**Theorem 35** (Derivatives correct)**.** *The following are mutually true*

1. $F[\![\partial^{\rho}_{d} T^{\circ}]\!] = \{\pi : d\pi \in F[\![T^{\circ}]\!]_{\rho}\}$
2. $P[\![\partial^{\rho}_{d} T^{\circ}]\!] = \{\pi : d\pi \in P[\![T^{\circ}]\!]_{\rho}\}$
3. $F[\![\partial_{d} T]\!] = \{\pi : d\pi \in F[\![T]\!]\}$
4. $P[\![\partial_{d} T]\!] = \{\pi : d\pi \in P[\![T]\!]\}$

The key lemma is the correctness of our nullability function, which follows from a simple induction.

**Lemma 36** (Nullability)**.** $\nu(T^{\circ}) = \epsilon \iff \epsilon \in F[\![T^{\circ}]\!]_{\rho}$

Thus assuming Lemma 36, each case of Theorem 35 has a straightforward proof except in the $\neg$ case, shown below (for the first proposition):

**Case** $T^{\circ} \equiv \neg T^{\circ\prime}$.

**Case** $H : \nu(\partial^{\rho}_{d} T^{\circ\prime}) = \epsilon$.

| | | |
|---|---|---|
| (1) | $F[\![\partial^{\rho}_{d} T^{\circ}]\!] = \emptyset$ | by computation |
| (2) | $\epsilon \in F[\![\partial^{\rho}_{d} T^{\circ\prime}]\!]$ | by H, lemma 36 |
| (3) | $d \in F[\![T^{\circ\prime}]\!]_{\rho}$ | by IH, (2) |

To show $\{\pi : d\pi \in F[\![T^{\circ}]\!]_{\rho}\} = \emptyset$, we suppose $\pi \in \neg F[\![T^{\circ\prime}]\!]_{\rho}$ and show $\pi \not\equiv d\pi'$:

**Case** $\pi \equiv d\pi'$.

Since $d \in F[\![T^{\circ\prime}]\!]_{\rho}$, by definition of $\neg$, contradiction.

**Otherwise.**

$\pi$ not prefixed by $d$

**Case** $H : \nu(\partial^{\rho}_{d} T^{\circ\prime}) = \bot$.

| | | |
|---|---|---|
| (1) | $\epsilon \notin F[\![\partial^{\rho}_{d} T^{\circ\prime}]\!]$ | by lemma 36 |

(2) $\{\pi : d\pi \in F[\![T^{\circ\prime}]\!]_\rho\} = F[\![\partial_d^\rho T^{\circ\prime}]\!]$ by IH
(3) $d \notin F[\![T^{\circ\prime}]\!]_\rho$ by (1), (2)
(4) Goal is $\neg\{\pi : d\pi \in F[\![T^{\circ\prime}]\!]_\rho\} = \{\pi : A\pi \in \neg F[\![T^{\circ\prime}]\!]_\rho\}$ by computation

We prove this goal by bi-containment:

**Case** $Hs : \pi \in \neg\{\pi : d\pi \in F[\![T^{\circ\prime}]\!]_\rho\}$.

(1) $\forall \pi' \in \{\pi : A\pi \in F[\![T^{\circ\prime}]\!]_\rho\} \setminus \{\epsilon\}. \pi' \not\leq \pi$ by *Hs* and inversion
(2) Suppose $\pi' \in F[\![T^{\circ\prime}]\!]_\rho$
(3) $\pi' \not\leq A\pi$ by (1), (2), prefix cancellation
(4) $\pi \in \neg F[\![T^{\circ\prime}]\!]_\rho$ by (3)

**Case** $Hs : \pi \in \neg F[\![T^{\circ\prime}]\!]_\rho$.

(1) $\forall \pi' \in F[\![T^{\circ\prime}]\!]_\rho \setminus \{\epsilon\}. \pi' \not\leq A\pi$ by *Hs*, inversion
(2) Suppose $\pi' \in \{\pi : A\pi \in F[\![T^{\circ\prime}]\!]_\rho\} \setminus \{\epsilon\}$
(3) $A\pi' \in F[\![T^{\circ\prime}]\!]_\rho$ by (2)
(4) $\pi' \not\leq \pi$ by (1), (3), prefix cancellation
(5) $\pi \in \neg\{\pi : d\pi \in F[\![T^{\circ\prime}]\!]_\rho\}$ by (4)

Paths relate to repeated derivation:

**Corollary 37.** $\pi \in P[\![T]\!] \iff \nu(\partial_\pi T) = \epsilon$

An additional surprising property of this semantics is that triple negation is single negation, but double negation is a separate beast. The first note is that double negation has the property in 38. For convenience, let's define the following helper.

$$done? : Temporal \to \mathbb{B}$$
$$done?(T) = \nu(T) \stackrel{?}{=} \epsilon$$

**Theorem 38** (DN kills future)**.** $\partial_d\neg\neg T = done?(\partial_d T) \to ..., \bot$

*Proof.*

$$\begin{aligned}
\partial_d\neg\neg T &= done?(\partial_d\neg T) \to \bot, \neg\partial_d\neg T \\
&= done?(done?(\partial_d T) \to \bot, \neg\partial_d T) \to \bot, \neg\partial_d\neg T \\
&= done?(\partial_d T) \to && \textit{[if lift]} \\
&\quad (done?(\bot) \to \bot, \neg\partial_d\neg T), \\
&\quad (done?(\neg\partial_d T) \to \bot, \neg\partial_d\neg T) \\
&= done?(\partial_d T) \to \neg\partial_d\neg T, \bot && \textit{[def. } \nu\textit{]} \\
&= done?(\partial_d T) \to \neg done?(\partial_d T) \to \bot, \neg\partial_d T, \bot \\
&= done?(\partial_d T) \to \neg\bot, \bot && \textit{[case hyp.]} \\
&= done?(\partial_d T) \to ..., \bot
\end{aligned}$$

□

In other words, since negation restricts any lookahead ability to one event, a double negation limits the applicability of a temporal contract to only its first allowed event.

Triple negation has the flip quality of double negation: if the derivative is nullable, then fail, otherwise all is permissible.

**Lemma 39.** $\partial_d \neg\neg\neg T = done?(\partial_d T) \to \bot, ...$

*Proof.*

$$\begin{aligned}
\partial_d \neg\neg\neg T &= done?(\partial_d \neg\neg T) \to \bot, \neg\partial_d \neg\neg T \\
&= done?(done?(\partial_d T) \to ..., \bot) \to \bot, \neg\partial_d \neg\neg T && \text{[Theorem 38]} \\
&= done?(\partial_d T) \to && \text{[if lift]} \\
&\qquad (done?(...) \to \bot, \neg\partial_d \neg\neg T), \\
&\qquad (done?(\bot) \to \bot, \neg\partial_d \neg\neg T) \\
&= done?(\partial_d T) \to \bot, \neg\partial_d \neg\neg T && \text{[def. done?]} \\
&= done?(\partial_d T) \to \bot, \neg(done?(\partial_d T) \to ..., \bot) && \text{[Theorem 38]} \\
&= done?(\partial_d T) \to \bot, \neg\bot && \text{[case hyp.]} \\
&= done?(\partial_d T) \to \bot, ...
\end{aligned}$$

□

Finally, four negations squashes back to two.

**Theorem 40** (QNE). $\partial_d \neg\neg\neg\neg T = \partial_d \neg\neg T$

*Proof.*

$$\begin{aligned}
\partial_d \neg\neg\neg\neg T &= done?(\partial_d \neg\neg\neg T) \to \bot, \neg\partial_d \neg\neg\neg T \\
&= done?(done?(\partial_d T) \to \bot, ...) \to \bot, \neg\partial_d \neg\neg\neg T && \text{[Lemma 39]} \\
&= done?(\partial_d T) \to && \text{[if lift]} \\
&\qquad (done?(\bot) \to \bot, \neg\partial_d \neg\neg\neg T), \\
&\qquad (done?(...) \to \bot, \neg\partial_d \neg\neg\neg T) \\
&= done?(\partial_d T) \to \neg\partial_d \neg\neg\neg T, \bot && \text{[def. done?]} \\
&= done?(\partial_d T) \to \neg(done?(\partial_d T) \to \bot, ...), \bot && \text{[Lemma 39]} \\
&= done?(\partial_d T) \to \neg\bot, \bot && \text{[case hyp.]} \\
&= done?(\partial_d T) \to ..., \bot \\
&= \partial_d \neg\neg T
\end{aligned}$$

□

Actions should only be visible to a temporal contract monitor if the action affects that monitor. This is not an issue in DFM's semantics, since they consider only one module at a time. For each `tmon` redex, our semantics creates a fresh runtime monitor for the contract; we store the state of these monitors in a global environment $\tau$, where the freshness comes from a space of timelines. I call these different keys "timelines" since time is relative to each module. We use timelines to distinguish where different function contracts will send their actions to be checked. Thus, as functions cross module boundaries, they also shift timelines: the more boundaries a function crosses, the more timelines will be aware of the calls made to it (due to nested wrappings). The specifics of timelines are discussed in Section 7.3.

## 7.3 THE SEMANTICS IN *limp*

Temporal monitors capture values to compare for object identity. The monitors can thus create space leaks if not implemented carefully. Concrete space leaks imply to abstract precision leaks. This section not only shows how we model the semantics of temporal contracts in *Limp*, but also shows how to use weak references to stop monitoring "dead" values.

We employ abstract garbage collection to precisely analyze programs utilizing temporal contracts. Weak references are necessary for blame freedom predictions when a monitor is waiting to blame when a given would-be dead function is called, for example. Recall to the sort example's "don't call *cmp* after return" contract, and consider a program that calls the *sort* function more than once with the same comparator. The abstract semantics can use stale addresses for temporally contracting functions, so object identity checks between temporally contracted *cmp* functions become useless **May** results. A newly contracted *cmp* can be safely called during sorting in the concrete, but in the abstract the "new" *cmp* is indistinguishable from the "old" *cmp*. The "old" *cmp* shouldn't be called, so a call to the "new" *cmp* leads to both a blame and the comparison. Without collecting would-be dead functions in monitors, the analysis emits a spurious blame.

**SCHEME SYNTAX IN *limp*** The Scheme-like language's core expressions in the *Limp* language syntax are

```
[(e) Expr (app Expr Exprs)
          x
          (lam xs Expr)
          (smon ℓ ℓ ℓ Ssyn e)
          (tmon e Tsyn)
          (begine Expr Exprs)
          (letrece LClauses Expr)
          (ife Expr Expr Expr)
          TCon-syntax
          (primop Primops)
          Datum
          #:bounded]
```

```
[(Tsyn) TCon-syntax (bind Expr) (tpred Expr) (¬ Tsyn) (kl Tsyn)
                    (· Tsyn Tsyn)
                    (∪ Tsyn Tsyn) (∩ Tsyn Tsyn)
                    (⊥) (⊤) (ϵ)]
[(Ssyn) SCon-syntax (flat Expr) (--> Ssyns Ssyn ℓ Expr) (any/c)
                    (cons/c Ssyn Ssyn)]
```

As an implementation shortcut, the matching semantics of temporal contracts in the previous sections are deferred to user-level functions manipulating event values. This gives the semantics extra power that goes unchecked, like the ability to create unboundedly many temporal contracts. We limit all examples to a restricted coding style that is non-recursive and only uses predicates, `if`s, projections and equality checking. Thus, instead of a binding temporal contract form, `bind` instead expects an expression that evaluates to a function that takes an event value and produces "the rest" of the temporal contract. Similarly, the `tpred` form expects an expression that evaluates to a function that takes an event value and produces a truth value indicating success or failure to match.

The list forms in the syntax use trusted lists (will not be store-allocated)

```
[TList (#:Λ X (#:U (Nil) (TCons X (#:inst TList X))))
       #:trust-construction]
[(es) Exprs (#:inst TList Expr)]
[(xs) Names (#:inst TList Name)]
[(Ssyns) SCon-syntaxes (#:inst TList Ssyn)]
```

Names, labels, primitive operators' names and data are all external spaces.

```
  [(x) #:external Name #:syntax identifier?]
  [(ℓ) #:external Label #:syntax recognize-label]
  [#:external Primop #:syntax
    (λ (s)
       (memv (syntax-e s)
             '(cons car cdr pair? null?
               not box? make-box unbox
               call? call-label call-fn call-args
               ret? ret-fn ret-label ret-value
               boolean? real? equal? set-box!
               add1 sub1 = <= zero? + *
               new-timeline)))]
```

Labels must use the syntax of quoted symbols:

```
(define (recognize-label stx)
  (syntax-case stx (quote)
    [(quote x) (identifier? #'x) #t]
    [_ #f]))
```

```
(letrece
    (LC fact
        (lam (TCons n (Nil))
          (ife (app zero? (TCons n (Nil)))
               1
               (app *
                (TCons n
                 (TCons (app fact
                  (TCons (app sub1 (TCons n (Nil)))
                         (Nil)))
                  (Nil))))))
    (NLC))
  fact)
```

Figure 47: Factorial in our Schemey AST

The Datum external space is more involved, since it evaluates the syntax to get the quoted form.

```
[#:external Datum
 #:syntax
 (λ (s)
   (with-handlers ([values (λ _ #f)])
     (define ev
       (parameterize
           ([sandbox-eval-limits (list 1 1)])
         (make-evaluator 'racket/base)))
         (define x
           (call-in-sandbox-context
             ev
             (λ () (eval-syntax s))))
         (or (symbol? x)
             (boolean? x)
             (number? x)
             (string? x)
             (null? x)
             (void? x))))]
```

By attaching syntax recognizers to external spaces, we can write terms that will get tagged with the appropriate external space's name. Figure 47 shows how we write factorial in this little language:

MACHINE REPRESENTATION IN *limp* The abstract machine we use has thirteen (13) kinds of states.

```
[State (ans v) (ev e ρ eκ) (coe eκ v)
       (ap fnv vs eκ)
       (blame ℓ S v) (tblame ℓ T event)
       (ev-syn Ssyn ρ sκ)
```

```
    (check ℓ ℓ S v cκ)
    (check-app (#:inst TList S) vs Blessed vs eκ)
    (send T event ℓ η tκ)
    (cod tκ T) (cos sκ S) (coc cκ v)]
```

The answer (`ans`), expression eval (`ev`), continue expression eval (here `coe` but usually `co`) and apply (`ap`) states should be familiar. The other states are for assigning structural contract blame (`blame`), assigning temporal contract blame (`tblame`), constructing a structural contract (`ev-syn`), checking a value against a structural contract (`check`), checking an arrow contract against a call (`check-app`), sending an event to a temporal monitor (`send`), continuing a temporal derivative computation (`cod`), continuing a structural contract construction (`cos`), and continuing a structural contract check (`coc`).

The four continue states correspond to the four classifications of continuations. The continuations in the machine can be visualized as striped in four (4) different colors.

1. An expression continuation (`EKont`) expects a value;
2. a structural contract checking continuation (`CKont`) expects a value;
3. a temporal contract derivative contract (`TKont`) expects a temporal contract value; and
4. a structural contract constructing continuation (`SKont`) expects a blessed structural contract.

All but the expression continuation are constructed in an expression context (an `EKont` in the tail), or in their own context.

```
[(eκ) EKont (Halt) (ECons eφ eκ) (PCons pφ tκ)
            (VCons vφ cκ) (ACons aφ sκ)]
[(cκ) CKont (CCons cφ cκ) (HCons hφ eκ)]
[(tκ) TKont (τCons dφ tκ) (LCons lφ eκ)]
[(sκ) SKont (SCons sφ sκ) (BCons bφ eκ)]
```

All the different modes of execution need expression evaluation at some point, so expression continuations have different constructors to carry the different modes' continuation types.

A value is a constructed temporal contract, a timeline, a primitive (datum or operation), an event, a function, a blessed function, a cons, or a `letrec` cell.

```
[(fnv) Proc-Value (primop Primop) (Clo xs Expr Env) Blessed]
[Blessed (Clo/blessed ℓ ℓ (#:inst TList SCon) S ℓ η fnv)]
[(v) Value
     fnv T η
     event
     (LR-cell (#:addr #:expose #:identity))
     Primop Datum (cons Value Value)]
```

The interesting rules are for checking contracts and sending events to the temporal monitor. When we apply a function that has been blessed with a temporal arrow contract, we first check the arguments:

```
[#:--> #:name wrap-app
       (ap (#:name fn (Clo/blessed _ _ Svs- _ _ _ _)) vs κ)
       (check-app Svs- vs fn (Nil) κ)]
```

The `check-app` form components are the argument contracts, argument values to check, the function called, the (reverse) of all checked/blessed values, and the continuation. If there are contracts and values still to check, switch over to the `check` state and remember the rest of the contracts and arguments to check.

```
[#:--> (check-app (TCons Sv- Svs-) (TCons v0 vs-to-check)
       (#:name fn (Clo/blessed ℓ- ℓ+ _ _ _ _ _)) vs-checked κ)
       (check ℓ- ℓ+ Sv- v0
              (HCons (ch∗k Svs- fn vs-to-check vs-checked) κ))]
```

If there are no more contracts and values to check, send the call event with the blessed arguments to the temporal monitor. After the contract on the timeline derives against the call event, we will call `fn` with the appropriate arguments. If the derivation invalidates the contract, we remember the event that causes temporal blame.

```
[#:--> (check-app (Nil)
                  (Nil)
                  (#:name fn (Clo/blessed ℓ- ℓ+ _ sv+ ℓ η clv))
                  vs-checked κ)
       (send (#:cast TCon (#:lookup a)) ev ℓ- η
             (LCons (blcall fn args-checked ev) κ))
       [#:where (timeline a) η]
       [#:where args-checked
        (#:call reverse #:inst [Value] vs-checked)]
       [#:where ev (call fn args-checked)]]
```

The `send` state does some of the temporal contract derivation, and `cod` continues it.

```
[#:--> (send T ev ℓ η κ)
       (#:match T
         [(ϵ) (cod κ (⊥))]
         [(⊥) (cod κ (⊥))]
         [(⊤) (cod κ (⊤))]
         [(bindv v) (ap v (TCons ev (Nil)) (PCons (mk-tcon) κ))]
         [(klv T∗) (send T∗ ev ℓ η (τCons (seqk T) κ))]
         [(¬v T∗) (send T∗ ev ℓ η (τCons (negt) κ))]
         [(·v T_0 T_1)
          (send T_0 ev ℓ η (τCons φ κ))
          [#:where φ (#:if (#:call ν?v T_0)
```

```
                            (seq2k T_1 ev η ℓ)
                            (seqk T_1))]]
           [(∪v T_0 T_1)
            (send T_0 ev ℓ η (τCons (∪_0 T_1 ev η ℓ) κ))]
           [(∩v T_0 T_1)
            (send T_0 ev ℓ η (τCons (∩_0 T_1 ev η ℓ) κ))]
           [(tpredv v)
            (ap v (TCons ev (Nil)) (PCons (pred-to-T) κ))])]
```

The sequencing rule chooses up front if it will need to do the second derivative, since we only need to derive the right contract if the left contract is nullable:

$$\partial^{\rho}_{d} T^{\circ}_{0} \cdot T^{\circ}_{1} = \cup\{\partial^{\rho}_{d} T^{\circ}_{0} \cdot (T^{\circ}_{1}, \rho),\ \nu(T^{\circ}_{0}) \cdot \partial^{\rho}_{d} T^{\circ}_{1}\}$$

The `(mk-tcon)` and `(pred-to-T)` frames direct the expression evaluation back to temporal contract derivation:

```
[#:--> (coe (PCons φ κ) v)
       (#:match φ
          [(mk-tcon)
           (cod κ (#:cast TCon v))]
          [(pred-to-T)
           (cod κ (#:if v (ϵ) (⊥)))])]
```

The result of the `bindv` function on the event *is* the derivative of the `bindv` against the event. A predicate contract denotes to a single event string, so if the predicate succeeds, then the derivative is the empty string. If the predicate fails, the contract derives to failure.

Continuing a derivation is fairly straightforward.

```
[#:--> (cod (τCons φ κ) v)
       (#:match φ
         [(negt) (cod κ (#:if (#:call ν?v v*)
                              (⊥)
                              (#:call mk¬v v*)))]
         [(seqk T_1) (cod κ (#:call mk·v v* T_1))]
         [(seq2k T_1 ev η ℓ-)
          (send T_1 ev ℓ- η (τCons (∪_1 (#:call mk·v v* T_1)) κ))]
         [(∪_0 T ev η ℓ-) (send T ev ℓ- η (τCons (∪_1 v*) κ))]
         [(∩_0 T ev η ℓ-) (send T ev ℓ- η (τCons (∩_1 v*) κ))]
         [(∪_1 T) (cod κ (#:call mk∪v T v*))]
         [(∩_1 T) (cod κ (#:call mk∩v T v*))])
       [#:where v* (#:cast TCon v)]]
```

The endpoints of derivation are to kick off a blessed function call (waiting to check the output contract) or finally return from a blessed function call with its result.

```
[#:-->
 (cod (LCons φ κ) v)
 (#:match φ
   [(blcall (#:name fn (Clo/blessed ℓ- ℓ+ _ Sv+
                                     ℓ (timeline a) clv))
            vs ev)
    (#:if (#:call μ?v v)
          (tblame ℓ- (#:cast TCon (#:lookup a)) ev)
          (#:let ([#:update a v])
            (ap clv vs (ECons (chret fn) κ))))]
   [(blret (#:name ev (ret (Clo/blessed _ ℓ+ _ _ _
                                          (timeline a) _)
           rv)))
    (#:if (#:call μ?v v)
          (tblame ℓ+ (#:cast TCon (#:lookup a)) ev)
          (#:let ([#:update a v])
            (coe κ rv)))])]
```

Whenever the contract is in obvious failure, the semantics blames. Obvious failure means algebraically ⊥ (see Figure 48), which is sufficient for denotationally ⊥. Being algebraically ⊥ is not sufficient for denotationally ⊥. It is undecidable to determine if a contract is denotationally ⊥ because `(predv v)` is denotationally ⊥ only if `v` is contextually equivalent to `(λ (ev) ff)`. The temporal contract constructors apply algebraic simplifactions to make these decisions faster and represent fewer distinct yet equivalent contracts in the state space.

Checking structural contracts is simple: an arrow contract of $n$ arguments checked against a function of $n$ arguments creates a monitor around the function that carries the contracts to check on call/return, and the parties involved. The wrapped function is called after all arguments are checked and the call event is accepted by the temporal monitor. Conses check contracts structurally, reconstructing the conses of checked/blessed values. Flat contracts return the original value if the predicate does not evaluate to `ff`.

```
[#:--> (check ℓ+ ℓ- S v κ)
       (#:match S
        [(-->/blessed Svs- Sv+ ℓ η)
         (#:match v
          [(#:name v∗ (Clo args _ _))
           (coc κ (Clo/blessed ℓ- ℓ+ Svs- Sv+ ℓ η v∗))
           [#:when (#:call eq-len args Svs-)]]
          [(#:name v∗ (Clo/blessed _ _ args _ _ _ _))
           (coc κ (Clo/blessed ℓ- ℓ+ Svs- Sv+ ℓ η v∗))
           [#:when (#:call eq-len args Svs-)]])]
          [(cons/c A D)
           (#:match v
            [(cons Av Dv)
```

```
(μ?v : (TCon → #:boolean)
 [(μ?v (⊥)) (#:external boolean #t)]
 [(μ?v (∪v T_0 T_1)) (#:if (#:call μ?v T_0)
                            (#:call μ?v T_1)
                            (#:external boolean #f))]
 [(μ?v (∩v T_0 T_1)) (#:if (#:call μ?v T_0)
                            (#:external boolean #t)
                            (#:call μ?v T_1))]
 [(μ?v (·v T_0 T_1)) (#:if (#:call μ?v T_0)
                           (#:external boolean #t)
                           (#:if (#:call ν!?v T_0)
                                 (#:call μ?v T_1)
                                 (#:external boolean #f)))]
 [(μ?v _) (#:external boolean #f)])
```

The `ν!?v` function decides if a temporal contract is algebraically $\epsilon$.

```
(ν!?v : (TCon → #:boolean)
 [(ν!?v (ε)) (#:external boolean #t)]
 [(ν!?v (klv T)) (#:call ν!?v T)]
 [(ν!?v (¬v T)) (#:call μ?v T)]
 [(ν!?v (∪v T_0 T_1)) (#:if (#:call ν!?v T_0)
                             (#:call ν!?v T_1)
                             (#:external boolean #f))]
 [(ν!?v (·v T_0 T_1)) (#:if (#:call ν!?v T_0)
                            (#:call ν!?v T_1)
                            (#:external boolean #f))]
 [(ν!?v (∩v T_0 T_1)) (#:if (#:call ν!?v T_0)
                             (#:external boolean #t)
                             (#:call ν!?v T_1))]
 [(ν!?v _) (#:external boolean #f)])
```

Figure 48: Algebraic ⊥-ness decision as *Limp* metafunction

```
            (check ℓ+ ℓ- A Av (CCons (chDk ℓ+ ℓ- D Dv) κ))]
           [_ (blame ℓ+ S v)])]
         [(any/c) (coc κ v)]
         [(#:name Sp (predv fn))
          (ap fn (TCons v (Nil)) (VCons (flatk v Sp ℓ-) κ))])]
```

The continuing checking state has two continuation forms. The first is for checking the cdr contract and finally constructing the blessed cons.

```
[#:--> (coc (CCons φ κ) v)
       (#:match φ
        [(chDk ℓ+ ℓ- D Dv)
         (check ℓ+ ℓ- D Dv (CCons (consk v) κ))]
        [(consk Av) (coc κ (cons Av v))])]
```

The second is for continuing checking function call arguments, sending the return event after the return value passes the output contract, and finishing an smon evaluation with the contracted value.

```
[#:--> (coc (HCons φ κ) v)
       (#:match φ
        [(ch*k Svs- fn vs-to-check vs-checked)
         (check-app Svs- vs-to-check fn (TCons v vs-checked) κ)]
        [(sret (#:name fn (Clo/blessed ℓ- ℓ+ _ _ ℓ η _)))
         (send (#:cast TCon (#:lookup a)) event ℓ+ η
                      (LCons (blret event) κ))
         [#:where (timeline a) η]
         [#:where event (ret fn v)]]
        [(Checking) (coe κ v)])]
```

Structural contract construction switches to expression evaluation to both construct flat contracts and evaluate the timeline component of a temporal arrow contract. The timeline is the final component of a temporal arrow contract, so the arrk frame carries the other components to finally construct the contract. The argument contracts are reversed, since they are checked in order but accumulated in reverse.

```
[#:--> (coe (ACons φ κ) v)
       (#:match φ
        [(mkflat)
         (cos κ (predv pred))
         [#:where (#:has-type fnv pred) v]]
        [(arrk Svs Sv ℓ)
         (cos κ (-->/blessed (#:call reverse #:inst [S] Svs)
              Sv ℓ η))
         [#:where (#:has-type Timeline η) v]])]
```

Temporal contract derivation switches to expression evaluation when it reaches a bindv to evaluate or a tpredv to check. Once those function calls finish evaluating, the bindv result is taken to mean the derivative, and the `tpredv` result determines if the checked event is in the singleton string of events that `tpredv` denotes.

```
[#:--> (coe (PCons φ κ) v)
       (#:match φ
        [(mk-tcon)
         (cod κ (#:cast TCon v))]
        [(pred-to-T)
         (cod κ (#:if v (ϵ) (⊥)))])]
```

Finally, when a flat contract finishes evaluating, a truish result returns the checked value. A false result means contract failure, meaning the semantics should blame.

```
[#:--> (coe (VCons (flatk vc _ _) κ) v)
       (#:if v
             (coc κ vc)
             (blame ℓ− Sp vc))]
```

The full semantics is in the software artifact[29].

[29] *Available at https://github.com/ianj/limp/blob/master/thocon.rkt*

## 7.4 evaluation

The temporal contract case study pushed the limits of the *Limp* system. Bugs were found and fixed, limited expressiveness was expanded, and "the AAM transformation" is closer to having a technical meaning (partially completed work undeveloped in this document). The *Limp* system has plenty of room to grow.

Whereas one could write the pushdown abstraction rules given access to the store object, the language does not linguistically support the construction. Without a pushdown abstraction, spurious backward flows from an arrow's return value contract lead to repeated temporal events and thus spurious blame. If weak references are not adequately collected in temporal monitors, repeated calls to contracted functions can lead to spurious blame. Garbage collection is expensive in terms of the potential state space size. We need to be able to reuse portions of the analysis when irrelevant parts of the state have changed, via sparse techniques.

# 8 RELATED WORK

Program analysis is a rich area where no single analysis is an island. We all stand on the shoulders of giants. This chapter on related work is separated into sections as they relate to the different chapters of this dissertation.

## 8.1 ENGINEERING ENGINEERED SEMANTICS (OPTIMIZING AAM)

ABSTRACTING ABSTRACT MACHINES This work clearly closely follows Van Horn and Might's original papers on abstracting abstract machines [96, 97], which in turn is one piece of the large body of research on flow analysis for higher-order languages (see Midtgaard [64] for a thorough survey). The AAM approach sits at the confluence of two major lines of research: (1) the study of abstract machines [57] and their systematic construction [78], and (2) the theory of abstract interpretation [20, 21].

ABSTRACT INTERPRETATION In the framework of abstract interpretation, the accepted method of gathering information about a program's execution is to manipulate the semantics to do some extra task. For example, it can build an environment at each "control point"[30] that maps variables to an over-approximation of all the values they can take — the so-called constant propagation analysis. The modified semantics is called the *non-standard semantics*, and can take any form you want, so long as it remains sound. A programming language semantics is treated extensionally as the set of all execution traces that the language deems valid. This viewpoint, while powerful, is so general it is easy to get lost trying to apply it.

[30] *Think of a node in a control-flow graph.*

AAM provides a focused viewpoint. Instead of bothering with a platonic set of all traces, it instead deals with single (computable) steps of an abstract machine.

FRAMEWORKS FOR FLOW ANALYSIS OF HIGHER-ORDER PROGRAMS Besides the original AAM work, the analysis most similar to that presented in section 3.2 is the infinitary control-flow analysis of Nielson and Nielson [70] and the unified treatment of flow analysis by Jagannathan and Weeks [38]. Both are parameterized in such a way that in the limit, the analysis is equivalent to an interpreter for the language, just as is the case here. What is different is that both give a constraint-based formulation of the abstract semantics rather than a finite machine model.

ABSTRACT COMPILATION Boucher and Feeley [11] introduced the idea of abstract compilation, which used closure generation [30] to improve the performance of control flow analysis. We have adapted the closure generation technique from compositional evaluators to abstract machines and applied it to similar effect.

CONSTRAINT-BASED PROGRAM ANALYSIS FOR HIGHER-ORDER LANGUAGES Constraint-based program analyses (e.g. [70, 105, 62, 90]) typically compute sets of abstract values for each program point. These values approximate values arising at run-time for each program point. Value sets are computed as the least solution to a set of (inclusion or equality) constraints. The constraints must be designed and proved as a sound approximation of the semantics. Efficient implementations of these kinds of analyses often take the form of worklist-based graph algorithms for constraint solving, and are thus quite different from the interpreter implementation. The approach thus requires effort in constraint system design and implementation, and the resulting system require verification effort to prove the constraint system is sound and that the implementation is correct.

This effort increases substantially as the complexity of the analyzed language increases. Both the work of maintaining the concrete semantics and constraint system (and the relations between them) must be scaled simultaneously. However, constraint systems, which have been extensively studied in their own right, enjoy efficient implementation techniques and can be expressed in declarative logic languages that are heavily optimized [12]. Consequently, constraint-based analyses can be computed quickly. For example, Jagannathan and Wright's polymorphic splitting implementation [105] analyses the Vardoulakis and Shivers benchmark about 5.5 times faster than the fastest implementation considered here. These analyses compute very different things, so the performance comparison is not apples-to-apples.

The AAM approach, and the state transition graphs it generates, encodes temporal properties not found in classical constraint-based analyses for higher-order programs. Such analyses (ultimately) compute judgments on program terms and contexts, e.g., at expression $e$, variable $x$ may have value $v$. The judgments do not relate the order in which expressions and context may be evaluated in a program, e.g., it has nothing to say with regard to question like, "Do we always evaluate $e_1$ before $e_2$?"The state transition graphs can answer these kinds of queries, but evaluation demonstrated this does not come for free.

## 8.2 PUSHDOWN ANALYSIS

PUSHDOWN MODELS AND MEMOIZATION The idea of relating pushdown automata with memoization is not new. In 1971, Stephen Cook [18] devised a transformation to simulate 2-way (on a fixed input) *deterministic* pushdown automata in time linear in the size of the input, that uses the same "context irrelevance" idea to skip from state q seen before to a corresponding first state that produced a smaller stack than q was seen with. Such a state is an instance of what are called *terminator* states. A *terminator* state is simply a state that performs a pop operation. Six years later, Neil D. Jones[45] simplified the transformation instead to *interpret* a stack machine program to work *on-the-fly* still on a deterministic machine, but with the same idea of using memo tables to remember corresponding terminator states. Thirty-six years after that, at David Schmidt's Festschrift, Robert Glück extended the technique to two-way *non-deterministic* pushdown automata, and showed that the technique can be used to recognize context-free languages in the standard $\mathcal{O}(n^3)$ time [35]. Glück's technique (correctness unproven at time of writing) uses the meta-language's stack with a deeply recursive interpretation function to preclude the use of a frontier and something akin to $\Xi$[1]. By exploring the state space *depth-first*, the interpreter can find all the different terminators a state can reach one-by-one by destructively updating the memo table with the "latest" terminator found. The trade-offs with this technique are that it does not obviously scale to first-class control, and the search can overflow the stack when interpreting moderate-sized programs. We have not performed an extensive evaluation to test the latter point, however. A minor disadvantage is that it is also not a fair evaluation strategy when allocation is unbounded. The technique can nevertheless be a viable alternative for languages with simple control-flow mechanisms. It has close similarities to "Big-CFA2" in Vardoulakis' dissertation [98].

In 1981, Sharir and Pnueli [84] proposed a "functional approach" to interprocedural program analysis that first captured the notion of summarization. Summaries themselves look like memo table entries. The specifics of the technique limited its use to first-order programming languages until CFA2 generalized the approach to higher-order programs written in continuation-passing-style (CPS).

In 1994, Andersen and Jones [4] took the insight of memoization Jones used on 2-way pushdown automata and applied it to imperative stack programs. They transform a program to insert textual pushes and pops in order to run programs faster, using more memory. This work was for concrete execution, but it has a close lineage to the techniques used in the abstract by this dissertation, Sharir and Pnueli, and the following related work.

**CFA2 AND PDCFA** The immediately related work is that of PDCFA [28, 29], CFA2 [103, 102], and AAM [95], the first two of which we recreated in full detail. The version of CFA2 that handles `call/cc`

1 See `gluck.rkt` in supplementary materials for a lambda calculus analysis in Glück's style does not handle composable control, is dependent on a restricted CPS representation, and has untunable precision for first-class continuations. Our semantics adapts to `call/cc` by removing the metacontinuation operations, and thus this work supersedes theirs. The extended version of PDCFA that inspects the stack to do garbage collection [29] also fits into our model; the addresses that the stack keeps alive can be accumulated by "reading through" the continuation table, building up the set of addresses in each portion of the stack that we come across.

STACK INSPECTION Stack inspecting flow analyses also exist, but operate on pre-constructed regular control-flow graphs [7], so the CFGs cannot be trimmed due to the extra information at construction time, leading to less precision. Backward analyses for stack inspection also exist, with the same prerequisite [15].

ANALYSIS OF PUSHDOWN AUTOMATA Pushdown models have existed in the first-order static analysis literature [68, Chapter 7][77], and the first-order model checking literature [10], for some time. The important difference when we move higher-order is that the model construction to feed these methods is an additional problem—the one we solve here. Additionally, the algorithms employed in these works expect a complete description of the model up front, rather than work with a modified `step` function (also called `post`), such as in "on-the-fly" model-checking algorithms for finite state systems [82].

DERIVATION FROM ABSTRACT MACHINES The trend of deriving static analyses from abstract machines does not stop at flow analyses. The model-checking community showed how to check temporal logic queries for collapsible pushdown automata (CPDA), or equivalently, higher-order recursion schemes, by deriving the checking algorithm from the Krivine machine [81]. The expressiveness of CPDAs outweighs that of PDAs, but it is unclear how to adapt higher-order recursion schemes (HORS) to model arbitrary programming language features. The method is strongly tied to the simply-typed call-by-name lambda calculus and depends on finite sized base-types.

The finite-sized base types restriction is close to AAM's restriction on base types. Particularly, it is a weak and entirely natural restriction that takes a change of perspective to look past. In AAM, states are explored if their exact representation has not been seen before[31], so if there are an unbounded number of representations for base types' *abstractions*, then the analysis will not terminate. Abstract interpretation's notion of *widening* is the tool we would use in AAM to force convergence on abstractions that have an unbounded representation space. In HORS, there is work on counter-example-guided abstraction refinement (CEGAR) [53] to break coarse abstractions down to finer pieces in order to verify properties. If the process of abstraction-refinement is well-founded, the "finite types" restriction is not violated.

[31] *With garbage collection, exactness can be replaced with a store subsumption check.*

## 8.3 SEMANTICS OF ABSTRACT MACHINES

The term "abstract machine" in the general computer science sense can mean any theoretical model of computation. In the programming languages discipline, however, the term "abstract machine" is understood well enough to show "functional correspondences" between evaluation functions for language terms and equivalent abstract machines for those languages [2]. The work in functional correspondence is still "by hand," and never defines the concept of "abstract machine." On the one hand, by leaving the term "abstract machine" open to an informal style of semantics specification, we leave the community room for innovation and creative freedom. On the other hand, we don't even have a working definition that encompasses *enough* known abstract machines to be an interesting object of study. A matter of culture would allow the innovative-yet-definition-defying "abstract machine" constructions to influence and grow our *formal* understanding of the term.

I could argue that the contextual rewriting semantics [51] of PLT Redex [32] provides a more than suitable foundation for abstract machines. The focus of this dissertation has been on *environment* machines – machines that maintain administrative data structures so that recursive decompositions are unnecessary. The novel semantics of PLT Redex is therefore not quite what we need. Perhaps an abstract machine is simply a conditional term rewriting system (CTRS)? Not exactly – abstract machines are clumsily expressed in all of Klop's characterizations of CTRSs [52]. Conditions depend on notions of convertability with respect to the relation being defined, joinability (reduction to a common term), evaluation (with the defined reduction relation) to specific ground terms, or an external first-order logical system. TRSs are not concerned with non-term data structures. Pattern matching could be expressed in the first two characterizations, but with additional unnecessary power.

The K framework [83] has a similar goal of expressing just the concrete semantics and getting "for free" a program analysis, but the group has so-far made light on their progress.

The specifics of store refinements and store updates in Chapter 6 are a combination of *strong updates* and *weak updates*. A *strong update* is safe if the address is fresh. If an address is not fresh, the semantics falls back on the more common *weak update*. The combination of weak and strong updates is called *conservative updates*, originally from Chase et al..

### 8.3.1 *Synthesizing correct analyses*

The language I developed in Chapter 6 is not an isolated incident of analysis synthesis. The common distinguishing characteristic of each synthesis in this section is that none of them provides an executable concrete semantics.

A remarkable example of a well-designed analysis synthesis tool is Flow Logic [72]. The language for flow logic is efficiently implemented, and is natural for anyone familiar to constraint-based approaches to program analysis. The constraint framework is tailored to regular analyses, so proper call/return matching is outside its grasp.

Rhodium [58] computes the least fixed point of programmer-provided flow facts on top of a single language model. The novelty of the tool is that it can guarantee soundness of transformations justified by analysis results, as long as its generated proof obligations can be discharged. The programmer-provided flow facts are also automatically checked for soundness. The rules that are definable on control-flow graph edges cannot introduce new edges, so higher-order analyses are not within the framework.

The abstract domains that Chapter 6 supports are open-ended, but not well-exercised. Strong abstract domains are similarly automatically constructable, as shown by Thakur et al. [91]. This approach finds the most precise inductive invariants of its inputs using a "bilateral approach" and an SMT solver. A bilateral approach is the combination of both forward and backward analysis to converge on high precision answers. The primary downside to the approach is its basis in interprocedural control-flow graphs (ICFGs) instead of a description of a concrete semantics. Not only is there no executable model, but the foundation is unsuited for higher-order analysis due to the inability to add behavior based on calculated facts.

# 9 CONCLUSION AND FUTURE WORK

I have addressed the itemized problems in the introduction:

- **Unsoundness:** All of the techniques in this dissertation are designed to be drop-in replacements for both the concrete implementation and analysis, where the two are separated by a single parameter: the memory allocator. An advantage to this approach is that the two semantics can be run in parallel and checked against each other to be confident in their behavior.
- **Imprecision and state-space explosion:** I have shown how precisely handling call and return sites of function calls leads to better predictions and performance. I have also shown how the precision of allocation and machine states themselves can be modularly tuned without affecting the correctness of the core analysis.
- **Non-termination:** When any component of a program state can be nested without bound, it is an almost certainty that state-space exploration will not terminate in general. I developed a language for expressing abstract machines in the natural way with recursive constructions. The language's support for allocation enables AAM-like store-allocation to prevent unbounded nesting. The only source of new values is with memory allocation. A finite allocation strategy is easier to control than an entire state space - any allocation strategy is sound.

The techniques developed in this dissertation are not only lightweight, they are relatively easy to prove correct. The key takeaway from this work is that abstraction should be an *external* input to an analysis framework. When the inputs that guide abstraction can only weaken predictions and not correctness, we get a single, more trustworthy framework that doubles as a language interpreter and a program analyzer.

I have shown implementation techniques that are all rigorous and orders of magnitude better than a naive translation of "math" to code. Indeed, the result of the techniques is still "math," just structured in a way more amenable to efficient implementation.

## 9.1 FUTURE WORK

A STATIC SEMANTICS FOR AAM I have an unfinished language, called *Limp*, on top of the core metalanguage in Chapter 6. Synthesizing an allocation function is a chore, which I conjecture can be mitigated by "heapification" annotations on types. The annotations themselves can be inferred for recursive types. The heapification annotations are treated as the annotated type for purposes of subtyping, but any removal or addition of a heapification annotation during subtyping is treated as an explicit coercion. For example, a stack is a list of frames:

$$\mu \mathit{List}. \cup\{\texttt{(Nil)},\texttt{(Cons}\ \varphi\ \mathit{List}\texttt{)}\}$$

where we can determine by type structure that *List* is a recursive reference that should thus be heapified. When a semantics pushes a frame on the stack, the *List* type we have in hand must be coerced to a heapified *List* type. Adding a heapification annotation translates to a store-allocation. Removing a heapification annotation translates to a store lookup.

When all recursive constructions are identified and translated to explicit address manipulation, the result is an "AAM-ified" semantics. Indeed I believe that a type-directed transformation gives algorithmic meaning to "the AAM transformation."

IMPROVED *limp* PERFORMANCE The *Limp* implementation is ripe for mechanizing the systematic implementation strategies of the first part of the dissertation. The software artifact at the time of writing contains no data specialization or imperative strategies to efficiently represent and explore the state space. Ideally, I would like to write a *Limp* to Racket compiler so that abstract machines in *Limp* are implemented exactly how I would write them by hand, or better. Metafunction evaluation is especially lacking in smarts. For functions with simple recursion schemes on trusted data structures, expensive memoization to prevent non-termination is unnecessary.

The run-time dependency analysis that TAJS performs to do sparse analysis appears to be general enough to apply to *Limp*'s evaluation model. Sparse analyses are crucial to analysis scalability.

IMPROVED *limp* EXPRESSIVENESS If *Limp* could express the notion of a *context*, and mark a state component as "the stack," then it could import the pushdown techniques in Chapter 4. I suspect that determining which component is "stack-like" is a fairly simple analysis. Since the reduction relation requires knowing the type of a state, the analysis could search to see if each state variant contains a component isomorphic to a list to designate as "the stack." Further, the stack must be treated in a "stack-like" fashion - destruction that never drops "the rest" of the stack, and bounded construction that always keeps "the rest" of the stack within the output stack.

Pushdown treatment of stack-capturing semantics requires a more advanced analysis. If the component marked as "the stack" from analysis of the state type ever flows to the store (an update expression's type contains a supertype of "the stack"), we have to introduce a context approximation function. I am less sure how the other interactions with the stack would be managed.

A glaring omission from this whole dissertation is treatment of unknowns - black-hole values that can be arbitrarily manipulated by an adversarial (AKA demonic) context. Automatically generating sound demonic contexts might be easy, but the predictions may be too conservative. Expressing language-provided program invariants impervious to manipulation by any context is a topic of deep research in logical relations and bisimulations. On one hand, this appears to be a big opportunity for abstract interpretation. Kripke logical relations for verifying imperative and concurrent programs include "protocols" that appear to be small abstract machines [94, 27, 93]. On the other hand, these logical relations are highly specialized to specific language features and may be difficult to automatically generate and apply in the context of abstract abstract machines.

IMPROVED *limp* PRECISION The forward-execution model in *Limp* limits its usefulness and precision. Forward analyses start with an under-approximation of the state space; we can only trust that the final result is sound because there is nothing more to add to the approximation. Therefore a long-running forward analysis cannot be stopped mid-stream to extract a sound result. Backward analyses start with an over-approximation that gets trimmed until nothing more can be soundly removed. The over-approximations can be too coarse to analyze efficiently - imagine a higher-order program where all calls to first-class functions are treated as calls to *all* functions that exist in the program. The first-order abstract interpretation community has known this since the beginning (Cousot and Cousot's debut AI paper [20] discussed "dual approximation methods"). After an extensive literature search, I found no higher-order analysis methods that apply these dual methods.

Backwards-executing *only* approaches exist for higher-order control-flow analyses [9, 74, 88]. Their commonality is in the use of the set-constraint formulation of 0CFA as a logic program that can "run backwards." Some of the difficulty is definitely in procedure-call boundaries, where backward execution requires knowing all the reaching closures to the call. There is definitely some synergy to be gained from combining forward and backward analyses.

# Part III

# APPENDIX

# NOTATIONAL CONVENTIONS

This dissertation follows a strict set of notational rules with respect to data and control-flow representation, scoping rules, and computational versus propositional relations. This appendix is meant to be used as a "legend" to help read the mathematical constructions of this document.

## 1 META RULES

The following rules are for notations themselves. If a binary relation has an overset '?', then it is meant as a decision procedure for membership in the relation. In the cases decision problems arise, the procedures should be obvious from the relation definition. The simplest decision we will see is $x \stackrel{?}{=} y$, where $x$ and $y$ are some type $A$ with decidable equality (for example, natural numbers or containers whose members are of a type with decidable equality).

I will give spaces of data as equalities or itemizations in EBNF. When I write

$$\begin{aligned} e \in \mathit{Expr} &::= \mathbf{Var}(x) \mid \mathbf{App}(e, \mathbf{e}) \mid \mathbf{Lam}(\mathbf{x}, e) \\ x &\in \mathit{Name} \end{aligned}$$

I mean

1. $e$ is a metavariable that, by convention, will be of *Expr* type (also $e$ with primes, subscripts and superscripts will be *Expr* type);
2. $\mathbf{e} : \mathit{Expr}^*$ emboldened is a list of the type of the metavariable's type by convention. Here we have a list of expressions; and
3. *Expr* is a closed inductive data type whose variants (named injections) are **Var**, **App**, and **Lam** with types

$$\begin{aligned} \mathbf{Var} &: \mathit{Name} \to \mathit{Expr} \\ \mathbf{App} &: \mathit{Expr} \to \mathit{Expr}^* \to \mathit{Expr} \\ \mathbf{Lam} &: \mathit{Name}^* \to \mathit{Expr} \to \mathit{Expr} \end{aligned}$$

I use two notations for list metavariables. As above, $\mathbf{e}$ is a list that is indexable - its $i^{th}$ element is written $e_i$. The metavariable $\overline{e}$ is also an element of $\mathit{Expr}^*$, but is interpolatable as $e\ldots$ following the rules of Kohlbecker and Wand [54]. For example, if $\overline{e} = \langle 1, 2, 3\rangle$ and $f : \mathbb{N} \to \mathbb{N} \to \mathbb{N} \to X$, then

$$f(e\ldots) = f(1, 2, 3)$$

Similarly, the metavariable $\tilde{e}$, is an element of $\wp(\mathit{Expr})$ (the powerset of *Expr*) and is also interpolatable in contexts that are associative and commutative (the order of interpolation doesn't matter). Any metavariable $x$ explicitly in $\wp(E)$ or $E^*$ is interpolatable ($x\ldots$) and indexable ($x_i$).

## 2 DATA

- $\langle e_0, \ldots, e_{n-1} \rangle$: a tuple of $n$-many elements.
- If **t** is in scope, $t_i$ is the $i^{th}$ element of the **t** list.

## 3 CONDITIONALS

The most general form of conditional expression I use in the dissertation is *pattern matching*. A datatype $T$ with $n$-many variants $V_i(t^i_j, \ldots)$ is primitively eliminated with a `case` expression. If $e : T$ and $\mathit{fv}(t^i_j)\ldots \vdash \mathit{rhs}^i : A$ for each $i \in \{0 \ldots n-1\}$, then the following expression is type $A$:

$$\begin{aligned} &\texttt{case}\ e\ \texttt{of} \\ &V_0(t^0_j \ldots) : \mathit{rhs}^0 \\ &\qquad\vdots \\ V_{n-1}&(t^{n-1}_j \ldots) : \mathit{rhs}^{n-1} \end{aligned}$$

Incomplete matches are partial definitions. I may also provide fall-through cases via **else** or _, meaning I don't care about $e$'s value, or I will give an arbitrary binder (say, $x$) if I wish to refer to $e$'s value. In uncommon cases, I will use the same binder in a variant pattern to state a side condition that the corresponding subterms must be equal (and the equality will be decidable). In rare occasions I will give explicit side conditions that must evaluate to `tt` in order for a case to "match."

Another form is the 'if' expression. I use the Dijkstra notation for 'if', where *guard* → *then*, *else* desugars to

$$\begin{aligned} &\texttt{case}\ \mathit{guard}\ \texttt{of} \\ &\quad \texttt{tt} : \mathit{then} \\ &\quad \texttt{ff} : \mathit{else} \end{aligned}$$

## 4 QUANTIFICATION AND SCOPE

Set and map comprehensions are commonplace in the dissertation. The scoping and implicit quantification rules are important for understanding the formal meanings.

A comprehension is of the form

*left-delimiter element-expression* (optional $\in$ *domain-expression*) **:** *variable-constraints right-delimiter*

The delimiters are { } for set construction, and [ ] for finite function (map) construction. The free variables in *element-expression* are universally quantified in their domain of discourse (shadowing the local context), and in scope in *variable-constraints*. The free variables in *variable-constraints* that are bound in the local context reference the local context. The free variables in *variable-constraints* that are **not** bound in the local context are existentially quantified. The entire comprehension is read as "the largest collection of *element-expression* that further satisfy *variable-constraints*."

Explicit quantifiers such as $\exists$ and $\forall$ have scope extending to the farthest right extent with balanced parentheses.

I use $x \equiv S(y, \ldots)$ to mean "$x$ matches $S(y, \ldots)$;" alternatively, there exist elements $y \ldots$ such that $x = S(y, \ldots)$.

## 5 LIFTING AND ORDERING

Most of the dissertation depends on partial orders and implicit structural lifting. This means that if I define a function $f : A \to B$, and we come across a container $c$ containing $A$, then $f(c)$ is a generic map over $c$ that applies $f$ to elements of type $A$. Additionally, if $A$ has a defined order $\sqsubseteq$, then all containers of $A$ are ordered by a pointwise lifting. For instance,

- **Sets:** Given $S, S' : \wp(A)$, if $\forall a \in S. \exists a' \in S'. a \sqsubseteq a'$, then $S \sqsubseteq S'$.
- **Partial functions:** Given $f, g : B \rightharpoonup A$, if $\forall b \in \mathbf{dom}(f). f(b) \sqsubseteq g(b)$ then $f \sqsubseteq g$.
- **Datatypes:** (by example of 2-3 trees) Inductively,
  1. if $a \sqsubseteq a'$ then $\mathbf{Leaf}(a) \sqsubseteq \mathbf{Leaf}(a')$
  2. if $t_0 \sqsubseteq t'_0$ and $t_1 \sqsubseteq t'_1$ then $\mathbf{Two}(t_0, t_1) \sqsubseteq \mathbf{Two}(t'_0, t'_1)$.
  3. if $t_0 \sqsubseteq t'_0$, $t_1 \sqsubseteq t'_1$, and $t_2 \sqsubseteq t'_2$ then $\mathbf{Three}(t_0, t_1, t_2) \sqsubseteq \mathbf{Three}(t'_0, t'_1, t'_2)$.

  Generally, variants of the datatype must align, and their subterms must be covariantly ordered.

$\lceil \hat{t} \rceil_{\hat{\sigma}}$ : lift a term $\hat{t}$ to be an abstract term; if $\hat{t}$ is a delayed dereference, use $\hat{\sigma}$ to dereference.

## 6 LISTS

I use tuple notation, string notation, and cons notation for lists. String notation does not name more than the first character A list of numbers 1, 2, 3, in that order can be written as

- Tuple notation: $\langle 1, 2, 3\rangle$
- String notation: $1\ell$ where $\ell$ is a metavariable denoting the list of numbers 2, 3
- Cons notation: $1 : \ell$ with similar $\ell$, and additionally $1 : 2 : 3 : \epsilon$

In tuple notation the empty list is written as $\langle\rangle$. In string and cons notation, the empty list is $\epsilon$.

The append operation for lists is written differently for the different list notations. For lists $\ell$ denoting 1, 2, 3 and $\ell'$ denoting 4, 5, 6:

- Tuple notation: $(++\ell, \ell')$, $\langle \ell \ldots, \ell' \ldots\rangle$, $\langle 1, 2, 3, \ell' \ldots\rangle$, and $\langle \ell \ldots, 4, 5, 6\rangle$.
- String notation: $\ell\ell'$
- Cons notation: $(++\ell, \ell')$, $\ell + +\ell'$.

## 7 SETS

Union ($\cup$), intersection ($\cap$), membership ($\in$), subset ($\subseteq$), proper subset ($\subset$) are all standard. Any comma-separated big operations are nested big operations. For example,

$$\bigcup_{a \in A, b \in B(a)} P(a, b) = \bigcup_{a \in A} \bigcup_{b \in B(a)} P(a, b)$$

The powerset of a set $A$ is written $\wp(A)$. The set of finite subsets of $A$ is written $\wp_{\text{fin}}(A)$.

## 8 RECORDS

A record is human-friendly way to write large tuples: instead of positions there are field names. If a record $r$ has a field $f$, the notation to get the value in $f$ is $r.f$. To create a new record where $f$ is set to some value $v$, we write $r[f := v]$. Multiple fields can be updated with comma-separated $:=$ directives, *e.g.*, $r[f_0 := v_0, f_1 := v_1]$.

Sometimes I will treat tuples as records where the field names are the metavariables used to define the tuple in the BNF grammar. For example, if I wrote $\varsigma \in \mathit{State} ::= \langle e, \rho, \sigma, \kappa\rangle$ in the grammar, and I have a state $\varsigma' \equiv \langle e', \rho', \sigma', \kappa'\rangle$, then I can refer to $\sigma'$ by writing $\varsigma.\sigma$, and I can write $\varsigma[\sigma := \sigma'']$ to mean $\langle e', \rho', \sigma'', \kappa'\rangle$. I do not use this notation if a tuple's definition has two of the same kind of metavariable, like a function application $(e\ e)$.

## 9 FUNCTIONS

A partial function space from $A$ to $B$ is denoted $A \rightharpoonup B$.
A finite function space from $A$ to $B$ is denoted $A \underset{\text{fin}}{\rightharpoonup} B$. I sometimes use $A \rightharpoonup B$ when finiteness is implicit from context for notational brevity.
If A and B are partially ordered, then $A \overset{\text{mono}}{\to} B$ is the function space for monotonic functions ($A \overset{\text{mono}}{\rightharpoonup} B$ for monotonic partial functions). A monotonic function $f : A \overset{\text{mono}}{\to} B$ satisfies the property (say A and B are ordered via $\preceq$ and $\sqsubseteq$ respectively)

$$\forall a, a' : A. a \preceq a' \implies f(a) \sqsubseteq f(a').$$

An antitonic function $g : A \overset{\text{anti}}{\to} B$ satisfies the property

$$\forall a, a' : A. a \preceq a' \implies g(a') \sqsubseteq g(a).$$

An injective function $f : A \rightarrowtail B$ satisfies the property

$$\forall a, a' : A. f(a) = f(a') \implies a = a'$$

A surjective function $f : A \twoheadrightarrow B$ satisfies the property

$$\forall b : B. \exists a : A. f(a) = b$$

Two partially ordered types A and B can be adjoined with a Galois connection: $A \underset{\alpha}{\overset{\gamma}{\leftrightarrows}} B$. The adjoint condition the functions $\gamma : B \to A$ and $\alpha : A \to B$ are

$$\forall a : A. a \preceq \gamma(\alpha(a))$$
$$\forall b : B. \alpha(\gamma(b)) \sqsubseteq b$$

which is equivalent to

$$\alpha(a) \preceq b \iff a \sqsubseteq \gamma(b)$$

I use $\lambda$ notation and bracket notation for constructing functions.

- $[a \mapsto b]$: A map of one key/value pair. Call it f. The semantics is $f(a) = b$ and $f(a') = \bot$ for $a \neq a'$.
- **dom**$(f)$: the domain of f. The set of points S where $\forall x \in S. f(x) \neq \bot$.
- $f[a \mapsto b]$: an overwriting update for a function: $\lambda a'. a \overset{?}{=} a' \to b, f(a')$.
- $f \triangleleft g$: a right-biased map extension: $\lambda a. a \overset{?}{\in} \mathbf{dom}(g) \to g(a), f(a)$.
- $\hat{\sigma} \blacktriangleleft \delta$: refine $\hat{\sigma}$ with $\delta$: $\hat{\sigma} \triangleleft \lambda \hat{a}. \lceil \delta(\hat{a}) \rceil_{\hat{\sigma}}$

# OAAM SUPPLEMENTALS

The entire definitions of each intermediate semantics were not shown in the chapter due to the mundanity of the differences. I thus give a reference here to the complete semantics of each other machine to use as a basis for the proofs in the following appendix.

## 9.1 λ**IF** *with store-allocated results*

Machine configuration space:

$$
\begin{aligned}
\varsigma \in \mathit{State} &= \mathtt{ev}^{t}(e, \rho, \sigma, \kappa) \mid \mathtt{co}\;(\kappa, v, \sigma) \mid \mathtt{ap}^{t}_{\ell}(v, a, \sigma, \kappa) \mid \mathtt{ans}\;(\sigma, v) \\
\kappa \in \mathit{Kont} &::= \mathtt{halt} \mid \mathbf{arg}^{t}_{\ell}(e, \rho, a, a, a) \mid \mathbf{fun}^{t}_{\ell}(a, a, a) \mid \mathtt{ifk}^{t}_{\ell}(e, e, \rho, \kappa) \\
\ell \in \mathit{Label} &\text{ an infinite set} \\
v \in \mathit{Value} &::= \mathtt{l} \mid o \mid \mathtt{clos}\;(x, e, \rho) \\
s \in \mathit{Storeable} &::= \mathtt{l} \mid o \mid \mathtt{clos}\;(x, e, \rho) \mid \kappa \\
\rho \in \mathit{Env} &= \mathit{Var} \underset{\text{fin}}{\rightharpoonup} \mathit{Addr} \\
\sigma \in \mathit{Store} &= \mathit{Addr} \underset{\text{fin}}{\rightharpoonup} \wp(\mathit{Storeable})
\end{aligned}
$$

Reduction semantics:

$$
\begin{aligned}
\mathtt{ev}\,(x\,,\rho,\sigma,\kappa) &\longmapsto \mathtt{co}\,(\kappa,v,\sigma) \text{ if } v \in \sigma(\rho(x))\\
\mathtt{ev}\,(\mathtt{lit}\,(l),\rho,\sigma,\kappa) &\longmapsto \mathtt{co}\,(\kappa,l,\sigma)\\
\mathtt{ev}\,(\lambda\ \ x.\,e,\rho,\sigma,\kappa) &\longmapsto \mathtt{co}\,(\kappa,\mathtt{clos}\,(x,e,\rho),\sigma)\\
\mathtt{ev}^t((e_0\ e_1)^\ell,\rho,\sigma,\kappa) &\longmapsto \mathtt{ev}^t(e_0,\rho,\sigma',\mathbf{arg}^t_\ell(e_1,\rho,a_\kappa))\\
&\quad \text{where } a_\kappa = \mathit{allockont}^t\ell(\sigma,\kappa)\\
&\quad\quad \sigma' = \sigma \sqcup [a_\kappa \mapsto \{\kappa\}]\\
\mathtt{ev}^t(\mathtt{if}^\ell(e_0,e_1,e_2),\rho,\sigma,\kappa) &\longmapsto \mathtt{ev}^t(e_0,\rho,\sigma',\mathtt{ifk}^t(e_1,e_2,\rho,a))\\
&\quad \text{where } a_\kappa = \mathit{allockont}^t\ell(\sigma,\kappa)\\
&\quad\quad \sigma' = \sigma \sqcup [a_\kappa \mapsto \{\kappa\}]\\
\mathtt{co}\,(\mathtt{halt},v,\sigma) &\longmapsto \mathtt{ans}\,(\sigma,v)\\
\mathtt{co}\,(\mathbf{arg}^t_\ell(e,\rho,a_\kappa),v,\sigma) &\longmapsto \mathtt{ev}^t(e,\rho,\sigma',\mathbf{fun}^t_\ell(a_f,a_\kappa))\\
&\quad \text{where } \begin{array}{ll} a_f & = \mathit{alloc}(\varsigma)\\ \sigma' & = \sigma \sqcup [a_f \mapsto \{v\}] \end{array}\\
\mathtt{co}\,(\mathbf{fun}^t_\ell(a_f,a_\kappa),v,\sigma) &\longmapsto \mathtt{ap}^t_\ell(u,v,\kappa,\sigma) \text{ where } \kappa \in \sigma(a_\kappa), u \in \sigma(a_f)\\
\mathtt{co}\,(\mathtt{ifk}^t(e_0,e_1,\rho,a),\mathtt{tt},\sigma) &\longmapsto \mathtt{ev}^t(e_0,\rho,\sigma,\kappa) \text{ where } \kappa \in \sigma(a)\\
\mathtt{co}\,(\mathtt{ifk}^t(e_0,e_1,\rho,a),\mathtt{ff},\sigma) &\longmapsto \mathtt{ev}^t(e_1,\rho,\sigma,\kappa) \text{ where } \kappa \in \sigma(a)\\
\mathtt{ap}^t_\ell(\mathtt{clos}\,(x,e,\rho),v,\sigma,\kappa) &\longmapsto \mathtt{ev}^{t'}(e,\rho',\sigma',\kappa)\\
&\quad \text{where } \begin{array}{ll} a & = \mathit{alloc}(\varsigma)\\ \rho' & = \rho[x \mapsto a]\\ \sigma' & = \sigma \sqcup [a \mapsto \{v\}] \end{array}\\
\mathtt{ap}^t_\ell(o,v,\sigma,\kappa) &\longmapsto \mathtt{co}\,(\kappa,v',\sigma) \text{ where } v' \in \Delta(o,v)
\end{aligned}
$$

$$
\begin{aligned}
\mathit{inject}(e) &= \mathtt{ev}^{t_0}(e,\bot,\bot,\mathtt{halt})\\
\mathit{reache} &= \{\varsigma \mid \mathit{inject}(e) \longmapsto^* \varsigma\}
\end{aligned}
$$

## 9.2 *Store-allocated results with lazy nondeterminism*

Machine configuration space:

$$
\begin{aligned}
\varsigma \in \mathit{State} &= \mathtt{ev}^t(e,\rho,\sigma,\kappa) \mid \mathtt{co}\,(\kappa,v,\sigma) \mid \mathtt{ap}^t_\ell(v,v,\sigma,\kappa) \mid \mathtt{ans}\,(\sigma,v)\\
v \in \mathit{Value} &::= l \mid o \mid \mathtt{clos}\,(x,e,\rho) \mid \mathtt{addr}\,(a)
\end{aligned}
$$

*Kont*, *Storeable*, *Env* and *Store* are defined the same as previously.

Reduction semantics:

$$\begin{aligned}
\mathtt{ev}\ (x\ ,\rho,\sigma,\kappa) &\longmapsto \mathtt{co}\ (\kappa,\mathtt{addr}\ (\rho(x)),\sigma)\\
\mathtt{ev}\ (\mathtt{lit}\ (l),\rho,\sigma,\kappa) &\longmapsto \mathtt{co}\ (\kappa,l,\sigma)\\
\mathtt{ev}\ (\lambda\ \ x.\ e,\rho,\sigma,\kappa) &\longmapsto \mathtt{co}\ (\kappa,\mathtt{clos}\ (x,e,\rho),\sigma)\\
\mathtt{ev}^t((e_0\ e_1)^\ell,\rho,\sigma,\kappa) &\longmapsto \mathtt{ev}^{t'}(e_0,\rho,\sigma',\mathbf{arg}^t_\ell(e_1,\rho,a))\\
&\qquad \text{where } a_\kappa = \mathit{allockont}^t\ell(\sigma,\kappa)\\
&\qquad\qquad \sigma' = \sigma\sqcup[a_\kappa\mapsto\{\kappa\}]\\
\mathtt{ev}^t(\mathtt{if}^\ell(e_0,e_1,e_2),\rho,\sigma,\kappa) &\longmapsto \mathtt{ev}^{t'}(e_0,\rho,\sigma',\mathtt{ifk}^t(e_1,e_2,\rho,a))\\
&\qquad \text{where } a_\kappa = \mathit{allockont}^t\ell(\sigma,\kappa)\\
&\qquad\qquad \sigma' = \sigma\sqcup[a_\kappa\mapsto\{\kappa\}]
\end{aligned}$$

$$\begin{aligned}
\mathtt{co}\ (\mathtt{halt},v,\sigma) &\longmapsto \mathtt{ans}\ (\sigma,u) \text{ where } u\in\mathit{force}(\sigma,v)\\
\mathtt{co}\ (\mathbf{arg}^t_\ell(e,\rho,a),v,\sigma) &\longmapsto \mathtt{ev}^{t'}(e,\rho,\sigma',\mathbf{fun}^t_\ell(a_f,a))\\
&\qquad \text{where } \begin{array}{ll} a_f & = \mathit{alloc}(\varsigma)\\ \sigma' & = \sigma\sqcup[a_f\mapsto\mathit{force}(\sigma,v)]\end{array}\\
\mathtt{co}\ (\mathbf{fun}^t_\ell(a_f,a),v,\sigma) &\longmapsto \mathtt{ap}^{t'}_\ell(u,v,\kappa,\sigma) \text{ where } \kappa\in\sigma(a),u\in\sigma(a_f)\\
\mathtt{co}\ (\mathtt{ifk}^t(e_0,e_1,\rho,a),\mathtt{tt},\sigma) &\longmapsto \mathtt{ev}^{t'}(e_0,\rho,\sigma,\kappa) \text{ where } \kappa\in\sigma(a)\\
\mathtt{co}\ (\mathtt{ifk}^t(e_0,e_1,\rho,a),\mathtt{ff},\sigma) &\longmapsto \mathtt{ev}^{t'}(e_1,\rho,\sigma,\kappa) \text{ where } \kappa\in\sigma(a)
\end{aligned}$$

$$\begin{aligned}
\mathtt{ap}^t_\ell(\mathtt{clos}\ (x,e,\rho),v,\sigma,\kappa) &\longmapsto \mathtt{ev}^{t'}(e,\rho',\sigma',\kappa)\\
&\qquad \text{where } \begin{array}{ll} a & = \mathit{alloc}(\varsigma)\\ \rho' & = \rho[x\mapsto a]\\ \sigma' & = \sigma\sqcup[a\mapsto\mathit{force}(\sigma,v)]\end{array}\\
\mathtt{ap}^t_\ell(o,v,\sigma,\kappa) &\longmapsto \mathtt{co}\ (\kappa,v',\sigma) \text{ where } u\in\mathit{force}(\sigma,v),v'\in\Delta(o,u)
\end{aligned}$$

$$\begin{aligned}
\mathit{force}(\sigma,\mathtt{addr}\ (a)) &= \sigma(a)\\
\mathit{force}(\sigma,v) &= \{v\}
\end{aligned}$$

$$\begin{aligned}
\mathit{inject}(e) &= \mathtt{ev}^{t_0}(e,\bot,\bot,\mathtt{halt})\\
\mathit{reache} &= \{\varsigma \mid \mathit{inject}(e)\longmapsto^*\varsigma\}
\end{aligned}$$

## 9.3 *Lazy nondeterminism with abstract compilation*

Machine configuration space:

$$\begin{aligned}
\varsigma\in\mathit{State} &= \mathtt{co}\ (\kappa,v,\sigma) \mid \mathtt{ap}^t_\ell(v,v,\sigma,\kappa) \mid \mathtt{ans}\ (\sigma,v)\\
k\in\mathit{Compiled} &= (\mathit{Env}\times\mathit{Store}\times\mathit{Kont}\times\mathit{Time})\to\mathit{State}\\
\kappa\in\mathit{Kont} &::= \mathtt{halt} \mid \mathbf{arg}^t_\ell(k,\rho,a) \mid \mathbf{fun}^t_\ell(a,a) \mid \mathtt{ifk}^t_\ell(k,k,\rho,a)\\
v\in\mathit{Value} &::= l \mid o \mid \mathtt{clos}\ (x,k,\rho) \mid \mathtt{addr}\ (a)\\
\mathit{Storeable} &::= l \mid o \mid \mathtt{clos}\ (x,k,\rho) \mid \kappa
\end{aligned}$$

*Store* and *Env* are defined the same as previously.

We write $\lambda^t(args\ldots).body$ (and without superscript) to mean $\lambda(args\ldots t).body$ and $k^t(\rho,\sigma,\kappa)$ to mean $k(\rho,\sigma,\kappa,t)$ for notational consistency.

Abstract compilation function:

$$\begin{aligned}
[\![_]\!] &: \mathit{Expr} \to \mathit{Compiled} \\
[\![x\ ]\!] &= \lambda(\rho,\sigma,\kappa).\mathtt{co}\ (\kappa,\mathtt{addr}\ (\rho(x)),\sigma) \\
[\![\mathtt{lit}\ (l)]\!] &= \lambda(\rho,\sigma,\kappa).\mathtt{co}\ (\kappa,l,\sigma) \\
[\![\lambda\ \ x.\ e]\!] &= \lambda(\rho,\sigma,\kappa).\mathtt{co}\ (\kappa,\mathtt{clos}\ (x,[\![e]\!],\rho),\sigma) \\
[\![(e_0\ e_1)^\ell]\!] &= \lambda^t(\rho,\sigma,\kappa).[\![e_0]\!]^{t'}(\rho,\sigma',\mathbf{arg}^t_\ell([\![e_1]\!],\rho,a)) \\
&\quad \text{where } a_\kappa = \mathit{allockont}^t\ell(\sigma,\kappa) \\
&\quad\quad \sigma' = \sigma\sqcup[a_\kappa\mapsto\{\kappa\}] \\
[\![\mathtt{if}^\ell(e_0,e_1,e_2)]\!] &= \lambda^t(\rho,\sigma,\kappa).[\![e_0]\!]^{t'}(\rho,\sigma',\mathtt{ifk}^t([\![e_1]\!],[\![e_2]\!],\rho,a)) \\
\text{where } a_\kappa &= \mathit{allockont}^t\ell(\sigma,\kappa) \\
&\quad\quad \sigma' = \sigma\sqcup[a_\kappa\mapsto\{\kappa\}]
\end{aligned}$$

Reduction semantics:

$$\begin{aligned}
\mathtt{co}\ (\mathtt{halt},v,\sigma) &\longmapsto \mathtt{ans}\ (\sigma,u) \text{ where } u\in \mathit{force}(\sigma,v) \\
\mathtt{co}\ (\mathbf{arg}^t_\ell(k,\rho,a),v,\sigma) &\longmapsto k^t(\rho,\sigma',\mathbf{fun}^t_\ell(a_f,a)) \\
\text{where } a_f &= \mathit{alloc}(\varsigma) \\
\sigma' &= \sigma\sqcup[a_f\mapsto \mathit{force}(\sigma,v)] \\
\mathtt{co}\ (\mathbf{fun}^t_\ell(a_f,a),v,\sigma) &\longmapsto \mathtt{ap}^t_\ell(u,v,\sigma,\kappa) \text{ where } \kappa\in\sigma(a), u\in\sigma(a_f) \\
\mathtt{co}\ (\mathtt{ifk}^t(k_0,k_1,\rho,a),\mathtt{tt},\sigma) &\longmapsto k^t_0(\rho,\sigma,\kappa) \text{ where } \kappa\in\sigma(a) \\
\mathtt{co}\ (\mathtt{ifk}^t(k_0,k_1,\rho,a),\mathtt{ff},\sigma) &\longmapsto k^t_1(\rho,\sigma,\kappa) \text{ where } \kappa\in\sigma(a)
\end{aligned}$$

$$\begin{aligned}
\mathtt{ap}^t_\ell(\mathtt{clos}\ (x,k,\rho),v,\sigma,\kappa) &\longmapsto k^{t'}(\rho',\sigma',\kappa) \\
\text{where } a &= \mathit{alloc}(\varsigma) \\
\rho' &= \rho[x\mapsto a] \\
\sigma' &= \sigma\sqcup[a\mapsto \mathit{force}(\sigma,v)] \\
\mathtt{ap}\ (o,v,\sigma,\kappa) &\longmapsto \mathtt{co}\ (\kappa,v',\sigma) \\
&\text{where } \kappa\in\sigma(a) \text{ and } u\in \mathit{force}(\sigma,v), v'\in\Delta(o,u)
\end{aligned}$$

$$\begin{aligned}
\mathit{inject}(e) &= [\![e]\!]^{t_0}(\bot,\bot,\mathtt{halt}) \\
\mathit{reach} e &= \{\varsigma \mid \mathit{inject}(e)\longmapsto^* \varsigma\}
\end{aligned}$$

## 9.4 *Widened abstract compilation*

Machine configuration space:

$$\begin{aligned}
\varsigma\in \mathit{State} &= \mathtt{co}\ (\kappa,v) \mid \mathtt{ap}^t_\ell(v,v,\kappa) \mid \mathtt{ans}\ (v) \\
\mathit{System} &= (\wp(\mathit{State}\times \mathit{Store})\setminus\{\varnothing\})\times\wp(\mathit{State})\times \mathit{Store}
\end{aligned}$$

$$\begin{aligned}
nw(\mathtt{co}\ (\kappa, v, \sigma)) &= \mathtt{co}\ (\kappa, v), \sigma \\
nw(\mathtt{ap}^t_\ell(u, v, \sigma, \kappa)) &= \mathtt{ap}^t_\ell(u, v, \kappa), \sigma \\
nw(\mathtt{ans}\ (\sigma, v)) &= \mathtt{ans}\ (v), \sigma \\
wn(\mathtt{co}\ (\kappa, v), \sigma) &= \mathtt{co}\ (\kappa, v, \sigma) \\
wn(\mathtt{ap}^t_\ell(u, v, \kappa), \sigma) &= \mathtt{ap}^t_\ell(u, v, \sigma, \kappa) \\
wn(\mathtt{ans}\ (v), \sigma) &= \mathtt{ans}\ (\sigma, v)
\end{aligned}$$

Reduction semantics:

$$\begin{aligned}
&\mathit{inject}(e) = (\{(\varsigma', \sigma)\}, \{\varsigma'\}, \sigma) \\
&\text{where } \varsigma = [\![e]\!]^{t_0}(\bot, \bot, \mathtt{halt}) \\
&\qquad \varsigma', \sigma = nw(\varsigma) \\
&\mathit{reach}e = \{wn(\varsigma, \sigma') \mid \mathit{inject}(e) \longmapsto^* (S, F, \sigma), (\varsigma, \sigma') \in S\} \\
&(S, F, \sigma) \longmapsto (S \cup S', F', \sigma') \\
&\text{where } I = \{nw(\varsigma^*) \mid \varsigma \in F, wn(\varsigma, \sigma) \longmapsto \varsigma^*, nw(\varsigma^*) \notin S\} \\
&\qquad F' = \{\varsigma \mid \exists \sigma.(\varsigma, \sigma) \in S'\} \\
&\qquad \Sigma = \{\sigma \mid \exists \varsigma.(\varsigma, \sigma) \in S'\} \\
&\qquad \sigma' = \bigsqcup_{\sigma \in \Sigma} \sigma \\
&\qquad S' = \{(\varsigma, \sigma') \mid \varsigma \in F'\}
\end{aligned}$$

## 9.5 *Abstract compilation with store deltas*

All previous machines had a trivial widening operator for the store that would expand states without stores to states with stores, reduce with the written semantics, and then remove the resulting stores and join them again so that there is one store shared amongst all states. Here we have a different widening that accumulates store changes so that entire stores need not be joined each step - just their changes. Machine configuration space:

$$\begin{aligned}
\varsigma \in \mathit{State} &= \mathtt{co}\ (\kappa, v) \mid \mathtt{ap}^t_\ell(v, v, \kappa) \mid \mathtt{ans}\ (v) \\
k \in \mathit{Compiled} &= (\mathit{Env} \times \mathit{Store} \times \mathit{Store}' \times \mathit{Kont} \times \mathit{Time}) \to (\mathit{State} \times \mathit{Store}') \\
\kappa \in \mathit{Kont} &::= \mathtt{halt} \mid \mathbf{arg}^t_\ell(k, \rho, a) \mid \mathbf{fun}^t_\ell(a, a) \mid \mathtt{ifk}^t_\ell(k, k, \rho, a) \\
\xi \in \mathit{Store}' &= (\mathit{Addr} \times \wp(\mathit{Storeable}))^*
\end{aligned}$$

*Storeable*, *Store*, *Env* and *Value* are defined the same as previously.

Abstract compilation function:

$$\begin{aligned}
\Delta[\![_]\!] &: \mathit{Expr} \to \mathit{Compiled} \\
\Delta[\![x\ ]\!] &= \lambda(\rho,\sigma,\xi,\kappa).(\mathtt{co}\ (\kappa, \mathtt{addr}\ (\rho(x)),\sigma),\bot) \\
\Delta[\![\mathtt{lit}\ (l)]\!] &= \lambda(\rho,\sigma,\xi,\kappa).(\mathtt{co}\ (\kappa, l,\sigma),\bot) \\
\Delta[\![\lambda\ \ x.\ e]\!] &= \lambda(\rho,\sigma,\xi,\kappa).(\mathtt{co}\ (\kappa, \mathtt{clos}\ (x,\Delta[\![e]\!],\rho),\sigma),\bot) \\
\Delta[\![(e_0\ e_1)^\ell]\!] &= \lambda^t(\rho,\sigma,\xi,\kappa).\Delta[\![e_0]\!]^{t'}(\rho,\sigma,\xi',\mathbf{arg}^t_\ell(\Delta[\![e_1]\!],\rho,a)) \\
&\text{where } a_\kappa = \mathit{allockont}^t\ell(\sigma,\kappa) \\
&\qquad \sigma' = \sigma \sqcup [a_\kappa \mapsto \{\kappa\}] \\
\Delta[\![\mathtt{if}^\ell(e_0,e_1,e_2)]\!] &= \lambda^t(\rho,\sigma,\xi,\kappa).\Delta[\![e_0]\!]^{t'}(\rho,\sigma,\xi',\mathtt{ifk}^t(\Delta[\![e_1]\!],\Delta[\![e_2]\!],\rho,a)) \\
&\text{where } a_\kappa = \mathit{allockont}^t\ell(\sigma,\kappa) \\
&\qquad \sigma' = \sigma \sqcup [a_\kappa \mapsto \{\kappa\}]
\end{aligned}$$

Reduction semantics helper (write $\varsigma \longmapsto^\xi_\sigma \varsigma', \xi'$ to mean $((\varsigma,\sigma,\xi),(\varsigma',\xi')) \in \longmapsto$):

$$\begin{aligned}
\longmapsto &\subseteq (\mathit{State} \times \mathit{Store} \times \mathit{Store\acute{}}) \times (\mathit{State} \times \mathit{Store\acute{}}) \\
\mathtt{co}\ (\mathtt{halt}, v) &\longmapsto^\xi_\sigma \mathtt{ans}\ (u), \xi \text{ where } u \in \mathit{force}(\sigma, v) \\
\mathtt{co}\ (\mathbf{arg}^t_\ell(k,\rho,a), v) &\longmapsto^\xi_\sigma k^t(\rho, \mathbf{fun}^t_\ell(a_f, a)), \xi \\
\text{where } a_f &= \mathit{alloc}(\varsigma) \\
\xi' &= (a_f, \mathit{force}(\sigma, v)){:}\xi \\
\mathtt{co}\ (\mathbf{fun}^t_\ell(a_f, a), v) &\longmapsto^\xi_\sigma \mathtt{ap}^t_\ell(u, v, \kappa), \xi \text{ where } \kappa \in \sigma(a), u \in \sigma(a_f) \\
\mathtt{co}\ (\mathtt{ifk}^t(k_0,k_1,\rho,a), \mathtt{tt}) &\longmapsto^\xi_\sigma k_0^{t'}(\rho,\sigma,\bot,\kappa), \xi \text{ where } \kappa \in \sigma(a) \\
\mathtt{co}\ (\mathtt{ifk}^t(k_0,k_1,\rho,a), \mathtt{ff}) &\longmapsto^\xi_\sigma k_1^{t'}(\rho,\sigma,\bot,\kappa), \xi \text{ where } \kappa \in \sigma(a) \\
\mathtt{ap}^t_\ell(\mathtt{clos}\ (x,k,\rho), v, \kappa) &\longmapsto^\xi_\sigma k^{t'}(\rho',\sigma,\xi,\kappa), \xi' \\
\text{where } a &= \mathit{alloc}(\varsigma) \\
\rho' &= \rho[x \mapsto a] \\
\xi' &= (a, \mathit{force}(\sigma, v)){:}\xi \\
\mathtt{ap}\ (o, v, \kappa) &\longmapsto^\xi_\sigma \mathtt{co}\ (\kappa, v'), \xi \\
&\text{where } \kappa \in \sigma(a) \text{ and } u \in \mathit{force}(\sigma, v), v' \in \Delta(o, u)
\end{aligned}$$

Reduction semantics:

$$\begin{aligned}
\mathit{inject}(e) &= (\{(\varsigma,\sigma)\},\{\varsigma\},\sigma)\\
\text{where } \varsigma,\xi &= \Delta[\![e]\!]^{t_0}(\bot,\bot,\bot,\mathtt{halt})\\
\sigma &= \mathit{replay}(\xi,\bot)\\
\mathit{reache} &= \{\mathit{wn}(\varsigma,\sigma') \mid \mathit{inject}(e) \longmapsto^* (S,F,\sigma),(\varsigma,\sigma') \in S\}\\
(S,F,\sigma) &\longmapsto (S\cup S',F',\sigma')\\
\text{where } (F',\xi') &= \mathit{step}^*(\varnothing,F,\sigma,\epsilon)\\
\sigma' &= \mathit{replay}(\xi',\sigma)\\
S' &= \{(c,\sigma') \mid c \in F'\}\\
\mathit{step}^*(F',\varnothing,\xi) &= (F',\xi)\\
\mathit{step}^*(F',\{c\}\cup F,\xi) &= \mathit{step}^*(F'\cup \mathit{cs}^*,F,\xi^*)\\
\mathit{cs}^* &= \{c' \mid (c,\sigma,\xi) \longmapsto_{\sigma\xi} (c',\xi^c)\}\\
\xi^* &= \mathit{appendall}(\{\xi^c \mid (c,\sigma,\xi) \longmapsto_{\sigma\xi} (c',\xi^c)\})\\
\mathit{appendall}(\varnothing) &= \epsilon\\
\mathit{appendall}(\{\xi\}\cup\Xi) &= \mathit{append}(\xi,\mathit{appendall}(\Xi))
\end{aligned}$$

## 9.6 *Store deltas with timestamped store*

$$\begin{aligned}
\mathit{inject}(e) &= (S_0,\{\varsigma\},\mathit{replay}(\xi,\bot),0)\\
\text{where } \varsigma,\xi &= \Delta[\![e]\!]^{t}(\bot,\bot,\bot,\mathtt{halt})\\
S_0 &= \lambda\varsigma'.\begin{cases} 0 & \text{if } \varsigma' = \varsigma\\ \epsilon & \text{otherwise}\end{cases}\\
\mathit{reachable}(e) &= \{\mathit{wn}(\varsigma,\Sigma(n)) \mid \mathit{inject}(e) \longmapsto^* (S,F,\Sigma,n'),\mathit{hd}(S(\varsigma)) = n\}
\end{aligned}$$

$$\begin{aligned}
\mathit{System} &= (\mathit{State} \to \mathbb{N}^*) \times \wp(\mathit{State}) \times \mathit{Store}^* \times \mathbb{N}\\
(S,F,\Sigma,n) &\longmapsto (S',F\cup F',\sigma'{:}\Sigma,n')\\
\text{where } \sigma &= \mathit{hd}(\Sigma)\\
I &= \{(\varsigma',\xi) \mid \varsigma \in F,\varsigma \longmapsto_\sigma \varsigma',\xi\}\\
\sigma',\mathit{updated?} &= \forall\mathit{replay}\Delta(\{\xi \mid \exists\varsigma.(\varsigma,\xi) \in I\},\sigma,\mathtt{ff})\\
n' &= \begin{cases} n+1 & \text{if } \mathit{updated?}\\ n & \text{otherwise}\end{cases}\\
F' &= \{\varsigma \mid \exists\xi.(\varsigma,\xi) \in I, n' \neq \mathit{hd}(S(\varsigma))\}\\
S' &= \lambda\varsigma.\begin{cases} n'S(\varsigma) & \text{if } \varsigma \in F'\\ S(\varsigma) & \text{otherwise}\end{cases}
\end{aligned}$$

$$\begin{aligned}
\forall \mathit{replay}\Delta(\{\xi\} \cup \Xi, \sigma, \mathit{updated?}) &= \forall \mathit{replay}\Delta(\Xi, \sigma', \mathit{updated?} \vee \mathit{join?}) \\
\text{where } \sigma', \mathit{join?} &= \mathit{replay}\Delta(\xi, \sigma, \mathtt{ff}) \\
\forall \mathit{replay}\Delta(\varnothing, \sigma, \mathit{updated?}) &= \sigma, \mathit{updated?} \\
\mathit{replay}\Delta(\xi[a \mapsto S], \sigma, \mathit{join?}) &= \mathit{replay}\Delta(\xi, \sigma', \mathit{join?} \vee \mathit{join?}') \\
\text{where } S' &= S \sqcup \sigma(a) \\
\mathit{join?}' &= \sigma(a) \overset{?}{=} S' \\
\sigma' &= \sigma[a \mapsto S'] \\
\mathit{replay}\Delta(\bot, \sigma, \mathit{join?}) &= \sigma, \mathit{join?}
\end{aligned}$$

The *pop* function I hinted at for a memoizing $CESIK^*_t\Xi$ machine is defined here.

$$\begin{aligned}
\mathit{pop} &: \mathit{LKont} \times \widehat{\mathit{Kont}} \times \mathit{KStore} \times \mathit{Memo} \times \mathit{Relevant} \to \times \mathit{Memo}\wp(\mathit{LKont} \times \widehat{\mathit{Kont}}) \\
\mathit{pop}(\iota, \hat{\kappa}, \Xi, M, r) &= \mathit{pop}^*(\iota, \hat{\kappa}, \Xi, r, \emptyset)(M)
\end{aligned}$$

where $\mathit{pop}^*$ is written with a variant of the *State*[*Memo*] monad to monotonically grow M. Instead of *get* and *put*, we just have *join*:

$$\begin{aligned}
\mathit{State}[a, b] &= a \to a \times b \\
\mathit{return} &: b \to \mathit{State}[a, b] \\
\mathit{return}(S) &= \lambda M.\langle M, S\rangle \\
\mathit{bind} &: \mathit{State}[a, b] \times (b \to \mathit{State}[a, c]) \to \mathit{State}[a, c] \\
\mathit{bind}(s, f) &= \lambda M.f(b)(M') \\
&\qquad \text{where } \langle M', b\rangle = s(M) \\
\mathit{join} &: \mathit{Context} \times \mathit{Relevant} \to \mathit{State}[\mathit{Memo}, ()] \\
\mathit{join}(\tau, r, f) &= \lambda M.\langle M \sqcup [\tau \mapsto \{r\}], ()\rangle
\end{aligned}$$

Now we can nicely write $\mathit{pop}^*$:

$$\begin{aligned}
\mathit{pop}^*(\epsilon, \epsilon, \Xi, r, G) &= \mathit{return}(\emptyset) \\
\mathit{pop}^*(\phi{:}\iota, \hat{\kappa}, \Xi, r, G) &= \mathit{return}(\{(\phi, \iota, \hat{\kappa})\}) \\
\mathit{pop}^*(\epsilon, \tau, \Xi, r, G) &= \texttt{do}\ \ \mathit{join}(\tau, r) \\
&\qquad \mathit{foldM}(\lambda S, \tau'.\mathit{bind}(\mathit{pop}^*(\epsilon, \tau', \Xi, r, G \cup G'), \\
&\qquad\qquad \lambda S'.\mathit{return}(S \cup S')), \\
&\qquad\quad \{(\phi, \iota, \hat{\kappa}) \ : \ (\phi{:}\iota, \hat{\kappa}) \in \Xi(\tau)\}, \\
&\qquad\quad G') \\
&\text{where } G' = \{\tau' \ : \ (\epsilon, \tau') \in \Xi(\tau)\} \setminus G
\end{aligned}$$

The *foldM* is like a $\bigcup$ comprehension over the $\tau' \in G'$, but allows the M to flow through and grow.

And the argument evaluation rule (indeed any popping rule) uses it the following way:

$$\begin{aligned}
\langle v, \sigma, \iota, \hat{\kappa}\rangle, \Xi, M \longmapsto\ & \langle e, \rho', \sigma, \mathbf{appR}(v, \rho){:}\iota', \hat{\kappa}'\rangle, \Xi, M' \\
& \text{if } \mathbf{appL}(e, \rho'), \iota', \hat{\kappa}' \in K \\
& \text{where } \langle M', K\rangle = \mathit{pop}(\iota, \hat{\kappa}, \Xi, M, \langle v, \sigma\rangle)
\end{aligned}$$

## 10 CONTEXT CONGRUENCE WITH $inv_\Xi$

$$\frac{}{inv_\Xi(e_{pgm}, \bot)}$$

$$\frac{inv_\Xi(e_{pgm}, \Xi) \qquad \epsilon \in K \implies inject(e_{pgm}) \longmapsto^*_{CESK_t} extend(\tau, \epsilon) \quad \forall \phi{:}\hat{\kappa}_c \in K, \kappa.A(\tau, \kappa) \implies extend(\tau, \kappa) \longmapsto^*_{CESK_t} extend(\tau, \phi{:}\kappa)}{inv_\Xi(e_{pgm}, \Xi[\tau \mapsto K])}$$

# PROOFS FOR OAAM

## 10.1 *Soundness of lazy-nondeterminism*

$$\begin{aligned}
\alpha(\varsigma) &= \varsigma \\
\gamma(\mathsf{ev}^t(e,\rho,\sigma,\kappa)) &= \{\mathsf{ev}^t(e,\rho,\sigma,\kappa)\} \\
\gamma(\mathsf{co}\ (\kappa,v,\sigma)) &= \{\mathsf{co}\ (\kappa,v',\sigma) \mid v' \in \mathit{force}(\sigma,v)\} \\
\gamma(\mathsf{ap}^t_\ell(v,u,\sigma)) &= \{\mathsf{ap}^t_\ell(v',u',\sigma) \mid v' \in \mathit{force}(\sigma,v), u' \in \mathit{force}(\sigma,u)\} \\
\gamma(\mathsf{ans}\ (\sigma,v)) &= \{\mathsf{ans}\ (\sigma,v') \mid v' \in \mathit{force}(\sigma,v)\mathsf{sto},\mathit{force}(v)\} \\
\alpha^*(C) &= \{\alpha(\varsigma) \mid \varsigma \in C\} \\
\gamma^*(A) &= \bigcup_{\hat\varsigma \in A} \gamma(\hat\varsigma)
\end{aligned}$$

$$\frac{\mathbf{dom}(\sigma) = \mathbf{dom}(\sigma') \qquad \forall a.\sigma(a) \subseteq \sigma'(a)}{\sigma \sqsubseteq \sigma'}$$

$$\frac{\sigma \sqsubseteq \sigma'}{\mathsf{ev}^t(e,\rho,\sigma,\kappa) \sqsubseteq \mathsf{ev}^t(e,\rho,\sigma',\kappa)} \qquad \frac{\mathit{force}(\sigma,v) \subseteq \mathit{force}(\sigma',v') \qquad \sigma \sqsubseteq \sigma'}{\mathsf{co}\ (\kappa,v,\sigma) \sqsubseteq \mathsf{co}\ (\kappa,v',\sigma')}$$

$$\frac{\mathit{force}(\sigma,v) \subseteq \mathit{force}(\sigma',v') \qquad \mathit{force}(\sigma,u) \subseteq \mathit{force}(\sigma',u') \qquad \sigma \sqsubseteq \sigma'}{\mathsf{ap}^t_\ell(v,u,\sigma) \sqsubseteq \mathsf{ap}^t_\ell(v',u',\sigma')}$$

$$\frac{\mathit{force}(\sigma,v) \subseteq \mathit{force}(\sigma',v') \qquad \sigma \sqsubseteq \sigma'}{\mathsf{ans}\ (\sigma,v) \sqsubseteq \mathsf{ans}\ (\sigma',v')} \qquad \frac{\forall \hat\varsigma \in S, \exists \hat\varsigma' \in S'.\hat\varsigma \sqsubseteq \hat\varsigma'}{S \sqsubseteq S'}$$

By definitions of $\alpha,\gamma,\mathit{force}$, $\gamma^* \circ \alpha^* = 1_C$. Also $\alpha^* \circ \gamma^* \leqslant 1_A$ is straightforward to prove. This forms a Galois connection.

**THEOREM 3** If $\varsigma \longmapsto \varsigma'$ and $\alpha(\varsigma) \sqsubseteq \hat\varsigma$ then $\exists \hat\varsigma'.\hat\varsigma \longmapsto \hat\varsigma'$.

*Proof.* By assumption, $\gamma^*(\alpha^*(\{\varsigma\})) \sqsubseteq \gamma^*\{\hat\varsigma\}$. By the above property, and definition of $\gamma^*$, $\varsigma \in \gamma(\hat\varsigma)$. Since $\alpha$ does not introduce $\mathsf{addr}\ ()$ values, most cases follow by definition.

By cases on $\varsigma \longmapsto \varsigma'$:

**Case** $\mathsf{ev}\ (x\ ,\rho,\sigma,\kappa) \longmapsto \mathsf{co}\ (\kappa,v,\sigma)$**.**

where $v \in \sigma(\rho(x))$

By assumption, $\hat\varsigma \equiv \mathsf{ev}\ (x\ ,\rho,\sigma',\kappa)$ such that $\sigma \sqsubseteq \sigma'$.

(1) Let $\hat\varsigma' = \mathsf{co}\ (\kappa,\mathsf{addr}\ (\rho(x)),\sigma')$

(2) $\{\varsigma'\} \sqsubseteq \gamma(\hat\varsigma')$ — by def. $\gamma$

(3) $\alpha(\varsigma') \sqsubseteq \hat\varsigma'$ — by def. $\alpha$, $\alpha^* \circ \gamma^* \leqslant 1_A$

**Case** $\mathtt{ev}\ (\mathtt{lit}\ (l), \rho, \sigma, \kappa) \longmapsto \mathtt{co}\ (\kappa, l, \sigma)$.

By assumption, $\hat{\varsigma} \equiv \mathtt{ev}\ (x\ , \rho, \sigma', \kappa)$ such that $\sigma \sqsubseteq \sigma'$. Let $\hat{\varsigma}' = \mathtt{co}\ (\kappa, l, \sigma')$. Conclusion holds by definition of $\alpha, \sqsubseteq, \longmapsto$.

**Case** $\mathtt{ev}\ (\lambda\ \ x.\ e, \rho, \sigma, \kappa) \longmapsto \mathtt{co}\ (\kappa, \mathtt{clos}\ (x, e, \rho), \sigma)$.

By assumption, $\hat{\varsigma} \equiv \mathtt{ev}\ (x\ , \rho, \sigma', \kappa)$ such that $\sigma \sqsubseteq \sigma'$. Let $\hat{\varsigma}' = \mathtt{co}\ (\kappa, \mathtt{clos}\ (x, e, \rho), \sigma')$. Conclusion holds by definition of $\alpha, \sqsubseteq, \longmapsto$.

**Case** $\mathtt{ev}^t((e_0\ e_1)^\ell, \rho, \sigma, \kappa) \longmapsto \mathtt{ev}^t(e_0, \rho, \sigma', \mathbf{arg}^t_\ell(e_1, \rho, a))$.

where $a_\kappa = \mathit{allockont}^t\ell(\sigma, \kappa)\ \sigma' = \sigma \sqcup [a_\kappa \mapsto \{\kappa\}]$

By assumption, $\hat{\varsigma} \equiv \mathtt{ev}\ (x\ , \rho, \sigma^*, \kappa)$ such that $\sigma \sqsubseteq \sigma^*$. Let $\hat{\varsigma}' = \mathtt{ev}^t(e_0, \rho, \sigma^*_1, \mathbf{arg}^t_\ell(e_1, \rho, a))$ where $\sigma^*_1 = \sigma^*[a_\kappa \mapsto \{\kappa\}]$. Conclusion holds by definition of $\alpha, \sqsubseteq, \longmapsto$.

**Case** $\mathtt{ev}^t(\mathtt{if}^\ell(e_0, e_1, e_2), \rho, \sigma, \kappa) \longmapsto \mathtt{ev}^t(e_0, \rho, \sigma', \mathtt{ifk}^t(e_1, e_2, \rho, a))$.

where $a_\kappa = \mathit{allockont}^t\ell(\sigma, \kappa)\ \sigma' = \sigma \sqcup [a_\kappa \mapsto \{\kappa\}]$

By assumption, $\hat{\varsigma} \equiv \mathtt{ev}\ (\mathtt{if}^\ell(e_0, e_1, e_2), \rho, \sigma^*, \mathtt{ifk}^t(e_1, e_2, \rho, a))$ such that $\sigma \sqsubseteq \sigma^*$. Let $\hat{\varsigma}' = \mathtt{ev}^t(e_0, \rho, \sigma^*_1, \mathtt{ifk}^t(e_1, e_2, \rho, a))$ where $\sigma^*_1 = \sigma^*[a_\kappa \mapsto \{\kappa\}]$. Conclusion holds by definition of $\alpha, \sqsubseteq, \longmapsto$.

**Case** $\mathtt{co}\ (\mathtt{halt}, v, \sigma) \longmapsto \mathtt{ans}\ (\sigma, v)$.

By assumption, $\hat{\varsigma} \equiv \mathtt{co}\ (\mathtt{halt}, v', \sigma^*)$ where $\sigma \sqsubseteq \sigma^*$ and $\mathit{force}(\sigma, v) \subseteq \mathit{force}(\sigma^*, v')$.

By definition of $\mathit{force}, \sqsubseteq$, $v \in \mathit{force}(\sigma^*, v')$. Let $\hat{\varsigma}' = \mathtt{ans}\ (\sigma^*, v)$. Conclusion holds by definition of $\longmapsto, \mathit{force}$.

**Case** $\mathtt{co}\ (\mathbf{arg}^t_\ell(e, \rho, a), v, \sigma) \longmapsto \mathtt{ev}^t(e, \rho, \sigma', \mathbf{fun}^t_\ell(a_f, a))$.

where $a_f = \mathit{alloc}(\varsigma), \sigma' = \sigma \sqcup [a_f \mapsto \mathit{force}(\sigma, v)]$.

**Case** $\hat{\varsigma} \equiv \mathtt{co}\ (\mathbf{arg}^t_\ell(e, \rho, a), \mathtt{addr}\ (c), \sigma^*)$.

where $\sigma \sqsubseteq \sigma^*$

Let $\hat{\varsigma}' = \mathtt{ev}^t(e, \rho, \sigma^*, \mathbf{fun}^t_\ell(a_f, a))$ where $a_f = \mathit{alloc}(\varsigma), \sigma* = \sigma \sqcup [a_f \mapsto \mathtt{addr}\ (c)]$. Conclusion holds by definition of $\alpha, \sqsubseteq, \longmapsto$.

**Otherwise.**

$\hat{\varsigma} \equiv \mathtt{co}\ (\mathbf{arg}^t_\ell(e, \rho, a), v, \sigma^*)$ where $\sigma \sqsubseteq \sigma^*$ and $v \not\equiv \mathtt{addr}\ (c)$.

Let $\hat{\varsigma}' = \mathtt{ev}^t(e, \rho, \sigma^*_1, \mathbf{fun}^t_\ell(a_f, a))$ where $a_f = \mathit{alloc}(\varsigma), \sigma^*_1 = \sigma^* \sqcup [a_f \mapsto \mathit{force}(\sigma, v)]$ Conclusion holds by definition of $\alpha, \sqsubseteq, \longmapsto$.

**Case** $\mathtt{co}\ (\mathbf{fun}^t_\ell(a_f, a), v, \sigma) \longmapsto \mathtt{ap}^t_\ell(u, v, \kappa, \sigma)$.

where $\kappa \in \sigma(a), u \in \sigma(a_f)$

By assumption, $\hat{\varsigma} \equiv \mathtt{co}\ (\mathbf{fun}^t_\ell(a_f, a), v', \sigma^*)$ where $\sigma \sqsubseteq \sigma^*$ and $\mathit{force}(\sigma, v) \subseteq \mathit{force}(\sigma^*, v')$.

By definition of $\sqsubseteq$, *force*, $\kappa \in \sigma^*(a)$ and $u \in \sigma^*(a_f)$. Thus letting $\hat{\varsigma}' = \mathtt{ap}^t_\ell(u, v, \kappa, \sigma^*)$. Conclusion holds by definition of $\alpha, \longmapsto$.

**Case** $\mathtt{co}\ (\mathtt{ifk}^t(e_0, e_1, \rho, a), \mathtt{tt}, \sigma) \longmapsto \mathtt{ev}^t(e_0, \rho, \sigma, \kappa)$.

where $\kappa \in \sigma(a)$

By assumption, $\hat{\varsigma} \equiv \mathtt{co}\ (\mathtt{ifk}^t(e_0, e_1, \rho, a), \mathtt{tt}, \sigma^*)$ where $\sigma \sqsubseteq \sigma^*$.

Let $\hat{\varsigma}' = \mathtt{ev}^t(e_0, \rho, \sigma^*, \kappa)$. Conclusion holds by definition of $\alpha, \longmapsto$.

**Case** $\mathtt{co}\ (\mathtt{ifk}^t(e_0, e_1, \rho, a), \mathtt{ff}, \sigma) \longmapsto \mathtt{ev}^t(e_1, \rho, \sigma, \kappa)$.

where $\kappa \in \sigma(a)$

By assumption, $\hat{\varsigma} \equiv \mathtt{co}\ (\mathtt{ifk}^t(e_0, e_1, \rho, a), \mathtt{ff}, \sigma^*)$ where $\sigma \sqsubseteq \sigma^*$.

Let $\hat{\varsigma}' = \mathtt{ev}^t(e_1, \rho, \sigma^*, \kappa)$. Conclusion holds by definition of $\alpha, \longmapsto$.

**Case** $\mathtt{ap}^t_\ell(\mathtt{clos}\ (x, e, \rho), v, \sigma, \kappa) \longmapsto \mathtt{ev}^{t'}(e, \rho', \sigma', \kappa)$.

where $a = \mathit{alloc}(\varsigma), \rho' = \rho[x \mapsto a], \sigma' = \sigma \sqcup [a \mapsto \{v\}]$

By assumption, $\hat{\varsigma} \equiv \mathtt{ap}^t_\ell(\mathtt{clos}\ (x, e, \rho), v', \sigma^*, \kappa)$ where $\sigma \sqsubseteq \sigma^*$ and $\mathit{force}(\sigma, v) \subseteq \mathit{force}(\sigma^*, v')$.

Let $\hat{\varsigma}' = \mathtt{ev}^{t'}(e, \rho', \sigma^*_1, \kappa)$ where $\rho' = \rho[x \mapsto a], \sigma^*_1 = \sigma^* \sqcup [a \mapsto \mathit{force}(\sigma, v')]$. Conclusion holds by definition of $\alpha, \sqsubseteq, \longmapsto$.

**Case** $\mathtt{ap}^t_\ell(o, v, \sigma, \kappa) \longmapsto \mathtt{co}\ (\kappa, v', \sigma)$.

where $v' \in \Delta(o, v)$

By assumption, $\hat{\varsigma} \equiv \mathtt{ap}^t_\ell(o, v^*, \sigma^*, \kappa)$ where $\sigma \sqsubseteq \sigma^*$ and $\mathit{force}(\sigma, v) \subseteq \mathit{force}(\sigma^*, v^*)$. By definition of *force*, $v \in \mathit{force}(\sigma^*, v')$.

Let $\hat{\varsigma}' = \mathtt{co}\ (\kappa, v', \sigma^*)$. Conclusion holds by definition of $\alpha, \longmapsto$.

□

## 10.2 *Semantic equivalence with abstract compilation*

We show that in the presence of abstract compilation, even though there are fewer represented states in the reduction relation, that there is a bisimulation between the two. Particularly, the compiled semantics is a WEB refinement (defined in pages 57-64 of Manolios [59]) of the non-compiled semantics. We equate states that "commit" to non-$\mathtt{ev}$ () states.

To differentiate the two states spaces, denote the machine configuration space from the abstractly-compiled machine as $[\![\mathit{State}]\!]$. We additionally denote the reduction relation as $[\![\longmapsto]\!]$.

$$\begin{aligned}
commit(\mathtt{ev}^t(e,\rho,\sigma,\kappa)) &= commitev(t,\rho,\sigma,\kappa,e) \\
commit(\varsigma) &= \varsigma \quad \text{otherwise} \\
commitev(t,\rho,\sigma,\kappa,x\ ) &= \mathtt{co}\ (\kappa,\mathtt{addr}\ (\rho(x)),\sigma) \\
commitev(t,\rho,\sigma,\kappa,\mathtt{lit}\ (l)) &= \mathtt{co}\ (\kappa,\mathtt{lit}\ (l),\sigma) \\
commitev(t,\rho,\sigma,\kappa,\lambda\ \ x,e,\rho.\ ) &= \mathtt{co}\ (\kappa,\mathtt{clos}\ (x,e,\rho),\sigma) \\
commitev(t,\rho,\sigma,\kappa,(e_0\ e_1)^\ell) &= commitev(t,\rho,\sigma',\mathbf{arg}^t_\ell(e_1,\rho,a),e_0) \\
\text{where } a &= allockont^t\ell(\sigma,\kappa) \\
\sigma' &= \sigma \sqcup [a \mapsto \{\kappa\}] \\
commitev(t,\rho,\sigma,\kappa,\mathtt{if}^\ell(e_0,e_1,e_2)) &= commitev(t,\rho,\sigma',\mathtt{ifk}^t_\ell(e_1,e_2,\rho,a),e_0) \\
\text{where } a &= allockont^t\ell(\sigma,\kappa) \\
\sigma' &= \sigma \sqcup [a \mapsto \{\kappa\}]
\end{aligned}$$

Next, the refinement map from non-compiled to compiled states.

$$\begin{aligned}
r &: State \to [\![State]\!] \\
r(\mathtt{ev}^t(e,\rho,\sigma,\kappa)) &= r(commit(\mathtt{ev}^t(e,\rho,\sigma,\kappa))) \\
r(\mathtt{co}\ (\kappa,v,\sigma)) &= \mathtt{co}\ (r(\kappa),r(v),r(\sigma)) \\
r(\mathtt{ap}^t_\ell(u,v,\sigma,\kappa)) &= \mathtt{ap}^t_\ell(r(u),r(v),r(\sigma),r(\kappa)) \\
r(\mathtt{ans}\ (\sigma,v)) &= \mathtt{ans}\ (r(\sigma),r(v)) \\
r(\sigma) &= \lambda a.\{r(v) \mid v \in \sigma(a)\} \\
r(\mathtt{halt}) &= \mathtt{halt} \\
r(\mathbf{arg}^t_\ell(e,\rho,a)) &= \mathbf{arg}^t_\ell([\![e]\!],\rho,a) \\
r(\mathbf{fun}^t_\ell(a_f,a)) &= \mathbf{fun}^t_\ell(a_f,a) \\
r(\mathtt{ifk}^t(e_0,e_1,\rho,a)) &= \mathtt{ifk}^t([\![e_0]\!],[\![e_1]\!],\rho,a) \\
r(o) &= o \\
r(l) &= l \\
r(\mathtt{addr}\ (a)) &= \mathtt{addr}\ (a) \\
r(\mathtt{clos}\ (x,e,\rho)) &= \mathtt{clos}\ (x,[\![e]\!],\rho)
\end{aligned}$$

$r$ terminates because the non-structurally descreasing call case is guaranteed to not happen unless on a structurally smaller counterpart. This is because *commit* never returns an $\mathtt{ev}\ ()$ state.

Next we relate states across the different machines with an equivalence relation $B$ on the two state spaces $S = State \cup [\![State]\!]$, such that $\forall \varsigma \in State.sBr(s)$. Let $B$ be the reflexive, symmetric closure of $B^*$:

$$\frac{r(s) = s'}{sB^*s'}$$

Finally, we show that $B$ is a WEB on the transition system $\langle S, \Rightarrow \rangle$ where $\Rightarrow = \longmapsto \cup [\![\longmapsto]\!]$.

Let $\langle W, \lessdot \rangle$ be the well-ordered set of expressions (ordered structurally) with a bottom element $\bot$.

$$\begin{aligned} \mathit{erankt}(\mathtt{ev}^t(e, \rho, \sigma, \kappa)) &= e \\ \mathit{erankt}(\varsigma) &= \bot \quad \text{otherwise} \\ \mathit{erankl}(s, s') &= 0 \quad \mathit{Unnecessary} \end{aligned}$$

We need one lemma:

LEMMA 8 [COMPILE/COMMIT] For all $t, e, \rho, \sigma, \kappa$, $[\![e]\!]^t(\rho, r(\sigma), r(\kappa)) = r(\mathtt{ev}^t(e, \rho, \sigma, \kappa))$.

*Proof.* By induction on $e$.

**Case** Base: $x$ **.**

By definitions of $[\![_]\!], r, \mathit{commit}$.

**Case** Base: $\mathtt{lit}\ (l)$**.**

By definitions of $[\![_]\!], r, \mathit{commit}$.

**Case** Induction step: $\lambda\ \ x.\ e'$**.**

By definitions of $[\![_]\!], r, \mathit{commit}$.

**Case** Induction step: $(e_0\ e_1)^\ell$**.**

By IH, $[\![e_0]\!]^t(\rho, r(\sigma'), r(\kappa')) = r(\mathtt{ev}^t(e_0)\rho, \sigma', \kappa')$ where $\kappa' = \mathbf{arg}^t_\ell(e_1, \rho, a)$ and $a_\kappa = \mathit{allockont}^t\ell(\sigma, \kappa)$ $\sigma' = \sigma \sqcup [a_\kappa \mapsto \{\kappa\}]$ Thus holds by definitions of $r, \mathit{commit}, [\![_]\!]$.

**Case** Induction step: $\mathtt{if}\ (e_0, e_1, e_2)$**.**

By IH, $[\![e_0]\!]^t(\rho, r(\sigma'), r(\kappa')) = r(\mathtt{ev}^t(e_0)\rho, \sigma', \kappa')$ where $\kappa' = \mathtt{ifk}^t_\ell(e_1, e_2, \rho, a)$ and $a_\kappa = \mathit{allockont}^t\ell(\sigma, \kappa)$ $\sigma' = \sigma \sqcup [a_\kappa \mapsto \{\kappa\}]$ Thus holds by definitions of $r, \mathit{commit}, [\![_]\!]$.

□

THEOREM 9 $B$ is a WEB on the transition system $\langle S, \Rightarrow \rangle$.

*Proof.* Let $s, u, w \in S$ be arbitrary such that $sBw$ and $s \Rightarrow u$. If $w = s$, the first case of WEB trivially holds with witness $u$. We assume $w \neq s$. Thus $w = r(s)$. By cases on $s \Rightarrow u$:

**Case** $\mathtt{ev}\ (x\ , \rho, \sigma, \kappa) \longmapsto \mathtt{co}\ (\kappa, \mathtt{addr}\ (\rho(x)), \sigma)$**.**

Since $w = r(s)$, $w = r(u)$ by definition of $r$. The second case of WEB holds by definition of *erankt*, $\lessdot$ and case analysis on $w$.

**Case** $\mathtt{ev}\ (\mathtt{lit}\ (l), \rho, \sigma, \kappa) \longmapsto \mathtt{co}\ (\kappa, l, \sigma)$**.**

Since $w = r(s)$, $w = r(u)$ by definition of $r$. The second case of WEB holds by definition of *erankt*, $\lessdot$ and case analysis on $w$.

**Case** $\mathtt{ev}\ (\lambda\ \ x.\ e, \rho, \sigma, \kappa) \longmapsto \mathtt{co}\ (\kappa, \mathtt{clos}\ (x, e, \rho), \sigma)$.

Since $w = r(s)$, $w = r(u)$ by definition of $r$. The second case of WEB holds by definition of *erankt*, $\lessdot$ and case analysis on $w$.

**Case** $\mathtt{ev}^{t}((e_0\ e_1)^{\ell}, \rho, \sigma, \kappa) \longmapsto \mathtt{ev}^{t}(e_0, \rho, \sigma', \mathbf{arg}^{t}_{\ell}(e_1, \rho, a))$.

where $a = \mathit{allockont}^{t}\ell(\sigma, \kappa)\ \sigma' = \sigma \sqcup [a \mapsto \{\kappa\}]$

By definition of *commit*, $r(u) = w$, thus $uBw$. By definition of $\lessdot$, $\mathit{erankt}(u) < \mathit{erankt}(s)$. Thus the second case of WEB holds.

**Case** $\mathtt{ev}^{t}(\mathtt{if}^{\ell}(e_0, e_1, e_2), \rho, \sigma, \kappa) \longmapsto \mathtt{ev}^{t}(e_0, \rho, \sigma', \mathtt{ifk}^{t}(e_1, e_2, \rho, a))$.

where $a = \mathit{allockont}^{t}\ell(\sigma, \kappa)\ \sigma' = \sigma \sqcup [a \mapsto \{\kappa\}]$

By definition of *commit*, $r(u) = r(s) = w$. By definition of $\lessdot$, $\mathit{erankt}(u) \lessdot \mathit{erankt}(s)$. Thus the second case of WEB holds.

**Case** $\mathtt{co}\ (\mathtt{halt}, v, \sigma) \longmapsto \mathtt{ans}\ (\sigma, u)$.

where $u \in \mathit{force}(\sigma, v)$

By definition of $\Rightarrow, [\![\longmapsto]\!]$, $w \Rightarrow r(u)$, satisfying the first case of WEB.

**Case** $\mathtt{co}\ (\mathbf{arg}^{t}_{\ell}(e, \rho, a), v, \sigma) \longmapsto \mathtt{ev}^{t}(e, \rho, \sigma', \mathbf{fun}^{t}_{\ell}(a_f, a))$.

where $a_f = \mathit{alloc}(\varsigma), \sigma' = \sigma \sqcup [a_f \mapsto \mathit{force}(\sigma, v)$

By definition of $\Rightarrow, [\![\longmapsto]\!]$, $w \Rightarrow [\![e]\!]^{t}(\rho, r(\sigma'), r(\mathbf{fun}^{t}_{\ell}(a_f, a)))$. By the compile/commit lemma, $w \Rightarrow r(u)$. Thus the first case of WEB holds with witness $r(u)$.

**Case** $\mathtt{co}\ (\mathbf{fun}^{t}_{\ell}(a_f, a), v, \sigma) \longmapsto \mathtt{ap}^{t}_{\ell}(u, v, \kappa, \sigma)$.

where $\kappa \in \sigma(a), u \in \sigma(a_f)$

By definition of $[\![\longmapsto]\!]$, $w\ [\![\longmapsto]\!]\ r(u)$, satisfying the first case of WEB.

**Case** $\mathtt{co}\ (\mathtt{ifk}^{t}(e_0, e_1, \rho, a), \mathtt{tt}, \sigma) \longmapsto \mathtt{ev}^{t}(e_0, \rho, \sigma, \kappa)$.

where $\kappa \in \sigma(a)$

By definition of $\Rightarrow, [\![\longmapsto]\!]$, $w \Rightarrow [\![e_0]\!]^{t}(\rho, r(\sigma), r(\kappa))$. By the compile/commit lemma, $w \Rightarrow r(u)$. Thus the first case of WEB holds with witness $r(u)$.

**Case** $\mathtt{co}\ (\mathtt{ifk}^{t}(e_0, e_1, \rho, a), \mathtt{ff}, \sigma) \longmapsto \mathtt{ev}^{t}(e_1, \rho, \sigma, \kappa)$.

where $\kappa \in \sigma(a)$

By definition of $\Rightarrow, [\![\longmapsto]\!]$, $w \Rightarrow [\![e_0]\!]^{t}(\rho, r(\sigma), r(\kappa))$. By the compile/commit lemma, $w \Rightarrow r(u)$. Thus the first case of WEB holds with witness $r(u)$.

**Case** $\mathtt{ap}^{t}_{\ell}(\mathtt{clos}\ (x, e, \rho), v, \sigma, \kappa) \longmapsto \mathtt{ev}^{t'}(e, \rho', \sigma', \kappa)$.

where $a = \mathit{alloc}(\varsigma), \rho' = \rho[x \mapsto a], \sigma' = \sigma \sqcup [a \mapsto \mathit{force}(\sigma, v)]$

By definition of $\Rightarrow, [\![\longmapsto]\!], w \Rightarrow [\![e]\!]^{t'}(\rho', r(\sigma'), r(\kappa))$. By the compile/commit lemma, $w \Rightarrow r(u)$. Thus the first case of WEB holds with witness $r(u)$.

**Case** $\mathsf{ap}^t_\ell(o, v, \sigma, \kappa) \longmapsto \mathsf{co}\ (\kappa, v', \sigma)$.

where $u \in \mathit{force}(\sigma, v), v' \in \Delta(o, u)$

By definition of $[\![\longmapsto]\!]$, $w\ [\![\longmapsto]\!]\ r(u)$, satisfying the first case of WEB.

**Case** $s\ [\![\longmapsto]\!]\ u$.

Must be the case that $s = w$, thus the first case of WEB holds.

□

## 10.3 *Soundness of widened abstract compilation*

$$\mathit{prep}(S, F, \hat{\sigma}) = S \cup \{(\hat{\varsigma}, \hat{\sigma}) \mid \hat{\varsigma} \in F\}$$

$$\frac{\hat{\sigma} \sqsubseteq \hat{\sigma}'}{(\hat{\varsigma}, \hat{\sigma}) \sqsubseteq (\hat{\varsigma}, \hat{\sigma}')} \qquad \frac{\mathit{prep}(S, F, \hat{\sigma}) \sqsubseteq \mathit{prep}(S', F', \hat{\sigma}')}{(S, F, \hat{\sigma}) \sqsubseteq (S', F', \hat{\sigma}')}$$

$$\begin{aligned}
\alpha(\varsigma) &= (\{(\hat{\varsigma}, \hat{\sigma})\}, \{\hat{\varsigma}\}, \hat{\sigma}) \\
\text{where } \hat{\varsigma}, \hat{\sigma} &= \mathit{nw}(\varsigma) \\
\gamma((S, F, \hat{\sigma})) &= \{\mathit{wn}(\hat{\varsigma}, \sigma) \mid (\hat{\varsigma}, \hat{\sigma}') \in \mathit{prep}(S, F, \hat{\sigma}), \sigma \in \gamma(\hat{\sigma}')\} \\
\gamma(\hat{\sigma}) &= \{R' \mid R' \subseteq R, R' \text{ functional}, \mathbf{dom}(R') = \mathbf{dom}(\hat{\sigma})\} \\
\text{where } R &= \{(a, \hat{s}) \mid a \in \mathbf{dom}(\hat{\sigma}), \hat{s} \subseteq \hat{\sigma}(a)\} \\
\alpha^*(C) &= \{\alpha(\varsigma) \mid \varsigma \in C\} \\
\gamma^*(A) &= \bigcup_{\hat{\varsigma} \in A} \gamma(\hat{\varsigma})
\end{aligned}$$

**Lemma 41.** $\gamma^* \circ \alpha^* \geqslant 1_C$

*Proof.* This is immediate if $\hat{\sigma} \in \gamma(\hat{\sigma})$, which is true, since $\hat{\sigma}(a) \subseteq \hat{\sigma}(a)$. □

**Lemma 42.** $\alpha^* \circ \gamma^* \leqslant 1_A$

*Proof.* Let $A \subseteq \mathit{System}$ be arbitrary. It suffices to show that $\alpha^*(\gamma^*(A)) \sqsubseteq A$. Let $(S, F, \hat{\sigma}) \in A$ be arbitary. Let $(\hat{\varsigma}, \hat{\sigma}) \in S$ be arbitrary. By definition of $\gamma$, $\hat{\sigma} \in \gamma(\hat{\sigma})$. Thus $\mathit{wn}(\hat{\varsigma}, \hat{\sigma}) \in \gamma^*(A)$. By definition of $\alpha^*, \alpha$, $(\{(\hat{\varsigma}, \hat{\sigma})\}, \{\hat{\varsigma}\}, \hat{\sigma}) \in \alpha^*(\gamma^*(A))$. By definition of *prep*, $\sqsubseteq$, $(\{(\hat{\varsigma}, \hat{\sigma})\}, \{\hat{\varsigma}\}, \hat{\sigma}) \sqsubseteq (S, F, \hat{\sigma})$. Thus since $(S, F, \hat{\sigma})$ was arbitrary, $\alpha^*(\gamma^*(A)) \sqsubseteq A$. □

**Theorem 43.** *If* $\varsigma\ [\![\longmapsto]\!]\ \varsigma'$ *and* $\alpha(\varsigma) \sqsubseteq (S, F, \hat{\sigma})$ *then there exist* $S', F', \hat{\sigma}'$ *such that* $(S, F, \hat{\sigma}) \longmapsto (S', F', \hat{\sigma}')$ *and* $\alpha(\varsigma') \sqsubseteq (S', F', \hat{\sigma}')$

*Proof.* By definition of $\alpha, \sqsubseteq$, there exists a $\hat{\varsigma} \in F$ such that, with $\hat{\varsigma}^*, \hat{\sigma}^* = nw(\varsigma)$, $\hat{\varsigma} = \hat{\varsigma}^*$ and $\hat{\sigma}^* \sqsubseteq \hat{\sigma}$.

Thus, by definition of $\longmapsto, nw, wn, prep, \sqsubseteq$ (call the result $(S', F', \hat{\sigma}')$), letting $\hat{\varsigma}_1^*, \hat{\sigma}_1^* = nw(\varsigma')$, $\hat{\sigma}_1^* \sqsubseteq \hat{\sigma}'$ and $\{(\hat{\varsigma}_1^*, \hat{\sigma}')\} \sqsubseteq prep(S', F', \hat{\sigma}')$. Thus $\alpha(\varsigma') \sqsubseteq (S', F', \hat{\sigma}')$. □

## 10.4 *Semantic equivalence with locally log-based store deltas*

Here we show extensional equality of the relations. We will use $\longmapsto_{\sigma\xi}$ for the store delta semantics and $\Delta[\![_]\!]$ for its compilation function.

Let $\equiv \subseteq Expr$ be the reflexive, transitive, symmetric closure of $\equiv^*$ with structural lifting where non-*Expr* elements are compared with equality, and that lifted to functions $Addr \to \wp(Storeable)$.

$$\frac{}{\Delta[\![e]\!] \equiv^* [\![e]\!]}$$

**Lemma 44** (Compile store independence)**.** *Let* $[\![e]\!]^t(\rho, \sigma, \kappa) = wn(\hat{\varsigma}, \sigma')$. $\exists \xi. \sigma' = replay(\xi, \sigma)$.

*Proof.* By induction on $e$

**Case** Base $x$ **.**

Witness $\epsilon$

**Case** Base `lit` $(l)$**.**

Witness $\epsilon$

**Case** Induction step $\lambda\ x.\ e$**.**

Witness $\epsilon$

**Case** Induction step $(e_0\ e_1)$ **.**

Let $a = allockont_\ell^t(\sigma, \kappa)$. Let $\sigma'' = \sigma \sqcup [a \mapsto \{\kappa\}]$. Let $\kappa' = \mathbf{arg}_\ell^t([\![e_1]\!], \rho, a)$. By IH with $e_0, t', \sigma'', \kappa'$, $\exists \xi. \sigma' = replay(\xi, \sigma'')$. Thus the witness is $(a, \{\kappa\}){:}\xi$ by definitions of $[\![_]\!]$, *replay*.

**Case** Induction step `if` $(e_0, e_1, e_2)$**.**

Let $a = allockont_\ell^t(\sigma, \kappa)$. Let $\sigma'' = \sigma \sqcup [a \mapsto \{\kappa\}]$. Let $\kappa' = \mathtt{ifk}^t([\![e_1]\!], [\![e_2]\!], \rho, a)$. By IH with $e_0, t', \sigma'', \kappa', \exists \xi. \sigma' = replay(\xi, \sigma'')$. Thus the witness is $(a, \{\kappa\}){:}\xi$ by definitions of $[\![_]\!]$, *replay*.

□

We need an additional property on *allockont* such that if $\sigma \equiv \sigma'$ and $\kappa \equiv \kappa'$, then $allockont_\ell^t(\sigma, \kappa) = allockont_\ell^t(\sigma', \kappa')$, which is a very reasonable assumption.

**Lemma 45** (*replay* and *append*)**.** $replay(\xi, replay(\xi', \sigma)) = replay(append(\xi, \xi'), \sigma)$

*Proof.* By induction on $\xi$. □

**Lemma 46** (Compile coherence). *For all* $e, t, \rho, \sigma, \xi, \kappa, \sigma^*, \kappa^*, \xi^*$,
*let* $(\hat{\varsigma}, \xi') = \Delta[\![e]\!]^t(\rho, \sigma, \xi, \kappa)$,
*let* $\mathit{nw}(\hat{\varsigma}', \sigma^{*\prime}) = [\![e]\!]^t(\rho, \sigma^*, \kappa^*)$.
*If* $\sigma \equiv \sigma^*$, $\xi \equiv \xi^*$, *and* $\kappa \equiv \kappa^*$ *then* $\hat{\varsigma}' \equiv \hat{\varsigma}$ *and there exists an* $\xi''$ *such that* $\mathit{replay}(\xi'', \mathit{replay}(\xi^*, \sigma^*)) \equiv \mathit{replay}(\xi', \sigma)$.

*Proof.* By induction on $e$

**Case** Base $x$ **.**

By definitions of $\Delta[\![_]\!], [\![_]\!], \mathit{replay}, \mathit{nw}$, witness is $\epsilon$.

**Case** Base $\mathtt{lit}\ (l)$**.**

By definitions of $\Delta[\![_]\!], [\![_]\!], \mathit{replay}, \mathit{nw}$, witness is $\epsilon$.

**Case** Induction step $\lambda\ \ x.\ e$**.**

By definitions of $\Delta[\![_]\!], [\![_]\!], \mathit{replay}, \mathit{nw}$, witness is $\epsilon$.

**Case** Induction step $(e_0\ e_1)$ **.**

Let $\kappa' = \mathbf{arg}^t_\ell(\Delta[\![e_1]\!], \rho, a)$ By definition of $\Delta[\![_]\!]$, Let $\Delta[\![e_0]\!]^{t'}(\rho, \sigma, \xi'_1, \kappa') = (\hat{\varsigma}, \xi')$ where $a = \mathit{allockont}^t\ell(\sigma, \kappa)$ $\xi'_1 = (a, \{\kappa\}){:}\xi$ Let $\kappa^{*\prime} = \mathbf{arg}^t_\ell([\![e_1]\!], \rho, a)$. By definition of $[\![_]\!]$, $[\![e_0]\!]^{t'}(\rho, \sigma^*_1, \kappa^{*\prime}) = \mathit{wn}(\hat{\varsigma}', \sigma^{*\prime})$ where $\sigma^*_1 = \sigma^* \sqcup [a \mapsto \{\kappa^*\}]$.

By lemma 44, there exists a $\xi^*_1$ such that $\sigma^{*\prime} = \mathit{replay}(\xi^*_1, \sigma^*_1)$

Let $\xi^{*\prime} = (a, \{\kappa^*\}){:}\xi^*_1$. By definition of *replay*, $\sigma^{*\prime} = \mathit{replay}(\xi^{*\prime}, \sigma^*)$.

By IH, with $e_0, t', \rho, \sigma, \xi'_1, \kappa', \sigma^*, \kappa^{*\prime}, \xi^{*\prime}$, $\hat{\varsigma} \equiv \hat{\varsigma}'$ and there exists an $\xi''_1$ such that $\mathit{replay}(\xi'', \mathit{replay}(\xi^{*\prime}, \sigma^*)) \equiv \mathit{replay}(\xi', \sigma)$. Thus the witness is $\mathit{append}(\xi''_1, (a, \{\kappa^*\}){:}\epsilon)$ by lemma 45 and associativity of *append*.

**Case** Induction step $\mathtt{if}\ (e_0, e_1, e_2)$**.**

Similar to above case.

□

**Theorem 47.** *If* $S \equiv S^*, F \equiv F^*, \sigma \equiv \sigma^*$, *then* $(S, F, \sigma) \longmapsto (S', F', \sigma')$ *iff* $\exists S^*_1, F^*_1, \sigma^*_1 . S' \equiv S^*_1 \wedge F' \equiv F^*_1 \wedge \sigma' \equiv \sigma^*_1 \wedge (S^*, F^*, \sigma^*) \longmapsto_{\sigma\xi} (S^*_1, F^*_1, \sigma^*_1)$

*Proof.* By definitions of $\longmapsto, \longmapsto_{\sigma\xi}$, *replay*, *appendall*, commutativity and associativity of $\sqcup$, and the previous lemma. □

## 10.5 *Semantic equivalence of log-based updates to a timestamped store*

Because the store is monotonically increasing, we know that $\sqsubseteq$ forms a total order on stores in the system. We use this information to sort and index the stores. Call the timestamped reduction relation $\longmapsto_n$.

$$\begin{aligned}
\alpha((S,F,\sigma)) &= (\alpha(S,\Sigma),F,\Sigma,|\Sigma|-1) \\
\text{where } \Sigma &= \lambda n.\mathit{sorted}_n \\
\mathit{sorted} &= \mathit{sort}(\{\sigma \mid (_,\sigma) \in \mathit{prep}(S,F,\sigma)\},\sqsubseteq) \\
\alpha(S,\Sigma) &= \{(\varsigma, \mathit{map}((\lambda\sigma.|\Sigma| - \mathit{indexof}(\Sigma,\sigma) - 1), \\
&\qquad \mathit{sort}(\{\sigma \mid (\varsigma,\sigma) \in S\},\sqsupseteq))) \mid (\varsigma,_,\in)S\} \\
\gamma((S,F,\Sigma,n)) &= (\gamma(S,\Sigma),F,\mathit{hd}(\Sigma)) \\
\gamma(S,\Sigma) &= \{(\varsigma,\Sigma(\ell_i)) \mid S(\varsigma) = \ell, 0 \leqslant i < |\ell|\}
\end{aligned}$$

$$\begin{aligned}
\sigma_\Xi \in \mathit{Store}_\Xi &= \mathit{Addr} \to \mathit{ValStack} \\
V \in \mathit{ValStack} &= (\mathit{Time} \times \wp(\mathit{Storeable}))^* \\
\alpha_\sigma(\epsilon) &= \alpha_\sigma(\lambda a.\varnothing) = \lambda a.\epsilon \\
\alpha_\sigma(\sigma) &= \lambda a.(0,\sigma(a)) \\
\alpha_\sigma(\sigma\sigma'\ldots) &= \mathit{merge}(\sigma,\alpha_\sigma(\sigma'\ldots)) \\
\mathit{merge}(\sigma,\sigma_\Xi) &= \lambda a.\begin{cases} (t+1, vs \sqcup \sigma(a)){:}\sigma_\Xi(a) & \text{if } vs \neq \sigma(a) \\ \sigma_\Xi(a) & \text{otherwise} \end{cases} \\
\text{where } t &= \mathit{size}(\sigma_\Xi) \\
vs &= \pi_1(\mathit{hd}(\sigma_\Xi(a))) \\
\mathit{size}(\bot) &= -1 \\
\mathit{size}(\sigma_\Xi) &= \max\{t \mid \sigma_\Xi(a) \equiv (t,vs){:}V\} \\
\gamma_\sigma(\sigma_\Xi) &= \mathit{snapshot}(\sigma_\Xi,n)\ldots_n \text{ where } n = \mathit{size}(\sigma_\Xi) \text{ down to } 0 \\
\mathit{snapshot}(\sigma_\Xi,n) &= \lambda a.\mathit{firstunder}(\sigma_\Xi(a),n) \\
\mathit{firstunder}(\epsilon,n) &= \varnothing \\
\mathit{firstunder}((t,vs){:}V,n) &= \begin{cases} vs & \text{if } t \leqslant n \\ \mathit{firstunder}(V,n) & \text{otherwise} \end{cases}
\end{aligned}$$

Of course we rely on the timestamps being sequential from 0 to $n$ (including each number in the range), so we add a well-formedness condition on $\mathit{Store}_\Xi$:

$$\mathit{wf}_\Xi(\sigma_\Xi) = (\forall a.\mathit{ordered}(\sigma_\Xi(a))) \wedge \forall 0 \leqslant i \leqslant \mathit{size}(\sigma_\Xi).\exists a,j.\pi_0(\sigma_\Xi(a)_j) = i$$

$$\frac{}{\mathit{ordered}(\epsilon)} \qquad \frac{vs \neq \varnothing}{\mathit{ordered}((t,vs))}$$

$$\frac{\mathit{ordered}((t,vs){:}V) \qquad t' > t \qquad vs' \sqsupset vs}{\mathit{ordered}((t',vs'){:}(t,vs){:}V)}$$

Additionally, the stack of stores we deal with must be in order, different, and greatest to least.

$$\frac{}{\mathit{wf}(\epsilon)} \qquad \frac{}{\mathit{wf}(\sigma)} \qquad \frac{\mathit{wf}(\sigma{:}\Sigma) \qquad \sigma' \sqsupset \sigma}{\mathit{wf}(\sigma'{:}\sigma{:}\Sigma)}$$

**Lemma 48** (Snapshot order)**.** *If* $wf_{\Xi}(\sigma_{\Xi})$ *then for all* $1 \leqslant n \leqslant size(\sigma_{\Xi})$, $snapshot(\sigma_{\Xi}, n) \sqsupset snapshot(\sigma_{\Xi}, n-1)$.

*Proof.* By cases on $n$.

**Case** $0$.

Vacuously true

**Case** $n+1$.

Let $a, j$ be the witnesses of the second well-formedness condition with $i = n$. By definition of *ordered*, either $j = |\sigma_{\Xi}(a)| - 1$ and thus $\pi_1(\sigma_{\Xi}(a)_0) \sqsupset snapshot(\sigma_{\Xi}, n-1)(a) = \varnothing$, or the third rule of *ordered* applies and $\sqsupset$ holds outright.

□

**Lemma 49.** *If* $wf_{\Xi}(\sigma_{\Xi})$ *and* $\sigma \sqsupseteq snapshot(\sigma_{\Xi}, size(\sigma_{\Xi}))$ *then* $\mathrm{wf}_{\Xi}(merge(\sigma, \sigma_{\Xi}))$

**Lemma 50** (Wellformedness (a))**.** $\gamma_\sigma : \{\sigma_{\Xi} \mid \sigma_{\Xi} \in Store_{\Xi}, wf_{\Xi}(\sigma_{\Xi})\} \to \{\Sigma \mid \Sigma \in Store^*, wf(\Sigma)\}$

*Proof.* Let $\sigma_{\Xi}$ be arbitrary such that $wf_{\Xi}(\sigma_{\Xi})$. By lemma 48, and straightforward induction on $size(\sigma_{\Xi})$. □

**Lemma 51** (Wellformedness (b))**.**

$$\alpha_\sigma : \{\Sigma \in Store^* \;:\; wf(\Sigma)\} \to \{\sigma_{\Xi} \in Store_{\Xi} \;:\; wf_{\Xi}(\sigma_{\Xi})\}$$

*Proof.* Let $\Sigma$ be arbitrary such that $wf(\Sigma)$. By induction on $\Sigma$:

**Case** Base $\epsilon$ or $\lambda a.\varnothing$.

Vacuously true by definitions of $\alpha_\sigma$, *ordered*, *size* and $wf_{\Xi}$.

**Case** Base $\sigma \sqsupset \lambda a.\varnothing$.

By definitons of $\alpha_\sigma$, *ordered*, *size* for first condition. Second condition witnesses are $a$ such that $\sigma(a) \neq \varnothing$ (exists by assumption) and 0.

**Case** Induction step $\sigma{:}\sigma'{:}\Sigma'$ such that $\sigma \sqsupset \sigma'$.

By IH, $wf_{\Xi}(\alpha_\sigma(\sigma'{:}\Sigma'))$. Let $a$ be arbitrary.

**Case** $a$ is such that $\sigma(a) \sqsupset \sigma'(a)$.

Let $\sigma'_{\Xi} = \alpha_\sigma(\sigma'{:}\Sigma)$. By definitions of $\alpha_\sigma$, *merge*, $\alpha_\sigma(\Sigma)(a) = (t+1, \sigma(a)){:}\sigma'_{\Xi}$, where $t = size(\sigma'_{\Xi})$. By definition of *ordered* and $wf_{X_i}$, $ordered(\alpha_\sigma(\Sigma)(a))$. For the second condition, $i \leqslant t$ is handled by IH. Otherwise, the witnesses are $a$ and 0 (and $a$ must exist by assumption).

**Otherwise.**

First condition holds by IH. Second by previous reasoning.

□

$$\begin{aligned}
&\mathit{prop}(S, \sigma) = \mathit{totally\text{-}ordered}(\Sigma, \sqsubseteq) \wedge \sigma \text{ an upper bound of } \Sigma \\
&\quad \text{where } \Sigma = \{\sigma \mid (_, \sigma) \in S\} \\
&\mathit{prop}^*(S, \Sigma, n) = (\forall \varsigma, i.0 \leqslant i < |S(\varsigma)| \implies 0 \leqslant \Sigma(\varsigma)_i < |\Sigma|) \wedge \\
&\qquad (\forall i.0 \leqslant i < |\Sigma| - 1 \implies \Sigma(i) \sqsubseteq \Sigma(i+1)) \wedge \\
&\qquad n = |\Sigma| - 1
\end{aligned}$$

**Lemma 52** (Monotone store collection)**.** *If* $\mathit{prop}(S, \sigma)$ *and* $(S, F, \sigma) \longmapsto (S', F', \mathit{msto}')$ *then* $\mathit{prop}(S', \sigma')$

*Proof.* Since $\forall \Xi, \sigma. \forall \mathit{replay}(\Xi, \sigma) \sqsupseteq \sigma$, this is trivial. □

**Lemma 53** (Monotone store timestamps)**.** *If* $\mathit{prop}^*(S, \Sigma, n)$ *and* $(S, F, \Sigma, n) \longmapsto (S', F', \Sigma', n')$ *then* $\mathit{prop}^*(S', \Sigma', n')$

*Proof.* Since $\forall \Xi, \sigma.\text{let } \sigma', \mathit{updated?} = \forall \mathit{replay}\Delta(\Xi, \sigma) \text{ in } \sigma' \sqsupseteq \sigma$, this is trivial. □

**Lemma 54** (Change is change)**.** *For all* $\xi, \sigma, \mathit{join?}$, *let* $\langle \sigma', \mathit{join?}' \rangle = \mathit{replay}\Delta(\xi, \sigma, \mathit{join?})$ *for* $\mathit{join?} \vee (\mathit{join?}' \iff \sigma \neq \sigma')$ *and* $\sigma' = \mathit{replay}(\xi, \sigma)$.

*Proof.* By induction on $\xi$.

**Case** Base $\perp$**.**

By definition of $\mathit{replay}\Delta, \mathit{replay}$, $\sigma = \sigma' = \mathit{replay}(\perp, \sigma)$ and $\mathit{join?} = \mathit{join?}'$.

**Case** Induction step $\xi'[a \mapsto \hat{s}]$**.**

Let $\mathit{replay}\Delta(\xi', \sigma^*, \mathit{join?} \vee \mathit{join?}^*) = \sigma^*_1, \mathit{join?}^*_1$ where $\hat{s}' = \hat{s} \sqcup \sigma(a)$, $\sigma^* = \sigma[a \mapsto \hat{s}']$, $\mathit{join?}^* = \hat{s}' \stackrel{?}{=} \sigma(a)$. If $\mathit{join?}^*$, then $\sigma^*_1 \neq \sigma$ and $\mathit{join?}^*_1$ because $\mathit{replay}\Delta$ monotonically increases $\sigma$ and $\mathit{join?}$. Otherwise, by IH, if $\mathit{join?}$, then $\mathit{join?}^*_1$; otherwise, $\sigma^*_1 \neq \sigma^* \iff \mathit{join?}^*_1$. Also by IH, $\sigma^*_1 = \mathit{replay}(\xi', \sigma^*)$. Thus by definition of $\mathit{replay}$, $\sigma' = \mathit{replay}(\xi, \sigma)$.

□

**Lemma 55** (Change all is change)**.** *For all* $\Xi$ *finite*, $\sigma, \mathit{updated?}$, *let* $\langle \sigma', \mathit{updated?}' \rangle = \forall \mathit{replay}\Delta(\Xi, \sigma, \mathit{updated?})$ *for* $\mathit{updated?} \vee (\mathit{updated?}' \iff \sigma \neq \sigma')$ *and* $\sigma' = \forall \mathit{replay}(\Xi, \sigma)$

*Proof.* By induction on $\Xi$.

**Case** Base $\varnothing$**.**

By definition of $\mathit{replay}\Delta$, $\mathit{replay}$, $\sigma = \sigma' = \mathit{replay}(\varnothing, \sigma)$ and $\mathit{updated?} = \mathit{updated?}'$.

**Case** Induction step $\{\xi\} \cup \Xi'$**.**

Let $\forall\mathit{replay}\Delta(\Xi', \sigma^*, \mathit{updated?} \vee \mathit{updated?}^*) = \sigma_1^*, \mathit{updated?}_1^*$ where $\mathit{replay}\Delta(\xi, \sigma, \mathit{updated?}) = \sigma^*, \mathit{updated?}^*$. By the previous lemma, $\mathit{updated?} \vee (\mathit{updated?}^* \iff \sigma \neq \sigma^*)$ and $\sigma^* = \mathit{replay}(\xi, \sigma)$.. If $\mathit{updated?}^*$ then $\sigma \neq \mathit{msto}_1^*$ and $\mathit{updated?}_1^*$ because $\forall\mathit{replay}\Delta$ monotonically increases $\sigma$ and $\mathit{updated?}$. Otherwise, by IH, if $\mathit{updated?}$, then $\mathit{updated?}_1^*$; otherwise $\sigma_1^* \neq \sigma^* \iff \mathit{updated?}_1^*$. Also by IH, $\sigma_1^* = \forall\mathit{replay}(\Xi', \sigma^*)$. Thus, by definition of $\forall\mathit{replay}$, $\sigma' = \forall\mathit{replay}(\Xi, \sigma)$.

□

**Theorem 56.** *If* $(S, F, \sigma) \longmapsto (S', F', \sigma')$ *and* $\mathit{prop}(S, \sigma)$ *and* $\alpha((S, F, \sigma)) \sqsubseteq (S^*, F, \Sigma, n)$ *then there exist* $S_1^*, \Sigma', n'$ *such that* $(S^*, F, \Sigma, n) \longmapsto_n (S_1^*, F', \Sigma', n')$ *and* $\alpha((S', F', \sigma')) \sqsubseteq (S_1^*, F', \Sigma', n')$.

*Proof.* By definition of $\alpha$, and lemma 52, $\Sigma(n) = \sigma$. By definitions of $\longmapsto_n$, $\longmapsto$, $\alpha$ and the previous lemma, $\sigma' = \Sigma'(n')$. By definitions of $\longmapsto_n$, $\longmapsto$, $\alpha$, and the previous statement, $\alpha(S') = S_1^*$. By definition of $\alpha$ and lemma 52, $(S^*, F, \Sigma, n) \longmapsto_n (S_1^*, F', \Sigma', n')$ and $\alpha((S', F', \sigma')) \sqsubseteq (S_1^*, F', \Sigma', n')$. □

**Theorem 57.** *If* $(S, F, \Sigma, n) \longmapsto_n (S', F', \Sigma', n')$ *and* $\gamma((S, F, \Sigma, n)) \sqsubseteq (S^*, F, \sigma)$ *then there exist* $S_1^*, \sigma'$ *such that* $(S^*, F, \sigma) \longmapsto (S_1^*, F', \sigma')$ *and* $\gamma((S', F', \Sigma', n')) \sqsubseteq (S_1^*, F', \sigma')$.

*Proof.* By definition of $\gamma$, and lemma 53, $\Sigma(n) = \sigma$. By definitions of $\longmapsto_n$, $\longmapsto$, $\gamma$ and the previous lemma, $\sigma' = \Sigma'(n')$. By definitions of $\longmapsto_n$, $\longmapsto$, $\gamma$, and the previous statement, $\gamma(S', \Sigma') = S_1^*$. By definition of $\gamma$ and lemma 53, $(S^*, F, \sigma) \longmapsto (S_1^*, F', \sigma')$ and $\gamma((S', F', \Sigma', n')) \sqsubseteq (S_1^*, F', \sigma')$. □

## 11 PROOFS FOR SECTION 4.2

$$\frac{}{ht_\kappa(\kappa,\kappa)} \qquad \frac{ht_\kappa(\kappa,\kappa')}{ht_\kappa(\phi{:}\kappa,\kappa')} \qquad \frac{ht_\kappa(\kappa,\kappa')}{ht_{\hat{\varsigma}}((\langle e,\hat{\rho},\kappa\rangle,\sigma,t),\kappa')}$$

$$\frac{}{ht(\epsilon,\kappa)} \qquad \frac{ht(\pi\varsigma,\kappa) \quad \varsigma \longmapsto \varsigma' \quad ht_{\hat{\varsigma}}(\varsigma',\kappa)}{ht(\pi\varsigma\varsigma',\kappa)}$$

$$\begin{aligned}
rt_\kappa(\kappa,\kappa,\kappa') &= \kappa' \\
rt_\kappa(\phi{:}\kappa,\kappa',\kappa'') &= \phi{:}rt_\kappa(\kappa,\kappa',\kappa'') \\
rt_{\hat{\varsigma}}((\langle e,\hat{\rho},\kappa\rangle,\sigma,t),\kappa',\kappa'') &= \langle e,\hat{\rho},rt_\kappa(\kappa,\kappa',\kappa'')\rangle,\sigma,t \\
rt(\epsilon,\kappa,\kappa'') &= \epsilon \\
rt(\pi\varsigma,\kappa,\kappa') &= rt(\pi,\kappa,\kappa')rt_{\hat{\varsigma}}(\varsigma,\kappa,\kappa')
\end{aligned}$$

**Lemma 58** ($ht_\kappa$ implies $rt_\kappa$ defined)**.** $\forall\kappa,\kappa'.\ ht_\kappa(\kappa,\kappa') \implies \forall\kappa'' \in Kont.\ \exists\kappa'''.rt_\kappa(\kappa,\kappa',\kappa'') = \kappa'''$

*Proof.* By induction on $\kappa$:

**Case** Base: $\epsilon$**.**

By inversion on $ht_\kappa(\kappa,\kappa')$, $\kappa' = \epsilon$, so $rt_\kappa(\kappa,\kappa',\kappa'') = \kappa''$.

**Case** Induction step: $\phi{:}\kappa_{pre}$**.**

By cases on $ht_\kappa(\kappa,\kappa')$:

**Case** $\kappa = \kappa'$**.**

By definition $rt_\kappa(\kappa,\kappa',\kappa'') = \kappa''$

**Case** $ht_\kappa(\kappa_{pre},\kappa')$**.**

By let $\kappa_{IH}$ be the witness from the induction hypothesis. By definition $rt_\kappa(\kappa,\kappa',\kappa'') = \phi{:}\kappa_{IH}$.

□

**Lemma 59** (*ht* implies *rt* defined)**.** $\forall\pi, \in CESK_t{}^*, \kappa,\kappa' \in Kont.\ ht(\pi,\kappa) \implies \exists\pi'.rt(\pi,\kappa,\kappa') = \pi'$

*Proof.* By induction on $\pi$ and application of Lemma 58. □

CORRECTNESS: THEOREM 14 For all expressions $e_{pgm}$, if for all $\hat{\kappa}$ such that $\varsigma.\kappa \in \mathit{unroll}_{\Xi}(\hat{\kappa})$, both $\mathit{tick}_{CESK_t}(\varsigma) = \mathit{tick}_{CESK^*_t\Xi}(\varsigma\{\kappa := \hat{\kappa}\})$ and $\mathit{alloc}_{CESK_t}(\varsigma) = \mathit{alloc}_{CESK^*_t\Xi}(\varsigma\{\kappa := \hat{\kappa}\})$, then

- **Soundness:** if $\varsigma, \sigma, t \longmapsto_{CESK_t} \varsigma', \sigma', t'$ and $\varsigma.\kappa \in \mathit{unroll}_{\Xi}(\hat{\kappa})$, then there are $\Xi', \hat{\kappa}'$ such that $\varsigma\{\kappa := \hat{\kappa}\}, \sigma, t, \Xi \longmapsto_{CESK^*_t\Xi} \varsigma'\{\kappa := \hat{\kappa}'\}, \sigma', \Xi', t'$ and $\varsigma'.\kappa \in \mathit{unroll}_{\Xi'}(\hat{\kappa}')$

- **Local completeness:** if $\hat{\varsigma}, \sigma, t, \Xi \longmapsto_{CESK^*_t\Xi} \hat{\varsigma}', \sigma', t', \Xi'$ and $\mathit{inv}(\hat{\varsigma}, \sigma, t, \Xi)$, for all $\kappa$, if $\kappa \in \mathit{unroll}_{\Xi}(\hat{\varsigma}.\hat{\kappa})$ then there is a $\kappa'$ such that $\hat{\varsigma}\{\hat{\kappa} := \kappa\}, \sigma, t \longmapsto_{CESK_t} \hat{\varsigma}'\{\hat{\kappa} := \kappa'\}, \sigma', t'$ and $\kappa' \in \mathit{unroll}_{\Xi}(\hat{\varsigma}'.\hat{\kappa})$.

*Proof.* Soundness follows by cases on $\longmapsto_{CESK_t}$:

**Case** $\langle x, \rho, \kappa\rangle, \sigma, t \longmapsto_{CESK_t} \langle v, \kappa\rangle, \sigma, u$.

where $v \in \sigma(\rho(x))$.

The witnesses are $\Xi, \hat{\kappa}$. The step is constructible with the lookup rule and the tick assumption.

**Case** $\langle (e_0\ e_1), \rho, \kappa\rangle, \sigma, t \longmapsto_{CESK_t} \langle e_0, \rho, \mathbf{appL}(e_1, \rho){:}\kappa\rangle, \sigma, u$.

The witnesses are $\Xi \sqcup [\tau \mapsto \hat{\kappa}]$ and $\mathbf{appL}(e_1, \rho){:}\tau$, where $\tau = \langle (e_0\ e_1), \rho, \sigma\rangle$. The step is constructible with the application expression rule and the tick assumption.

**Case** $\langle v, \mathbf{appL}(e, \rho'){:}\kappa\rangle, \sigma, t, \longmapsto_{CESK_t} \langle e, \rho', \mathbf{appR}(v){:}\kappa\rangle, \sigma, u$.

$\hat{\kappa}$ must be of the form $\mathbf{appL}(e, \rho'){:}\hat{\kappa}'$, where $\kappa \in \mathit{unroll}_{\Xi}(\hat{\kappa}')$, by the definition of unrolling.

The witnesses are $\Xi$ and $\mathbf{appR}(:)\hat{\hat{\kappa}}'$. The step is constructible with the argument evaluation rule and the tick assumption.

**Case** $\langle v, \rho, \mathbf{appR}(\lambda\, x.\ e, \rho'){:}\kappa\rangle, \sigma, t \longmapsto_{CESK_t} \langle e, \rho'', \kappa\rangle, \sigma', u$.

where $\rho'' = \rho'[x \mapsto a]$, $\sigma' = \sigma \sqcup [a \mapsto v]$.

$\hat{\kappa}$ must be of the form $\mathbf{appR}(\lambda\, x.\ e, \rho'){:}\hat{\kappa}'$, where $\kappa \in \mathit{unroll}_{\Xi}(\hat{\kappa}')$, by the definition of unrolling. The witnesses are thus $\Xi$ and $\hat{\kappa}'$. The step is constructible with the function call rule and the alloc and tick assumptions.

Completeness follows by cases on $\longmapsto_{CESK^*_t\Xi}$:

**Case** $\langle x, \rho, \sigma, \hat{\kappa}\rangle_t, \Xi \longmapsto_{CESK^*_t\Xi} \langle v, \sigma, \hat{\kappa}\rangle_u, \Xi$.

where $v \in \sigma(\rho(x))$

The witness is $\kappa$. The step is constructible with the lookup rule and the tick assumption.

**Case** $\langle (e_0\ e_1), \rho, \sigma, \hat{\kappa}\rangle_t, \Xi \longmapsto_{CESK^*_t\Xi} \langle e_0, \rho, \sigma, \mathbf{appL}(e_1, \rho){:}\tau\rangle_u, \Xi'$.

where $\tau = \langle (e_0\ e_1), \rho, \sigma\rangle_t$, $\Xi' = \Xi \sqcup [\tau \mapsto \hat{\kappa}]$.

The witness is $\mathbf{appL}(e_, \rho){:}\kappa$ by definition of unrolling. The step is constructible with the application expression rule and the tick assumption.

**Case** $\langle v, \sigma, \mathbf{appL}(e, \rho'){:}\tau\rangle_t, \Xi \longmapsto_{CESK_t^*\Xi} \langle e, \rho', \sigma, \mathbf{appR}(v){:}\tau\rangle_u, \Xi$.

The given $\kappa$ must be of the form $\mathbf{appL}(e, \rho'){:}\kappa'$ by definition of unrolling. The witness is $\mathbf{appR}(v){:}\kappa'$. The step is constructible with the argument evaluation rule and the tick assumption.

**Case** $\langle v, \rho, \sigma, \mathbf{appR}(\lambda x.\ e, \rho'){:}\tau\rangle_t, \Xi \longmapsto_{CESK_t^*\Xi} \langle e, \rho'', \sigma', \hat{\kappa}\rangle_u, \Xi$.

where $\hat{\kappa} \in \Xi(\tau)$, $\rho'' = \rho'[x \mapsto a]$, $\sigma' = \sigma \sqcup [a \mapsto v]$.

The given $\kappa$ must be of the form $\mathbf{appR}(\lambda x.\ e, \rho'){:}\kappa'$ by definition of unrolling. The witness is $\kappa'$. The step is constructible with the function application rule and the tick and alloc assumptions.

□

CORRECTNESS THEOREM 15 For all $e_0$, let $\varsigma_0 = \langle e_0, \bot, \epsilon\rangle, \bot, t_0$ in $\forall n \in \mathbb{N}, \varsigma, \varsigma' \in \mathit{CESK}_t$:

- if $(\varsigma, \varsigma') \in \mathit{unfold}(\varsigma_0, \longmapsto_{CESK_t}, n)$ then there is an $m$ such that $\varsigma \longmapsto_{\mathit{reify}(\mathcal{F}_{e_0}^m(\bot))} \varsigma'$
- if $\varsigma \longmapsto_{\mathit{reify}(\mathcal{F}_{e_0}^n(\bot))} \varsigma'$ then there is an $m$ such that $(\varsigma, \varsigma') \in \mathit{unfold}(\varsigma_0, \longmapsto_{CESK_t}, m)$

*Proof.* By induction on $n$.

**Case** $0$.

Both vacuously true.

**Case** $i + 1$.

First bullet: If $(\varsigma, \varsigma')$ is not newly added at $i + 1$, then holds by IH. Otherwise, we have a step $\varsigma \longmapsto_{CESK_t} \varsigma'$ by definition of $\mathit{unfold}_1$. By IH, in $i$ steps $\varsigma$ is reachable, in the reified system. By cases on the rule that added $(\varsigma, \varsigma')$ to *unfold*. Reasoning follows the same as soundness bullet of Theorem 14.

Second bullet: Let $\mathcal{S} = \mathit{reify}(\mathcal{F}_{e_0}^{i+1}(\bot))$ If the step is not newly added at $i + 1$, then holds by IH. Otherwise, we have a pair $\varsigma, \varsigma' \in \mathcal{S}.R$ that was extended by a step $\varsigma, \Xi \longmapsto \hat{\varsigma}', \Xi'$ where $\Xi' \sqsubseteq \mathcal{S}.\Xi$ and $\mathit{inv}_\Xi(\Xi')$. Reasoning follows the same as the local completeness bullet of Theorem 14.

□

## 12 PROOFS FOR SECTION 4.4

For *alloc* in $\longmapsto_{SR}$ and $\widehat{alloc}$ in $\longmapsto_{SRS\chi_t}$, the two "behave" if

$$\frac{\exists a, a'. \forall \kappa \sqsubseteq \mathit{unroll}_{\Xi_{\hat{\kappa}},\chi}(\hat{\kappa}). \qquad \forall C \sqsubseteq \mathit{unrollC}_{\Xi_{\hat{\kappa}},\Xi_{\hat{C}},\chi}(.) \quad (a, a') = \widehat{alloc}(\mathtt{ev}\ (\mathtt{shift}\ x.e, \rho, \hat{\sigma}, \chi, \hat{\kappa}, \hat{C}), \Xi_{\hat{\kappa}}, \Xi_{\hat{C}})}{\mathit{alloc}(\mathtt{ev}\ (\mathtt{shift}\ x.e, \rho, \sigma, \kappa, C)) = a}$$

$$\frac{\exists a. \forall \kappa \sqsubseteq \mathit{unroll}_{\Xi_{\hat{\kappa}},\chi}(\hat{\kappa}). \quad \forall C \sqsubseteq \mathit{unrollC}_{\Xi_{\hat{\kappa}},\Xi_{\hat{C}},\chi}(.) \qquad (\mathbf{fun}(\lambda\, x.\ e, \rho), \tau) \in \mathit{pop}(\Xi_{\hat{\kappa}}, \chi, \hat{\kappa})}{\mathit{alloc}(\mathtt{co}\ (\mathbf{fun}(\lambda\, x.\ e, \rho){:}\kappa\kappa, C, v, \sigma)) = \widehat{alloc}(\mathtt{co}\ (\hat{\kappa}, \hat{C}, \hat{v}, \hat{\sigma}, \chi), \Xi_{\hat{\kappa}}, \Xi_{\hat{C}})}$$

**Lemma 60** ($\mathbb{A}$ is sound)**.** *If* $\kappa \sqsubseteq \mathit{unroll}_{\Xi_{\hat{\kappa}},\chi}(\hat{\kappa})$ *then for* $(\chi', \tilde{\kappa}) = \mathbb{A}(\chi, a, \hat{\kappa})$, $\kappa \sqsubseteq \mathit{unroll}_{\Xi_{\hat{\kappa}},\chi'}(\tilde{\kappa})$

*Proof.* By routine case analysis on $\hat{\kappa}$. □

SOUNDNESS THEOREM 23 If $\varsigma \longmapsto_{SR} \varsigma'$, and $\varsigma \sqsubseteq \hat{\varsigma}, \Xi_{\hat{\kappa}}, \Xi_{\hat{C}}$ and the allocation functions behave, then there are $\hat{\varsigma}', \Xi'_{\hat{\kappa}}, \Xi'_{\hat{C}}$ such that $\hat{\varsigma}, \Xi_{\hat{\kappa}}, \Xi_{\hat{C}} \longmapsto \hat{\varsigma}', \Xi'_{\hat{\kappa}}, \Xi'_{\hat{C}}$ and $\varsigma' \sqsubseteq \hat{\varsigma}', \Xi'_{\hat{\kappa}}, \Xi'_{\hat{C}}$.

*Proof.* By cases on the concrete step:

**Case** $\mathtt{ev}\ (\mathtt{reset}\ e, \rho, \sigma, \kappa, C) \longmapsto \mathtt{ev}\ (e, \rho, \sigma, \epsilon, \kappa \circ C)$.

By assumption, we must have some $\hat{\sigma}, \chi, \hat{\kappa}, \hat{C}$ such that $\hat{\varsigma} = \mathtt{ev}\ (\mathtt{reset}\ e, \rho, \hat{\sigma}, \chi, \hat{\kappa}, \hat{C})$ with the appropriate ordering in $\hat{\varsigma}$. The step is then to

$$\mathbf{ev}\langle e, \rho, \hat{\sigma}, \chi, \epsilon, \gamma\rangle, \Xi_{\hat{\kappa}}, \Xi'_{\hat{C}}$$
$$\text{where } \gamma = \langle e, \rho, \hat{\sigma}, \chi\rangle$$
$$\Xi'_{\hat{C}} = \Xi_{\hat{C}} \sqcup [\gamma \mapsto \{(\hat{\kappa}, \hat{C})\}]$$

Where the ordering is trivial.

**Case** $\mathtt{co}\ (\epsilon, \kappa \circ C, v, \sigma) \longmapsto \mathtt{co}\ (\kappa, C, v, \sigma)$.

We must have $\hat{v}, \hat{\sigma}, \chi, \gamma$ such that

$$\hat{\varsigma} = \mathtt{co}\ (\epsilon, \gamma, \hat{v}, \hat{\sigma}, \chi)$$
$$\kappa \circ C \sqsubseteq \mathit{unrollC}_{\Xi_{\hat{\kappa}},\Xi_{\hat{C}},\chi}(\gamma)$$
$$v \sqsubseteq_{\Xi_{\hat{\kappa}},\chi} \hat{v}$$

By decomposing the unroll ordering, we get our hands on the appropriate $(\hat{\kappa}, \hat{C}) \in \Xi_{\hat{C}}(\gamma)$ so that the step is to

$$\mathtt{co}\ (\hat{\kappa}, \hat{C}, \hat{v}, \hat{\sigma}, \chi)$$

The ordering is by assumption.

**Case** $\mathtt{ev}\ (\mathtt{shift}\ x.e, \rho, \sigma, \kappa, C) \longmapsto \mathtt{ev}\ (e, \rho[x \mapsto a], \sigma', \epsilon, C)$**.**

where $a = alloc(\varsigma)$, $\sigma' = \sigma \sqcup [a \mapsto \mathbf{comp}(\kappa)]$.

We must hav $\hat{\sigma}, \chi, \hat{\kappa}, \hat{C}$ such that

$$\hat{\sigma} = \mathtt{ev}\ (\mathtt{shift}\ x.e, \rho, \hat{\sigma}, \chi, \hat{\kappa}, \hat{C})$$
$$\sigma \sqsubseteq_{\Xi_{\hat{\kappa}}, \chi} \hat{\sigma}$$
$$\kappa \sqsubseteq unroll_{\Xi_{\hat{\kappa}}, \chi}(\hat{\kappa})$$
$$C \sqsubseteq unrollC_{\Xi_{\hat{\kappa}}, \Xi_{\hat{C}}, \chi}(\hat{C})$$

By the $\widehat{alloc}$ assumption there is a $a'$ such that

$$(a, a') = \widehat{alloc}(\hat{\varsigma}, \Xi_{\hat{\kappa}}, \Xi_{\hat{C}})$$

Let $(\chi', \tilde{\kappa}) = \mathbb{A}(\chi, a', \hat{\kappa})$. By Lemma 60, the step to

$$\mathtt{ev}\ (e, \rho[x \mapsto a], \hat{\sigma} \sqcup [a \mapsto \{\tilde{\kappa}\}], \chi', \epsilon, \hat{C}), \Xi_{\hat{\kappa}}, \Xi_{\hat{C}}$$

is correctly ordered.

**Case** $\mathtt{co}\ (\mathbf{fun}(\mathbf{comp}(\kappa')){:}\kappa, C, v, \sigma) \longmapsto \mathtt{co}\ (\kappa', \kappa \circ C, v, \sigma)$**.**

The ordering assumption makes this trivial.

**Case** variable lookup**.**

Trivial.

**Case** closure creation**.**

Trivial.

**Case** application expression**.**

Trivial.

**Case** argument evaluation**.**

Trivial.

**Case** function call**.**

Same argument as for standard pushdown, using the alloc assumption.

□

## 13 PROOFS FOR SECTION 4.5

For the completeness result in this global system, we need that $inv_{\Xi}$ is maintained over the system's $\Xi$. The primary difference is about maintenance through join. Each trace guaranteed by the invariant is independent of the table, so we can add each mapping of a table in whatever order.

**Lemma 61.** *If* $inv_\Xi(e_{pgm}, \Xi)$ *and* $inv_\Xi(e_{pgm}, \Xi')$*, then* $inv_\Xi(e_{pgm}, \Xi \sqcup \Xi')$*.*

*Proof.* By induction on the proof of $inv_\Xi(e_{pgm}, \Xi')$. □

**Lemma 62.** *If* $inv_M(M)$ *and* $inv_\Xi(e_{pgm}, M')$*, then* $inv_M(M \sqcup M')$*.*

*Proof.* By induction on the proof of $inv_M(M')$. □

The state invariant entirely for program $e_{pgm}$ is

$$\begin{aligned} inv(e_{pgm}, \varsigma, \Xi, M) = {} & inv_\Xi(e_{pgm}, \Xi) \wedge inv_M(M) \wedge \mathbf{dom}(M) \subseteq \mathbf{dom}(\Xi) \\ & \wedge \varsigma.\hat{\kappa} \equiv \epsilon \implies \langle e_{pgm}, \bot, \bot, \epsilon\rangle \longmapsto^*_{CESK_t} \varsigma \\ & \wedge \varsigma.\hat{\kappa} \equiv \phi{:}\tau \implies \\ & \quad \tau \in \mathbf{dom}(\Xi) \wedge \\ & \quad \forall \kappa. A(\tau, \kappa) \implies extend(\tau, \kappa) \longmapsto^*_{CESK_t} \varsigma\{\hat{\kappa} := \kappa\} \end{aligned}$$

**Lemma 63** (Memo invariant)**.** *If* $inv(e_{pgm}, \varsigma, \Xi, M)$ *and* $\varsigma, \Xi, M \longmapsto_{CESK^*_t\Xi M} \varsigma', \Xi', M'$ *then* $inv(e_{pgm}, \varsigma', \Xi', M')$*.*

*Proof.* The $inv_\Xi$ component is the same as before except in the memo use rule. The piecewise traces based on the current state's continuation are simple. I focus on the $inv_M$ component. By cases on the step:

**Case** $\langle x, \rho, \sigma, \hat{\kappa}\rangle_t, \Xi, M \longmapsto \langle v, \sigma, \hat{\kappa}\rangle_u, \Xi, M$**.**

where $v \in \sigma(\rho(x))$

By assumption.

**Case** $\langle (e_0\ e_1), \rho, \sigma, \hat{\kappa}\rangle_t, \Xi, M \longmapsto \langle e_0, \rho, \sigma, \mathbf{appL}(e_1, \rho){:}\tau\rangle_u, \Xi', M$**.**

where $\tau = \langle (e_0\ e_1), \rho, \sigma\rangle_t$, $\Xi' = \Xi \sqcup [\tau \mapsto \hat{\kappa}]$, and $\tau \notin \mathbf{dom}(M)$

By assumption.

**Case** $\langle (e_0\ e_1), \rho, \sigma, \hat{\kappa}\rangle_t, \Xi, M \longmapsto \langle e', \rho', \sigma', \hat{\kappa}\rangle_u, \Xi', M$**.**

where $\tau = \langle (e_0\ e_1), \rho, \sigma\rangle_t$, $\Xi' = \Xi \sqcup [\tau \mapsto \hat{\kappa}]$, and $\langle e', \rho', \sigma'\rangle \in M(\tau)$.

The $inv_\Xi(e_{pgm}, \Xi)$ path comes from path concatenation with the path from $inv_M$.

**Case** $\langle v, \sigma, \mathbf{appL}(e, \rho'){:}\tau\rangle_t, \Xi, M \longmapsto \langle e, \rho', \sigma, \mathbf{appR}(v){:}\tau\rangle_u, \Xi, M$**.**

By assumption.

**Case** $\langle v, \rho, \sigma, \mathbf{appR}(\lambda x.\ e, \rho'){:}\tau\rangle_t, \Xi, M \longmapsto \langle e, \rho'', \sigma', \hat{\kappa}\rangle_u, \Xi, M'$**.**

where

$$\begin{aligned} \hat{\kappa} &\in \Xi(\tau) \\ \rho'' &= \rho'[x \mapsto a] \\ \sigma' &= \sigma \sqcup [a \mapsto v] \\ M' &= M \sqcup [\tau \mapsto \{\langle e, \rho'', \sigma'\rangle\}] \end{aligned}$$

Let $\kappa$ be arbitrary. We must show $\mathit{extend}(\tau, \kappa) \longmapsto^{*}_{CESK_t} \mathit{plug}(\langle e, \rho'', \sigma' \rangle, \kappa)$. By the *inv* assumption, there is a path from the starting state of the continuation to the left-hand state. The function call rule is immediately applicable, and the invariant holds for this addition to M.

$\square$

## 14 WEAK EQUALITY PROOFS

Our abstract term equality is an *exact approximation* if we can show the relationship:

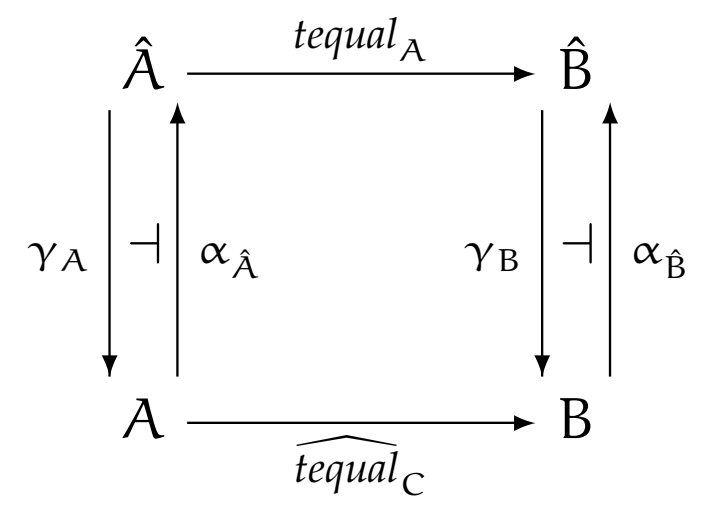

Specifically, we want our hands on the equation:

$$tequal_A = \alpha_{\hat{B}} \circ \widehat{tequal}_C \circ \gamma_A$$

The As and Bs are named as such to illustrate the relationship. Informally, this diagram says that $tequal_A$ does the best it can to mimic $\widehat{tequal}_C$'s behavior within the abstract domain.

In our case $\hat{A}$ and $\hat{B}$ are

$$\hat{A} = \widehat{Store} \times Count \times Term_A \times Term_A$$
$$\hat{B} = \widehat{Equality}$$

The concrete term equality function is hatted because we lift *tequal* over the powerset of its input type:

$$\begin{aligned} &\widehat{tequal}_C : A \to B \\ &\text{where } A = \wp(Store \times Term_C \times Term_C) \text{ non-empty} \\ &\qquad B = \{\{\mathtt{tt}\}, \{\mathtt{ff}\}, \{\mathtt{tt}, \mathtt{ff}\}\} \\ &\widehat{tequal}_C(S) = \{tequal(\sigma)(t_0, t_1) \ : \ \langle \sigma, t_0, t_1 \rangle \in S\} \end{aligned}$$

The relationship between a non-empty set of booleans and $\widehat{Equality}$ is the obvious Galois connection (also isomorphism):

$$\begin{aligned} \alpha_{\hat{B}}(\{\mathtt{tt}\}) &= \textbf{Equal} & \gamma_B(\textbf{Equal}) &= \{\mathtt{tt}\} \\ \alpha_{\hat{B}}(\{\mathtt{ff}\}) &= \textbf{Unequal} & \gamma_B(\textbf{Unequal}) &= \{\mathtt{ff}\} \\ \alpha_{\hat{B}}(\{\mathtt{tt}, \mathtt{ff}\}) &= \textbf{May} & \gamma_B(\textbf{May}) &= \{\mathtt{tt}, \mathtt{ff}\} \end{aligned}$$

The remaining pieces are $\alpha_{\hat{A}}$ and $\gamma_A$.

USER-DEFINED ABSTRACTION The address spaces $Addr_A$ and $Addr_C$ are user-defined but (their powersets) must have a Galois *insertion* (abstraction is surjective). It is sufficient for us to require a user-provided surjective address abstraction function, $\alpha$, that we pointwise lift:

$$\begin{aligned}
\alpha &: Addr_C \twoheadrightarrow Addr_A \\
\overline{\alpha} &: \wp(Addr_C) \twoheadrightarrow \wp(Addr_A) \\
\overline{\alpha}(\{a \ldots\}) &= \{\alpha(a) \ldots\} \\
\overline{\gamma} &: \wp(Addr_A) \rightarrow \wp(Addr_C) \\
\overline{\gamma}(S)&\{a' \ : \ \hat{a} \in S, \alpha(a') = \hat{a}\}
\end{aligned}$$

The user-supplied $\alpha$ is sufficient to build the Galois insertion:

$$\langle \wp(Addr_C), \subseteq \rangle \underset{\overline{\alpha}}{\overset{\overline{\gamma}}{\leftrightarrows}} \langle \wp(Addr_A), \subseteq \rangle$$

POINTWISE ABSTRACTION The abstraction function is a pointwise abstraction with counting:

$$\alpha_{\hat{A}}(S) = \{\langle \hat{\sigma}, \mu, \dot{\alpha}(t_0), \dot{\alpha}(t_1) \rangle \ : \ \langle \sigma, t_0, t_1 \rangle \in S, \langle \hat{\sigma}, \mu \rangle = \alpha_S(\sigma)\}$$

The $\dot{\alpha}$ function is a structural lifting of $\alpha$ over concrete terms[32]. The $\hat{\alpha}$ function is $\dot{\alpha}$ lifted over a set of concrete terms:

[32] $\dot{\alpha}$ *defers to* $E.\gamma$*'s right adjoint on external terms.*

$$\begin{aligned}
\hat{\alpha} &: \wp(Term_C) \rightarrow Term_A \\
\hat{\alpha}(S) &= \bigsqcup_{t \in S} \dot{\alpha}(t)
\end{aligned}$$

The $\alpha_S$ function abstracts and counts addresses in a concrete store:

$$\begin{aligned}
\alpha_S &: Store \rightarrow \widehat{Store} \times Count \\
\alpha_S(\sigma) &= \langle \hat{\sigma}, \mu \rangle \\
\hat{\sigma} &= \bigsqcup_{a \in \mathbf{dom}(\sigma)} [\alpha(a) \mapsto \dot{\alpha}(\sigma(a))] \\
\mu &= \lambda \hat{a}.0 \oplus \bigoplus_{a \in \mathbf{dom}(\sigma)} [\alpha(a) \mapsto 1]
\end{aligned}$$

The $\oplus$ operator is an abstract plus in $\hat{\mathbb{N}}$, lifted above to maps:

$$0 \oplus \hat{n} = n \qquad \hat{n} \oplus 0 = n \qquad \hat{n} \oplus \hat{n}' = \omega \text{ otherwise}$$

CONCRETIZING THE STORE I've come across some misunderstandings of abstract counting, so I'm going to suggest and disspell a couple of false starts:

1. we can say an address $\hat{a}$ is fresh or concretely identifiable if $|\overline{\gamma}(\{\hat{a}\})| = 1$.
   This intepretation is wrong because the Galois connection is unchanging. Most abstract addresses will always concretize to an infinite number of concrete addresses they can stand for. Thus while this criterion is technically correct, it is largely inapplicable.

2. The infinite set of concrete addresses $\overline{\gamma}$ returns can be trimmed with the additional context of the store. We might then say $\hat{a}$ is fresh if its corresponding concrete store only binds one of its concrete meanings:

   $$|\mathbf{dom}(\sigma) \cap \overline{\gamma}(\{\hat{a}\})| = 1$$

   But whence the concrete store, $\sigma$? A naive interpretation is a pointwise concretization of the abstract store (which depends on concretizing terms via some $\gamma_T : \mathit{Term}_A \to \wp(\mathit{Term}_C)$):

   $$\begin{aligned} \gamma_S &: \widehat{\mathit{Store}} \to \wp(\mathit{Store}) \\ \gamma_S(\bot) &= \{\bot\} \\ \gamma_S(\hat{\sigma}[\hat{a} \mapsto \hat{t}]) &= \{\sigma[a \mapsto t] \ : \ \sigma \in \gamma_S(\hat{\sigma}), \alpha(a) = \hat{a}, \hat{t} \in \gamma_T(\hat{t})\} \end{aligned}$$

   Which, first of all, isn't even right. Here $\gamma_S$ creates one concrete store entry per abstract address, yet potentially infinitely many such stores for all the concretizations of an abstract address. An abstract address can denote *unboundedly many* concrete addresses, though. This definition should really be making *infinitely many* stores with all non-empty subsets of the $\alpha^{-1}(\hat{a})$ in their domains. Infinitely many is too many. We have $\mu$ to tell us we *do* know how many concrete addresses an abstract address denotes. But $\mu$ is not utilized at all here.

These false starts illuminate that concretization must take freshness information into account. The concretization $\gamma_A$ not only concretizes the an abstract store and count, but also two abstract terms. A term is understood in the context of a store, so we first focus on concretizing the store, which we call $\gamma_S$:

$$\gamma_S : \widehat{\mathit{Store}} \times \mathit{Count} \to \wp(\mathit{Store})$$

and then we focus on the term concretization function $\gamma_T$:

$$\gamma_T : \wp(\mathit{Addr}_C) \times \mathit{Term}_A \to \wp(\mathit{Term}_C).$$

The above attempt to define $\gamma_S$ failed to properly understand addresses. If an address is used, we have no idea which or how many of its concrete allocations could be mapped. Therefore, each used address represents a set of sets of addresses; each individual set is

$$\begin{aligned}
\gamma_S &: \widehat{Store} \times Count \to \wp(Store) \\
\gamma_S(\hat{\sigma}, \mu) &= \bigcup_{D \in P(\mathbf{dom}(\hat{\sigma}))} \gamma_D(D, \hat{\sigma}) \\
\text{where} Ds &: Addr_A \to \wp(\wp(Addr_C)) \\
Ds(\hat{a}) &= \texttt{case}\ \mu(\hat{a})\ \texttt{of} \\
&\qquad \omega : \wp(\alpha^{-1}(\hat{a})) \setminus \{\emptyset\} \\
&\qquad 1 : \{\{a\} \ :\ a \in \alpha^{-1}(\hat{a})\} \\
&\qquad 0 : \{\emptyset\} \\
P &: \wp(Addr_A) \to \wp(\wp(Addr_C)) \\
P(\emptyset) &= \{\emptyset\} \\
P(\{\hat{a}\} \cup \hat{A}) &= \{A \cup A' \ :\ A \in Ds(\hat{a}), A' \in P(\hat{A})\} \\
\gamma_D &: \wp(Addr_C) \times \wp(Addr_C) \times \widehat{Store} \to \wp(Store) \\
\gamma_D(\emptyset, D, \hat{\sigma}) &= \{\bot\} \\
\gamma_D(\{a\} \cup D_{rec}, D, \hat{\sigma}) &= \{\sigma[a \mapsto t] : \sigma \in \gamma_D(D_{rec}, D, \hat{\sigma}), \\
&\qquad t \in \gamma_T(D)(\hat{\sigma}(\alpha(a)))\}
\end{aligned}$$

Figure 49: Store concretization

the slice of a concrete store's domain that all map through $\alpha$ to the one used address. If an address is fresh, we still don't know which concrete address it stands for, just that there is exactly one of them.

An *NDTerm* in the store may refer to other addresses. As such, concretization needs the entire scope of a concrete store it's building *before* it concretizes any terms. With the set of all concrete store domains, one-by-one we concretize each term with respect to the domain.

The definition of $\gamma_S$ is in Figure 49. We implement the previous informal description with functions *Ds*, P and $\gamma_D$. The *Ds* function builds the domain slices that an abstract address gives rise to. The P function produces the big product of these slices into whole domains of a concrete store. The $\gamma_D$ function is mapped over each domain to produce all the possible concretizations of each term in the abstract store, as scoped to the concrete store's domain.

Even though the fresh addresses produce many stores, we view stores with an equivalence relation that identifies "$\alpha$-equivalent" stores. The $\alpha$-equivalence treats addresses in the store domain as binding positions, and addresses in the codomain as reference positions.

TERM CONCRETIZATION A term is a well-founded data structure, but we sometimes understand an address as its mapping in the store. Conflating an address with its contents in the store can lead to infinite (ill-founded) terms as we saw in the `(cons b b)` example before.

We separate the concerns of understanding the address and the concretization of a term by always concretizing an abstract address to a set of concrete addresses. The important piece of the definition in Figure 50 is that each address is concretized to a set of concrete addresses that must be in D. We don't want to produce ill-formed terms that have dangling pointers.

$\gamma_T : \wp(\mathit{Addr}_C) \to \mathit{Term}_A \to \wp(\mathit{Term}_C)$

$$
\begin{aligned}
\gamma_T(D)(\mathbf{EAddr}(\hat{a})) &= \{\mathbf{EAddr}(a) \ : \ a \in \alpha^{-1}(\hat{a}) \cap D\} \\
\gamma_T(D)(\mathbf{IAddr}(\hat{a}, \mathit{lm})) &= \{\mathbf{IAddr}(a, \mathit{lm}) \ : \ a \in \alpha^{-1}(\hat{a}) \cap D\} \\
\gamma_T(D)(\mathbf{Delay}(\hat{a})) &= \{\mathbf{Delay}(a) \ : \ a \in \alpha^{-1}(\hat{a}) \cap D\} \\
\gamma_T(D)(\mathbf{External}(E, \hat{v})) &= \{\mathbf{External}(E, v) \ : \ v \in E.\gamma(\hat{v})\} \\
\gamma_T(D)(\mathbf{NDT}(\mathit{ts}, \mathit{Es})) &= \bigcup_{\hat{t} \in \mathit{Choose}(\mathbf{NDT}(\mathit{ts}, \mathit{Es}))} \gamma_T(D)(\hat{t}) \\
\gamma_T(D)(\mathbf{Variant}(n, \hat{\mathbf{t}})) &= \mathit{each}(\hat{\mathbf{t}}, \langle\rangle) \\
\text{where } \mathit{each}(\langle\rangle, \mathbf{t}) &= \{\mathbf{Variant}(n, \mathbf{t})\} \\
\mathit{each}(\langle \hat{t}_0 \hat{t}_i \ldots \rangle, \langle t \ldots \rangle) &= \bigcup_{t_0 \in \gamma_T(D)(\hat{t}_0)} \mathit{each}(\hat{t}_i \ldots, \langle t \ldots \ t_0 \rangle)
\end{aligned}
$$

I will write $\gamma_\sigma$ to mean $\gamma_T(\mathbf{dom}(\sigma))$.

Figure 50: Term concretization

An important property we need later is that smaller refinements mean larger concretizations. This means if you restrict the store less, it is free to mean more. The set $\mathit{Refinements}(\hat{\sigma}, \mu)$ is all the possible refinements: $\{\delta \ : \ \mathit{refines}(\delta, \hat{\sigma}, \mu)\}$.

**Lemma 64** (Restrictive overwriting is antitonic). *For functions* $f, g, g' : A \rightharpoonup B$ *where* $B$ *is ordered by* $\sqsubseteq$, *if* $g \sqsubseteq g' \sqsubseteq f$ *(discretely) then* $f \triangleleft g \sqsupseteq f \triangleleft g'$.

*Proof.* Let $a \in A$ be arbitrary. By cases on $a \in \mathbf{dom}(g)$:

**Case** $a \in \mathbf{dom}(g)$.

then so must $a \in \mathbf{dom}(g')$, and $g(a) = g'(a)$, so $f \triangleleft g'(a) \sqsubseteq f \triangleleft g(a)$.

**Case** $a \notin \mathbf{dom}(g)$.

So, by cases on $a \in \mathbf{dom}(g')$:

**Case** $a \in \mathbf{dom}(g')$.

$f \triangleleft g(a) = f(a) \sqsupseteq g'(a) = f \triangleleft g'(a)$

**Case** $a \notin \mathbf{dom}(g')$.

$f \triangleleft g(a) = f(a) \sqsupseteq f(a) = f \triangleleft g'(a)$

□

**Lemma 65** (Refinement is antitonic). *If* $\delta, \delta' \in \mathit{Refinements}(\hat{\sigma}, \mu)$ *and* $\delta \sqsubseteq \delta'$, *then* $\gamma_S(\hat{\sigma} \blacktriangleleft \delta, \mu) \supseteq \gamma_S(\hat{\sigma} \blacktriangleleft \delta', \mu)$.

*Proof.* Follows from Lemma 64 and the fact that $\lceil \bullet \rceil_{\hat{\sigma}} \circ \delta' \sqsubseteq \hat{\sigma}$. □

The Galois insertion of addresses additionally implies a Galois insertion of terms (provided external descriptors have a Galois insertion). First we need a couple auxiliary functions.

Let $\mathit{fa} : \mathit{Term}_X \to \wp(\mathit{Addr}_X)$ be be the set of "free addresses" (all addresses) in a term, structurally lifted as necessary.

**Lemma 66** (Term abstraction is a Galois insertion). *For all* $D \in \wp(\mathit{Addr}_C)$, *if* $\hat{\alpha}, \gamma_T(D)$ *is a Galois insertion on external descriptors, then it is a Galois insertion on terms. For all* $\hat{t} \in \mathit{Term}_A$ *where* $\mathit{fa}(\hat{t}) \subseteq \alpha_A(D)$, $(\hat{\alpha} \circ \gamma_\sigma)(\hat{t}) = \hat{t}$ *and for all* $T \subseteq \mathit{Term}_C$ *where* $\mathit{fa}(T) \subseteq D$, $(\gamma_T(D) \circ \hat{\alpha})(T) \sqsupseteq T$

The Galois insertion property is important for reasoning about fresh addresses.

EQUALITY CORRECTNESS The order we use on intermediate results ensures that equality's constructors are incomparable. This ensures that, even if the term pair sets are overapproximate, the ultimate output of $\widehat{\mathit{tequal}}$ is exact.

The "so-far" result type is ordered against $\mathit{EqRes}_A$ via $\sqsubseteq_{op}$:

$$\frac{}{\mathbf{Unequal}(\Delta) \sqsubseteq_{op} \mathbf{None}} \qquad \frac{A \subseteq A'}{\mathbf{May}(A) \sqsubseteq_{op} \mathbf{Some}(A')}$$

Let $\ddot{\gamma}_\sigma : \widehat{\mathit{Pairs}} \to \wp(\mathit{Pairs})$ be

$$\ddot{\gamma}_\sigma(\emptyset) = \{\emptyset\}$$
$$\ddot{\gamma}_\sigma(\{\langle \hat{t}_0, \hat{t}_1 \rangle\} \cup \hat{A}) = \{A \cup (T_0 \times T_1) \ : \ A \in \ddot{\gamma}_\sigma(\hat{A}), T_0 \in \tilde{\gamma}_\sigma(\hat{t}_0), T_1 \in \tilde{\gamma}_\sigma(\hat{t}_1)\}$$

where $\tilde{\gamma}_\sigma : \mathit{Term}_A \to \wp(\wp(\mathit{Term}_C))$ gives the set of sets of possible concretizations of an abstract term from $\hat{A}$ that could appear in $\hat{A}$:

$$\tilde{\gamma}_\sigma(\hat{t}) = \wp(\gamma_\sigma(\hat{t})) \setminus \{\emptyset\}$$

We remove the empty set of terms because each term has at least one concretization (since $\alpha$ is total).

Let $\alpha_{eq} : \wp(\mathit{EqRes}_C) \to \mathit{EqRes}_A$ be

$$\alpha_{eq}(S) = \mathit{add\text{-}none}(S, \bigsqcup_{\mathbf{Some}(A) \in S} \mathbf{Equal}(\alpha(A)))$$

$$\mathit{add\text{-}none}(S, \mathbf{Unequal}) = \mathbf{Unequal}(\emptyset)$$
$$\mathit{add\text{-}none}(\{\mathbf{None}, _ \ldots\}, \mathbf{Equal}(A)) = \mathbf{May}(A)$$
$$\mathit{add\text{-}none}(S, \mathbf{Equal}(A)) = \mathbf{Equal}(\emptyset, A) \text{ otherwise}$$

The refinement sets are empty because the concrete world has perfect information; no state splitting is necessary.

Let $\ddot{\alpha} : \mathit{Pairs} \to \widehat{\mathit{Pairs}}$ be

$$\ddot{\alpha}(A) = \{\langle \alpha(t_0), \alpha(t_1) \rangle \ : \ \langle t_0, t_1 \rangle \in A\}$$

[TERM ABSTRACTION IS A GALOIS INSERTION] Lemma 66
If $\hat{\alpha}, \gamma_\sigma$ is a Galois insertion on external descriptors, then it is a Galois insertion on terms.
For all $\hat{t} \in \widehat{Term}$,

$$fa(\hat{t}) \subseteq \overline{\alpha}(\mathbf{dom}(\sigma)) \implies (\hat{\alpha} \circ \gamma_\sigma)(\hat{t}) = \hat{t}$$

for all $T \subseteq Term_C$

$$fa(T) \subseteq \mathbf{dom}(\sigma) \implies (\gamma_\sigma \circ \hat{\alpha})(T) \sqsupseteq T$$

*Proof.* First part: induct on $\hat{t}$.

**Case IAddr**$(\hat{a}, lm)$**.**

By surjectivity of $\alpha$, $\alpha^{-1}(\hat{a})$ is non-empty.

By assumption, $\hat{a} \in \overline{\alpha}(\mathbf{dom}(\sigma))$.

Thus $\gamma_\sigma(\hat{t}) = \{\mathbf{IAddr}(a, lm) \ : \ a \in \alpha^{-1}(\hat{a})\}$ is non-empty.

By definition of $\hat{\alpha}$, the goal holds.

**Case EAddr**$(\hat{a})$**.**

Same as previous case.

**Case Delay**$(\hat{a})$**.**

Same as previous case.

**Case NDT**$(ts, Es)$**.**

By definition, $\gamma_\sigma(\mathbf{NDT}(ts, Es)) = \bigcup\limits_{\hat{t}' \in Choose(\hat{t})} \gamma_\sigma(\hat{t}')$

By IH for each $\hat{t}' \in ts$, $\hat{\alpha}(\gamma_\sigma(\hat{t}')) = \hat{t}'$.

Since $\hat{\alpha}$ is structural, $\hat{\alpha}(\gamma_\sigma(\hat{t})) = \bigsqcup\limits_{t \in \gamma_\sigma(\hat{t})} \dot{\alpha}(t)$.

By definition this equals $\hat{\alpha}(\gamma_\sigma(\hat{t}))$.

**Case External**$(E, \hat{v})$**.**

By assumption.

**Case Variant**$(n, \hat{\mathbf{t}})$**.**

We prove a lemma with nested induction on the recursion scheme of *each*:

$$\hat{\alpha}(each(n)(\hat{\mathbf{t}}, \langle t \ldots \rangle)) = \mathbf{Variant}(n, \langle \dot{\alpha}(t) \ldots \rangle +\!\!+ \hat{\mathbf{t}})$$

**Case** $\langle \rangle, \mathbf{t}$**.**

By definition.

**Case** $\langle \hat{t}_0 \hat{t}_i \ldots \rangle, \langle t \ldots \rangle$**.**

By the definitions of $\hat{\alpha}$ and $\dot{\alpha}$,

$$\hat{\alpha}(each(n)(\langle \hat{t}_0 \hat{t}_i \ldots \rangle, \langle t \ldots \rangle)) = \bigsqcup_{t_0 \in \gamma_\sigma(\hat{t}_0)} \hat{\alpha}(each(n)(\hat{t}_i \ldots, \langle t \ldots \ t_0 \rangle))$$

$$\begin{aligned}
&= \text{[By inner IH]} \\
&\quad \bigsqcup_{t_0 \in \gamma_\sigma(\hat{t}_0)} \mathbf{Variant}(n, \langle \dot{\alpha}(t) \ldots \; \dot{\alpha}(t_0) \rangle {+}{+} \langle \hat{t}_i \ldots \rangle) \\
&= \text{[By structural definition of } \dot{\alpha}] \\
&\quad \mathbf{Variant}(n, (\langle \dot{\alpha}(t) \ldots \rangle {+}{+} \langle \hat{\alpha}(\gamma_\sigma(\hat{t}_0)) \rangle) {+}{+} \langle \hat{t}_i \ldots \rangle) \\
&= \text{[By outer IH]} \\
&\quad \mathbf{Variant}(n, (\langle \dot{\alpha}(t) \ldots \rangle {+}{+} \langle \hat{t}_0 \rangle) {+}{+} \langle \hat{t}_i \ldots \rangle)
\end{aligned}$$

By associativity of append, the conclusion holds.

Instantiate the lemma with $\hat{\mathbf{t}}, \langle\rangle$.

The second part is a simple structural induction on an arbitrary $t \in T$. □

## 14.1 *Correctness*

We have to take special care with the term pair set - the higher specificity of $\widehat{tequalaux}_S$ over $\widehat{tequalaux}$ means that any pair set we get back will be a subset of what $\widehat{tequalaux}$ might produce. An equality result P is processed into a *possible* **Both** answer in the following way:

$$strength(P) = \mathbf{Both}(\bigsqcup_{\delta' \in \mathbf{dom}(P), \mathbf{Equal}(dp) = P(\delta')} dp, \bigsqcup_{\delta' \in \mathbf{dom}(P), \mathbf{Unequal}(\Delta) = P(\delta')} \Delta)$$

**Theorem 67** (Approximation ordering)**.** *If $\hat{\varsigma} \sqsubseteq \hat{\varsigma}'$, $\hat{\sigma} \sqsubseteq \hat{\sigma}'$, $\delta' \sqsubseteq \delta$ possible refinements, $\widehat{tequal}_S(\hat{\varsigma}, \hat{\sigma}, t_0, t_1, \delta) \sqsubseteq \widehat{tequal}_S(\hat{\varsigma}', \hat{\sigma}', t_0, t_1, \delta')$*

*Proof.* Straightforward structural induction. □

We need an ordering that makes equality and inequality results incomparable in the upcoming proof.

$$\frac{}{\mathbf{None} \sqsubseteq_A \mathbf{None}} \qquad \frac{ps \subseteq ps'}{\mathbf{Some}(ps) \sqsubseteq_A \mathbf{Some}(ps')}$$

We can concretize an equality answer given a store with $\gamma_{EqRes}$:

$$\begin{aligned}
\gamma_{EqRes} &: \widehat{Store} \to \widehat{EqRes} \to \wp(EqRes) \\
\gamma_{EqRes}(\hat{\sigma})(\widehat{eq}) &= \{\mathbf{None}\} \\
\gamma_{EqRes}(\hat{\sigma})(\mathbf{Must}(dp)) &= \{\mathbf{Some}(ps) \mid \delta \in \mathbf{dom}(dp), \sigma \in \gamma_S(\hat{\sigma} \blacktriangleleft \delta), \\
&\qquad ps \in \ddot{\gamma}_\sigma(dp(\delta))\} \\
\gamma_{EqRes}(\hat{\sigma})(\mathbf{May}(\widehat{ps})) &= \{\mathbf{None}\} \cup \{\mathbf{Some}(ps) \; : \; \sigma \in \gamma_S(\hat{\sigma}), ps \in \ddot{\gamma}_\sigma(\widehat{ps})\}
\end{aligned}$$

An $\widehat{Equality}$ approximates a set of Booleans the following way:

$$\begin{aligned}\gamma_{\widehat{Equality}}(\mathbf{Equal}) &= \{\mathtt{tt}\}\\ \gamma_{\widehat{Equality}}(\mathbf{Unequal}) &= \{\mathtt{ff}\}\\ \gamma_{\widehat{Equality}}(\mathbf{May}) &= \{\mathtt{tt},\mathtt{ff}\}\end{aligned}$$

We can concretize the input to equality with $\gamma^e$:

$$\begin{aligned}\gamma^e &: (\hat\sigma : \widehat{Store}) \times \widehat{Term} \times \widehat{Term} \times \mathit{Refinements}(\hat\sigma)\\ &\to \wp(\mathit{Store} \times \mathit{Term} \times \mathit{Term})\\ \gamma^e(\hat\sigma, \hat t_0, \hat t_1, \delta) &= \{\langle \sigma, t_0, t_1\rangle \ : \ \sigma \in \gamma_S(\hat\sigma \blacktriangleleft \delta), t_0 \in \gamma_\sigma(\hat t_0), t_1 \in \gamma_\sigma(\hat t_1)\}\end{aligned}$$

We can concretize the input to equality's auxiliary function with $\gamma^\sharp$:

$$\begin{aligned}\gamma^\sharp &: (\hat\sigma : \widehat{Store}) \times \widehat{Term} \times \widehat{Term} \times \mathit{Refinements}(\hat\sigma) \times \widehat{Pairs}\\ &\to \wp(\mathit{Store} \times \mathit{Term} \times \mathit{Term} \times \mathit{Pairs})\\ \gamma^\sharp(\hat\sigma, \hat t_0, \hat t_1, \delta, \widehat{ps}) &= \{\langle \sigma, t_0, t_1, ps\rangle \ : \ \sigma \in \gamma_S(\hat\sigma \blacktriangleleft \delta), t_0 \in \gamma_\sigma(\hat t_0), t_1 \in \gamma_\sigma(\hat t_1), ps \in \ddot\gamma_\sigma(\widehat{ps})\}\end{aligned}$$

**Theorem 68** (Non-splitting equality is an exact approximation). *$\gamma_{\widehat{Equality}} \circ \widehat{tequal} = \widehat{tequal}_C \circ \gamma^e$ provided that external descriptors' equality is an exact approximation.*

*Proof.* We prove a lemma that has the goal as a corollary. In particular (let $\widehat{tequal}^*_C = \mathit{lift}(\mathit{tequal}^*_C)$),

$$\begin{aligned}&\gamma_{\mathit{EqRes}}(\hat\sigma) \circ \widehat{tequalaux}(\hat\sigma)(\hat t_0, \hat t_1)(\delta, \widehat{ps})\\ &\sqsupseteq_A (\widehat{tequal}^*_C \circ \gamma^\sharp)(\hat\sigma, \hat t_0, \hat t_1, \delta, \widehat{ps})\end{aligned}$$

Let $\hat\sigma$ be arbitrary.

By induction on $\widehat{tequalaux}$'s recursion scheme (a "larger" $\widehat{ps}$ is a smaller obligation since $\widehat{ps}$ is bounded by the number of subterms that exist in the finite store and given terms). For ease of proof, we split the *resolve* rules into direct recursive calls unless we have a **NDT**.

**Case EAddr**$(\hat a)$, **EAddr**$(\hat a)$.

By cases on $\mu(\hat a) \overset{?}{\leqslant} 1$:

**Case** `tt`.

Every concretization of $\hat\sigma$ will produce only one $a$ such that $\alpha(a) = \hat a$. Thus, concrete equality will always return **Some**$(ps)$ for the concretized $ps$ set. Abstract equality returns **Must**$(\widehat{ps})$, which concretizes to **Some**$(ps)$ for each concretization $\sigma \in \gamma_S(\hat\sigma \blacktriangleleft \delta), ps \in \ddot\gamma_\sigma$.

**Case** `ff`**.**

By assumption, for each $\sigma \in \gamma_S(\hat{\sigma} \blacktriangleleft \delta)$, $|\gamma(\hat{a}) \cap \mathbf{dom}(\sigma)| > 1$. Thus there is a **None** answer for the different addresses, and each **Some** answer for the pairs set concretizations.

**Case IAddr**$(\hat{a}, _), _$**.**

By cases on $\hat{a} \in \mathbf{dom}(\delta)$:

**Case** `tt`**.**

By IH with $\widehat{tequalaux}(ctx)(\delta(\hat{a}), \hat{t}_1)(\delta, \widehat{ps} \cup \{\langle \delta(\hat{a}), \hat{t}_1 \rangle\})$

**Case** `ff`**.**

By cases on $\mu(\hat{a}) \overset{?}{\leqslant} 1$:

**Case** `tt`**.**

By definition of *select*, we recur with $\delta' = \delta[\hat{a} \mapsto \hat{t}]$ for each $\hat{t} \in Choose(\hat{\sigma}.h(\hat{a}))$. By cases on the result of the join:

**Case Equal**$(dp)$**.**

Each mapping in *dp* comes from a recursive call. The concretizations line up by IH.

**Case Unequal.**

All recursive calls must be unequal. Holds by IH.

**Case May**$(\widehat{ps}')$**.**

All pairs come from recursive calls. Both equality and inequality are represented. The concretizations line up by IH.

**Case** `ff`**.**

Similar to above, without changing $\delta$.

**Case** $_,$ **IAddr**$(\hat{a}, _)$**.**

Similar to previous case.

**Case Delay**$(a), _$**.**

Similar to previous case.

**Case** $_,$ **Delay**$(a)$**.**

Similar to previous case.

**Case External**$(E, \hat{v}_0)$, **External**$(E, \hat{v}_1)$**.**

By assumption.

**Case NDT**$(ts, Es), _$**.**

By cases on the join:

**Case Equal**$(dp)$**.**

All choices were equal and each entry from $dp$ comes from the answer of at least one choice. Composes with IH.

**Case Unequal.**

All choices were unequal. Composes with IH.

**Case May**$(\widehat{ps})$**.**

Both equal and unequal results are possible. Whatever pairs we come up with will contain the pair sets of the **May** or **Equal** results from the recursive calls that contributed to $\widehat{ps}$, so the conclusion holds.

**Case** $_, \mathbf{NDT}(ts, Es)$**.**

Similar to previous case.

**Case** $\mathbf{Variant}(n, \mathbf{t}), \mathbf{Variant}(n, \mathbf{t}')$**.**

By induction on both $\mathbf{t}$ and $\mathbf{t}'$,

$$\gamma_{EqRes}(\hat{\sigma}) \circ V_A \sqsupseteq_A V_C \circ \overline{\gamma}^{\sharp}(\hat{\sigma})$$

where $\overline{\gamma}$ is like $\gamma^{\sharp}$, but mapped over lists of terms:

$$\begin{aligned} \overline{\gamma}(\hat{\sigma})(\hat{\mathbf{t}}, \hat{\mathbf{t}}')(\delta, \widehat{ps}) &= \{\langle \sigma, \mathbf{t}, \mathbf{t}', ps \rangle : \sigma \in \gamma_S(\hat{\sigma} \blacktriangleleft \delta), \mathbf{t} \in \overline{\gamma}_{\sigma}(\hat{\mathbf{t}}), \mathbf{t}' \in \overline{\gamma}_{\sigma}(\hat{\mathbf{t}}'), ps \in \ddot{\gamma}_{\sigma}(\widehat{ps})\} \\ \overline{\gamma}_{\sigma}(\langle\rangle) &= \{\langle\rangle\} \\ \overline{\gamma}_{\sigma}(\hat{t} : \hat{\mathbf{t}}) &= \{t : \mathbf{t} \; : \; t \in \gamma_{\sigma}(\hat{t}), \mathbf{t} \in \overline{\gamma}_{\sigma}(\hat{\mathbf{t}})\} \end{aligned}$$

We restrict the domain of $\overline{\gamma}^{\sharp}$ so that its term pair sets must contain the current $\widehat{ps}$ set. This way, we can use the outer IH within this induction.

**Case** $\langle\rangle, \langle\rangle, \delta', \widehat{ps}'$**.**

Same concretizations.

**Case** $\hat{t} : \hat{\mathbf{t}}, \hat{t}' : \hat{\mathbf{t}}', \delta', \widehat{ps}'$**.**

By cases on $\widehat{tequalaux}(\hat{\sigma})(\hat{t}, \hat{t}')(\delta', \widehat{ps'})$:

**Case Equal**$(dp)$**.**

By outer IH, the equality has the same concretization, so we can continue with the same strength. By cases on the joined recursive calls:

**Case Equal**$(dp')$**.**

Each entry of $dp'$ comes from a recursive call, so we can use the inner IH to show the concretization is overapproximate.

**Case Unequal.**

Both sides are $\{$**None**$\}$.

**Case May**$(\widehat{ps}'')$**.**

By the inner IH, this is overapproximate thus satisfies the goal.

**Case May**$(\widehat{ps}'')$**.**

By cases on the joined recursive calls:

**Case Unequal.**

The result must be **Unequal**, so the concrete semantics will return **None**.

**Otherwise.**

The result weakens to a **May** with an overapproximate pair set.

**Case Unequal.**

By the ordering, $V_C$'s output must be **None**.

**Otherwise.**

Unequal lengths, so both sides are {**None**}.

**Otherwise.**

The remaining terms are structurally incompatible, so $\widehat{tequalaux}$ produces **Unequal** and $\widehat{tequal}^*_C$ produces {**None**}. Since $\alpha'(\{\mathbf{None}\}) =$ **Unequal**, the conclusion holds.

This statement about the helper functions is then easily translated to the $\widehat{Equality}$ domain. □

[CONCRETIZATION SPLIT] Theorem 27
For all C such that $Cut(C, Refinements(\hat{\sigma}))$, $\gamma(\hat{\sigma}) = \bigcup\{\gamma(\hat{\sigma} \blacktriangleleft \delta) : \delta \in C\}$

*Proof.* The $\supseteq$ direction is fairly obvious, so we focus on $\subseteq$. Let $\sigma \in \gamma(\hat{\sigma})$ be arbitrary. Let $\delta \in Refinements(\hat{\sigma})$ be such that $\mathbf{dom}(\delta) = \mathbf{dom}(\hat{\sigma}.h)$, and that for all $\hat{a} \in \mathbf{dom}(\delta)$, $\sigma(a) \in \gamma(\delta(\hat{a}))$ (let $a$ be the unique address where $\mathbf{dom}(\sigma) \cap \overline{\gamma}(\hat{a}) = \{a\}$). Such a $\delta$ must exist since $\sigma(a)$ is concretized from a choice from abstract stores' mappings. In fact $\sigma \in \gamma(\langle\hat{\sigma} \blacktriangleleft \delta\rangle)$ (*).

By definition of *Cut*, there is a $\delta' \in C$ such that $\delta' \circledast \delta$. If $\delta' \sqsupseteq \delta$, then $\delta' = \delta$ by the fact that $\delta$ is largest, which makes $\sigma \in \gamma(\hat{\sigma} \blacktriangleleft \delta')$. If $\delta' \sqsubseteq \delta$, then by Lemma 65 and (*), $\sigma \in \gamma(\hat{\sigma} \blacktriangleleft \delta')$. □

[WORTHWHILE COMPOSITION] Lemma 28
Given total $P, P' : Refinements(\hat{\sigma}) \to Equality$, if $worthwhile(C, P)$, $worthwhile(C', P')$ and $\neg conflicting(C, P, C', P')$ then $worthwhile(C \sqcup C', P \sqcup P')$.

*Proof.* First, a simple fact of order theory gives us that the pointwise join of antitonic functions is antitonic. Let $f, g : A \to B$ be antitonic functions where $\langle A, \preceq\rangle$ and $\langle B, \sqsubseteq\rangle$ are join-semilattices. Let $a \preceq a'$ be arbitrary elements of A. By antitonicity, $f(a) \sqsupseteq f(a')$ and $g(a) \sqsupseteq g(a')$. We then must show $f(a) \sqcup g(a) \sqsupseteq f(a') \sqcup g(a')$. Since $f(a') \sqsubseteq f(a) \sqsubseteq f(a) \sqcup g(a)$ and $g(a') \sqsubseteq g(a) \sqsubseteq f(a) \sqcup g(a)$, the least upper bound property of join implies $f(a') \sqcup g(a') \sqsubseteq f(a) \sqcup g(a)$.

Next, we show that $C \sqcup C'$ is a cut of $\mathit{Refinements}(\hat{\sigma})$. By definition of $C \sqcup C'$, no two elements are comparable. Let $\delta \in \mathit{Refinements}(\hat{\sigma})$ be arbitrary. Since $C$ and $C'$ are both cuts, both have comparable refinements, $\delta'$ and $\delta''$. If $\delta' \sqsubseteq \delta''$, then $\delta'' \in C \sqcup C'$. If $\delta'' \sqsubseteq \delta'$ then $\delta' \in C \sqcup C'$. Otherwise, both are in $C \sqcup C'$ and either choice is adequate. Thus $\mathit{Cut}(C \sqcup C', \mathit{Refinement}(\hat{\sigma}))$.

Finally, we must show that for any $\delta \in C \sqcup C'$, $(P \sqcup P')(\delta) \sqsubset \mathbf{May}$. The only troublesome case is when $\delta \in \mathbf{dom}(P) \sqcap \mathbf{dom}(P')$, because the other cases are handled by the corresponding worthwhile cuts. If $\delta \in C$, then let $\delta' \in C'$ be the refinement guaranteed by the cut property of $C'$: $\delta \circledast \delta'$. By the non-conflicting hypothesis, $P(\delta) \sqcup P'(\delta') \neq \mathbf{May}$. By definition of $C \sqcup C'$, $\delta' \sqsubseteq \delta$. By antitonicity, $P'(\delta) \sqsubseteq P'(\delta')$. By property of the *Equality* lattice, $P(\delta) = P'(\delta') = P'(\delta)$. Therefore $(P \sqcup P')(\delta) \neq \mathbf{May}$, and since $\mathbf{May}$ is $\top$, $(P \sqcup P')(\delta) \sqsubset \mathbf{May}$.

The argument is symmetric if $\delta \in C'$. □

[CONFLICTING COMPOSITION NEVER WORTHWHILE] Lemma 29
If $\mathit{conflicting}(C, P, C', P')$, then for all $C''$, $\neg \mathit{worthwhile}(C'', P \sqcup P')$.

*Proof.* The goal restated in simpler terms is all cuts of $P \sqcup P'$ have a refinement that maps to $\mathbf{May}$. By the conflict hypothesis, there are $\delta \in C$, $\delta' \in C'$ such that $\delta \circledast \delta'$ and $P(\delta) \sqcup P'(\delta') = \mathbf{May}$. By antitonicity, the larger of the two maintains the same answer in the *Equality* join-semilattice. Without loss of generality, let $\delta$ be the larger. There must be a $\delta'' \circledast \delta$ in $C''$ since it is a cut. If $\delta'' \sqsubseteq \delta$, then $P(\delta'') \sqsupseteq P(\delta)$ and $P'(\delta'') \sqsupseteq P'(\delta)$, meaning the mapping is the same or $\mathbf{May}$. We already know the mapping at $\delta$ joins to $\mathbf{May}$, so $(P \sqcup P')(\delta'') = \mathbf{May}$. If $\delta \sqsubseteq \delta''$, then $P(\delta'') \sqsubseteq P(\delta)$, meaning $P(\delta'') = P(\delta)$ and $P'(\delta'') = P'(\delta)$, so again $(P \sqcup P')(\delta'') = \mathbf{May}$. Thus, since $C''$ was arbitrary, there is no worthwhile cut of $P \sqcup P'$. □

The next lemma tells us that once an equality produces a collection of refinements, we can replay the equality given each refinement and get the same single refinement back.

**Lemma 69** (Answers don't split). *If* $\widehat{\mathit{tequalaux}}_S(\mathit{ctx})(\hat{t}_0, \hat{t}_1)(\delta, \mathit{ps}) = \mathbf{Both}(\mathit{dp}, \Delta)$, *then*

- *for all* $\delta' \in \mathbf{dom}(\mathit{dp})$, $\widehat{\mathit{tequalaux}}_S(\mathit{ctx})(\hat{t}_0, \hat{t}_1)(\delta', \mathit{ps}) = \mathbf{Both}([\delta' \mapsto \mathit{dp}(\delta')], \emptyset)$
- *for all* $\delta' \in \Delta$, $\widehat{\mathit{tequalaux}}_S(\mathit{ctx})(\hat{t}_0, \hat{t}_1)(\delta', \mathit{ps}) = \mathbf{Both}(\bot, \{\delta'\})$

*if the same holds for external descriptors.*

*Proof.* By induction on $\widehat{tequalaux}_S$'s recursion scheme.

**Case EAddr**$(\hat{a})$, **EAddr**$(\hat{a})$**.**

By cases on $\mu(\hat{a}) \overset{?}{\leqslant} 1$

**Case** $\mu(\hat{a}) = 1$**.**

The goal holds by computation.

**Otherwise.**

Vacuously true, since not a **Both** answer.

**Case** *resolvable*, _**.**

If an already refined address, the IH applies. Otherwise, we combine several results.

By cases on the result of the join.

**Case May**$(A')$**.**

Vacuously true.

**Case Both**$(R, D)$**.**

Each refinement in **dom**$(R)$ and $D$ come from one of the recursive calls' answers. By IH, the goal holds.

**Case** _, *resolvable*, _**.**

Like above.

**Case Variant**$(n, \mathbf{t})$, **Variant**$(n, \mathbf{t}')$**.**

By nested induction on the recursion scheme of $V_S$. The combination logic of the resolve case is similar.

**Case External**$(E, v_0)$, **External**$(E, v_1)$**.**

By assumption.

**Otherwise.**

Structurally unequal, so **Both**$(\bot, \{\delta\})$ is the answer. The goal holds since the only $\delta' \in \Delta$ is $\delta$.

□

We have a new concretization for $\widehat{EqRes}_S$:

$$\begin{aligned}
\gamma_{EqRes_S} &: \widehat{Store} \to \widehat{EqRes}_S \to \wp(EqRes) \\
\gamma_{EqRes_S}(\hat{\sigma})(\mathbf{Both}(dp, \Delta)) &= \{\mathbf{None} \ : \ \exists \delta \in \Delta\} \\
&\quad \cup \{\mathbf{Some}(ps) \ : \ \delta \in \mathbf{dom}(dp), \sigma \in \gamma_S(\hat{\sigma} \blacktriangleleft \delta), ps \in \ddot{\gamma}_\sigma(dp(\delta))\} \\
\gamma_{EqRes_S}(\hat{\sigma})(\mathbf{May}(\widehat{ps})) &= \{\mathbf{None}\} \cup \{\mathbf{Some}(ps) \ : \ \sigma \in \gamma_S(\hat{\sigma}), ps \in \ddot{\gamma}_\sigma(\widehat{ps})\}
\end{aligned}$$

Another concretization for $\widehat{Equality}_S$:

$$\begin{aligned}
\gamma_{\widehat{Equality}_S}(\mathbf{May}) &= \{\mathtt{tt},\mathtt{ff}\}\\
\gamma_{\widehat{Equality}_S}(\mathbf{Split}(_,_)) &= \{\mathtt{tt},\mathtt{ff}\}\\
\gamma_{\widehat{Equality}_S}(\mathbf{Equal}) &= \{\mathtt{tt}\}\\
\gamma_{\widehat{Equality}_S}(\mathbf{Unequal}) &= \{\mathtt{ff}\}
\end{aligned}$$

**Theorem 70** (Splitting equality is an exact approximation)**.** $\gamma_{\widehat{Equality}_S} \circ \widehat{tequal}_S = \widehat{tequal}_C \circ \gamma^\sharp$ *provided that external descriptors' equality is an exact approximation.*

*Proof.* Follows the same reasoning as the non-splitting version. □

**Theorem 71** (Splitting equality worthwhile)**.**

$$\begin{aligned}
\widehat{tequalaux}_S(\hat{\varsigma},\hat{\sigma})(\hat{t}_0,\hat{t}_1)(\delta,ps) \in\ & \mathtt{if}\ \tilde{P} \stackrel{?}{=} \emptyset\ \mathtt{then}\\
& \quad \{\mathbf{May}(ps') \ :\ \mathbf{May}(ps') \sqsubseteq_A \widehat{tequalaux}(\hat{\varsigma},\hat{\sigma})(\hat{t}_0,\hat{t}_1)(\delta,\widehat{ps})\}\\
& \mathtt{else}\ \{strength(P) \ :\ P \in \tilde{P}\}
\end{aligned}$$

*where*

$$\begin{aligned}
P &= [\delta' \mapsto \widehat{tequalaux}(\hat{\varsigma},\hat{\sigma})(\hat{t}_0,\hat{t}_1)(\delta',ps) \ :\ \delta' \in Refinements(\hat{\sigma}), \delta \sqsubseteq \delta']\\
\tilde{P} &= \{P|_C \ :\ worthwhile(C, \widehat{equality} \circ P)\}
\end{aligned}$$

*if external descriptors satisfy the same.*

*Proof.* Fix $\hat{\varsigma}, \hat{\sigma}$ and induct on the recursion scheme of $\widehat{tequalaux}_S(\hat{\varsigma},\hat{\sigma})$.

**Case EAddr**$(\hat{a})$**, EAddr**$(\hat{a})$**.**

By cases on $\mu(\hat{a}) \stackrel{?}{\leqslant} 1$

**Case** $\mu(\hat{a}) = 1$**.**

No further refinement necessary. The cut is a singleton of the bottom element, $\delta$.

**Otherwise.**

No refinement possible. Not worthwhile, so **May** is correct.

**Case** *resolvable*, _**.**

If an already refined address, the IH applies. Otherwise, we combine several results.

By cases on the result of the join.

**Case May**$(A')$**.**

Either a term did not have worthwhile cut, or a conflict lead to the jump to **May**. In the first case, we use fruitless extension Lemma 30. If the second case, we use the conflicts are never worthwhile Lemma 29.

**Case Both**$(R, D)$**.**

All choices must lead to strong results that do not conflict. By IH, each individual term has a worthwhile cut. By Lemma 28, their combination is worthwhile. The goal space represents all worthwhile answers.

**Case** $_$*, resolvable,* $_$**.**

Like above.

**Case Variant**$(n, \mathbf{t})$**, Variant**$(n, \mathbf{t}')$**.**

By nested induction on the recursion scheme of $V_S$.

**Case External**$(E, v_0)$**, External**$(E, v_1)$**.**

By assumption.

**Otherwise.**

Structurally unequal, so all $\widehat{tequalaux}$ results are **Unequal**, and the current refinement is ample justification.

□

## 15 WEAK MATCHING PROOFS

Non-refining matching functions is similarly definable. Generalize *worthwhile* to allow **May** and **Equal** to carry arbitrary payloads for the following. The *strength* operation additionally generalizes.

$$\begin{aligned} \mathit{strength}(P) = \mathbf{Both}([\delta' \mapsto \{\rho \ldots\} \;:\; \mathbf{Strongly}(\{\rho \ldots\}) = P(\delta')], \\ \{\delta' \;:\; \mathbf{Unequal} = P(\delta')\}) \end{aligned}$$

[NON-REFINING MATCHING AN EXACT APPROXIMATION] Theorem 31
$\gamma' \circ \hat{M} = M \circ \gamma$ where $\gamma$ is the structural concretization of $\hat{M}$'s inputs, and $\gamma'$ is the concretization of $\mathit{Res}[\widehat{MEnv}]$.

*Proof.* Simple induction following the same reasoning as equality. □

**Lemma 72** (Match answers don't split)**.** *If* $\hat{M}^*_S(ctx)(p, t, \rho)(\delta) = \mathbf{Both}(de, \Delta)$, *then*

- *for all* $\delta' \in \mathbf{dom}(de)$, $\hat{M}^*_S(ctx)(p, t, \rho)(\delta') = \mathbf{Both}([\delta' \mapsto de(\delta')], \emptyset)$
- *for all* $\delta' \in \Delta$, $\hat{M}^*_S(ctx)(p, t, \rho)(\delta') = \mathbf{Both}(\bot, \{\delta'\})$

*and if* $V_{\hat{M}}(ctx)(\overline{p}, \overline{t}, \rho)(\delta) = \mathbf{Both}(de, \Delta)$*, then*

- *for all* $\delta' \in \mathbf{dom}(de)$, $V_{\hat{M}}(ctx)(\overline{p}, \overline{t}, \rho)(\delta') = \mathbf{Both}([\delta' \mapsto de(\delta')], \emptyset)$
- *for all* $\delta' \in \Delta$, $V_{\hat{M}}(ctx)(\overline{p}, \overline{t}, \rho)(\delta') = \mathbf{Both}(\bot, \{\delta'\})$

*if the same holds for external descriptors.*

*Proof.* By induction on the recursion scheme.

**Case** $\mathbf{Name}(x, p), t$**.**

By cases on $x \overset{?}{\in} \mathbf{dom}(\rho)$:

**Case** $x \in \mathbf{dom}(\rho)$**.**

By cases on $\widehat{tequal}_S(ctx)(\rho(x), t, \delta)$:

**Case Equal.**

By IH.

**Case Unequal.**

By definition.

**Case** $\mathbf{Split}(\Delta_=, \Delta_{\neq})$**.**

By cases on the result of the join:

**Case** $\mathbf{Both}(de, \Delta')$**.**

By IH, Lemma 69.

**Case** $\mathbf{May}(U)$**.**

Vacuously true.

**Case May.**

If the result is a failure, use IH. Otherwise the result is **May** and the goal vacuously holds.

**Case** $x \notin \mathbf{dom}(\rho)$**.**

By cases on the result. If **Both**, then the positive answers are separable by refinement by definition of $\sqcup$ on *Refmap*. If **May**, then vacuously true.

**Case** $\mathbf{Wild}, t$**.**

By definition.

**Case** $\mathbf{Is\text{-}Addr}, \mathbf{EAddr}(_)$**.**

By definition.

**Case** $\mathbf{Is\text{-}External}(E), \mathbf{Ex}(E, _)$**.**

By definition.

**Case** $\mathbf{V}(n, \overline{p}), \mathbf{V}(n, \overline{t})$**.**

By induction on $V_{\hat{M}}$'s recursion scheme.

**Case** $\langle\rangle, \langle\rangle$**.**

By definition.

**Case** $p_0\overline{p}, t_0\overline{t}$**.**

If join is **Both**, we appeal to outer IH for $p_0, t_0$, and inner IH for the rest.

**Otherwise.**

By definition.

**Case** $p$, *resolvable*.

Same reasoning as $x \notin \mathbf{dom}(\rho)$ case.

**Otherwise.**

By definition.

□

[CORRECTNESS OF SPLITTING MATCHING] Theorem 32
$\hat{M}^*_S(\hat{\varsigma}, \hat{\sigma})(p, t, \delta, \rho)$ is in

$$\begin{aligned}
&\texttt{if } \tilde{P} \stackrel{?}{=} \emptyset \texttt{ then} \\
&\quad \{\hat{M}(\hat{\varsigma}, \hat{\sigma})(p, t, \rho)(\delta)\} \\
&\texttt{else} \\
&\quad \{\mathbf{Both}([\delta \mapsto U \ : \ P(\delta) = \mathit{return}(\delta, U)], P^{-1}(\mathbf{Fail})) \ : \ P \in \tilde{P}\}
\end{aligned}$$

$$\begin{aligned}
\text{where } P &= [\delta \mapsto \hat{M}(\hat{\varsigma}, \hat{\sigma})(p, t, \rho)(\delta') \ : \ \delta' \in \mathit{Refinements}(\hat{\sigma}), \delta \sqsubseteq \delta'] \\
\tilde{P} &= \{P|_C \ : \ \mathit{worthwhile}'(C, P)\}
\end{aligned}$$

*Proof.* By induction on $\hat{M}^*_S$'s recursion scheme. Cases below are by pattern and term since other arguments are constant.

**Case** $\mathbf{Name}(x, p), t$**.**

By cases on $x \stackrel{?}{\in} \mathbf{dom}(\rho)$:

**Case** $x \in \mathbf{dom}(\rho)$**.**

By cases on $\widehat{\mathit{tequal}}_S(\mathit{ctx})(\rho(x), t, \delta)$:

**Case Equal.**

By IH.

**Case Unequal.**

By definition.

**Case Split**$(\Delta_=, \Delta_{\neq})$**.**

By cases on the join:

**Case Both**$(de', \Delta')$**.**

All choices must lead to strong results that do not conflict. By IH, each individual term has a worthwhile cut. By Lemma 28, their combination is worthwhile. By Lemma 72, the goal space has the expected shape, and represents the worthwhile answers.

**Case May**$(U)$**.**

Either a match did not have worthwhile cut, or a conflict lead to the jump to **May**. In the first case, we use fruitless extension Lemma 30. If the second case, we use the conflicts are never worthwhile Lemma 29.

**Case May.**

By Theorem 71, there is no worthwhile cut to show equality. If the match fails in the recursive call, use IH.

**Case** $x \notin$ **dom**$(\rho)$**.**

Same reasoning as above for joins.

**Case Wild**, $t$**.**

By definition.

**Case Is-Addr**, **EAddr**$(_)$**.**

By definition.

**Case Is-External**$(E)$, **Ex**$(E, _)$**.**

By definition.

**Case V**$(n, \overline{p})$, **V**$(n, \overline{t})$**.**

By nested induction. Reasoning for joins follows previous cases.

**Case** $p$, *resolvable***.**

If an already resolved address, apply IH. Otherwise case split on the result of the join and use above reasoning.

**Otherwise.**

By definition.

□

# 16 WEAK EVALUATION PROOFS

**Lemma 73** (Cut composition)**.** *Let* $\langle P, \sqsubseteq \rangle$ *be a finite poset with a bottom element* $b$*. If* $Cut(C, P)$ *and* $\forall a \in C$ *there is a* $C_a$ *such that* $Cut(C_a, \{c \in P : a \sqsubseteq c\})$*, then* $Cut(\bigsqcup_a C_a, P)$*.*

*Proof.* Let $c \in P$ be arbitrary. We must show there is a $c' \in \bigsqcup_{a} C_a$ such that $c \circledast c'$. Let $d \in C$ be the cut element for $c$. By inversion on $d \circledast c$:

**Case** $d \sqsubseteq c$**.**

$C_d$ cuts the space $\{c_d \in P \ : \ d \sqsubseteq c_d\}$, which $c$ is in. Since $C$ is a cut, $d$ is incomparable to all other elements of $C$. Therefore, $d \in \bigsqcup_{a} C_a$.

**Case** $c \sqsubseteq d$**.**

Every element of $C_d$ is greater than $c$. Choose the largest in $\bigsqcup_{a} C_a$.

□

## 17 DENOTATIONS

FULL IN PREFIX Lemma 34 $F[\![T^\circ]\!]_\rho \subseteq P[\![T^\circ]\!]_\rho$

*Proof.* By induction on $T^\circ$. □

PREFIX CLOSED Theorem 33 $prefixes(P[\![T^\circ]\!]_\rho) = P[\![T^\circ]\!]_\rho$

*Proof.* By induction on $T^\circ$:

**Case** $A$.

Only traces are matching actions and $\epsilon$. Holds by definition.

**Case** $!A$.

Only traces are non-matching actions and $\epsilon$. Holds by definition.

**Case** $\epsilon$.

Holds by definition.

**Case** $\neg T^\circ$.

$P[\![\neg T^\circ]\!]_\rho = \neg F[\![T^\circ]\!]_\rho$.
We prove a generalized property that $prefixes(\neg(\Pi)) = \neg(\Pi)$: Let $\pi \in \neg(\Pi)$ and $\pi' \in Trace$ be arbitrary such that $\pi' \leqslant \pi$. We must show that $\pi' \in \neg(\Pi)$. If $\pi = \epsilon$, then $\pi' = \epsilon$ and we're done. By definition, there is no prefix of $\pi$ in $\Pi \setminus \{\epsilon\}$. Thus, since $\pi'$ is a prefix of $\pi$, it is not in $\Pi \setminus \epsilon$ and must therefore be in $\neg(\Pi)$.

**Case** $T^\circ \cdot T^\circ$.

Let $\pi \in P[\![T^\circ_0 \cdot T^\circ_1]\!]_\rho$ and $\pi' \leqslant \pi$ be arbitrary. If $\pi \in P[\![T^\circ_0]\!]_\rho$ then by IH, we're done. Otherwise, $\pi \equiv \pi_0\pi_1$ where $\pi_0 \in F[\![T^\circ_0]\!]_\rho$ and $\pi_1 \in P[\![T^\circ_1]\!]_\rho$ if $\pi' \leqslant \pi_0$, then by Lemma 34 and IH, we're done. Otherwise, $\pi' \equiv \pi_0\pi'_1$, and by IH, $\pi'_1 \leqslant \pi_1$, and we're done.

**Case** $T^{\circ *}$.

By IH.

**Case** $\cup\tilde{T}^\circ$.

By IH.

**Case** $\cap\tilde{T}^\circ$.

By IH.

**Case** $\langle A \rangle\ T^{\circ}$**.**

By IH and simple cases on empty and singleton traces.

□

## 18 DERIVATIVES

THEOREM 35 The following are mutually true

1. $F[\![\partial_d^{\rho} T^{\circ}]\!] = \{\pi \ : \ d\pi \in F[\![T^{\circ}]\!]_{\rho}\}$
2. $P[\![\partial_d^{\rho} T^{\circ}]\!] = \{\pi \ : \ d\pi \in P[\![T^{\circ}]\!]_{\rho}\}$
3. $F[\![\partial_d T]\!] = \{\pi \ : \ d\pi \in F[\![T]\!]\}$
4. $P[\![\partial_d T]\!] = \{\pi \ : \ d\pi \in P[\![T]\!]\}$

*Proof.* By mutual induction on the functional schemes, equational reasoning and Lemma 36. □

```haskell
{-# LANGUAGE FlexibleInstances, MultiParamTypeClasses,
          UndecidableInstances, GeneralizedNewtypeDeriving,
          NoMonomorphismRestriction, GADTs, KindSignatures,
          RankNTypes, ConstraintKinds #-}
import Data.Map hiding (fold)
import Data.Set hiding (fold)
import Data.Maybe
import Control.Monad.State
import Control.Monad.ConstrainedNormal
import qualified Data.Functor.Identity as Fid
import Test.HUnit hiding (State)
import qualified Data.Map as Map
import qualified Data.Set as Set

-- Monad magic

newtype MaybeT m a = MaybeT { runMaybeT :: m (Maybe a) }
bindMT :: (Monad m) => (MaybeT m a) -> (a -> MaybeT m b) -> MaybeT m b
bindMT x f = MaybeT $ runMaybeT x >>= maybe (return Nothing) (runMaybeT . f)

returnMT :: (Monad m) => a -> MaybeT m a
returnMT a = MaybeT $ return (Just a)

failMT :: (Monad m) => t -> MaybeT m a
failMT _ = MaybeT $ return Nothing

instance (Monad m) => Monad (MaybeT m) where
  return = returnMT
  (>>=) = bindMT
  fail = failMT

instance MonadTrans MaybeT where
  lift m = MaybeT (Just ‘liftM‘ m)

instance (MonadState s m) => MonadState s (MaybeT m) where
  get = lift get
  put k = lift (put k)

newtype MState a b = MS {
  runMS :: MaybeT (State a) b
  } deriving (Monad, MonadState a)

-- Concrete stuff

type Name = String

data AddrC = AddrC (String, [Int]) deriving (Eq,Ord,Show)
newtype AddrA = AddrA String deriving (Eq,Ord,Show)

-- Lookup mode
data LM = Deref | Delay | Resolve deriving (Eq,Ord,Show)
```

```
-- Match mode
data MM = Explicit | Implicit LM deriving (Eq,Ord,Show)
-- Equality mode
data EM = Structural | Identity deriving (Eq,Ord,Show)

-- A concrete term is a variant, a map, a qualified address or a delayed lookup
data TermC =
  VC Name [TermC]
  | QC AddrC MM EM
  | DC AddrC
    deriving (Eq, Ord, Show)

type Store = Map AddrC TermC

-- test data
ac0 = AddrC ("addr", [])
nilc = VC "nil" []
tc0 = VC "cons" [nilc, nilc]
tc1 = VC "mumble" [tc0, nilc]
s0 = Map.insert ac0 tc0 Map.empty

----------------------------------------
-- Concrete term equality
----------------------------------------
type Pairs = Set (TermC, TermC)
type EqResC = Maybe Pairs
tequalC :: Store -> TermC -> TermC -> Bool
tequalC s t0 t1 = case coindC s t0 t1 Set.empty of
              Just _ -> True
              Nothing -> False
-- The ps variable is the math's 'A'
coind :: (Functor f, Ord a) =>
       (Set (a,a) -> f (Set (a,a))) ->
       (a -> a -> Set (a,a) -> f (Set (a,a))) ->
       a -> a -> Set (a,a) -> f (Set (a,a))
coind ret f t0 t1 ps = if Set.member (t0,t1) ps then
                  ret ps
                else f t0 t1 (Set.insert (t0,t1) ps)
coindC s = coind Just (tequalauxC s)

tequalauxC :: Store -> TermC -> TermC -> Pairs -> EqResC
tequalauxC s (QC ac0 _ Identity) (QC ac1 _ Identity) ps | ac0 == ac1 = Just ps
-- Qualified and delayed terms just lookup
tequalauxC s (QC ac _ Structural) t1 ps = coindC s (s ! ac) t1 ps
tequalauxC s t0 (QC ac _ Structural) ps = coindC s t0 (s ! ac) ps
tequalauxC s (DC ac) t1 ps = coindC s (s ! ac) t1 ps
tequalauxC s t0 (DC ac) ps = coindC s t0 (s ! ac) ps
-- variants compared pointwise with eqvc
tequalauxC s (VC n0 ts0) (VC n1 ts1) ps | n0 == n1 = eqvc s ts0 ts1 ps
tequalauxC s _ _ ps = Nothing

equalTest0 = TestCase (assertBool "Reflexivity" (tequalC s0 tc0 tc0))
equalTest1 = TestCase (assertBool "Different" (not (tequalC s0 tc0 tc1)))

-- Are two lists of terms equal?
eqvc :: Store -> [TermC] -> [TermC] -> Pairs -> EqResC
eqvc s [] [] ps = Just ps
```

```
eqvc s (t0:ts0) (t1:ts1) ps = coindC s t0 t1 ps >>= eqvc s ts0 ts1
eqvc s _ _ ps = Nothing

----------------------------------------
-- Abstract terms
----------------------------------------

data Flat a = FTop | FVal a deriving (Eq,Ord,Show)

type Ctx t = (StateHat t, StoreHat t)

type AbsEq t = Ctx t -> ExtVal -> ExtVal -> EqResAM t
type SplitEq t = Ctx t -> ExtVal -> ExtVal
                -> EqResM t (ATerm t, ATerm t)
type ExtLess t = Ctx t -> ExtVal -> ExtVal -> LessRes t
type ExtJoin t = Ctx t -> ExtVal -> ExtVal -> JoinRes t ExtVal

-- shouldn't really derive for these, but this is a placeholder.
data ExtVal = EString (Flat String) | ENumber (Flat Int) deriving (Eq,Ord,Show)
data ExternalDescriptor t = ExternalDescriptor {
  name :: Name,
  equivA :: AbsEq t,
  equivS :: SplitEq t,
  less :: ExtLess t,
  ejoin :: ExtJoin t
  }
instance Eq (ExternalDescriptor t) where
  ExternalDescriptor {name=n} == ExternalDescriptor {name=n'} = n == n'
instance Ord (ExternalDescriptor t) where
  ExternalDescriptor {name=n} <= ExternalDescriptor {name=n'} = n <= n'
instance Show (ExternalDescriptor t) where
  show (ExternalDescriptor {name=n}) = "External: " ++ show n

stringEquiv :: AbsEq t
stringEquiv s (EString es) (EString es') = eq es es'
  where eq FTop _ d ps = May ps
        eq _ FTop d ps = May ps
        eq (FVal v) (FVal v') d ps = if v == v' then
                                       Equal (Map.singleton d ps)
                                     else
                                       Unequal (Set.singleton d)
stringEquiv s _ _ = \ d _ -> Unequal (Set.singleton d)

stringEquivS :: SplitEq t
stringEquivS s (EString es) (EString es') = eq es es'
  where eq FTop _ d ps = MayS ps
        eq _ FTop d ps = MayS ps
        eq (FVal v) (FVal v') d ps = if v == v' then
                                        Both (Map.singleton d ps) Set.empty
                                      else
                                        Both Map.empty (Set.singleton d)
stringEquivS s _ _ = \ d _ -> Both Map.empty (Set.singleton d)

fless :: (Eq a) => Flat a -> Flat a -> Bool
fless _ FTop = True
fless (FVal v) (FVal v') | v == v' = True
fless _ _ = False
```

```
fjoin :: (Eq a) => Flat a -> Flat a -> Flat a
fjoin fv@(FVal v) (FVal v') | v == v' = fv
fjoin _ _ = FTop

stringLess :: ExtLess t
stringLess s (EString es) (EString es') = if fless es es' then
                                    return ()
                                  else MS $ failMT ()

stringJoin :: ExtJoin t
stringJoin s (EString es) (EString es') = return $ EString $ fjoin es es'
stringJoin _ _ _ = error "Expected strings"

stringExt :: ExternalDescriptor t
stringExt = ExternalDescriptor {
  name = "String",
  equivA = stringEquiv,
  equivS = stringEquivS,
  less = stringLess,
  ejoin = stringJoin }

data STerm t =
  VA Name [TermA t]
  | QA AddrA MM EM
    deriving (Eq,Ord,Show)

data PreTerm t =
  STerm (STerm t)
  | EA (ExternalDescriptor t) ExtVal
  | DA AddrA
    deriving (Eq, Ord, Show)

data ATerm t =
  ASTerm (STerm t)
  | AEA (ExternalDescriptor t) ExtVal
    deriving (Eq, Ord, Show)

type ExtMap t = Map (ExternalDescriptor t) ExtVal

type AbsTerm t = (Set (STerm t), ExtMap t)
data TermA t = PreTerm (PreTerm t)
             | TAbs (AbsTerm t) deriving (Eq,Ord,Show)
type MapA t = Map (TermA t) (TermA t)

termbot :: TermA t
termbot = TAbs (Set.empty, Map.empty)

type HeapHat t = Map AddrA (AbsTerm t)
data StoreHat t = StoreHat (HeapHat t) Count deriving (Eq,Ord,Show)
type PairsHat t = Set (ATerm t, ATerm t)

-- test data
aa0 = AddrA "addr"
nila = VA "nil" []
unita = VA "unit" []
tnila = PreTerm $ STerm nila
```

```
tunita = PreTerm $ STerm unita
tblur0 = VA "blur" [TAbs (Set.insert nila (Set.singleton unita), Map.empty)]
tblur1 = VA "blur" [PreTerm $ STerm nila]

ma0 = Map.singleton (PreTerm (DA aa0)) tnila
ma1 = Map.insert (PreTerm (STerm tblur1)) tnila $
     Map.singleton (PreTerm (DA aa0)) tnila


----------------------------------------
-- TermA <= TermA
type LessRes t = MState (PairsHat t) ()

unwrapLess :: LessRes t -> (PairsHat t) -> Bool
unwrapLess lr ps = case runStateT (runMaybeT $ runMS lr) ps of
  Fid.Identity (Just _,_) -> True
  Fid.Identity (Nothing,_) -> False
-- Without an ML functor, all these functions take a StoreHat first. Ugh.

mkAbsTerm :: StoreHat t -> TermA t -> AbsTerm t
mkAbsTerm (StoreHat h _) (PreTerm (DA a)) = h ! a
mkAbsTerm s (PreTerm (EA ed v)) = (Set.empty, Map.singleton ed v)
mkAbsTerm s (PreTerm (STerm st)) = (Set.singleton st, Map.empty)
mkAbsTerm s (TAbs abs) = abs

-- Top level entry into term ordering.
termLess :: Ctx t -> TermA t -> TermA t -> Bool
termLess c@(_,s) t0 t1 = unwrapLess
                   (absTermLess c (mkAbsTerm s t0) (mkAbsTerm s t1))
                   Set.empty

absTermLess :: Ctx t -> AbsTerm t -> AbsTerm t -> LessRes t
absTermLess c (sts,es) (sts',es') = do stermsAllLess c sts sts'
                                 extsAllLess c es es'

extsAllLess :: Ctx t -> ExtMap t -> ExtMap t -> LessRes t
extsAllLess c es es' = Map.foldlWithKey (\ res ed v ->
                                 do res
                                   maybe
                                     (MS $ failMT ())
                                     (less ed c v)
                                     (Map.lookup ed es'))
                  (return ()) es

termAless :: Ctx t -> TermA t -> TermA t -> LessRes t
termAless c@(_,s) t0 t1 = absTermLess c (mkAbsTerm s t0) (mkAbsTerm s t1)

-- STerm <= STerm
stermLess :: Ctx t -> STerm t -> STerm t -> LessRes t
stermLess c@(_,s@(StoreHat h _)) st0 st1 =
  do ps <- get
     let pair = (ASTerm st0, ASTerm st1) in
       if Set.member pair ps then
         return ()
       else do put (Set.insert pair ps)
               case (st0,st1) of
                 (VA n0 ts0, VA n1 ts1) ->
```

```
              if n0 == n1 then
                foldM (\ res (t0,t1) -> do { termAless c t0 t1})
                ()
                (zip ts0 ts1)
              else MS $ failMT ()
             (QA _ _ Identity, QA _ _ Identity) -> if st0 == st1 then
                                                   return ()
                                                  else MS $ failMT ()
             (_, QA a _ Structural) -> stermOneLess c st0 (fst $ h ! a)
             (QA a _ Structural, _) -> absTermLessSTerm c (h ! a) st1
             (_,_) -> MS $ failMT ()

atermLessTop :: Ctx t -> ATerm t -> ATerm t -> Bool
atermLessTop s a0 a1 = unwrapLess (atermLess s a0 a1) Set.empty

atermLess :: Ctx t -> ATerm t -> ATerm t -> LessRes t
atermLess c (ASTerm st) (ASTerm st') = stermLess c st st'
atermLess c (AEA ed v) (AEA ed' v') | ed == ed' = less ed c v v'
atermLess c _ _ = MS $ failMT ()

absTermLessSTerm :: Ctx t -> AbsTerm t -> STerm t -> LessRes t
absTermLessSTerm c (sts,es) st = if Map.null es then
                                  stermAllLess c sts st
                                 else MS $ failMT ()

-- all t' in (Set STerm). t' <= t
stermAllLess :: Ctx t -> Set (STerm t) -> STerm t -> LessRes t
stermAllLess c ts t = Set.foldl (\ lr t' ->
                         do {lr; stermLess c t' t}) (return ()) ts
-- there is a t' in (Set STerm). t <= t'
stermOneLess :: Ctx t -> STerm t -> Set (STerm t) -> LessRes t
stermOneLess c t sts = do ps <- get
                    Set.foldl (\ lr t' ->
                             if unwrapLess lr ps then
                               lr
                              else stermLess c t t')
                     (MS $ failMT ()) sts
-- all t in ts0. there is a t' in ts1. t <= t'
stermsAllLess :: Ctx t -> Set (STerm t) -> Set (STerm t) -> LessRes t
stermsAllLess c ts0 ts1 = Set.foldl (\ lr t0 -> do {lr; stermOneLess c t0 ts1})
                    (return ()) ts1

----------------------------------------
-- End TermA <= TermA
----------------------------------------

----------------------------------------
-- TermA join TermA
----------------------------------------

--- XXX: joining goes through structure, so we need pairs!
-- JoinRes can't use state since we do backtracking search

type JoinRes t a = State (PairsHat t) a

joinTermTop :: Ctx t -> TermA t -> TermA t -> TermA t
joinTermTop c t0 t1 = case runStateT (joinTerm c t0 t1) Set.empty of
```

```
  Fid.Identity (t,_) -> t

joinTerm :: Ctx t -> TermA t -> TermA t -> JoinRes t (TermA t)
joinTerm c@(_,s) t0 t1 = joinAbsTerm c (mkAbsTerm s t0) (mkAbsTerm s t1)

joinAbsTerm :: Ctx t -> AbsTerm t -> AbsTerm t -> JoinRes t (TermA t)
joinAbsTerm c (sts,es) (sts',es') =
  do ps <- get
     case runStateT (joinSTermsSTerms c sts sts') ps of
       Fid.Identity (t,ps') ->
         do es'' <- joinExts c es es'
            case t of
              TAbs (sts'', _) -> return $ TAbs (sts'', es'')
              -- INVARIANT: pt can only be an STerm
              PreTerm (STerm st) -> return $ if Map.null es'' then
                                      t
                                    else TAbs (Set.singleton st, es'')
              _ -> error ("Uh oh " ++ show t)

joinAbsTermTerm :: Ctx t -> AbsTerm t -> TermA t -> JoinRes t (TermA t)
joinAbsTermTerm c@(_,s) abs t = joinAbsTerm c abs (mkAbsTerm s t)

joinExts :: Ctx t -> ExtMap t -> ExtMap t -> JoinRes t (ExtMap t)
joinExts c es es' = Map.foldlWithKey
               (\ jr ed v ->
                 do es'' <- jr
                    case Map.lookup ed es' of
                      Just v' -> do v'' <- ejoin ed c v v'
                                    return $ Map.insert ed v'' es''
                      Nothing -> return $ Map.insert ed v es'')
               (return es') es

joinSTermsSTerms :: Ctx t -> (Set (STerm t)) -> (Set (STerm t))
               -> JoinRes t (TermA t)
joinSTermsSTerms c sts0 sts1 = Set.foldl
                       (\ jr st0 ->
                         do t <- jr
                            case t of
                              TAbs abs -> joinSTermSTerms c st0 abs
                              _ -> error "Shouldn't be a non TAbs")
                       (return $ TAbs (sts1, Map.empty)) sts0
unwrapjoin jr ps = let (Fid.Identity j) = runStateT jr ps in j

-- If any sterms in sts join to a non-tabs, then replace that term
-- with the result. Otherwise, just add st to sts
joinSTermSTerms :: Ctx t -> STerm t -> AbsTerm t -> JoinRes t (TermA t)
joinSTermSTerms c st (sts,es) =
  do ps <- get
     case unwrapjoin (findJoin c st sts) ps of
       (Left sts',ps') -> do put ps'
                             return $ TAbs (sts',es)
       (Right sts',ps') -> do put ps'
                              return $ TAbs (Set.insert st sts', es)

-- Left = set with structurally joined terms
-- Right = sts rebuilt.
findJoin :: Ctx t -> STerm t -> Set (STerm t)
```

```
         -> JoinRes t (Either (Set (STerm t)) (Set (STerm t)))
findJoin c st sts =
  do ps <- get
     Set.foldl (help ps) (return $ Right Set.empty) sts
  where help ps jeither st' =
          do either <- jeither
             case either of
               Left sts' -> return $ Left (Set.insert st' sts')
               Right sts' -> case unwrapjoin (joinSTerm c st st') ps of
                 (TAbs _,_) -> return $ Right (Set.insert st' sts')
                 (PreTerm (STerm stj),ps') ->
                   do put ps'
                      return $ Left (Set.insert stj sts')
                 _ -> error "Bad join"

threadMap :: (a -> b -> (b, c)) -> b -> [a] -> (b,[c])
threadMap f acc [] = (acc,[])
threadMap f acc (x:xs) = let (acc',b) = f x acc in
  let (acc'',lst) = threadMap f acc' xs in
  (acc',b:lst)

-- Combine structurally if we can.
joinSTerm :: Ctx t -> STerm t -> STerm t -> JoinRes t (TermA t)
joinSTerm c@(_,s@(StoreHat h _)) st0 st1 =
  do ps <- get
     let pair = (ASTerm st0,ASTerm st1) in
      if Set.member pair ps then
       return $ twoSTerms st0 st1
      else do put $ Set.insert pair ps;
              case (st0,st1) of
                ((VA n0 ts0), (VA n1 ts1)) ->
                  if n0 == n1 && length ts0 == length ts1 then
                    do js <- zipWithM (joinTerm c) ts0 ts1
                       return $ PreTerm $ STerm $ VA n0 js
                  else return $ twoSTerms st0 st1
                ((QA _ _ Identity),(QA _ _ Identity)) ->
                  if st0 == st1 then
                    return $ PreTerm $ STerm $ st0
                  else return $ twoSTerms st0 st1
                (_, QA a _ Structural) ->
                  if st0 == st1 then
                    return $ PreTerm $ STerm $ st0
                  else joinAbsTermTerm c (h ! a) (PreTerm $ STerm st0)
                ((QA a _ Structural), _) -> joinSTermSTerms c st1 (h ! a)
                _ -> return $ twoSTerms st0 st1

twoSTerms :: STerm t -> STerm t -> TermA t
twoSTerms t0 t1 = TAbs (Set.insert t0 (Set.singleton t1), Map.empty)

-----------------------------------------
-- Term equality without splitting
-----------------------------------------
type Refinement t = (Map AddrA (ATerm t))
type Refinements t = Set (Refinement t)

-- C0 is just unmapped
data NatHat = C1 | Cinf deriving (Eq,Ord,Show)
```

```
type Count = Map AddrA NatHat

------------------------------------------------
-- Equality result type with join, return, bind
------------------------------------------------
data EqResA t = Unequal (Refinements t)
           | Equal (Map (Refinement t) (PairsHat t))
           | May (PairsHat t) deriving (Show)
type EqResAM t = Refinement t -> PairsHat t -> EqResA t

squash :: (Ord b) => (Map a (Set b)) -> (Set b)
squash dps = Set.unions $ Map.elems dps

joinA :: EqResA t -> EqResA t -> EqResA t
joinA (Equal dps) (Unequal _) = May $ squash dps
joinA (Equal dps) (Equal dps') = Equal (Map.unionWith Set.union dps dps')
joinA eq@(May ps) (Unequal _) = eq
joinA (May ps) (Equal dps) = May $ Set.union ps $ squash dps
joinA (May ps) (May ps') = May $ Set.union ps ps'
joinA (Unequal ps) (Unequal ps') = Unequal $ Set.union ps ps'
-- symmetric cases
joinA (Unequal _) (Equal dps) = May $ squash dps
joinA (Unequal _) eq@(May ps) = eq
joinA (Equal dps) (May ps) = May $ Set.union ps $ squash dps

returnA :: Refinement t -> PairsHat t -> EqResA t
returnA d ps = Equal (Map.singleton d ps)

failA d = Unequal (Set.singleton d)

weakenA :: EqResA t -> EqResA t
weakenA (Equal dps) = May $ squash dps
weakenA eq = eq

seqA :: EqResAM t -> EqResAM t -> EqResAM t
seqA r f = \ d ps -> case r d ps of
  eq@(Unequal _) -> eq
  May ps -> weakenA (f d ps)
  Equal dps -> case (Map.toList dps) of
    [] -> (f d Set.empty)
    (d,ps):rps' -> Prelude.foldl
                (\ eq (d',ps') -> joinA (f d' ps') eq)
                (f d ps) rps'

--------------------------
-- Abstract Term Equality
--------------------------
tequalauxA :: Ctx t -> TermA t -> TermA t -> EqResAM t
tequalauxA c@(_,s) t0 t1 = tequalAbsTermA c (mkAbsTerm s t0) (mkAbsTerm s t1)

tequalAbsTermA :: Ctx t -> AbsTerm t -> AbsTerm t -> EqResAM t
tequalAbsTermA c (sts,es) (sts',es') = seqA (stermsEqualA c sts sts') $
  extsEqualA c es es'

-- all st in sts, st' in sts'. st = st'
stermsEqualA :: Ctx t -> Set (STerm t) -> Set (STerm t) -> EqResAM t
stermsEqualA c sts sts' = Set.foldl (\ res st ->
```

```
                                   seqA res $ stermsAllEqualA c st sts)
                        returnA sts

stermsAllEqualA :: Ctx t -> STerm t -> Set (STerm t) -> EqResAM t
stermsAllEqualA c st sts = Set.foldl (\ res st' ->
                                    seqA res $ stermequal c st st')
                         returnA sts

-- all ed in dom(es). es(ed) = es'(ed)
-- and all ed' in dom(es') \ dom(es). es'(ed') =
extsEqualA :: Ctx t -> ExtMap t -> ExtMap t -> EqResAM t
extsEqualA c es es' = seqA (extsContainedA c es es')
                    (\ d ps -> if Map.keysSet es == Map.keysSet es' then
                                returnA d ps
                              else failA d)

extsContainedA :: Ctx t -> ExtMap t -> ExtMap t -> EqResAM t
extsContainedA c es es' = Map.foldlWithKey (\ res ed v ->
                                         seqA res $ \ d ps ->
                                         case Map.lookup ed es' of
                                           Just v' -> equivA ed c v v' d ps
                                           Nothing -> failA d)
                        returnA es


atermequalA :: Ctx t -> ATerm t -> ATerm t -> EqResAM t
atermequalA c a0 a1 d ps
  | Set.member pair ps = returnA d ps
  | otherwise = case (a0,a1) of
    (ASTerm s0, ASTerm s1) -> stermequal c s0 s1 d ps'
    (AEA ed v, AEA ed' v') -> if ed == ed' then
                               equivA ed c v v' d ps'
                             else
                               failA d
    _ -> failA d
  where pair = (a0,a1)
        ps' = Set.insert pair ps

stermequal :: Ctx t -> STerm t -> STerm t -> EqResAM t
stermequal c@(_,s@(StoreHat _ cnt)) st0 st1 d ps
  | Set.member pair ps = returnA d ps
  | otherwise =
    case (st0,st1) of
      (QA a0 _ Identity, QA a1 _ Identity) -> if a0 == a1 then
                                               case (cnt ! a0) of
                                                 Cinf -> May ps
                                                 _ -> returnA d ps
                                             else bad
      (VA n0 ts0, VA n1 ts1) -> if n0 == n1 then
                                  eqva c ts0 ts1 d ps
                                else bad
      (QA a0 _ Structural, _) -> (withResolveAddrA c a0 $ \ at0 ->
                                  atermequalA c at0 (ASTerm st1)) d ps
      (_, QA a1 _ Structural) -> (withResolveAddrA c a1 $ \ at1 ->
                                  atermequalA c at1 (ASTerm st0)) d ps
      _ -> bad
  where pair = (ASTerm st0, ASTerm st1)
```

```
      ps' = Set.insert pair ps
      bad = Unequal (Set.singleton d)

withResolveAddrA :: Ctx t -> AddrA -> (ATerm t -> EqResAM t) -> EqResAM t
withResolveAddrA c@(_,(StoreHat h cnt)) a f = \ d ps ->
  case Map.lookup a d of
   Just st -> f st d ps
   _ -> case Map.lookup a cnt of
     Just Cinf -> let (sts,es) = (h ! a) in
      let persts = Set.foldl (\ acc st ->
                         seqA acc (f $ ASTerm st)) returnA sts in
      Map.foldlWithKey (\ acc ed v ->
                    seqA acc (f $ AEA ed v)) persts es d ps
     _ -> let (sts,es) = h ! a in
      let persts = Set.foldl
                (\ acc st ->
                 seqA acc $ \ d' ->
                 f (ASTerm st) (Map.insert a (ASTerm st) d'))
                returnA sts in
      Map.foldlWithKey (\ acc ed v -> seqA acc $ \ d' ->
                    f (AEA ed v) (Map.insert a (AEA ed v) d'))
      persts es d ps

eqva :: Ctx t -> [TermA t] -> [TermA t] -> EqResAM t
eqva s [] [] = returnA
eqva s (t0:ts0) (t1:ts1) = seqA (tequalauxA s t0 t1)
                    (eqva s ts0 ts1)
eqva _ _ _ = \ d ps -> failA d

-------------------------------------------------
-- End abstract term equality without splitting
-------------------------------------------------

-----------------------------------------------------
-- Equality result with join for splitting equality
-----------------------------------------------------

-- decide all a in dom(d). d(a) <= d'(a)
refinementLess :: Ctx t -> Refinement t -> Refinement t -> Bool
refinementLess s d d' = Map.foldWithKey (\ addr t acc ->
                              acc &&
                              case Map.lookup addr d' of
                               Just t' -> atermLessTop s t t'
                               Nothing -> False)
                  True d

strictlyLessInSet s d ds =
  Set.fold (\ d' acc ->
         acc || (refinementLess s d d' && (not $ refinementLess s d' d)))
  False ds
-- Is d strictly less than some d' in dom(dps)? If so, return dps(d')
strictlyLessInKeys s d dps =
  Map.foldWithKey (\ d' ps acc ->
              if isJust acc then
               acc
              else if (refinementLess s d d' &&
                     (not $ refinementLess s d' d)) then
```

```
                     Just ps
                   else Nothing)
  Nothing dps

overlap s d d' = refinementLess s d d' || refinementLess s d' d
overlapInSet s d ds = Set.fold (\ d' acc -> acc || overlap s d d') False ds
overlapInMap s d dps = Map.foldWithKey (\ d' _ acc -> acc || overlap s d d')
                 False dps
bigoverlap s dps ds = Map.foldWithKey (\ d _ acc -> acc || overlapInSet s d ds)
                False dps

data EqResS t a = Both (Map (Refinement t) (Set a))
                 (Refinements t)
               | MayS (Set a) deriving (Show)
type EqResM t a = Refinement t -> (Set a) -> EqResS t a

joinS :: (Ord a) => Ctx t -> EqResS t a -> EqResS t a -> EqResS t a
joinS s (Both dps _) (MayS ps) = MayS $ Set.union ps $ squash dps
joinS s (MayS ps) (MayS ps') = MayS $ Set.union ps ps'
joinS s (Both dps ds) (Both dps' ds') =
  if bigoverlap s dps ds' || bigoverlap s dps' ds then
    MayS $ Set.union (squash dps) (squash dps')
  else
    Both (mergeKeysStrictlySmaller s dps dps')
    (joinCut s ds ds')
-- symmetric cases
joinS s (MayS ps) (Both dps _) = MayS $ Set.union ps $ squash dps

joinCut s ds ds' = removeAllStrictlySmaller s (Set.union ds ds')
-- Keep only the maximal refinements in a set.
-- { d : all d' in ds. overlap s d d' => refinementLess s d' d }
removeAllStrictlySmaller s ds = Set.fold (\ d acc ->
                                  if strictlyLessInSet s d ds then
                                    acc
                                  else Set.insert d acc) Set.empty ds

-- Keep only the maximal refinements in an equality justification,
-- but union togetherthe term pairs of comparable refinements.
-- Maintain invariant that domains are incomparable.
mergeKeysStrictlySmaller s dps dps' = combineSmaller s keyCut
                              (Map.unionWith Set.union dps dps')
  where keyCut = joinCut s (Map.keysSet dps) (Map.keysSet dps')
combineSmaller s ks dps = Set.fold
                    (\ d acc ->
                      case strictlyLessInKeys s d dps of
                        Just ps' ->
                          Map.insert d (Set.union ps' $
                                fromJust $ Map.lookup d dps) acc
                        Nothing -> Map.insert d
                              (fromJust $ Map.lookup d dps) acc)
                    Map.empty ks

weakenS :: (Ord a) => EqResS t a -> EqResS t a
weakenS eq@(Both dps ds) = if Map.null dps then eq else MayS $ squash dps
weakenS eq = eq

failS d = Both Map.empty (Set.singleton d)
```

```
badS ds = Both Map.empty ds

returnS :: Refinement t -> (Set a) -> EqResS t a
returnS d ps = Both (Map.singleton d ps) Set.empty

mustS dps = Both dps Set.empty

bindSM :: (Ord a) => Ctx t -> EqResM t a -> EqResM t a -> EqResM t a
bindSM c r f = \ d ps ->
  case r d ps of
   MayS ps -> weakenS $ f d ps
   eq@(Both dps ds) ->
     case Map.toList dps of
       [] -> eq -- no equalities. Stay failed.
       ((d,ps):rps) -> joinS c (badS ds)
                    $ Prelude.foldl (\ acc (d',ps') ->
                                   joinS c acc $ f d' ps')
                    (f d ps) rps

withResolveAddr :: (Ord a) => Ctx t -> AddrA -> (ATerm t -> EqResM t a)
              -> EqResM t a
withResolveAddr c@(_,(StoreHat h cnt)) a f = \ d ps ->
  case Map.lookup a d of
   Just st -> f st d ps
   _ -> case Map.lookup a cnt of
     Just Cinf -> let (sts,es) = (h ! a) in
       let persts = Set.foldl (\ acc st ->
                           bindSM c acc (f $ ASTerm st)) returnS sts in
       Map.foldlWithKey (\ acc ed v ->
                     bindSM c acc (f $ AEA ed v)) persts es d ps
     _ -> let (sts,es) = h ! a in
       let persts = Set.foldl
                  (\ acc st ->
                   bindSM c acc $ \ d' ->
                   f (ASTerm st) (Map.insert a (ASTerm st) d'))
                  returnS sts in
       Map.foldlWithKey (\ acc ed v -> bindSM c acc $ \ d' ->
                     f (AEA ed v) (Map.insert a (AEA ed v) d'))
       persts es d ps

-----------------------------------------
-- Abstract term equality with splitting
-----------------------------------------
data EqResTopS t =
  EqualS | UnequalS | MayTS | EqSplit (Refinements t) (Refinements t)
tequalS :: Ctx t -> Refinement t -> TermA t
        -> TermA t -> EqResTopS t
tequalS c d t0 t1 = case tequalauxS c t0 t1 d Set.empty of
  Both de ds -> if Map.null de then
              UnequalS
            else if Set.null ds then
                  EqualS
                else EqSplit (Map.keysSet de) ds
  MayS _ -> MayTS

tequalauxS :: Ctx t -> TermA t -> TermA t
          -> EqResM t (ATerm t, ATerm t)
```

```
tequalauxS c@(_,s) t0 t1 = tequalAbsTermS c (mkAbsTerm s t0) (mkAbsTerm s t1)

tequalAbsTermS :: Ctx t -> AbsTerm t -> AbsTerm t
              -> EqResM t (ATerm t, ATerm t)
tequalAbsTermS c (sts,es) (sts',es') = bindSM c (stermsEqualS c sts sts') $
  extsEqualS c es es'

-- all st in sts, st' in sts'. st = st'
stermsEqualS :: Ctx t -> (Set (STerm t)) -> (Set (STerm t))
            -> EqResM t (ATerm t, ATerm t)
stermsEqualS c sts sts' = Set.foldl (\ res st ->
                            bindSM c res $ stermsAllEqualS c st sts)
                   returnS sts

stermsAllEqualS :: Ctx t -> STerm t -> Set (STerm t)
              -> EqResM t (ATerm t, ATerm t)
stermsAllEqualS c st sts = Set.foldl (\ res st' ->
                             bindSM c res $ stermequalS c st st')
                    returnS sts

-- all ed in dom(es). es(ed) = es'(ed)
-- and all ed' in dom(es') \ dom(es). es'(ed') =
extsEqualS :: Ctx t -> ExtMap t -> ExtMap t -> EqResM t (ATerm t, ATerm t)
extsEqualS c es es' = bindSM c (extsContainedS c es es')
                (\ d ps -> if Map.keysSet es == Map.keysSet es' then
                          returnS d ps
                        else failS d)

extsContainedS :: Ctx t -> ExtMap t -> ExtMap t
              -> EqResM t (ATerm t, ATerm t)
extsContainedS c es es' = Map.foldlWithKey (\ res ed v ->
                                  bindSM c res $ \ d ps ->
                                  case Map.lookup ed es' of
                                    Just v' -> equivS ed c v v' d ps
                                    Nothing -> failS d)
                   returnS es


atermequalS :: Ctx t -> ATerm t -> ATerm t -> EqResM t (ATerm t, ATerm t)
atermequalS c a0 a1 d ps
  | Set.member pair ps = returnS d ps
  | otherwise = case (a0,a1) of
    (ASTerm s0, ASTerm s1) -> stermequalS c s0 s1 d ps'
    (AEA ed v, AEA ed' v') -> if ed == ed' then
                          equivS ed c v v' d ps'
                        else
                          failS d
    _ -> failS d
  where pair = (a0,a1)
     ps' = Set.insert pair ps

stermequalS :: Ctx t -> STerm t -> STerm t -> EqResM t (ATerm t, ATerm t)
stermequalS c@(_,(StoreHat h cnt)) (QA a0 _ Identity) (QA a1 _ Identity)
  | a0 == a1 = \ d ps -> case (cnt ! a0) of
    Cinf -> MayS ps
    _ -> returnS d ps
stermequalS c (VA n0 ts0) (VA n1 ts1) | n0 == n1 = eqvaS c ts0 ts1
```

```
stermequalS c (QA a0 _ Structural) st1 = withResolveAddr c a0
                               (atermequalS c $ ASTerm st1)
stermequalS c st0 (QA a1 _ Structural) = withResolveAddr c a1
                               (atermequalS c $ ASTerm st0)
stermequalS _ _ _ = \ d _ -> failS d

eqvaS :: Ctx t -> [TermA t] -> [TermA t] -> EqResM t (ATerm t, ATerm t)
eqvaS c [] [] = returnS
eqvaS c (t0:ts0) (t1:ts1) = bindSM c (tequalauxS c t0 t1) $ eqvaS c ts0 ts1
eqvaS _ _ _ = \ d ps -> failS d

-- Combine weak finds into a single proof.
forEachRefinement :: Refinements t -> (Refinement t -> Maybe (PairsHat t))
                -> Maybe (PairsHat t)
forEachRefinement ds f =
  Set.fold (\ d acc ->
          case f d of
           Just ps -> case acc of
             Just ps' -> Just (Set.union ps ps')
             Nothing -> Just ps
           Nothing -> acc)
  Nothing ds

findWeak :: Ctx t -> Refinement t -> TermA t -> [(PairsHat t, TermA t)]
         -> Maybe (PairsHat t)
findWeak s d v [] = Nothing
findWeak s d v ((ps,v'):pvs) =
  case tequalauxS s v v' d ps of -- XXX should be guardS
   Both dps ds -> if Map.null dps then
                findWeak s d v pvs
              else Just (squash dps)
   MayS ps -> Just ps

----------------------------------------
-- Pattern matching
----------------------------------------
newtype MVariable = MVariable String deriving (Eq,Ord,Show)
data Pattern t =
  PName MVariable (Pattern t)
  | PV Name [Pattern t]
  | PQ MM EM
  | PExt (ExternalDescriptor t)
  | PWild deriving (Show)
newtype MEnvC = MEnvC (Map MVariable TermC)
newtype MEnvA t = MEnvA (Map MVariable (TermA t))
                deriving (Eq,Ord,Show)

type MResC = Maybe MEnvC

matchC :: Store -> Pattern t -> TermC -> MEnvC -> MResC
matchC s (PName x PWild) t me@(MEnvC env) =
  case Map.lookup x env of
   Nothing -> case t of
    QC a mm _ ->
     case mm of
      Explicit -> Just $ MEnvC $ Map.insert x t env
      Implicit Delay -> Just $ MEnvC $ Map.insert x (DC a) env
```

```
        Implicit _ -> Just $ MEnvC $ Map.insert x (s ! a) env
      _ -> Just $ MEnvC $ Map.insert x t env
   Just t' -> if tequalC s t t' then Just me else Nothing
matchC s (PName x p) t me@(MEnvC env) =
  case Map.lookup x env of
   Nothing -> matchC s p t $ MEnvC $ Map.insert x t env
   Just t' -> if tequalC s t t' then matchC s p t me else Nothing
matchC s PWild t env = Just env
matchC s (PQ mm em) t@(QC _ mm' em') env | mm == mm' && em == em' = Just env
matchC s (PV n ps) (VC n' ts) env | n == n' = matchCmany s ps ts env

matchCmany :: Store -> [Pattern t] -> [TermC] -> MEnvC -> MResC
matchCmany s [] [] env = Just env
matchCmany s (p:ps) (t:ts) me = do env' <- matchC s p t me
                                   matchCmany s ps ts me
matchCmany _ _ _ _ = Nothing

----------------------------------------
-- Expressions
----------------------------------------
newtype Tag = Tag String deriving (Show)
data Expr t =
  ERef MVariable
  | EVariant Name Tag [Expr t]
  | EAlloc Tag
  | ELet [BU t] (Expr t)
  | ECall Name [Expr t]
  | ELookup (Expr t) LM deriving (Show)
data BU t = Where (Pattern t) (Expr t)
          | Update (Expr t) (Expr t)
          deriving (Show)
data Rule t = Rule (Pattern t) (Expr t) [BU t]
            deriving (Show)

data StuckFail a = Stuck | Fail | Fires(a)
newtype StuckFailT m a = StuckFailT { runStuckFailT :: m (StuckFail a) }
bindSFT :: (Monad m) => (StuckFailT m a) -> (a -> StuckFailT m b)
        -> StuckFailT m b
bindSFT x f = StuckFailT $ runStuckFailT x >>= \ sf -> case sf of
  Stuck -> return Stuck
  Fail -> return Fail
  Fires y -> runStuckFailT (f y)

returnSFT :: (Monad m) => a -> StuckFailT m a
returnSFT a = StuckFailT $ return $ Fires a

failSFT :: (Monad m) => t -> StuckFailT m b
failSFT _ = StuckFailT $ return Stuck

instance (Monad m) => Monad (StuckFailT m) where
  return = returnSFT
  (>>=) = bindSFT
  fail = failSFT

instance MonadTrans StuckFailT where
  lift m = StuckFailT (Fires `liftM` m)
```

```
instance (MonadState s m) => MonadState s (StuckFailT m) where
  get = lift get
  put k = lift (put k)

------------

newtype EvRes a = ER {
  runER :: MaybeT (State Store) a
  } deriving (Monad, MonadState Store)

newtype RuleRes a = RR {
  runRR :: StuckFailT (State Store) a
  } deriving (Monad, MonadState Store)


data Metafunction t =
  UserMF [Rule t]
  | ExternalMF ([TermC] -> EvRes TermC)

data Semantics t = Sem {
  rules :: [Rule t],
  metafunctions :: Map Name (Metafunction t),
  alloc :: Store -> Tag -> AddrC,
  mkV :: Name -> Tag -> [TermC] -> EvRes TermC
  }

maybeMT :: Maybe a -> (() -> b) -> (a -> b) -> b
maybeMT (Just a) fail good = good a
maybeMT Nothing fail good = fail ()

ev :: Semantics t -> Expr t -> MEnvC -> EvRes TermC
ev s (ERef x) (MEnvC env) = case Map.lookup x env of
  Just t -> ER $ returnMT t
  Nothing -> ER $ failMT ()
ev s (EVariant n tag es) me = do ts <- evmany s es me
                                 mkV s n tag ts
ev s (EAlloc tag) me = do { st <- get;
                          ER $ returnMT $
                          QC (alloc s st tag) Explicit Identity }
ev s (ELet bus e) me =
  do store <- get
     case runStateT (runStuckFailT $ runRR $ evbus s bus me) store of
       Fid.Identity (Fires me', store') ->
         do put store';
            ev s e me'
       Fid.Identity (_, _) -> ER $ failMT ()
ev s (ECall f es) me = do ts <- evmany s es me
                       evcall s f ts
ev s (ELookup e lm) me = do t <- ev s e me
                          case t of
                            QC a Explicit _ ->
                              do { st <- get ;
                                   ER $ returnMT $ st ! a }
                            _ -> ER $ failMT ()

evmany :: Semantics t -> [Expr t] -> MEnvC -> EvRes [TermC]
evmany s [] me = ER $ returnMT []
```

```
evmany s (e:es) me = do t <- ev s e me
                        ts <- evmany s es me
                        ER $ returnMT $ t:ts

evbu :: Semantics t -> BU t -> MEnvC -> RuleRes MEnvC
evbu s (Where p e) me =
  do store <- get
     case runStateT (runMaybeT $ runER $ ev s e me) store of
      Fid.Identity (Just t, store') ->
        case matchC store' p t me of
          -- side-effects happen only on success
          Just me' -> do { put store';
                           RR $ returnSFT me' }
          Nothing -> RR $ StuckFailT $ return Fail
      _ -> RR $ failSFT ()
evbu s (Update ea et) me =
  do store <- get
     case runStateT (runMaybeT $ runER $ ev s ea me) store of
      Fid.Identity (Just ta, store') ->
        case ta of
          QC a Explicit _ ->
            case runStateT (runMaybeT $ runER $ ev s et me) store' of
             Fid.Identity (Just tt, store'') ->
               do { put (Map.insert a tt store'');
                    return me }
             _ -> RR $ failSFT ()
          _ -> RR $ failSFT ()
      Fid.Identity (Nothing, _) -> RR $ failSFT ()


evbus :: Semantics t -> [BU t] -> MEnvC -> RuleRes MEnvC
evbus s [] me = RR $ returnSFT me
evbus s (bu:bus) me = evbu s bu me >>= evbus s bus

runEvRRes :: RuleRes TermC -> Store -> (StuckFail TermC, Store)
runEvRRes ev store = let (Fid.Identity p) =
                          runStateT (runStuckFailT $ runRR $ ev) store in
                p

evcall :: Semantics t -> Name -> [TermC] -> EvRes TermC
evcall s f ts = case Map.lookup f $ metafunctions s of
  Just mf ->
    case mf of
      UserMF rs ->
        do { store <- get;
             case runEvRRes (runInOrder s rs (VC f ts)) store of
              (Fires t, store') -> do { put store' ; ER $ returnMT t }
              _ -> ER $ failMT () }
      ExternalMF f -> f ts
  Nothing -> ER $ failMT ()

runRule :: Semantics t -> Rule t -> TermC -> RuleRes TermC
runRule s (Rule p e bus) t =
  do st <- get
     case matchC st p t (MEnvC Map.empty) of
      Just env ->
        do me <- evbus s bus env
```

```
        store <- get
        case runStateT (runMaybeT $ runER $ ev s e me) store of
          Fid.Identity (Just t, store') ->
            do { put store'; RR $ returnSFT t }
          -- evaluation got stuck, so the rule is stuck
          _ -> RR $ failSFT ()
      -- Rule didn't match. Fails
      Nothing -> RR $ StuckFailT $ return Fail

runInOrder :: Semantics t -> [Rule t] -> TermC -> RuleRes TermC
runInOrder s [] t = RR $ failSFT ()
runInOrder s (r:rs) t =
  do store <- get
     case runStateT (runStuckFailT $ runRR $ runRule s r t) store of
       Fid.Identity (Fires t', store') ->
         do { put store' ; RR $ returnSFT t' }
       Fid.Identity (Stuck, store') -> RR $ failSFT ()
       Fid.Identity (Fail, store') -> RR $ StuckFailT $ return Fail


----------------------------------------
-- Abstract evaluation
----------------------------------------
data Change t = Strong (TermA t)
              | Weak (TermA t)
              | Reset (TermA t)
              deriving (Eq,Ord,Show)
newtype StoreDelta t = StoreDelta (Map AddrA (Change t))
                   deriving (Eq,Ord,Show)

atermToTerm :: ATerm t -> TermA t
atermToTerm (ASTerm st) = PreTerm $ STerm st
atermToTerm (AEA ed v) = PreTerm $ EA ed v

applyRefinement :: StoreHat t -> Refinement t -> Set AddrA -> StoreHat t
applyRefinement (StoreHat h cnt) d disregard = StoreHat h'' cnt
  where h'' = Map.foldWithKey
          (\ a at h' ->
            if Set.member a disregard then
              h'
            else case at of
              ASTerm st -> Map.insert a (Set.singleton st, Map.empty) h'
              AEA ed v -> Map.insert a (Set.empty, Map.singleton ed v) h')
          h d

insertTerm :: AddrA -> TermA t -> HeapHat t -> HeapHat t
insertTerm a (PreTerm (DA a')) h = case Map.lookup a' h of
  Just abs -> Map.insert a abs h
  Nothing -> error ("Oh, bugger: " ++ show a')
insertTerm a (PreTerm (STerm st)) h = Map.insert a
                            (Set.singleton st, Map.empty) h
insertTerm a (PreTerm (EA ed v)) h = Map.insert a
                           (Set.empty, Map.singleton ed v) h
insertTerm a (TAbs abs) h = Map.insert a abs h

applyDelta :: Ctx t -> StoreDelta t -> StoreHat t
applyDelta c@(_,s@(StoreHat h cnt)) (StoreDelta pars) =
```

```
  Map.foldWithKey
  (\ a ch s'@(StoreHat h' cnt') ->
    case ch of
      Strong t -> StoreHat (insertTerm a t h') cnt'
      Weak t -> case Map.lookup a h of
      -- XXX: Can we join terms through addresses that might not be set yet?
        Just abs -> StoreHat (insertTerm a (joinTermTop c t (TAbs abs)) h') cnt'
        Nothing -> error ("Weak must be mapped: " ++ show a)
      Reset t -> StoreHat (insertTerm a t h') (Map.insert a Cinf cnt'))
  s pars

applychange :: Ctx t -> Refinement t -> StoreDelta t -> StoreHat t
applychange c d pars@(StoreDelta sd) = applyRefinement (applyDelta c pars) d
                                       (Map.keysSet sd)
  -- first apply changes, then refine non-changed addresses.


type ResS t a = EqResS t (a, StoreDelta t)
data RResS t a =
  FSU {fires :: Map (Refinement t) (Set (a, StoreDelta t)),
       stuck :: Refinements t,
       unapplicable :: Refinements t}
  | MayF (Set (a, StoreDelta t))

data MetafunctionHat t =
  UserAMF [Rule t]
  | ExternalAMF ([TermA t] -> EvResS t (TermA t))
data StateHat t = StateHat (TermA t) (StoreHat t) t deriving (Eq,Ord,Show)

data SemanticsHat t = SemHat {
  rulesH :: [Rule t],
  metafunctionsH :: Map Name (MetafunctionHat t),
  allocH :: Tag -> SRSD t AddrA,
  mkVH :: Name -> Tag -> [TermA t] -> EvResS t (TermA t),
  tickH :: MEnvA t -> SRSD t t
  }

newtype ResM t a = ResM (SemanticsHat t -> StateHat t -> Refinement t
                   -> StoreDelta t -> ResS t a)
newtype RResM t a = RResM (SemanticsHat t -> StateHat t -> Refinement t
                     -> StoreDelta t -> RResS t a)

weakenRS :: (Ord a) => ResS t a -> ResS t a
weakenRS eq@(Both r d) = if Map.null r then
                           eq
                         else MayS $ squash r
weakenRS r = r

joinResS :: (Ord a) => Ctx t -> ResS t a -> ResS t a -> ResS t a
joinResS s (Both dps _) (MayS ps) = MayS $ Set.union ps $ squash dps
joinResS s (MayS ps) (MayS ps') = MayS $ Set.union ps ps'
joinResS s (Both dps ds) (Both dps' ds') =
  if bigoverlap s dps ds' || bigoverlap s dps' ds then
    MayS $ Set.union (squash dps) (squash dps')
  else
    Both (mergeKeysStrictlySmaller s dps dps')
    (joinCut s ds ds')
```

```
-- symmetric cases
joinResS s (MayS ps) (Both dps _) = MayS $ Set.union ps $ squash dps

joinRResS :: (Ord a) => Ctx t -> RResS t a -> RResS t a -> RResS t a
joinRResS s (FSU dps _ _) (MayF ps) = MayF $ Set.union ps $ squash dps
joinRResS s (MayF ps) (MayF ps') = MayF $ Set.union ps ps'
joinRResS s (FSU f st u) (FSU f' st' u') =
  if bigoverlap s f st' || bigoverlap s f u'
     || bigoverlap s f' st || bigoverlap s f' u then
    MayF $ Set.union (squash f) (squash f')
  else
    FSU {fires=mergeKeysStrictlySmaller s f f',
         stuck=joinCut s st st',
         unapplicable=joinCut s u u'}
-- symmetric cases
joinRResS s (MayF ps) (FSU dps _ _) = MayF $ Set.union ps $ squash dps

----------------------------------------
-- Standard ResM monad actions
----------------------------------------

returnResAux :: (Ord a) => a -> Refinement t -> StoreDelta t -> ResS t a
returnResAux a d pars =
  Both (Map.singleton d (Set.singleton (a,pars))) Set.empty

returnRes :: (Ord a) => a -> ResM t a
returnRes a = ResM $ \ _ w d pars -> returnResAux a d pars

failRes :: (Ord a) => u -> ResM t a
failRes _ = ResM $ \ _ w d pars -> Both Map.empty (Set.singleton d)

bindRes :: (Ord a, Ord b) => ResM t a -> (a -> ResM t b) -> ResM t b
bindRes (ResM r) f = ResM $ \ sem w@(StateHat _ s _) d pars ->
  case r sem w d pars of
    Both dts ds -> case Map.toList dts of
      [] -> Both Map.empty ds
      (d',as):das ->
        let eachas d as =
              case Set.toList as of
                [] -> error ("Bad result at " ++ show d')
                (a,pars'):as' ->
                  Prelude.foldl (\ res (a,pars') ->
                                 joinResS (w,s) res (unResM (f a) sem w d pars'))
                  (unResM (f a) sem w d pars') as'
        in joinResS (w,s) (Both Map.empty ds) $
           Prelude.foldl (\ res (d',as') ->
                          joinResS (w,s) res $ eachas d' as')
           (eachas d' as) das
    MayS as -> case Set.toList as of
      [] -> error ("Bad result at " ++ show d)
      (a,pars'):as' ->
        Prelude.foldl (\ res (a,pars') ->
                       joinResS (w,s) res (unResM (f a) sem w d pars'))
        (unResM (f a) sem w d pars') as'

----------------------------------------
-- Standard RResM monad actions
```

```
---------------------------------------

returnRResAux :: (Ord a) => a -> Refinement t -> StoreDelta t -> RResS t a
returnRResAux a d pars = FSU {fires = Map.singleton d (Set.singleton (a,pars)),
                              stuck = Set.empty,
                              unapplicable = Set.empty}

returnRRes :: (Ord a) => a -> RResM t a
returnRRes a = RResM $ \ _ w d pars -> returnRResAux a d pars

failRRes :: (Ord a) => u -> RResM t a
failRRes _ = RResM $ \ _ w d pars -> FSU {fires = Map.empty,
                                         stuck = Set.empty,
                                         unapplicable = (Set.singleton d) }

bindRRes :: (Ord a, Ord b) => RResM t a -> (a -> RResM t b)
         -> RResM t b
bindRRes (RResM r) f = RResM $ \ sem w@(StateHat _ s _) d pars ->
  case r sem w d pars of
    FSU fires stuck unapplicable -> case Map.toList fires of
      [] -> FSU {fires=Map.empty, stuck= stuck, unapplicable=unapplicable}
      (d',as):das ->
        let eachas d as =
              case Set.toList as of
                [] -> error ("Bad rule result at " ++ show d')
                (a,pars'):as' ->
                  Prelude.foldl
                  (\ res (a,pars') ->
                    joinRResS (w,s) res (unRResM (f a) sem w d pars'))
                  (unRResM (f a) sem w d pars') as'
        in Prelude.foldl (\ res (d',as') ->
                            joinRResS (w,s) res $ eachas d' as')
             (eachas d' as) das
    MayF as -> case Set.toList as of
      [] -> error ("Bad rule result at " ++ show d)
      (a,pars'):as' ->
        Prelude.foldl (\ res (a,pars') ->
                          joinRResS (w,s) res (unRResM (f a) sem w d pars'))
        (unRResM (f a) sem w d pars') as'

---------------------------------------
-- Loewering operations
---------------------------------------

unResM :: ResM t a -> SemanticsHat t -> StateHat t ->
       Refinement t -> StoreDelta t -> ResS t a
unResM (ResM f) sem w d pars = f sem w d pars
runEvResS = unResM . lowerResM

unRResM :: RResM t a -> SemanticsHat t -> StateHat t -> Refinement t ->
        StoreDelta t -> RResS t a
unRResM (RResM f) sem w d pars = f sem w d pars
runEvRResS = unRResM . lowerRResM

ruleToExpr :: (Ord a) => EvRResS t a -> EvResS t a
ruleToExpr r = liftNM $ ResM $ \ sem w@(StateHat _ s _) d pars ->
  case runEvRResS r sem w d pars of
```

```
   FSU fires stuck unapplicable ->
     Both fires (joinCut (w,s) stuck unapplicable)
   MayF e -> MayS e

exprToRule :: (Ord a) => EvResS t a -> EvRResS t a
exprToRule r = liftNM $ RResM $ \ sem w d pars ->
  case runEvResS r sem w d pars of
   Both dts ds -> FSU {fires=dts, stuck=ds, unapplicable=Set.empty}
   MayS e -> MayF e


type EvResS t a = NM Ord (ResM t) a
type EvRResS t a = NM Ord (RResM t) a
lowerResM :: (Ord a) => EvResS t a -> ResM t a
lowerResM = lowerNM returnRes bindRes

lowerRResM :: (Ord a) => EvRResS t a -> RResM t a
lowerRResM = lowerNM returnRRes bindRRes

----------------------------------------
-- Special ResM monad actions
----------------------------------------
-- Update the store with the appropriate strength.
updateRes :: (Ord a) => AddrA -> TermA t -> EvRResS t a
          -> EvRResS t a
updateRes addr t next = liftNM $ RResM $ \ sem w d pars ->
  let (StateHat _ s@(StoreHat h cnt) _) = w in
  let (StoreDelta sd) = pars in
  -- Already changed?
  case Map.lookup addr sd of
   Just ch ->
     runEvRResS next sem w d $ StoreDelta $
     case ch of
       Strong _ -> Map.insert addr (Strong $ unSRSD (demand t) w d pars) sd
       Weak t' -> Map.insert addr (Weak $ joinTermTop (w,s) t t') sd
       Reset t' -> Map.insert addr (Reset $ joinTermTop (w,s) t t') sd
   Nothing -> runEvRResS next sem w d $ StoreDelta $ Map.insert addr
           (case Map.lookup addr cnt of
             -- Fresh. We can strongly update.
             Just C1 -> Strong t
             _ -> Weak t)
           sd

-- No abstraction yet
makeVariant :: Name -> Tag -> [TermA t] -> EvResS t (TermA t)
makeVariant n tag ts = withSemantics (\ s -> mkVH s n tag ts)

bumpAddr :: (Ord a) => AddrA -> EvResS t a -> EvResS t a
bumpAddr a next = liftNM $ ResM $ \ sem w@(StateHat _ (StoreHat _ cnt) _)
                           d pars@(StoreDelta sd) ->
  runEvResS next sem w d $ StoreDelta $
  case Map.lookup a sd of
   Just (Strong t) -> Map.insert a (Reset t) sd
   Just ch -> sd -- already bumped. Just run
   Nothing -> case Map.lookup a cnt of
     Just _ -> Map.insert a (Weak termbot) sd
     Nothing -> Map.insert a (Strong termbot) sd
```

```
allocA :: Tag -> EvResS t AddrA
allocA tag = withSemantics $ \ sem -> do addr <- liftSRSD $ allocH sem tag
                                         bumpAddr addr $ return addr

tickA :: (Ord t) => MEnvA t -> EvRResS t t
tickA me = withSemantics' $ \ sem -> liftSRSD' $ tickH sem me

withStateHat :: (Ord a) => (StateHat t -> EvResS t a)
            -> EvResS t a
withStateHat f = liftNM $ ResM $ \ sem w d pars -> runEvResS (f w) sem w d pars

withSemantics :: (Ord a) => (SemanticsHat t -> EvResS t a) -> EvResS t a
withSemantics f = liftNM $ ResM $ \ sem w d pars ->
  runEvResS (f sem) sem w d pars

getRefinement :: EvResS t (Refinement t)
getRefinement = liftNM $ ResM $ \ _ w d pars -> returnResAux d d pars

chooseRefinement :: (Ord a) => Refinements t -> EvResS t a -> EvResS t a
chooseRefinement ds next =
  case Set.toList ds of
    [] -> error "Empty set of refinements"
    d:ds' -> liftNM $ ResM $ \ sem w@(StateHat _ s _) d pars ->
      Prelude.foldl (\ res d ->
                  joinResS (w,s) res $ runEvResS next sem w d pars)
      (runEvResS next sem w d pars) ds'

-- If term is abstract, join the results of function applied to each term.
appeach :: (Ord a) => (TermA t -> EvResS t a) -> (ATerm t -> Refinement t)
        -> TermA t -> EvResS t a
appeach f upd t = liftNM $ ResM $ \ sem w@(StateHat _ s _) d pars ->
  case unSRSD (demand t) w d pars of
    TAbs (sts,es) ->
      case Set.toList sts of
        [] -> error ("Bad term: " ++ show t)
        st:sts' ->
         let withsts =
              Prelude.foldl
              (\ res st ->
               joinResS (w,s) res (runEvResS (f $ PreTerm $ STerm st)
                           sem w (upd $ ASTerm st) pars))
              (runEvResS (f $ PreTerm $ STerm st) sem w d pars) sts' in
         Map.foldlWithKey (\ res ed v ->
                    joinResS (w,s) res
                    (runEvResS (f $ PreTerm $ EA ed v)
                     sem w (upd $ AEA ed v) pars))
         withsts es
    t -> runEvResS (f t) sem w d pars

-- Lookup and refine, if the lookup mode asks for it.
slookup :: (Ord a) => AddrA -> LM -> (TermA t -> EvResS t a) -> EvResS t a
slookup a Delay f = f (PreTerm $ DA a)
slookup a lm f = liftNM $ ResM $
  \ sem w@(StateHat _ s@(StoreHat h cnt) _) d pars@(StoreDelta sd) ->
  case Map.lookup a sd of
    -- Already modified, so just use what we have here. Don't do any refinement
```

```
    Just ch ->
      let t = case ch of
            Strong t -> t
            Reset t -> t
            Weak t -> case Map.lookup a h of
              Just abs -> let Fid.Identity (t,_) =
                              runStateT (joinAbsTermTerm (w,s) abs t)
                              Set.empty in
                         t
              Nothing -> error ("Weak without mapping: " ++ show a)
              in
       case lm of
         Resolve -> runEvResS (appeach f (\ _ -> d) t) sem w d pars
         Deref -> runEvResS (f t) sem w d pars
    Nothing ->
      case Map.lookup a d of
        -- Refined, so use what we have.
        Just aterm -> runEvResS (f $ atermToTerm aterm) sem w d pars
        Nothing ->
          case Map.lookup a h of
            Just abs ->
              case lm of
                Resolve ->
                  runEvResS (appeach f
                           (case Map.lookup a cnt of
                              Just C1 -> \ st -> Map.insert a st d
                              _ -> \ _ -> d) $ TAbs abs)
                  sem w d pars
                Deref -> runEvResS (f (TAbs abs)) sem w d pars
            Nothing -> error ("Dangling pointer: " ++ show a)

resolve :: TermA t -> EvResS t (TermA t)
resolve (TAbs (sts,es)) = liftNM $ ResM $ \ _ w d pars ->
  Both (Map.foldlWithKey
          (\ acc ed v ->
            Map.insert d (Set.singleton (PreTerm $ EA ed v, pars)) acc)
          (fromsts d pars) es)
  Set.empty
  where fromsts d pars = Map.singleton d $
          Set.foldl (\ acc st ->
                      (Set.insert (PreTerm $ STerm st,pars) acc)
                     acc)
          Set.empty sts
resolve (PreTerm (DA a)) = slookup a Resolve return
resolve (PreTerm (STerm (QA a (Implicit lm) _))) = slookup a lm return
resolve t = return t

----------------------------------------
-- Demand a term (no resolutions)

-- read-only view of ResM
newtype SRSD t a = SRSD (StateHat t -> Refinement t
                             -> StoreDelta t -> a)
unSRSD (SRSD f) w d pars = f w d pars
returnSRSD a = SRSD (\ _ _ _ -> a)
bindSRSD f g = SRSD $ \ w d pars -> unSRSD (g (unSRSD f w d pars)) w d pars
liftSRSD :: (Ord a) => SRSD t a -> EvResS t a
```

```
liftSRSD f = liftNM $ ResM $ \ _ w d pars ->
  returnResAux (unSRSD f w d pars) d pars

liftSRSD' :: (Ord a) => SRSD t a -> EvRResS t a
liftSRSD' f = liftNM $ RResM $ \ _ w d pars ->
  returnRResAux (unSRSD f w d pars) d pars

withSemantics' :: (Ord a) => (SemanticsHat t -> EvRResS t a)
             -> EvRResS t a
withSemantics' f = liftNM $ RResM $ \ sem w d pars ->
  unRResM (lowerRResM (f sem)) sem w d pars


instance Monad (SRSD t) where
  return = returnSRSD
  (>>=) = bindSRSD

demand :: TermA t -> SRSD t (TermA t)
demand (PreTerm (DA a)) = flatlookup a
demand (PreTerm (STerm st)) = demandSTerm st
demand t@(TAbs sts) = return t

demandSTerm :: STerm t -> SRSD t (TermA t)
demandSTerm (VA n ts) = do ts <- Prelude.mapM demand ts
                           return $ PreTerm $ STerm $ VA n ts
demandSTerm t@(QA a Explicit _) = return $ PreTerm $ STerm t
demandSTerm (QA a _ _) = flatlookup a

flatlookup :: AddrA -> SRSD t (TermA t)
flatlookup a = SRSD $
  \ w@(StateHat _ s@(StoreHat h cnt) _) d pars@(StoreDelta sd) ->
  case Map.lookup a sd of
   Just ch ->
    case ch of
     Strong t -> t
     Reset t -> t
     Weak t -> case Map.lookup a h of
      Just abs -> let Fid.Identity (t,_) =
                     runStateT (joinAbsTermTerm (w,s) abs t)
                     Set.empty in
        t
      Nothing -> t
   Nothing ->
    case Map.lookup a d of
     Just aterm -> atermToTerm aterm
     Nothing ->
      case Map.lookup a h of
       Just sts -> TAbs sts
       Nothing -> error ("Dangling pointer: " ++ show a)
----------------------------------------
-- End demand
----------------------------------------

evS :: Expr t -> MEnvA t -> EvResS t (TermA t)
evS (ERef x) (MEnvA env) = case (Map.lookup x env) of
  Just t -> liftNM $ returnRes t
  Nothing -> error ("Unbound metavariable: " ++ show x)
```

```
evS (EVariant n tag es) env = (evMany es env) >>= makeVariant n tag
evS (EAlloc tag) env = do a <- withSemantics (\ s -> liftSRSD $ allocH s tag)
                       return $ PreTerm $ STerm $ QA a Explicit Identity
evS (ELet bus body) env = (ruleToExpr $ evBUsS bus env) >>= evS body
evS (ECall mf es) env = evMany es env >>= evMF mf
evS (ELookup e lm) env =
  do t <- evS e env
     t' <- resolve t
     case t' of
       PreTerm (STerm (QA a Explicit _)) -> slookup a lm return
       _ -> liftNM $ failRes ()

evMany :: [Expr t] -> MEnvA t -> EvResS t [TermA t]
evMany [] env = return []
evMany (e:es) env = do t <- evS e env
                       ts <- evMany es env
                       return (t:ts)

evBUS :: BU t -> MEnvA t -> EvRResS t (MEnvA t)
evBUS (Where p e) env = do t <- exprToRule $ evS e env
                           matchS p t env
evBUS (Update ea ev) env = do ta <- exprToRule $ evS ea env
                              ta' <- exprToRule $ resolve ta
                              case ta' of
                                PreTerm (STerm (QA a Explicit _)) ->
                                  do tv <- exprToRule $ evS ev env
                                     updateRes a tv (return env)

evBUsS :: [BU t] -> (MEnvA t) -> EvRResS t (MEnvA t)
evBUsS [] env = return env
evBUsS (bu:bus) env = evBUS bu env >>= evBUsS bus

data MatchResS t a = MFail | MSuccess (Set a)
                   | MSplit (Map (Refinement t) (Set a)) (Refinements t)
                   | MMay (Set a)

mresauxTomres :: (Ord a) => ResS t a -> StateHat t -> Refinement t
                -> StoreDelta t -> RResS t a
mresauxTomres (MayS s) w d pars = MayF s
mresauxTomres r@(Both dts ds) w d pars =
  if Map.null dts then
    FSU {fires= Map.empty, stuck=Set.empty, unapplicable= Set.singleton d}
  else if Set.null ds then
       FSU {fires = Map.singleton d $ squash dts,
            stuck =Set.empty,
            unapplicable=Set.empty}
     else FSU {fires = dts, unapplicable = ds, stuck = Set.empty}

matchS :: Pattern t -> TermA t -> MEnvA t -> EvRResS t (MEnvA t)
matchS p t env = liftNM $ RResM $ \ sem w d pars ->
  mresauxTomres (runEvResS (matchSaux p t env) sem w d pars) w d pars

resolvable :: TermA t -> Bool
resolvable (PreTerm (DA _)) = True
resolvable (PreTerm (STerm (QA _ (Implicit _) _))) = True
resolvable (TAbs _) = True
resolvable _ = False
```

```
matchSaux :: Pattern t -> TermA t -> MEnvA t -> EvResS t (MEnvA t)
matchSaux PWild t me = return me
matchSaux (PName x p) t me@(MEnvA env) = case Map.lookup x env of
  Just t' -> withStateHat $ \ w ->
    do d <- getRefinement
       case tequalS (let (StateHat _ s _) = w in (w,s)) d t t' of
         EqualS -> matchSaux p t me
         UnequalS -> liftNM $ failRes ()
         EqSplit eqs neqs -> liftNM $ ResM $ \ sem w' d' pars ->
           let (StateHat _ s' _) = w in
           joinResS (w',s') (Both Map.empty neqs) $
           (runEvResS (chooseRefinement eqs
                       (matchSaux p t me)) sem w' d' pars)
         MayTS -> liftNM $ ResM $ \ sem w' d' pars ->
           weakenRS $ runEvResS (matchSaux p t me) sem w' d' pars
  Nothing -> do t' <- resolve t
                matchSaux p t $ MEnvA $ Map.insert x t' env
matchSaux (PV n ps) (PreTerm (STerm (VA n' ts))) me
  | n == n' = matchMany ps ts me
matchSaux (PQ mm em) (PreTerm (STerm (QA _ mm' em'))) me
  | mm == mm' && em == em' = return me
matchSaux (PExt ed) (PreTerm (EA ed' _)) me | ed == ed' = return me
matchSaux p t me | resolvable t = do t' <- resolve t
                                     matchSaux p t' me
matchSaux p t me = liftNM $ failRes ()

matchMany :: [Pattern t] -> [TermA t] -> MEnvA t -> EvResS t (MEnvA t)
matchMany [] [] me = return me
matchMany (p:ps) (t:ts) me = matchSaux p t me >>= matchMany ps ts
matchMany _ _ _ = liftNM $ failRes ()

evRuleS :: Rule t -> TermA t -> MEnvA t -> EvRResS t (TermA t)
evRuleS (Rule p e bus) t me = do me' <- matchS p t me
                                 me'' <- evBUsS bus me'
                                 exprToRule $ evS e me''

maybefire :: (Ord a) => EvRResS t a -> EvResS t a -> EvResS t a
maybefire rres res =
  liftNM $ ResM $ \ sem w@(StateHat _ s _) d pars ->
  case runEvRResS rres sem w d pars of
    FSU fires stuck unapplicable ->
      if Set.null unapplicable then
        Both fires stuck
      else joinResS (w,s) (Both fires stuck) $
           runEvResS (chooseRefinement unapplicable res) sem w d pars

applyInOrder :: [Rule t] -> TermA t -> MEnvA t -> EvResS t (TermA t)
applyInOrder [] t me = liftNM $ failRes ()
applyInOrder (r:rs) t me = maybefire (evRuleS r t me) (applyInOrder rs t me)

tset :: Ctx t -> [StateHat t] -> Refinement t
      -> Set ((TermA t, t),StoreDelta t) -> [StateHat t]
tset c acc d tparss = Set.foldl (\ ws ((t,tk),pars) ->
                                 (StateHat t (applychange c d pars) tk):ws)
                acc tparss
```

```
finalize :: Ctx t -> RResS t (TermA t, t) -> [StateHat t]
finalize c (FSU fires _ _) = Map.foldlWithKey (tset c) [] fires
finalize c (MayF tparss) = tset c [] Map.empty tparss

evStepS :: (Ord t) => SemanticsHat t -> Rule t -> StateHat t -> [StateHat t]
evStepS sem (Rule p e bus) w@(StateHat term s tk) = finalize (w,s) $
  runEvRResS (do me <- matchS p term (MEnvA Map.empty)
              me’ <- evBUsS bus me
              t’ <- exprToRule $ evS e me’
              tk <- tickA me’
              return (t’, tk))
  sem w Map.empty (StoreDelta Map.empty)

evMF :: Name -> [TermA t] -> EvResS t (TermA t)
evMF mf ts = withSemantics $ \ s ->
  case Map.lookup mf $ metafunctionsH s of
    Just (UserAMF rs) -> applyInOrder rs (PreTerm (STerm (VA mf ts)))
                     (MEnvA Map.empty)
    Just (ExternalAMF f) -> f ts
    Nothing -> liftNM $ failRes ()

applyAll :: (Ord t) => SemanticsHat t -> [Rule t] -> StateHat t -> [StateHat t]
         -> [StateHat t]
applyAll sem [] w next = next
applyAll sem (r:rs) w next = applyAll sem rs w (evStepS sem r w ++ next)

stepHat :: (Ord t) => SemanticsHat t -> StateHat t -> [StateHat t]
stepHat sem w = applyAll sem (rulesH sem) w []

stepUntilFinished :: (Ord t, Num a) => SemanticsHat t -> [(a, StateHat t)] ->
                  [StateHat t] -> a -> [StateHat t] -> [(a, StateHat t)]
-- If there are no identified steps, return the set of states we ended up at.
stepUntilFinished sem stuck [] steps [] = stuck
stepUntilFinished sem stuck next steps [] =
  stepUntilFinished sem stuck [] (steps+1) next
stepUntilFinished sem stuck next steps (w:todo) =
  case stepHat sem w of
    [] -> stepUntilFinished sem ((steps,w):stuck) next steps todo
    next’ -> stepUntilFinished sem stuck (next’++next) steps todo

runProgram :: (Ord t, Num a) => SemanticsHat t -> t -> TermA t
           -> [(a,StateHat t)]
runProgram sem tk start = stepUntilFinished sem [] [] 0 [initial]
  where initial = (StateHat start (StoreHat Map.empty Map.empty) tk)
```